\documentclass[12pt,a4paper,twoside,openright]{book}

\usepackage[dvips]{epsfig}
\usepackage{fancybox}
\usepackage{subfigure}
\usepackage{booktabs}
\usepackage[latin1]{inputenc}
\usepackage{multirow}
\usepackage{array}
\usepackage{rotating}
\usepackage[british]{babel} 

\let\tmp\oddsidemargin
\let\oddsidemargin\evensidemargin
\let\evensidemargin\tmp
\reversemarginpar

\begin{document}


\pagestyle{headings}
\pagestyle{empty}

\begin{center}
\LARGE
{\sc Universidad Autónoma de Madrid} \\
\large {\sc Escuela Politécnica Superior}
\end{center}

\vspace{3cm}

\begin{center}
\LARGE {\sc ANALYSIS AND STUDY ON TEXT REPRESENTATION TO IMPROVE
THE ACCURACY OF THE NORMALIZED COMPRESSION DISTANCE}\\
\vspace{3cm} \LARGE
{\sc Thesis}\\
\end{center}

\vspace{3cm}

\begin{center}
\large
Ana Granados Fontecha\\
Madrid, 2012
\end{center}

\vspace{2cm}

\begin{tabular}{cl}
ADVISORS: & David Camacho Fernández \\
 & Francisco de Borja Rodríguez Ortiz \\
\end{tabular}

\newpage{\pagestyle{empty}\cleardoublepage}

Tribunal nombrado por el Mgfco. y Excmo. Sr. Rector de la
Universidad Autónoma de Madrid, el día ....... de
................................ de 2012.

\vspace{2cm}

Presidente: D. ..........................................\\

Vocal: D. ..........................................\\

Vocal: D. ..........................................\\

Vocal: D. ..........................................\\

Secretario: D. ..........................................\\

\vspace{2cm}

Realizado el acto de defensa y lectura de la Tesis el día ....... de
.................. de 2012 en
..............................................................

\vspace{1.5cm}

Calificación: ......................................................\\

\vspace{1cm}

\begin{tabular}{lr}
EL PRESIDENTE & ~~~~~~~~~~~~~~~~~~~~~~~~~~~~EL SECRETARIO\\
 & \\
 & \\
 & \\
 & \\
 & \\
 & \\
 & \\
LOS VOCALES & \\
\end{tabular}

\newpage{\pagestyle{empty}\cleardoublepage}

\begin{flushright}
{\it A mis padres.}
\end{flushright}

\newpage{\pagestyle{empty}\cleardoublepage}

%

\begin{flushright}
{\it After climbing a great hill, one only finds \\
that there are many more hills to climb.\\}
\vspace{1cm} {\it-Nelson Mandela }
\end{flushright}

\newpage{\pagestyle{empty}\cleardoublepage}

{\bf \LARGE Abstract}

\vspace{1cm}

The huge amount of information stored in text form makes methods
that deal with texts really interesting. This thesis focuses on
dealing with texts using compression distances. More specifically,
the thesis takes a small step towards understanding both the nature
of texts and the nature of compression distances. Broadly speaking,
the way in which this is done is exploring the effects that several
distortion techniques have on one of the most successful distances
in the family of compression distances, the Normalized Compression
Distance -NCD-.

The research carried out in this thesis can be divided into three
parts. The first part, which corresponds to Chapter \ref{Chapter:
Study on text distortion}, experimentally evaluates the impact that
several word removal techniques have on NCD-driven text clustering,
with the aim of better understanding of both the nature of
compression distances and the nature of textual information. This
goal is accomplished by analyzing how the information contained in
the documents and how the upper bound estimation of their Kolmogorov
complexity progress as words are removed from the documents. One of
the main conclusions that can be drawn from this analysis is that
the clustering accuracy can be improved by applying a specific word
removal technique. This distortion technique consists of removing
the most frequent words of the language preserving the previous text
structure.

The second part of the thesis, which corresponds to Chapter
\ref{Chapter: Relevance of the contextual information}, attempts to
shed light on the reasons why the application of such a distortion
technique can improve NCD-driven text clustering. The experimental
results show that the maintenance of both the previous text
structure and the remaining words structure have some relevance in
the clustering behavior.

The third part of the thesis, which corresponds to Chapter
\ref{Chapter: Application to Document Retrieval}, applies the above
mentioned distortion technique to NCD-driven document search. The
application of compression distances to document search is not
trivial due to the fact that they do not commonly perform well when
the compared objects have very different sizes. An NCD-based
document search engine that deals with that drawback by using
passage retrieval, is used in the third part of the thesis. The
results show that the search accuracy can be improved by applying
the distortion technique presented previously.

Summarizing, one of the distortion techniques explored in the thesis
has been found to be beneficial both in NCD-based document
clustering and in NCD-based document search.

\newpage

\frontmatter
\pagestyle{plain}
\tableofcontents
\listoffigures
\listoftables
\pagestyle{empty}
\mainmatter
\pagestyle{headings}

\makeatletter
\def\cleardoublepage{\clearpage\if@twoside \ifodd\c@page\else
    \hbox{}
    \thispagestyle{plain}
    \newpage
    \if@twocolumn\hbox{}\newpage\fi\fi\fi}
\makeatother \clearpage{\pagestyle{plain}\cleardoublepage}

%
%
%



\setcounter{chapter}{-1}

\chapter{Resumen de la Tesis}
\label{Chapter: Resumen}

Hoy en día, la mayoría de la información almacenada
electrónicamente, está almacenada en forma de texto. De hecho, si
reflexionamos sobre la cantidad de tiempo que cada día pasamos
delante del ordenador leyendo e-mails, noticias, artículos o
informes, nos daremos cuenta que, de hecho, la mayoría de la
información con la que trabajamos diariamente es texto. Esta
circunstancia hace que las áreas de investigación que estudian
diferentes aspectos relacionados con los datos textuales tengan más
importancia cada día.

Esta tesis se centra en el tratamiento de textos mediante el uso de
distancias basadas en compresión. Más específicamente, la tesis
pretende avanzar en la comprensión tanto de la naturaleza de la
información textual, como de las métricas basadas en compresión.

El fundamento teórico de las distancias basadas en compresión es la
complejidad de Kolmogorov \cite{Kolmogorov65}, la cual está
íntimamente relacionada con el concepto de entropía propuesto por
Shannon en el paper que dio lugar al nacimiento de la teoría de la
información \cite{Shannon48}.

En términos generales, la teoría desarrollada por Shannon cuantifica
la cantidad de información como la cantidad de sorpresa que la
información contiene al ser revelada. Una forma muy simple de
entender esta idea es pensar en la comunicación entre personas.

Por ejemplo, si una persona le dice a otra algo que la última ya
sabía, no habrá ninguna sorpresa en el mensaje, y por tanto, la
primera persona no habrá dado ninguna información a la segunda. Por
el contrario, si la primera persona le dice a la segunda algo que
ésta última no sabía, la primera persona le habrá dado a la última
algo de información.

Ahora bien, desde el punto de vista cuantitativo, la cantidad de
in\-for\-ma\-ción transmitida en el segundo ejemplo, dependerá de lo
probable que fuera el mensaje transmitido. No es lo mismo decir
"Acabo de asomarme a la ventana de mi casa y he visto pasar a una
persona por la calle", que decir, "Acabo de asomarme a la ventana de
mi casa y he visto pasar a la Reina de Inglaterra por la calle". De
esa manera, la información definida por Shannon es inversamente
proporcional a la probabilidad, es decir, cuanto menos probable sea
un mensaje, más información contendrá dicho mensaje.

Para cuantificar de manera formal la información asociada a un
sistema, Shannon definió el concepto de entropía como el promedio de
la ganancia de información de todos los eventos posibles del
sistema. Como cada evento puede ocurrir o no, con una cierta
probabilidad, la entropía creada por Shannon da un peso a la
información asociada a cada evento, en función a la probabilidad de
dicho evento.

El concepto de entropía se ha aplicado en numerosas áreas de
investigación desde su creación. En particular, la entropía es un
concepto básico en el área de la compresión de datos, ya que
proporciona un umbral teórico de la cantidad de compresión que se
puede alcanzar al comprimir una cadena
\cite{Balakrishnan04,OrnsteinW93,Salomon2004}. Este umbral teórico
coincide, aproximadamente, no sólo con la entropía de la cadena,
sino también con la complejidad de Kolmogorov de dicha cadena
\cite{Cover91}. Por tanto, ambos conceptos están directamente
relacionados.

La complejidad de Kolmogorov de una cadena, se define como la
longitud del programa más corto que puede generar la cadena en una
máquina universal de Turing \cite{Kolmogorov65,Li97,Turing36}. Una
cadena será más o menos compleja dependiendo de la naturaleza de la
misma. Por ejemplo, la cadena "0000000000000000" será menos compleja
que la cadena "0000111100001111", y a su vez, ésta será menos
compleja que la cadena "1011011100101010".

La definición de complejidad de Kolmogorov puede extenderse para
definir la complejidad condicional de Komogorov, la cual mide la
complejidad de una cadena $x$ relativa a otra cadena $y$. Esta
medida se define como la longitud del programa más corto que puede
generar la cadena $x$ teniendo la cadena $y$ como entrada a dicho
programa.

Li \emph{et al.} definieron una medida de similaridad entre dos
cadenas, llamada \emph{Normalized Information Distance} -NID-,
combinando los conceptos de complejidad de Komogorov y de
complejidad condicional de Kolmogorov \cite{Li04}.

Dado que la complejidad de Kolmogorov no es computable
\cite{Cover91}, la NID tampoco lo es. Sin embargo, Cilibrasi
\emph{et al.} propusieron una medida computable, llamada
\emph{Normalized Compression Distance} -NCD-, que utiliza algoritmos
de compresión para estimar cotas superiores de la complejidad de
Kolmogorov \cite{Cilibrasi05}. Puede encontrarse información
detallada sobre la NCD y la NID en la Sección \ref{State.
Compression Distances}.

La NCD en particular, y las métricas basadas en compresión en
general, se han aplicado a numerosas áreas de investigación debido a
su naturaleza libre de parámetros, a su efectividad y a su facilidad
de uso. Entre otras, las distancias basadas en compresión se han
utilizado en áreas de investigación, tales como el clustering de
documentos
\cite{Granados11eswa,Granados10tkde,Granados08,Granados10ideal,Helmer07,Telles07},
la re\-cu\-pe\-ra\-ción de documentos
\cite{Granados11tkde,Martinez08}, la clasificación de música
\cite{cilibrasi2004acm,GonzalezPardo10}, la minería de datos
\cite{cilibrasi07}, la seguridad de diferentes sistemas
computacionales \cite{Apel09,Bertacchini07,Wehner07}, la detección
de plagios \cite{Chen04}, la ingeniería del software
\cite{allen2001mca,arbuckle2007sdc,scott-new}, la bioinformática
\cite{Ferragina07,Kocsor06,krasnogor2004msp,Nykter08}, la química
\cite{Melville07}, la medicina \cite{Cohen09,Santos06} o incluso el
arte \cite{Svangard04}.

El hecho de que las distancias basadas en compresión se hayan
utilizado tanto, da una idea de lo útiles que son. Sin embargo, a
pesar de su amplio uso, se ha avanzado poco en la interpretación de
sus resultados o en la explicación de su comportamiento. Cada vez
que se lleva a cabo un trabajo analítico sobre las distancias
basadas en compresión, normalmente éste se centra en la manipulación
algebraica de conceptos algorítmicos de teoría de la información
\cite{Cilibrasi05,Li04,Zhang07}.

Uno de los propósitos de esta tesis es avanzar en el entendimiento
de las métricas basadas en compresión, para así poder mejorar el
rendimiento de este tipo de métricas. En particular, esta tesis se
centra en una de las más importantes distancias basadas en
compresión, la previamente mencionada NCD. El análisis llevado a
cabo en esta tesis es principalmente experimental. Por tanto, la
metodología de trabajo utilizada es la utilizada en ciencias
experimentales. Esta metodología se basa en perturbar el sistema
para observar las consecuencias que acarrea dicha perturbación en el
estado del sistema.

La hipótesis de partida es que se puede modificar la información
contenida en los textos, de manera que el compresor capture mejor la
estructura de los mismos, y por tanto, se pueda mejorar el
rendimiento de la NCD. La clave sería cambiar la representación de
los textos sin perder la información relevante, de forma que esa
nueva representación sea más favorable para que los compresores
capturen mejor las similitudes entre los textos.

Antes de describir los experimentos realizados a lo largo de esta
tesis y mostrar los correspondientes resultados, el Capítulo
\ref{Chapter: Related Work} presenta todos los conceptos necesarios
para comprender los contenidos de la tesis. Tras la presentación de
dichos conceptos, los Capítulos \ref{Chapter: Study on text
distortion} a \ref{Chapter: Application to Document Retrieval}
describen los experimentos realizados a lo largo de la tesis, y
muestran los resultados experimentales obtenidos. Cada uno de esos
capítulos tiene un objetivo claro marcado, y genera una serie de
contribuciones, las cuales se detallan siempre al comienzo de cada
capítulo.

La investigación correspondiente al Capítulo \ref{Chapter: Study on
text distortion} pretende avanzar en el entendimiento tanto de la
naturaleza de la información textual, como de la naturaleza de las
distancias basadas en compresión. Este avance se realiza evaluando
el impacto que tienen varias técnicas de distorsión basadas en la
eliminación de palabras sobre el rendimiento de la NCD.

En concreto, la investigación realizada tanto en el Capítulo
\ref{Chapter: Study on text distortion}, como en el Capítulo
\ref{Chapter: Relevance of the contextual information} utiliza el
método de clustering basado en la NCD desarrollado por los creadores
de la NCD \cite{complearn}, para medir el impacto que tienen las
técnicas de distorsión estudiadas sobre el rendimiento de la NCD. El
uso del método de clustering basado en la NCD como herramienta para
medir el rendimiento de la NCD, permite analizar cómo la información
contenida en los textos estudiados evoluciona a medida que las
palabras son eliminadas de los documentos.

En el Capítulo \ref{Chapter: Study on text distortion}, además de
estudiar cómo evoluciona la información contenida en los textos a
medida que avanza la distorsión de los mismos, se estudia cómo la
complejidad de los textos estudiados evoluciona a medida que se
eliminan más y más palabras de los textos
\cite{Granados10tkde,Granados08}.

Las principales contribuciones de ese capítulo pueden resumirme
brevemente como sigue:

\begin{itemize}
\item
Análisis y estudio de nuevas representaciones de datos textuales
para evaluar el comportamiento de la NCD.

\item
Una técnica de representación de los datos textuales, especialmente
diseñada para ser utilizada en herramientas que utilicen métricas
basadas en compresión, que reduce la complejidad de los documentos
mientras que mantiene la mayoría de la información relevante de los
mismos.

\item
Evidencia experimental de cómo refinar la representación de los
textos para permitir al compresor obtener similaridades más fiables,
y por tanto, permitir al método de clustering basado en la NCD
mejorar los resultados obtenidos al trabajar con los textos
originales, es decir, los textos sin distorsionar.
\end{itemize}

Una de las principales conclusiones que se pueden sacar del análisis
llevado a cabo en el Capítulo \ref{Chapter: Study on text
distortion} es que la precisión del clustering se puede mejorar
aplicando una de las técnicas analizadas. Esta técnica implica, no
sólo la eliminación de palabras, sino también la conservación de la
estructura contextual de los textos.

Esos resultados apuntan a que aunque la información más importante
de un texto esté contenida en las palabras más relevantes del mismo,
lo que rodea a esas palabras es importante también, ya que es el
sustrato que las soporta. La hipótesis sería que esa es la razón por
la cual, la técnica de distorsión que mantiene la información
relevante a la vez que preserva la información contextual, es la
mejor de todas las evaluadas.

El Capítulo \ref{Chapter: Relevance of the contextual information}
estudia si esa hipótesis es acertada o no, es decir, el capítulo
estudia en qué medida se ven afectados los resultados obtenidos
debido tanto a la preservación de la información relevante, como a
la preservación de la información contextual.

El concepto de información contextual se ha utilizado en numerosas
aplicaciones informáticas. Por ejemplo, se ha utilizado en áreas de
investigación como la recuperación de información
\cite{Ma07,Rhodes03,Sieg07,Speretta05,Tamine10}, los sistemas de
recomendación \cite{Adomavicius05,Kwon09,Su10,Weng09}, las
aplicaciones sensibles al contexto
\cite{Caus09,Driver08,Hegde09,Pascoe98,Schilit94}, la visión
artificial \cite{Acosta10,Belongie02,Chi08,Mori05,Mori06}, el
reconocimiento de voz \cite{Eronen06,Huang07,Lee90,Odell95thesis} o
el análisis del tráfico en redes \cite{Goodall06}, entre otros.

Cuando se está trabajando con información textual, la idea de
contexto es muy útil, ya que está intimamente relacionada con los
textos, debido a la naturaleza intrínseca de los mismos. Dado que
los textos no son sólo secuencias de palabras, sino que tienen una
estructura coherente \cite{Kozima93}, aplicar la idea de contexto al
manejo de textos surge de forma natural.

En esta tesis, la información contextual es un subproducto de la
a\-pli\-ca\-ción de la técnica de distorsión, presentada en el
Capítulo \ref{Chapter: Study on text distortion}, mencionada
anteriormente. El Capítulo \ref{Chapter: Relevance of the contextual
information} compara dicha técnica con tres nuevas técnicas creadas
a partir de la anterior, las cuales destruyen la información
contextual de diferentes maneras. Analizando los resultados
experimentales obtenidos, se puede observar que mantener la
información contextual es beneficioso en el campo del clustering de
textos basado en la NCD \cite{Granados11eswa,Granados10ideal}.

Las principales contribuciones del Capítulo \ref{Chapter: Relevance
of the contextual information} se resumen brevemente en los
siguientes puntos:

\begin{itemize}
\item
Evaluación experimental de la relevancia que la información
contextual tiene en el clustering de textos basado en la NCD, en un
escenario de eliminación de palabras.

\item
Nuevas perspectivas para la evaluación y el estudio del
comportamiento de las distancias basadas en compresión, en relación
a la información contextual.
\end{itemize}

Finalmente, el Capítulo \ref{Chapter: Application to Document
Retrieval} aplica los conocimientos adquiridos en los
Ca\-pí\-tu\-los \ref{Chapter: Study on text distortion} y
\ref{Chapter: Relevance of the contextual information} a la búsqueda
de documentos basada en la NCD. La aplicación de las distancias
basadas en compresión a la búsqueda de documentos no es trivial dado
que este tipo de distancias tienen un punto débil que tiene que
tenerse en cuenta si éstas se quieren aplicar en determinadas
circunstancias. Su punto débil es que cuando los objetos comparados
son muy diferentes en tamaño, las distancias obtenidas no son muy
fiables. Un método de búsqueda de documentos que aborda este
problema utilizando recuperación de pasajes se utiliza en la última
parte de la tesis.

Los resultados experimentales muestran que la búsqueda de documentos
se puede mejorar aplicando la técnica de distorsión presentada en la
primera parte de la tesis. Este hecho da mayor generalidad a los
resultados obtenidos en la primera parte de la tesis, ya que dicha
técnica ha resultado ser útil no sólo para el clustering de
documentos, sino también para la búsqueda de los mismos
\cite{Granados11tkde}.

Las principales contribuciones del Capítulo \ref{Chapter:
Application to Document Retrieval} se pueden resumir como sigue:

\begin{itemize}
\item
Aplicación práctica de las principales conclusiones sacadas de los
estudios llevados a cabo en las dos primeras partes de la tesis, a
la búsqueda de documentos textuales.

\item
Mejora en la representación de los documentos que permite obtener un
incremento considerable de la precisión en los resultados obtenidos
al buscar dichos documentos.
\end{itemize}


\chapter{Introduction}
\label{Chapter: Introduction}

Nowadays, most of the information stored electronically is stored in
text form. In fact, if we think of the time that we spend every day
reading e-mails, news, articles or reports, we will realize that
most of the information that we use every day is text. This fact
makes methods that deal with texts really interesting.

This thesis focuses on dealing with texts using compression
distances. More specifically, it takes a step towards understanding
both the nature of texts and the nature of compression distances.

The theoretical foundation of compression distances is the
Kolmogorov complexity, which is intimately related to the concept of
entropy proposed by Shannon in the paper that gave rise to
Information Theory \cite{Shannon48}.

Broadly speaking, the theory developed by Shannon quantifies the
amount of information as the amount of surprise that the information
contains when revealed. A very simple way of understanding this is
thinking of human communications.

For example, if one person tells another something that the latter
already knows, there is no surprise in the message, and therefore,
the first person has given the latter no information at all. On the
contrary, if a person tells another something that the latter does
not know, the first person has given the latter some information.

The amount of information transmitted in the second example, depends
on the likelihood of the transmitted message. For example, saying
``I have just looked out the window and I have seen a person walking
in the street'' gives less information than saying ``I have just
looked out the window and I have seen the Queen of England walking
in the street''. Thus, the information defined by Shannon is
inversely proportional to the probability, that is, the less
probable a message is, the more information it contains.

Shannon defined the concept of entropy with the aim of formally
quantifying the information associated with a system. Entropy was
defined as the average information gain from all possible events of
the system. Given that each event can occur with a certain
probability, the entropy created by Shannon weights the information
associated with each event, according to the probability of the
event.

The concept of entropy has been applied to numerous research areas.
In particular, entropy is a basic concept in the area of data
compression because it provides a theoretical bound on the amount of
compression that can be achieved
\cite{Balakrishnan04,OrnsteinW93,Salomon2004}. This theoretical
bound coincides approximately with not only the entropy of the
string, but also the Kolmogorov complexity of the string
\cite{Cover91}. Therefore, the concept of entropy is directly
related to the theoretical foundation of compression distances: the
Kolmogorov complexity.

The Kolmogorov complexity of a string is defined as the length of
the smallest program that can generate the string on a universal
computer \cite{Kolmogorov65,Li97}. A string would be more or less
complex depending on its nature. For example, the string
``0000000000000000'' would be less complex than the string
``0000111100001111'', and in turn, the latter would be less complex
than the string ``1011011100101010''.

The definition of Kolmogorov complexity can be extended to define
the conditional Kolmogorov complexity, which measures the complexity
of a string $x$ relative to another string $y$. This measure is
defined as the length of the smallest program that can generate the
string $x$ on a universal computer, having the string $y$ as input
to the program.

Li \emph{et al.} defined a measure of similarity between two
strings, called \emph{Normalized Information Distance} -NID-,
combining the concepts of Kolmogorov complexity, and conditional
Kolmogorov complexity \cite{Li04}.

Given that Kolmogorov complexity is non-computable \cite{Cover91},
NID is not computable either. However, Cilibrasi \emph{et al.}
proposed a computable measure, called \emph{Normalized Compression
Distance} -NCD-, that uses compression algorithms to estimate an
upper bound upon the Kolmogorov complexity \cite{Cilibrasi05}. More
detailed information on the NCD and the NID can be found in Section
\ref{State. Compression Distances}.

The NCD in particular and compression distances in general have been
applied to several research areas because of their parameter-free
nature, their wide applicability and their leading efficacy. Among
others, they have been applied to document clustering
\cite{Granados11eswa,Granados10tkde,Granados08,Granados10ideal,Helmer07,Telles07},
document retrieval \cite{Granados11tkde,Martinez08}, music
classification \cite{cilibrasi2004acm,GonzalezPardo10}, data mining
\cite{cilibrasi07}, security of computer systems
\cite{Apel09,Bertacchini07,Wehner07}, plagiarism detection
\cite{Chen04}, software engineering
\cite{allen2001mca,arbuckle2007sdc,scott-new}, bioinformatics
\cite{Ferragina07,Kocsor06,krasnogor2004msp,Nykter08}, chemistry
\cite{Melville07}, medicine \cite{Cohen09,Santos06} or even art
\cite{Svangard04}. The fact that compression distances have been so
widely used gives us an idea of how useful they are.

Despite their wide use, little has been done to interpret
compression distances results or to explain their behavior. Whenever
some analytical work on compression distances is carried out, it is
usually focused on the algebraic manipulation of algorithmic
information theory concepts \cite{Cilibrasi05,Li04,Zhang07}.

One of the objectives of this thesis is to make progress on the
understanding of compression distances in order to improve the
performance of these metrics. In particular, this thesis focuses on
one of the most important compression distances, the previously
mentioned NCD. The analysis carried out in this thesis is mainly
experimental. Therefore, the methodology used is the one used in
experimental sciences. This methodology is based on disturbing the
system to observe the consequences of the disturbance in the state
of the system.

The assumption is that the information contained in the texts can be
modified so that the compressor can better capture their structure,
and therefore, the obtained NCD-based clustering results can be
improved. The idea is to change the representation of the texts
without losing relevant information so that this new representation
is more suitable for compressors to better capture the similarities
between the texts.

Before describing the experiments carried out throughout the thesis,
Chapter \ref{Chapter: Related Work} presents all the concepts needed
to easily understand the contents of the thesis. After presenting
them, Chapters \ref{Chapter: Study on text distortion} to
\ref{Chapter: Application to Document Retrieval} describe the
experiments carried out throughout the thesis, and show the obtained
experimental results. Each of these chapters has a clear objective,
and generates a series of contributions, which are detailed always
at the beginning of each chapter.

The research that corresponds to Chapter \ref{Chapter: Study on text
distortion}, tries to take a step towards the understanding of both
the nature of textual information, and the nature of compression
distances. This purpose is accomplished by analyzing how the
information contained in the documents and how the upper bound
estimation of their Kolmogorov complexity progress as words are
removed from the documents. This is done by evaluating the impact
that different distortion techniques, based on word removal, have on
the NCD behavior \cite{Granados10tkde,Granados08}.

In particular, the research carried out in both Chapter
\ref{Chapter: Study on text distortion} and Chapter \ref{Chapter:
Relevance of the contextual information} uses the NCD-based
clustering method developed by the creators of the NCD
\cite{complearn}, to measure the impact that the explored distortion
techniques have on the NCD behavior. The use of the NCD-based
clustering method as a tool to measure the performance of the NCD,
allows analysis of how the information contained in the texts
progresses as words are removed from the texts.

The main contributions of this chapter can be briefly summarized as
follows:

\begin{itemize}
\item
Analysis and study of new representations of texts to evaluate the
behavior of the NCD.

\item
A technique to represent textual data, specially created to be used
with compression distances, that reduces the complexity of the
documents while preserving most of the relevant information.

\item
Experimental evidence of how to fine-tune the representation of
texts to allow the compressor to obtain more reliable similarities
and, therefore, to allow the compression-based clustering method to
improve the non-distorted clustering results.
\end{itemize}

One of the main conclusions that can be drawn from the analysis made
in Chapter \ref{Chapter: Study on text distortion}, is that the
accuracy of the clustering can be improved by applying a specific
word removal technique. That technique implies, not only the removal
of words, but also the maintenance of the previous text structure.

These results suggest that although the most important information
of a text is contained in the most relevant words thereof, the
information that surrounds these words is important too, because
that information is the substrate that supports them. The hypothesis
would be that this is the reason why the distortion technique that
maintains the relevant information while preserving the contextual
information is the best of all the evaluated distortion techniques.

Chapter \ref{Chapter: Relevance of the contextual information}
explores whether that hypothesis is correct or not. That is, the
chapter studies how the results are affected by both the maintenance
of the relevant information, and the maintenance of the contextual
information.

The concept of contextual information has been used in several
research areas. For example, it has been used in research areas such
as contextual information retrieval
\cite{Ma07,Rhodes03,Sieg07,Speretta05,Tamine10}, recommender systems
\cite{Adomavicius05,Kwon09,Su10,Weng09}, context-aware computing
applications \cite{Caus09,Driver08,Hegde09,Pascoe98,Schilit94},
computer vision \cite{Acosta10,Belongie02,Chi08,Mori05,Mori06},
speech recognition systems
\cite{Eronen06,Huang07,Lee90,Odell95thesis} or network traffic
analysis \cite{Goodall06}, among others.

In particular, in the management of textual data, the idea of
context is very useful because it is strongly bound to texts due to
their intrinsic nature. Since a text is not just a sequence of
words, but it has coherent structure \cite{Kozima93}, applying the
idea of context to text management arises naturally.

In this thesis, the contextual information is a byproduct of the
application of the distortion technique presented in Chapter
\ref{Chapter: Study on text distortion}, that can improve the
accuracy of the clustering. Chapter \ref{Chapter: Relevance of the
contextual information} compares that technique with three new
distortion techniques created from it, which destroy the contextual
information in different ways. Analyzing the obtained experimental
results, it can be observed that maintaining the contextual
information is beneficial in NCD-based text clustering
\cite{Granados11eswa,Granados10ideal}.

The main contributions of Chapter \ref{Chapter: Relevance of the
contextual information} can be briefly summarized as follows:

\begin{itemize}
\item
Experimental evaluation of the relevance that the contextual
information has in compression-based text clustering, in a word
removal scenario.

\item
New perspectives for the evaluation and explanation of the behavior
of compression distances, in relation to contextual information.
\end{itemize}

Finally, Chapter \ref{Chapter: Application to Document Retrieval},
applies the knowledge acquired in Chapters \ref{Chapter: Study on
text distortion} and \ref{Chapter: Relevance of the contextual
information} to NCD-based document search. The application of
compression distances to document search is not trivial due to their
having a weakness that must be taken into account if one wants to
apply them under particular circumstances. Their drawback is that
they do not commonly fit well when the compared objects have very
different sizes. A document search method that addresses this issue
by using passage retrieval is used in the last part of the thesis.

The experimental results show that the non-distorted document search
results can be improved by applying the distortion technique
presented in the first part of the thesis. This fact gives more
generality to the results obtained in the first part of the thesis,
since this technique has proven to be useful not only for document
clustering, but also for document search \cite{Granados11tkde}.

The main contributions of Chapter \ref{Chapter: Application to
Document Retrieval} can be briefly summarized as follows:

\begin{itemize}
\item
Practical application of the main conclusions taken from the studies
developed in the first two parts of the thesis to document search.

\item
Improvement in the representation of documents that allows
increasing the accuracy of the results obtained when searching
documents.
\end{itemize}

\chapter{Objectives}
\label{Objectives}

Broadly speaking, this thesis applies text distortion to
compression-based text clustering with the aim of taking a step
towards understanding the nature of compression distances, and the
nature of textual data. After that, it applies text distortion to
compression-based document retrieval with the aim of exploring a
possible practical application of the knowledge acquired in the
first study. These widespread objectives can be divided into more
specific goals:

\begin{itemize}
\item \emph{Objective 1. Providing new perspectives for
understanding the nature of textual data.}

The huge amount of information stored in text form makes the study
of the nature of texts really interesting. Many research areas
address several aspects of processing textual information in
different manners. This thesis uses compression distances to explore
how the application of different distortion techniques affects the
information contained in the evaluated texts.

\item \emph{Objective 2. Providing a technique to smoothly reduce
the complexity of the documents while preserving most of their
relevant information.}

Removing irrelevant parts of the data has been found to be
beneficial in data analysis. In fact, most of the research areas
that work with textual data apply that idea to text processing. This
thesis tries to provide a text distortion technique that reduces the
complexity of the texts while maintaining most of the relevant
information contained in them.

\item \emph{Objective 3. Giving experimental evidence of how to
fine-tune the text representation so that better results are
obtained when using NCD-driven text clustering.}

One of the purposes of this thesis is finding a text representation
that can improve the clustering results. This work explores
different distortion techniques with the aim of attaining this
objective. One of the explored techniques has been found to be
beneficial to NCD-based text clustering.

\item \emph{Objective 4. Giving new insights for the evaluation
and explanation of the behavior of the NCD.}

Compression distances have been widely used in knowledge discovery
and data mining due to their parameter-free nature, wide
applicability, and leading efficacy in several domains. However,
little has been done to interpret their results or to explain their
behavior. This thesis tries to shed light on this issue by
performing an experimental study on text distortion.

\item \emph{Objective 5. Experimentally evaluating the relevance
that the contextual information has in compression-based text
clustering, in a word removal scenario.}

The distortion technique that fine-tunes the text representation
implies not only the removal of words, but also the maintenance of
the previous text structure. Exploring the relevance of both factors
becomes necessary in order to better understand the results. This
research is carried out in the thesis as well.

\item \emph{Objective 6. Applying the main conclusions taken from
the studies developed in the first two parts of the thesis to
document search.}

A problem in which the distortion technique that fine-tunes the text
representation can be very useful is the search of texts. Applying
that technique to document search is one of the purposes of this
work.

\item \emph{Objective 7. Giving a representation of documents
that improves the non-distorted document search accuracy.}

Text representation plays an important role in document search.
Thus, good text representations can improve the accuracy of the
results, whereas bad ones can make the results get worse. Exploring
if the application of the above mentioned distortion technique can
lead to better document search results is the final goal of this
thesis.

\end{itemize}

\chapter{Thesis Overview}
\label{Thesis Overview}

The thesis is structured as follows:

\begin{itemize}
\item \emph{Chapter \ref{Chapter: Related Work}} presents and
discusses all the concepts needed to easily understand the
contents of the thesis.

\item
\emph{Chapter \ref{Chapter: Study on text distortion}} explores
several text distortion techniques based on word removal. It
analyzes how the information contained in the documents and how the
upper bound estimation of their Kolmogorov complexity progress as
the words are removed from the documents in different manners.

\item \emph{Chapter \ref{Chapter: Relevance of the contextual
information}} explores how the loss or the maintenance of the
contextual information affects the clustering accuracy. At the
same time, it explores how the loss or the preservation of the
remaining words structure affects the clustering.

\item \emph{Chapter \ref{Chapter: Application to Document
Retrieval}} applies the distortion technique that can lead to better
clustering results to document search.

\item
\emph{Chapter \ref{Chapter: Conclusions}} discusses the conclusions
drawn from the research carried out in the thesis.

\item \emph{Chapter \ref{Summary of Results}} summarizes each of
the contributions made from the work developed throughout the
thesis.

\item \emph{Chapter \ref{Chapter: Contributions}} presents the
papers created from the investigation carried out in the thesis.

\item \emph{Appendix \ref{Appendix Acronyms}} presents an index of
the acronyms used.

\item
\emph{Appendix \ref{Appendix Datasets}} contains the detailed
description of the datasets used throughout the thesis. In addition,
it shows, as a sample, a fragment of a document for each dataset.

\item
\emph{Appendix \ref{Appendix Queries}} contains the detailed
description of the queries used in the experiments carried out in
Chapter \ref{Chapter: Application to Document Retrieval}. It also
shows, as a sample, a fragment of a query for each dataset.

\item \emph{Appendix \ref{Appendix Detailed Experimental Results}}
contains all the detailed results obtained in the work presented
in Chapter \ref{Chapter: Study on text distortion}.
\end{itemize}


\chapter{Related Work}
\label{Chapter: Related Work}

This chapter presents and discusses all the concepts needed to
easily understand the contents of the thesis.

Compression distances have to be described in this chapter because
this thesis uses them to cluster and retrieve documents. Since,
compression distances are based on information theory concepts, the
latter have to be presented as well, in order to help and understand
compression distances. Furthermore, given that compression distances
use compression algorithms to calculate the similarity between two
objects, the compression algorithms explored in this thesis have to
be described.

Three compression algorithms are used in this thesis to calculate
compression distances. Each of them belongs to a different family of
compressors: LZMA, PPMZ, and BZIP2. LZMA compressor, is a
Lempel-Ziv-Markov chain algorithm \cite{lzmax}. PPMZ compressor is
an adaptive statistical data compression algorithm based on context
modeling and prediction \cite{ppmz}. BZIP2 compressor is a
block-sorting compressor based on the Burrows-Wheeler Transform,
Huffman codes, the Move-To-Front transform, and Run Length Encoding
\cite{Burrows94,Huffman52,Salomon2004,bzip2}. Among all the existing
compression algorithms, only these ones are reviewed in this
chapter.

In addition, given that the distortion techniques explored in this
thesis are based on word removal, the concept of word removal must
be presented as well. Moreover, since the most important distortion
technique used in the thesis maintains the contextual information
despite the removal, presenting how several research areas apply the
concept of contextual information is necessary.

The chapter contains a section for each of the concepts mentioned
above.

\section{Information Theory Concepts}
\label{State. Information Theory Concepts}

Information Theory -IT- is a branch of applied mathematics and
electrical engineering that focuses on the task of quantifying
information. The famous work by Claude Shannon in 1948
\cite{Shannon48} involved its creation. The research area of IT has
turned out to be one of the most influential ones because of its
wide applicability in many other domains \cite{Cover91}.

Roughly speaking, the theory developed by Shannon quantifies the
amount of information as the amount of surprise that the information
contains when revealed. A very simple way of understanding this is
thinking of human communications. For example, if one person tells
another something that the latter already knows, there is no
surprise in the message, and therefore, the first person has given
the latter no information at all. On the contrary, if a person tells
another something that the latter does not know, the first person
has given the latter some information.

The amount of information transmitted in the second example, depends
on the likelihood of the transmitted message. For example, saying
``I have just looked out the window and I have seen a person walking
in the street'' gives less information than saying ``I have just
looked out the window and I have seen the Queen of England walking
in the street''. Thus, the information should be proportional to the
probability, that is, the less probable a message is, the more
information it contains. The mathematical formulation of this idea
would be:

\begin{equation}
I(x) \sim \frac{1}{P(x)}
\end{equation}

where x is the event, and P(x) is the probability function.

Furthermore, independent information should be additive, that is, if
the first person tells the second one something more, then the first
has given the latter some more information, independent of, and
additional to, the information that the former gave the latter
previously.

Since the probability of independent events is the product of the
probabilities of the individual events, the function used to
represent the information should have the following property:

\begin{equation}
f(xy) = f(x) + f(y)
\end{equation}

The mathematical function that transforms a product in an addition
is the logarithm. That is the reason why logarithms are used to give
a measure of the information.

Therefore, from all the above, the following formal definition can
be derived:

\begin{equation}
I(x) = log_2~\frac{1}{P(x)}
\end{equation}

Several examples will be analyzed in order to better understand the
basis of IT. The simplest event that can be analyzed in order to
approach the basis of IT is the toss of a coin. Before the toss, the
result is uncertain, this uncertainty must be resolved by tossing
the coin. This will produce either a head or a tail. Thus, the
result of tossing the coin can be expressed with a single bit since
there are only two possibilities. Therefore, the information
contained in the result is one bit.

This strategy can easily be generalized to resolve more complex
problems. The idea is finding the minimum number of yes/no questions
that must be answered in order to resolve the uncertainty. The
number of questions will correspond to the number of bits needed to
express the information in the result because the information
contained in a yes/no question is a bit, since this kind of question
only produces two possible answers.

A more complex problem that intuitively introduces why the logarithm
is the mathematical function that quantifies information, is drawing
a card from a deck of 32 playing cards. This event can be thought as
guessing a number between 1 and 32. The minimum number of yes/no
questions needed to guess a number between 1 and 32 is given by the
binary search algorithm.

In computer science, a binary search locates an element in a sorted
array of elements. The algorithm works by comparing the searched
element with the element contained in the middle of the array. The
comparison determines whether the element is already found or must
be searched for again in the left half of the array or in the right
half of it. The asymptotical cost of this algorithm is $\log_2~{N}$,
N being the number of elements contained in the array. This
reasoning constitutes an alternative way of explaining why the
logarithm is the mathematical function that quantifies information.

Returning to the problem of guessing the card, the number of
questions that have to be answered to guess the card is 5, because
$\log_2~{32} = 5$.

One can easily interpret that result thinking of the bits needed to
codify the number and the suit of the card from a deck of 32 playing
cards. Since there are four possible suits, two bits will be
required to codify the suit. Similarly, since there are eight
possible numbers, three bits will be necessary to codify the number.
Again, this makes a total of 5 bits to codify the card:

\begin{itemize}
  \item \emph{Suit of the card:} 2 bits because there are 4 possible suits.
  \item \emph{Number of the card:} 3 bits because there are 8 possible
  numbers.
\end{itemize}

To quantify the information associated with a system, Shannon
defined the concept of entropy as the average information gain of
all possible system events \cite{Shannon48}. Since each event can
occur or not, with a certain probability, the entropy gives a weight
to the information associated with each event, according to the
probability of the event. Mathematically, the entropy $H(X)$ of a
discrete random variable $X$, with probability function $p(x)$ is
defined as follows:

\begin{equation}
H(X) = - \sum\limits_{x \in \chi} p(x)~log_2~p(x)
\end{equation}

Note that he entropy of X can also be interpreted as the expected
value of $log_2~\frac{1}{p(X)}$.\\

The expected value of a random variable g(X) is as follows:

\begin{equation}
E_p~g(X) = \sum\limits_{x \in \chi} p(x)~g(x)
\end{equation}

Therefore:\\

$E_p~log_2~\frac{1}{p(X)} = \sum\limits_{x \in \chi}
p(x)~log_2~\frac{1}{p(x)} = \sum\limits_{x \in \chi}
p(x)~log_2~p(x)^{-1}
= - \sum\limits_{x \in \chi} p(x)~log_2~p(x)$\\

Thus:

\begin{equation}
H(X) = E_p~log_2~\frac{1}{p(X)}
\end{equation}

The simple example discussed above, the toss of a coin, can be used
to clarify the concept of entropy. The probability of obtaining a
head or a tail is the same:

\[
  X = \left\{
  \begin{array}{l l}
    head &  \textrm{with probability $\frac{1}{2}$}\\
    & \\
    tail &  \textrm{with probability $\frac{1}{2}$}\\
  \end{array} \right.
\]

Then,\\

$H(X) = - \sum\limits_{x \in \chi} p(x)~log_2~p(x)$\\

$H(X) = - [\frac{1}{2}~log_2~\frac{1}{2} + \frac{1}{2}~log_2~\frac{1}{2} ] = -~log_2~\frac{1}{2} = -~log_2~2^{-1} = log_2~2 = 1~\textrm{bit.}$\\

Similarly, the entropy of the example of drawing a card from a deck
of 32 cards can be calculated, given that the probability of drawing
a card is the same for all the cards. That is: $p(x) =
\frac{1}{32}$.\\

$H(X) = - \sum\limits_{x \in \chi} p(x)~log_2~p(x) = -~[32~(\frac{1}{32}~log_2~\frac{1}{32})] = log_2~32 = 5~\textrm{bits.}$\\

Notice that these examples have an important characteristic: all the
possible events have the same probability of occurrence. In general,
in systems in which all the possible events have the same
probability of occurrence, the entropy is equivalent to the
logarithm of the number of possible events. Thus if N is the number
of possible events:

\begin{equation}
H(X) = - \sum\limits_{i = 1}^N \frac{1}{N}~log_2~\frac{1}{N} =
log_2~N~\textrm{for equally likely events}.
\end{equation}

Let us analyze a more complex example. The setting is the same as
the previous example, that is, drawing a card from a deck of 32
cards. However, in this example, the amount of uncertainty of an
event $E$ given another event $F$ is calculated. For example, given
the following events:

\begin{itemize}
 \item $E$ = The card drawn is the ace of hearts.
 \item $F$ = The card drawn is a heart.
\end{itemize}

The probability of $E$ given $F$ is:\\

$P(E/F) = \frac{P(E \cap F)}{P(F)} = \frac{P(E)}{P(F)}$ as $E
\subset F$\\

The probabilities of the events E and F are:
\begin{itemize}
 \item $P(E) = \frac{1}{32}$, since there is only one ace of hearts
 in the deck.
 \item $P(F) = \frac{1}{4}$, since there are four suits in the deck.
\end{itemize}

Therefore:\\

$H(E/F) = -~log_2~ P(E/F) = log_2~\frac{P(E)}{P(F)} =
log_2~\frac{32}{4} =
log_2~8 = 3~\textrm{bits.}$\\

This result can be easily interpreted. The fact that $F$ has
occurred determines the suit of the card, that is, determines two
bits, because as said previously, two bits are needed to codify the
suit of the card because there are four possible suits.
Consequently, specifying the card given that it is a heart, requires
only 5 - 2 = 3 bits. Thus, the uncertainty of $E$ has been reduced
thanks to the knowledge of $F$.

The main theorem proved by Shannon says that a message of $n$
symbols can, on average, be compressed down to $nH$ bits, but not
further. It also says that almost optimal compressors -called
\emph{entropy encoders}- exist \cite{Salomon2004}.

%

\subsection{Kolmogorov Complexity}
\label{State. Kolmogorov Complexity}

Directly related to the measure of information proposed by Shannon
is the Kolmogorov complexity of a string $x$, $K(x)$. Andréi
Kolmogórov defined the algorithmic complexity of an object $x$,
$K(x)$ as the length of the shortest program that can generate $x$
on a universal computer \cite{Kolmogorov65,Li97}. This definition
can be extended to define the conditional Kolmogorov complexity.
That is, the Kolmogorov complexity of a string $x$, relative to
another string $y$. The conditional Kolmogorov complexity $K(x|y)$
is the length of the smallest program that generates the string $x$
having the string $y$ as input to the program.

The most interesting result is that the expected length of the
shortest program of a random variable is approximately equal to its
entropy \cite{Cover91}.

The best way to assimilate the concept of Kolmogorov complexity is
intuitively analyzing some strings:

\begin{enumerate}
  \item 1010101010101010101010101010101010101010101010101010
  \item 1100001100000011001100000011000000110000001100110000
  \item 1000101011001011011000101111001010001011001100010101
\end{enumerate}

The question is: what is the shortest program that can generate each
of these strings?

Generating the first string with a program is simple because the
string could be generated using a for-loop that prints ``10'' in
each iteration.

The second string can be described as a ``11'', followed by $r_i$
repetitions of ``0'', where $r_i$ can be 2, 4 or 6. Therefore, the
shortest program that can generate such a string is more complex
than the previous one.

Finally, the shortest program that can generate the third string
should simply print all the bits of the sequence, because this
string cannot be expressed in any regular way. Consequently, this
program would be at least as big as the string itself. This program
would be definitely more complex than the previous ones.

The good news is that most binary strings used in practice to
represent texts are similar to the second string shown previously.
Therefore, they exhibit some regularity, and thus they can be
compressed \cite{Salomon2004}.

The concept of Kolmogorov complexity is directly related to this
thesis because it has been used to define a measure of similarity
between two strings, giving rise to the concept of \emph{Normalized
Information Distance} -NID- \cite{Li04}. The compression distance
used in this thesis is created from the NID. Both distances are
described in depth in Section \ref{State. Compression Distances}.

\section{Compression Algorithms}
\label{State. Compression Algorithms}

Data compression existed before the appearance of computers, as some
well-known codes, such as the Braille of 1825 or the Morse of 1838
show. Interesting approaches were used in both cases, as explained
below.

The Braille code is based on a communication method developed by
Charles Barbier in order to allow Napoleon's soldiers to communicate
silently and lightlessly. The Barbier method was rejected by the
military due to that it encoded each letter with a set of 12
embossed dots, making it too difficult for soldiers to read by
touch. However, it ended up leading to the creation of the extremely
important Braille code that has allowed blind people to read since
its creation.

The inception of the Braille code is due to the encounter between
Charles Barbier and Louis Braille in the National Institute for the
Blind in Paris in 1821. Braille, who only was 12 years old,
associated the failure of the method to the high number of dots used
to encode each letter. His hypothesis was that since the human
fingertip could not cover the whole symbol without moving, the
message could not be read efficaciously and efficiently. This led to
the creation of the Braille system, which encodes each symbol with 6
dots.

Each of the 6 dots in a symbol can be flat or raised, which means
that the information contained in a symbol is equivalent to 6
bits, which implies the possibility of coding $2^6 = 64$ different
symbols. Since the letters, digits, and punctuation marks do not
require the use of all the codes, the spare ones are used to code
common words, such as \emph{and}, \emph{for} and \emph{of}, and
common strings of letters, such as \emph{ound}, \emph{ation}, and
\emph{th}. Although this kind of data compression is modest, it is
important because books in Braille are usually very large due to
the room that each symbol takes up.

The first version of the Morse code, mentioned above, which dates
from 1832, allows the transmission of textual information as a
series of short and long dashes that represent numbers. A code book
or dictionary associates each number with a word. Thus, this first
version of the Morse code was a primitive form of data compression.

The famous Morse code used nowadays is the evolution of the
primitive Morse code. It allows the transmission of textual
information as a series of dots and dashes as well. It encodes
letters, digits, and some punctuation marks, using variable-size
codes to encode each symbol. This important feature of the Morse
code leads to a better efficiency because the length of each symbol
is approximately inversely proportional to its frequency of
occurrence in English. This is reminiscent of the basic idea of the
Huffman coding, which will be considered later.

These are some examples of data compression used before the
appearance of computers. After that, in the computer age, data
compression has become crucial, initially, to reduce the storage
needed for data, and later, after the appearance of the Internet, to
reduce transmission time.

Many compression strategies have been used since the emergence of
data compression as a research field, from primitive algorithms to
sophisticated algorithms that achieve very high compression rates.
Among these latter, most text compression methods are either
dictionary or statistical based. The next subsections explain in
more detail the characteristics of the most important text
compression algorithms.

\subsection{Statistical Methods}

Statistical compressors are based on developing statistical models
of the text. The model assigns probabilities to the input symbols,
and then, the symbols are coded based on these probabilities. The
model can be \emph{static} or \emph{dynamic} -also known as
\emph{adaptive}-.

\subsubsection{Huffman Coding}

David Huffman developed this entropy encoding algorithm in 1952
\cite{Huffman52}. This method uses variable-length codes for
encoding the symbols using bits. It assigns shorter codes to the
more frequent symbols and longer codes to the less frequent ones
to make the coding more efficient.

The method constructs a binary tree, with a symbol at each leaf,
which can be traversed to determine the codes of the symbols. Fig
\ref{STATE. Fig:huffman} shows an example of a Huffman tree
generation.

The process is as follows:

\begin{enumerate}
\item A list of nodes that contains the alphabet symbols is created
and sorted in increasing order of frequency.

\item Then, the tree is constructed from that list following these
steps:

\begin{enumerate}
\item Remove the two nodes of lowest frequency from the list.

\item Create a new node with these nodes as children and with
frequency equal to the sum of the children's frequencies.

\item Insert the new node into the ordered list of nodes.
\end{enumerate}

\item At the end of the process, a binary tree, which has a leaf
for each symbol of the alphabet, is obtained.
\end{enumerate}

\def\imagetop#1{\vtop{\null\hbox{#1}}}
\begin{figure*}
\begin{tabular}{cc}
\multicolumn{2}{c}{1st step} \\
\multicolumn{2}{c}{\imagetop{\includegraphics[width=8cm]{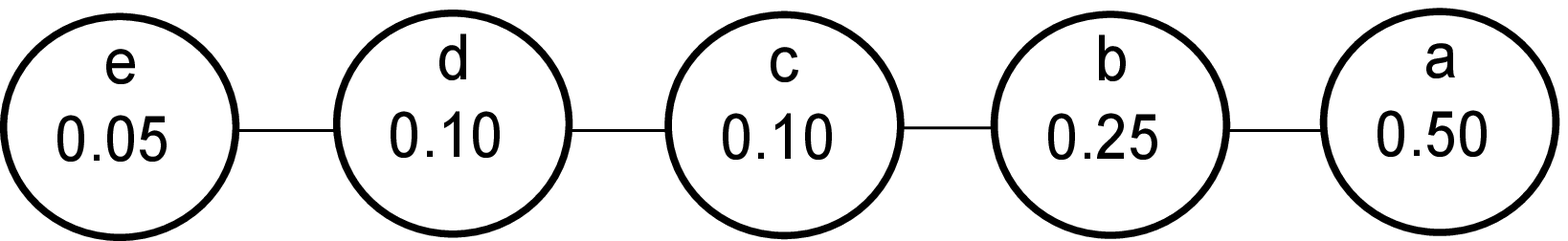}}} \\
 & \\
 2nd step & \texttt{~~~~3rd step} \\
\imagetop{\includegraphics[width=6.2cm]{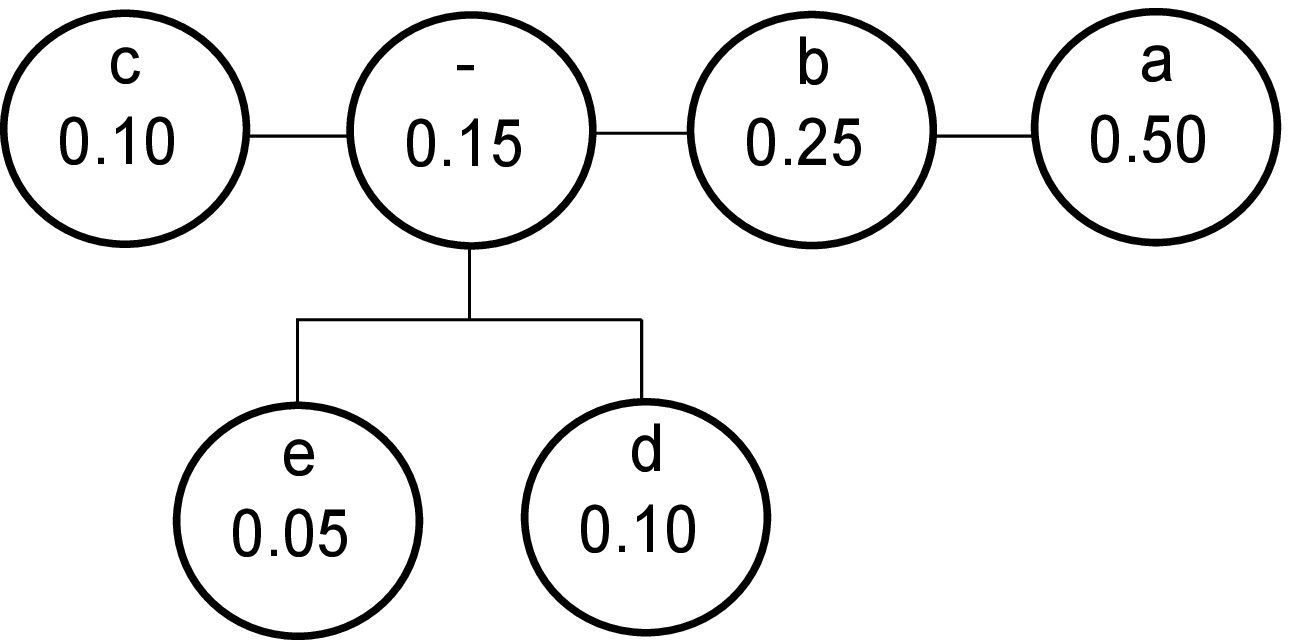}} & \imagetop{\includegraphics[width=5.4cm]{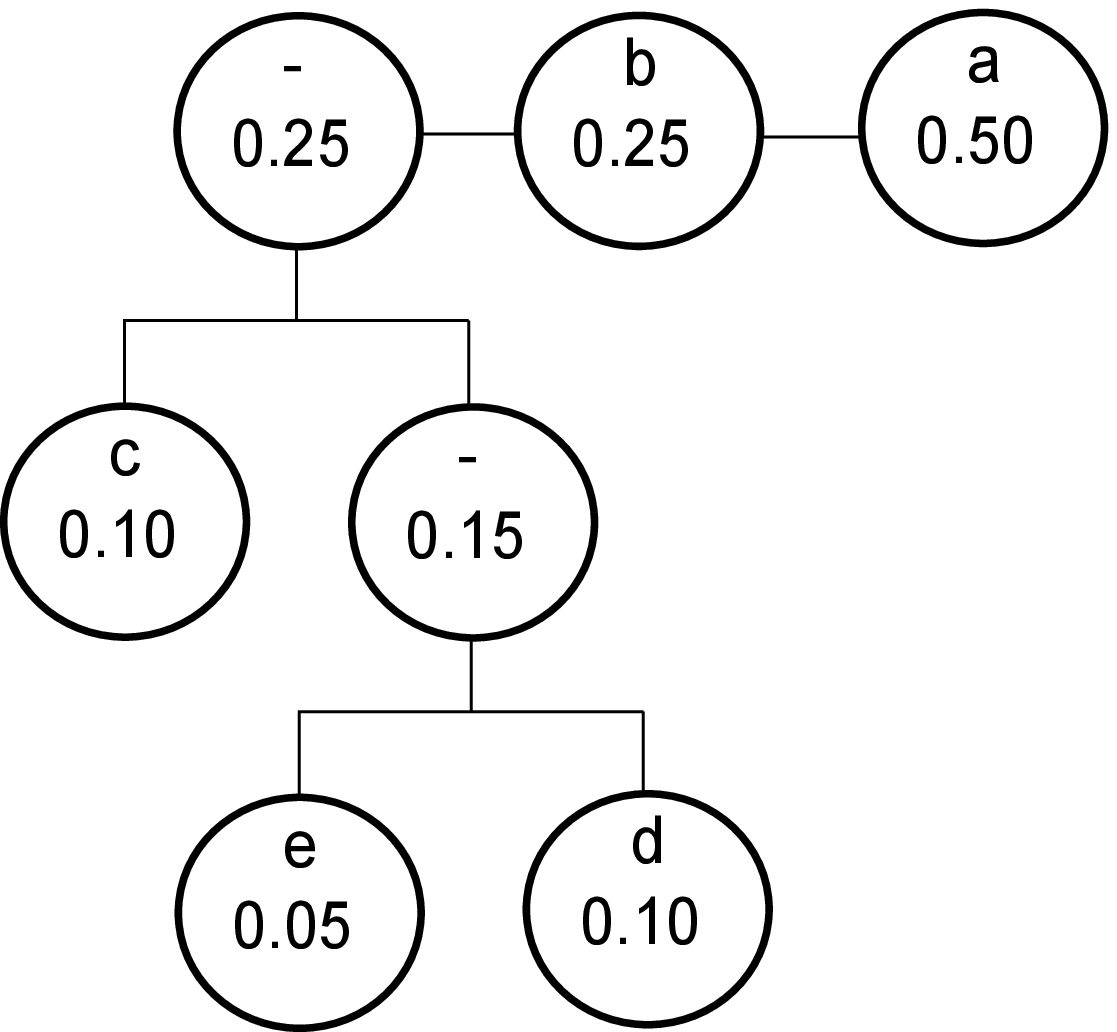}} \\
 & \\
 \texttt{~~~~~~~~4th step} & \texttt{~~~~~~~~5th step} \\
\imagetop{\includegraphics[width=4.5cm]{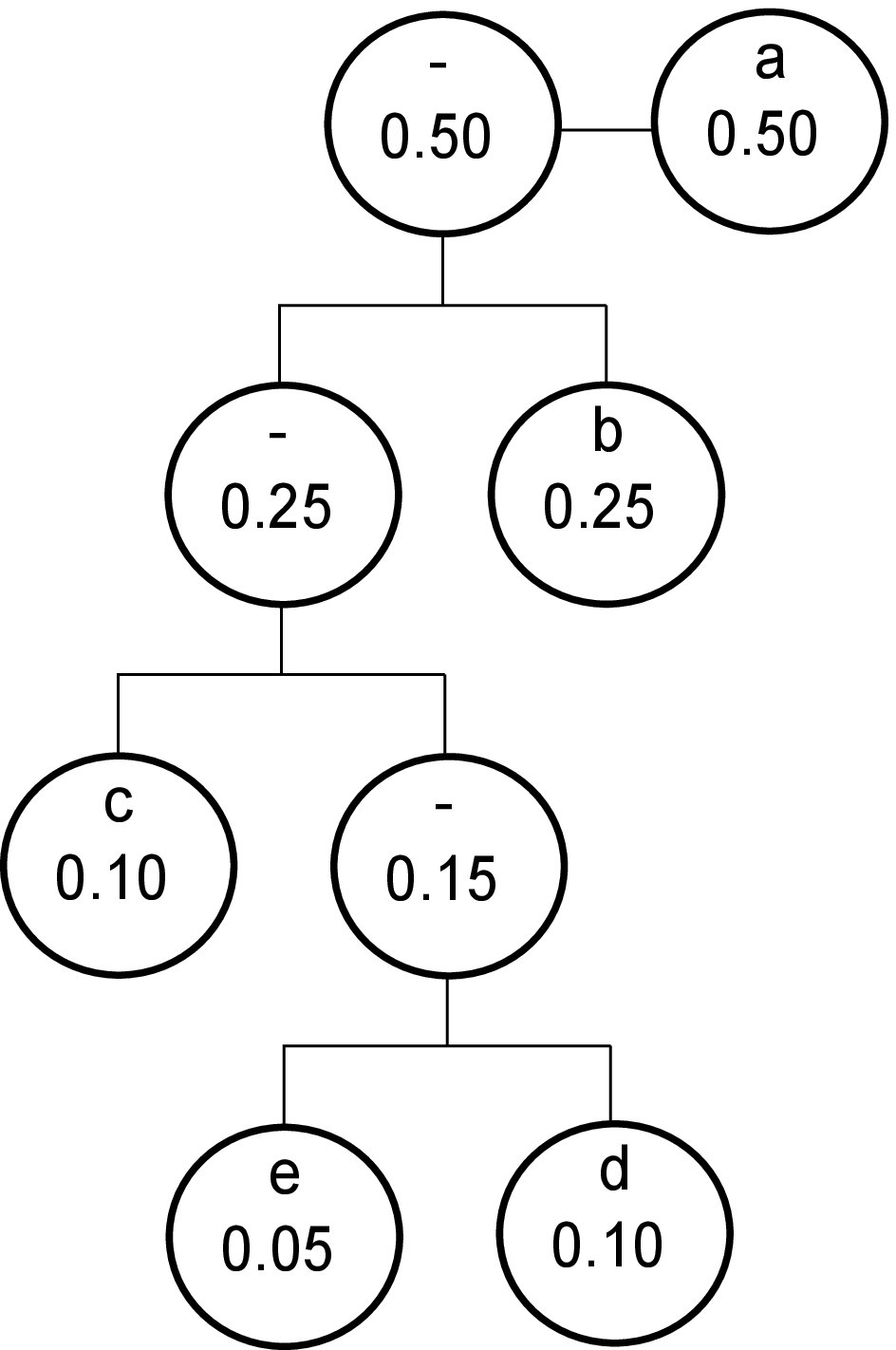}} & \imagetop{\includegraphics[width=4.5cm]{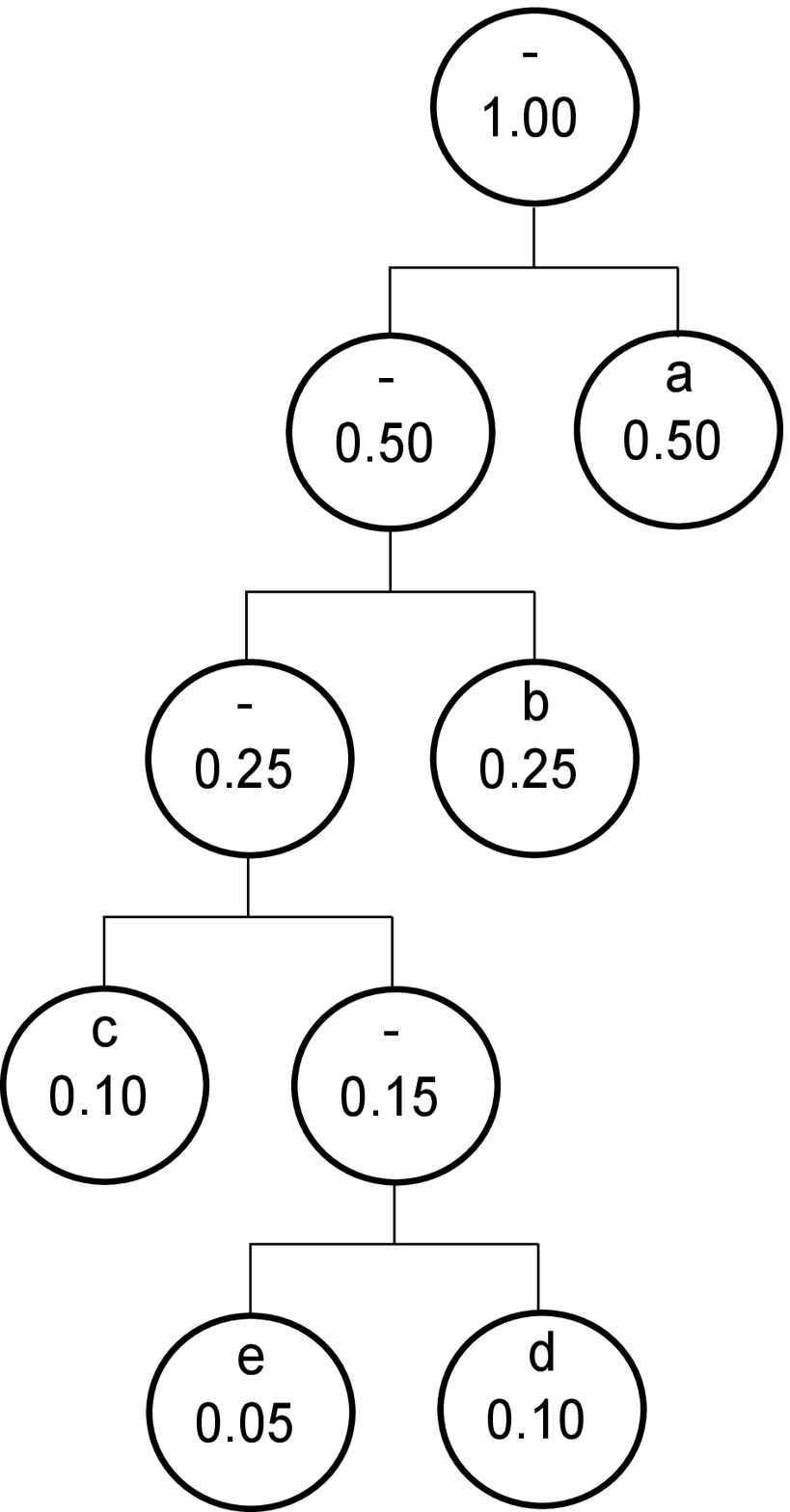}}   \\
\end{tabular}
\caption[Huffman tree generation.]{Huffman tree generation. The tree
is constructed from the list of nodes shown in the 1st step. That
list contains the alphabet symbols and their frequencies sorted in
increasing order of frequency. In each step, the list is updated by
removing its first two nodes, and then by inserting a new node that
has these nodes as children.} \label{STATE. Fig:huffman}
\end{figure*}

This binary tree is then used to assign the codes to the symbols by
traversing the tree from the root node to the leaf that contains the
symbol that is being coded. Since the tree is binary, there are two
possibilities of going from one node to the next one in the tree
traversal process. One is going through the left child of the node,
while the other is going through the right one. The coding process
assigns a different bit in every step depending on the edge used to
go from one level to the next. This implies that the Huffman coding
results in a prefix code, due to the fact that the bit string
representing some particular symbol is never a prefix of the bit
string representing any other symbol.

\subsubsection{PPM}

The PPM algorithm, whose name stands for \emph{Prediction with
Partial string Matching}, is an adaptive statistical data
compression technique based on an encoder that maintains a
statistical model of the text. It was originally developed by John
Clearly and Ian Witten \cite{Cleary84}, with extensions and an
implementation by Alistair Moffat \cite{Moffat90}.

There can be many statistical models depending on the way the input
data is treated. Thus, statistical models can take into account
separated symbols or groups of contiguous symbols. While the former
do not consider the context of the symbols because they treat them
separately, the latter do consider it because they take into account
the preceding symbols of each symbol. Because of that, they receive
the name of \emph{context-based} statistical models.

Depending on whether the probabilities are fixed or dynamic, that
is, updated as more data is being input, the modeler would be
\emph{static} or \emph{dynamic} -also known as \emph{adaptive}-. The
latter are more suitable because they adapt to the particularities
of the data contained in the file being compressed.

Although in principle, it can seem logical that a long context is
better than a small one because the longer retains information about
the nature of old data, experience shows that large data files
contain different distributions of symbols in different parts. Thus,
better compression can be achieved if the model takes into account
contexts of about 10 symbols \cite{Salomon2004}.

In general, an order-N adaptive context-based modeler considers the
N symbols preceding the symbol being processed. Although this
approach may sound good, there is a problem with it. The drawback is
that considering only order-N contexts can lead to no compression in
spite of the existence of smaller order instances which could be
used to compress the data. That is, when the encoder does not find
any order-N instance of a given symbol, it simply writes the symbol
on the compressed stream as a literal. However, the data could be
compressed using smaller contexts. The PPM method solves this
problem by switching to shorter contexts if necessary. Thus, the PPM
method uses smaller and smaller parts of the context in order to
achieve a better compression.

PPM uses sophisticated data structures and it usually achieves the
best performance of any real compressor although it is also usually
the slowest and most memory intensive \cite{Cilibrasi05}. One of the
data structures that can be used to implement the PPM algorithm is a
special type of tree called \emph{trie}.

Level 1 of a \emph{trie} contains the order-1 contexts, which means
that it contains one node for each symbol read so far. Level 2
contains all the order-2 contexts, and so on. In a \emph{trie}, each
context can be found by traversing the tree from the root to one of
the leaves.

\begin{figure*}
\begin{tabular}{ccc}
 1. b & 2. a & 3. n \\
\imagetop{\includegraphics[width=4cm]{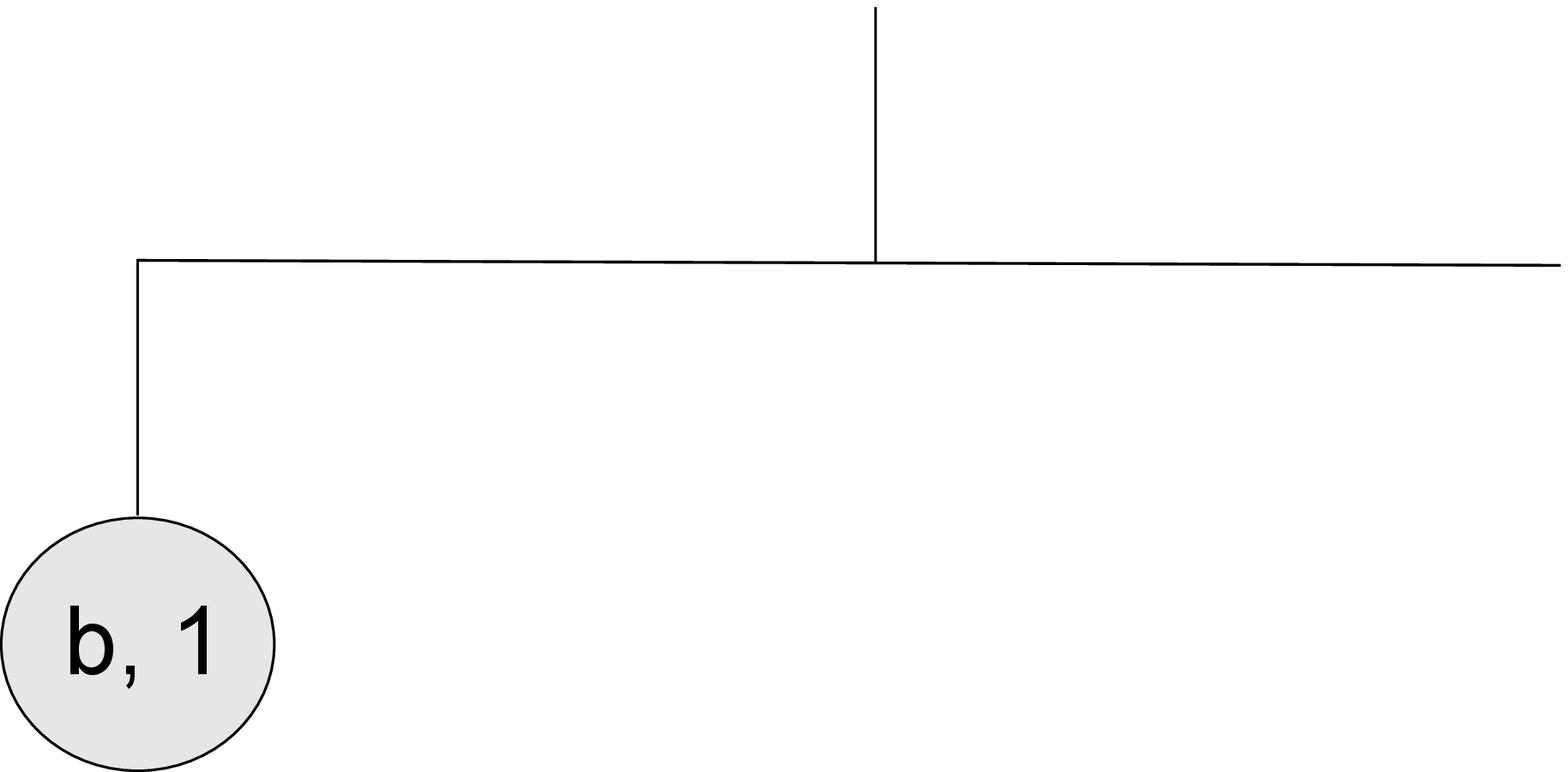}} & \imagetop{\includegraphics[width=4cm]{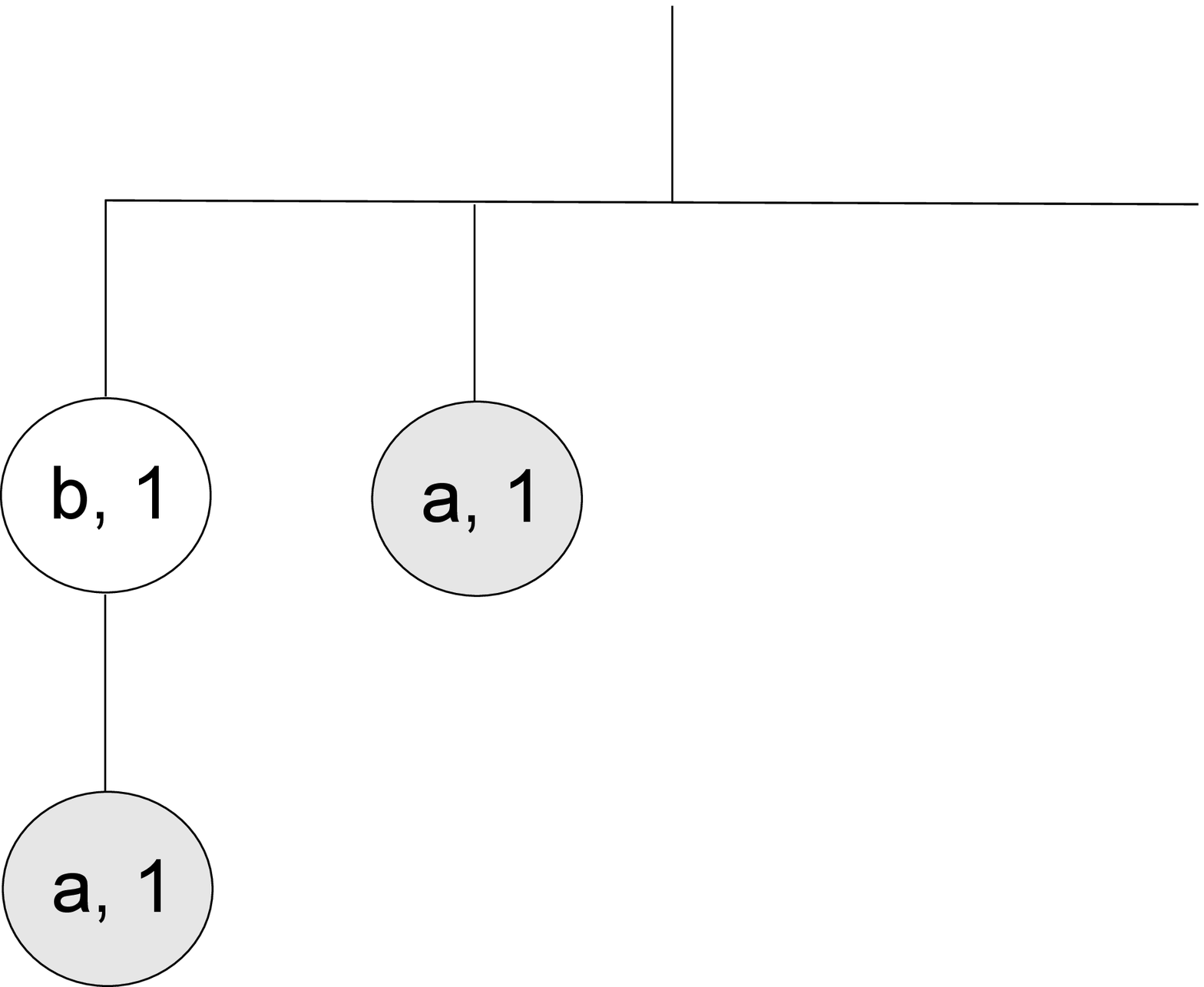}} & \imagetop{\includegraphics[width=4cm]{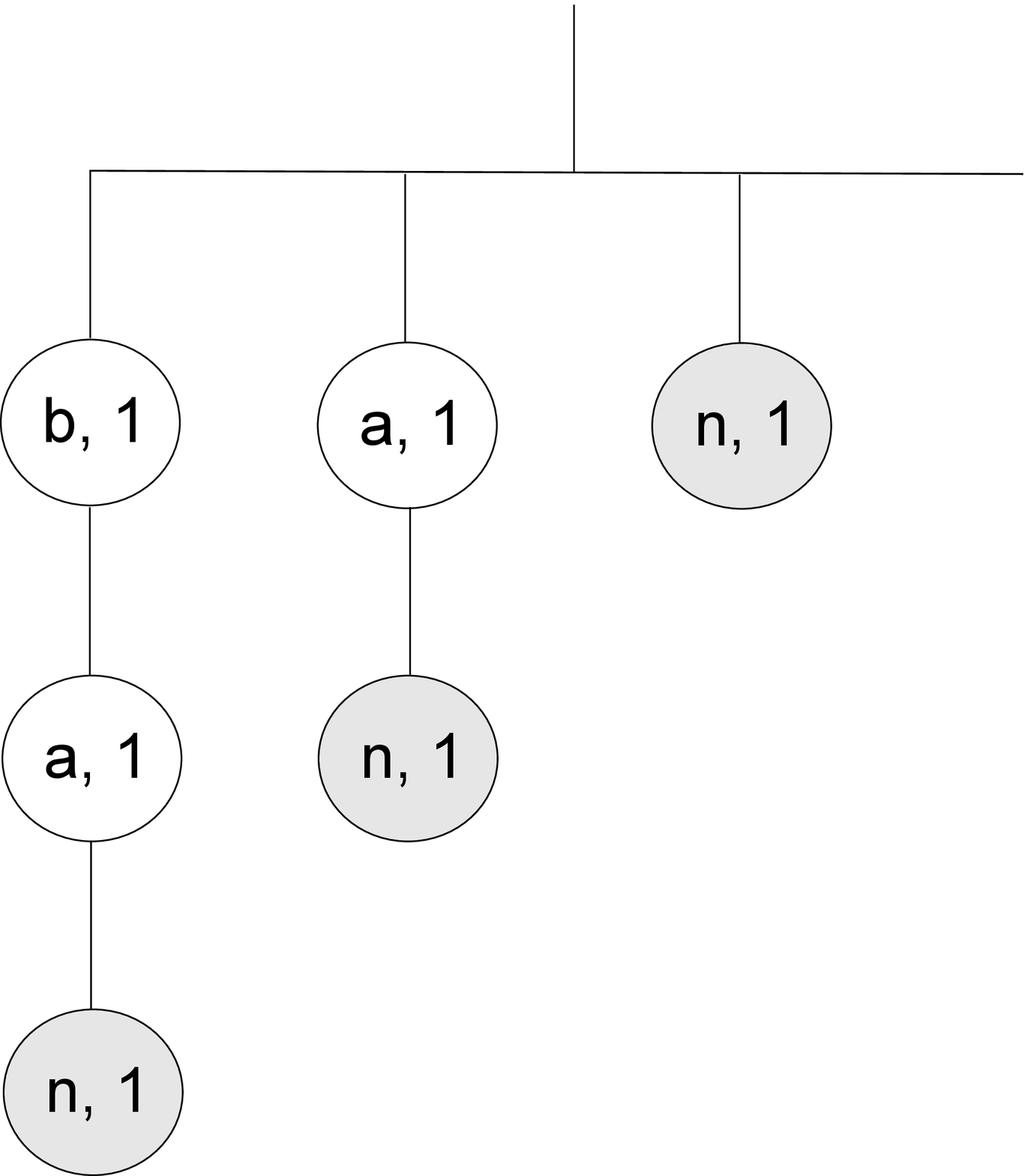}}  \\
 & & \\
 4. a & 5. n & 6. a \\
\includegraphics[width=4cm]{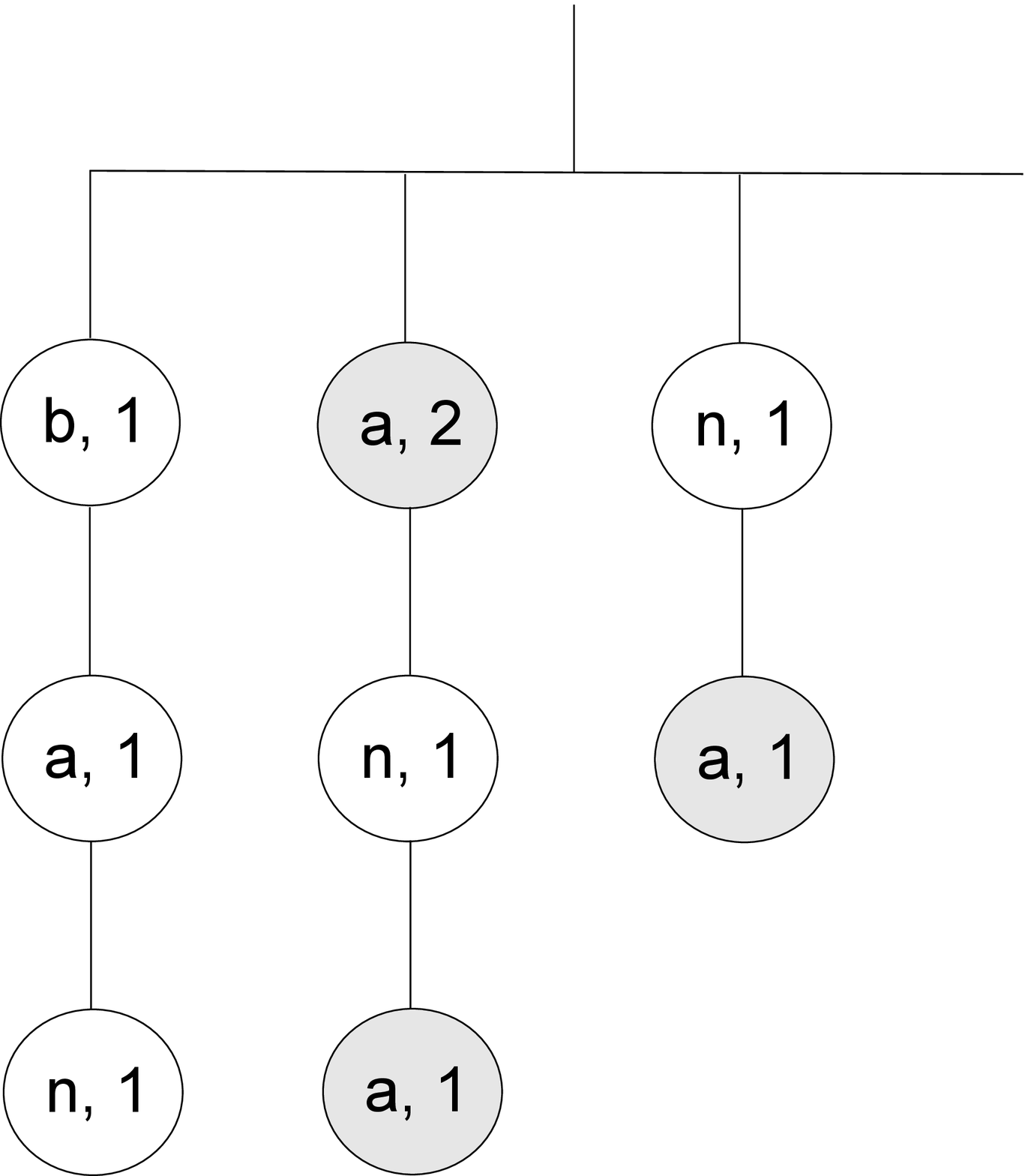} & \includegraphics[width=4cm]{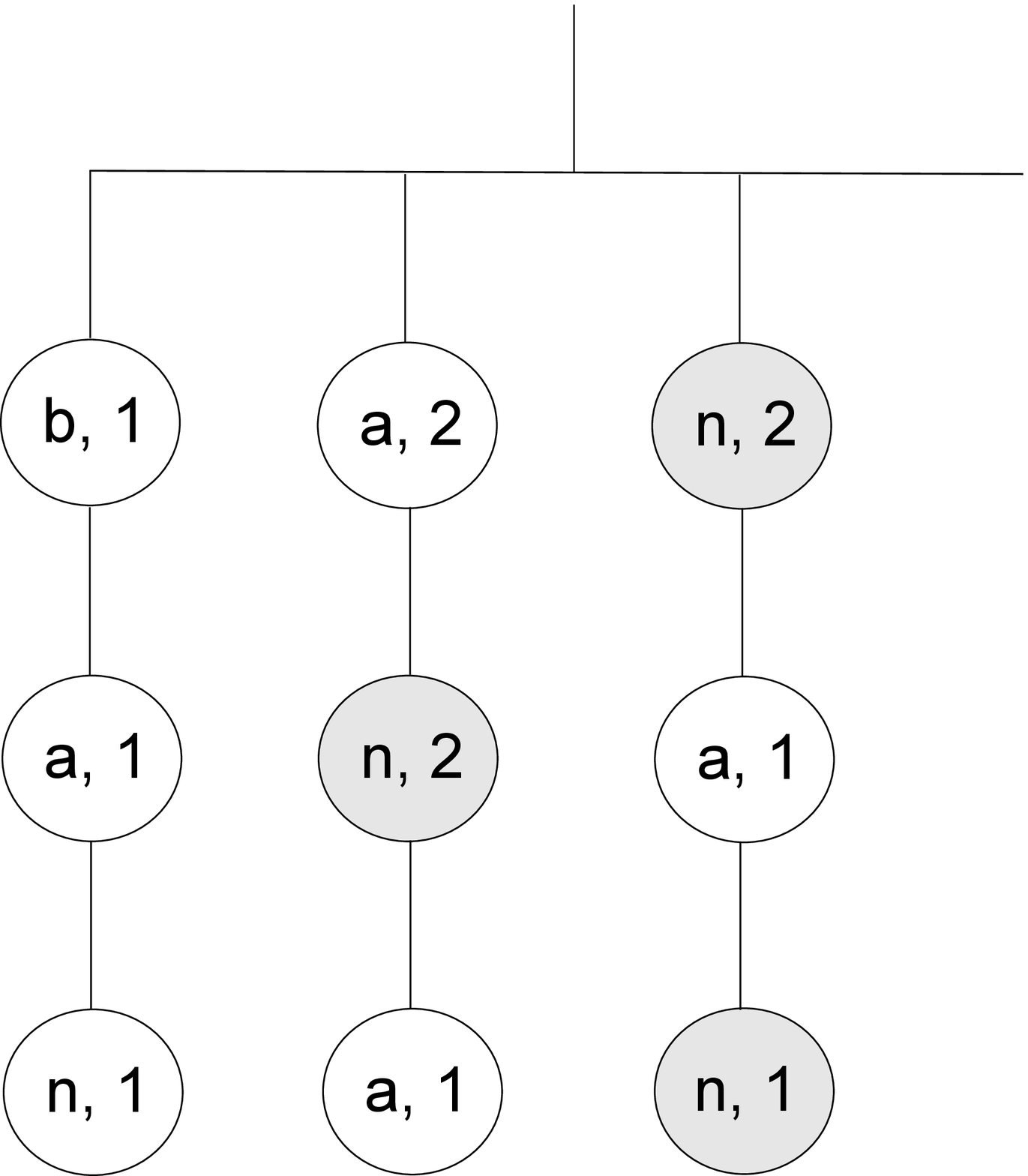} & \includegraphics[width=4cm]{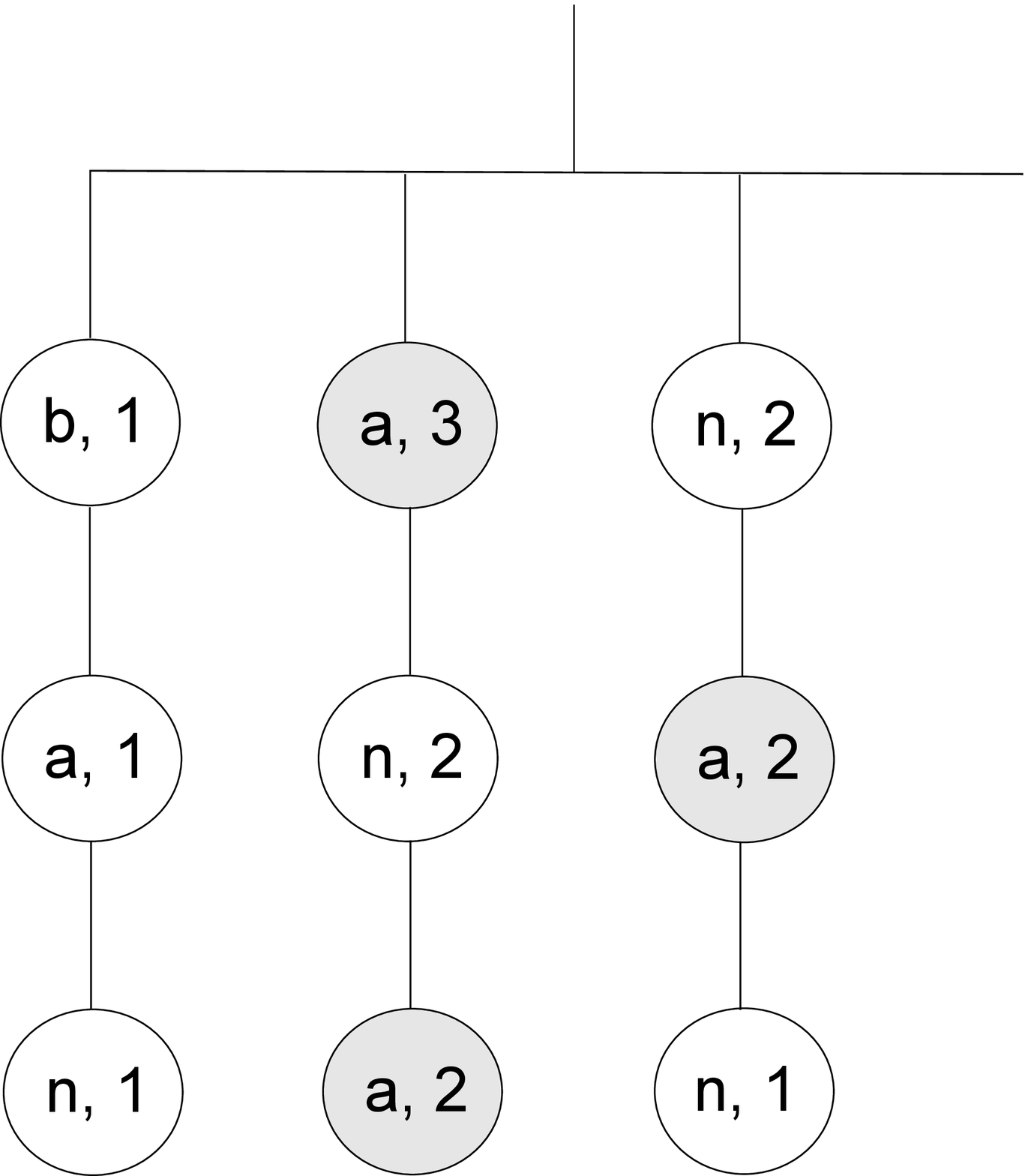} \\
 & & \\
 7. s & & \\
\includegraphics[width=4cm]{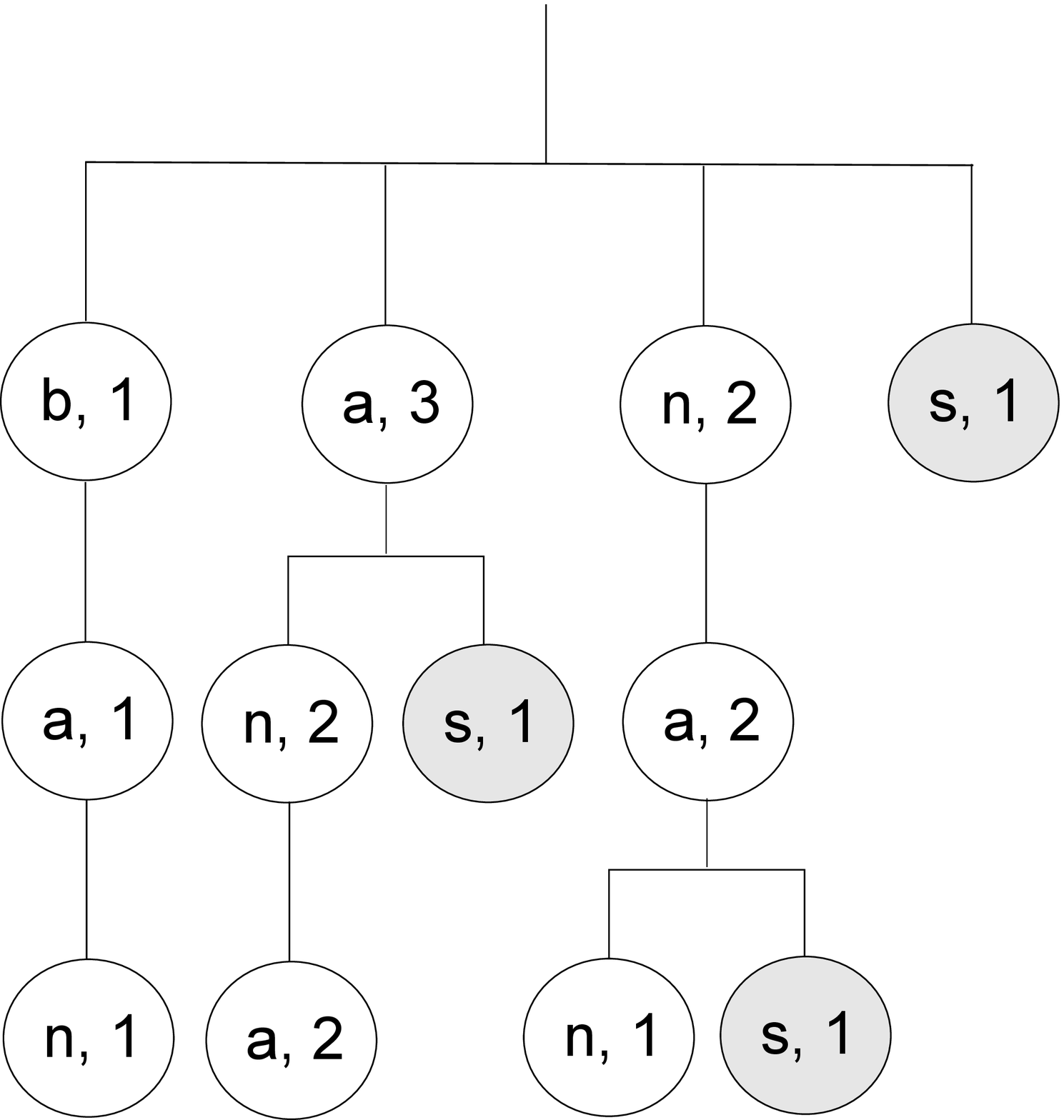} &  & \\
\end{tabular}
\caption[PPM: seven tries of ``bananas'' for a context of N =
2.]{PPM: seven \emph{tries} of ``bananas'' for a context of N = 2.
The characters are processed first to last. The numbers in the nodes
are context counts. Three nodes are involved in each step, except
the first two steps when the \emph{trie} has not yet reached its
final height. The nodes involved are shaded to ease the
understanding of the figure. For example, after reading the first
``n'', the tree is updated by adding three nodes. The node ``n,1'',
on level 1, means that the character ``n'' has occurred once, so
far. The node ``n,1'', on level 2, means that the sequence ``an''
has occurred once. Finally, the node ``n,1'', on level 3, means that
the sequence ``ban'' has occurred once.} \label{STATE. Fig.PPM}
\end{figure*}

Fig \ref{STATE. Fig.PPM} illustrates an example that helps to
understand the process of creation of a \emph{trie} and the meaning
of the nodes contained in it. The figure shows the seven steps
needed to construct the \emph{trie} for the string ``bananas'',
assuming N = 2. Note that the tree grows in width but not in depth.
In fact, it can be observed that its depth remains N + 1 regardless
of how many characters have been read.

The characters in the string are processed first to last, one at a
time. All the intermediate \emph{tries} shown in the figure
illustrate the state of the \emph{trie} after processing each
character. The numbers in the nodes are context counts. Notice that
three nodes are involved in each step, except the first two steps
when the \emph{trie} has not yet reached its final height. All the
nodes involved in each step are shaded to ease the understanding of
the figure.

The first tree contains only one node because only one character
(``b'') has been processed so far. The label ``b,1'' on the node
means that the ``b'' has occurred only once until that moment.

After reading the next symbol of the string (``a''), the tree is
updated by adding two nodes. The ``a,1'' on level 1 means that the
character ``a'' has occurred only once. The ``a,1'' that is on level
2, under the ``b,1'', means that the substring ``ba'' has occurred
only once.

After reading the next symbol (``n''), the tree is updated by adding
three nodes, one at each context:

\vspace{-0.1cm}
\begin{itemize}
\item
The node ``n,1'' on level 1 means that the character ``n'' has
occurred once.

\vspace{-0.1cm}

\item
The node ``n,1'' on level 2 means that the substring ``an'' has
occurred once.

\vspace{-0.1cm}

\item
Finally, the node ``n,1'' on level 3 means that the substring
``ban'' has been seen once.
\end{itemize}

\clearpage

The last \emph{trie} of Fig \ref{STATE. Fig.PPM}, that is, the 7th
one, can be analyzed to see the contexts that correspond to the
string ``bananas''. For example:

\begin{itemize}
\item
The ``a,3'' on level 1 of the tree means that the ``a'' occurs 3
times in the string ``bananas'':

\begin{itemize}
  \item b\underline{a}nanas
  \item ban\underline{a}nas
  \item banan\underline{a}s
\end{itemize}

\item
The ``n,2'' and ``s,1'' below it mean that these three occurrences
of ``a'' were followed by ``n'' twice, and by ``s'' once:

\begin{itemize}
  \item b\underline{an}anas
  \item ban\underline{an}as
  \item banan\underline{as}
\end{itemize}

\item
These two occurrences of ``an'', were followed always by ``a'', as
the node ``a,2'' on level 3 indicates.

\begin{itemize}
  \item b\underline{ana}nas
  \item ban\underline{ana}s
\end{itemize}

\end{itemize}

Many variants of the PPM algorithm have been implemented
\cite{Salomon2004}: PPMA, PPMB, PPMP, PPMX, PPMZ. However, the bases
of the method are always the ones explained above.

In this thesis, the variant called PPMZ is used. The PPMZ algorithm,
implemented by Charles Bloom \cite{Bloom98}, tries to improve the
PPM performance by handling features such as deterministic contexts,
unbounded-length contexts, and local order estimation, in an optimal
way \cite{Salomon2004}. Implementation details are difficult to
understand due to the code being very obscure. However, since PPMZ
belongs to the family of PPM algorithms, the basis of it are the
ones explained above.

\subsection{Dictionary Methods}

Dictionary compressors break the text into fragments that are saved
in a data structure called \emph{dictionary}. When a fragment of new
text is found to be identical to one of the dictionary entries, a
pointer to that entry is written on the compressed stream.

The simplest example of a dictionary compressor can be one that uses
an English dictionary to compress English texts, by coding each word
as its index in the dictionary, or by writing the word into the
output stream when the word is not found in the dictionary.
Obviously, this kind of approach is not a good choice for a
general-purpose compressor since the words contained in the
dictionary do not depend on the input.

The most famous dictionary compressors are the ones that belong to
the Lempel-Ziv family \cite{Salomon2004}. The origin of this
family of compressors is the LZ77, also known as LZ1, and the
LZ78, also known as LZ2, which were developed by Jacob Ziv and
Abraham Lempel \cite{Ziv78,Ziv77}.

\subsubsection{LZ77}

This algorithm uses as dictionary part of the input stream
previously seen. The method is based on a sliding window that the
encoder shifts as the strings of symbols are being encoded. That is
the reason why sometimes this method is called \emph{sliding
window}.

The window is divided into two parts, the first part, called the
\emph{search buffer}, is the current dictionary, while the second
part, called the \emph{look-ahead buffer} contains the text yet to
be encoded. It is important to point out that practical
implementations of this method use really long \emph{search buffers}
of thousands of bytes long, and small \emph{look-ahead buffers} of
tens of bytes long \cite {Salomon2004}.

The encoding algorithm works as follows:
\begin{enumerate}
\item It scans the \emph{search buffer} backwards looking for a
match to the first symbol in the \emph{look-ahead buffer}.

\item
Then, it calculates the length of the match by comparing the symbols
following the symbol found.

\item
After that, it keeps doing this in order to find longer matches.

\item
After the search process, it selects the longest one, or the last
one found in the event of a tie. This is done this way to avoid
having to memorize previously found matches.

\item Finally a token with three parts -offset, length of match,
and first symbol in the \emph{look-ahead buffer}- is written on the
output in this way:

\begin{enumerate}
\item If the backward search yields no match, a token with zero
offset, zero length, and the unmatched symbol is written on the
output. Then, the window is shifted to the right one position.

\item If there is a match, a token with the offset, the length of
match, and the symbol that follows the matched sequence in the
\emph{look-ahead buffer} is written on the output. Then, the window
is shifted to the right $L + 1$ positions, $L$ being the length of
match.
\end{enumerate}
\end{enumerate}

To sum up, the LZ77 encodes the input by generating tokens with
three parts: offset, length and next symbol in the \emph{look-ahead
buffer}. Table \ref{Table.LZ77} shows an example that helps to
understand the algorithm. It shows the evolution of the \emph{search
buffer} and the \emph{look-ahead buffer} for the input data
``the-abbess-and-the-abbot-are-in-the-abbey''.

Let us analyze some steps of the process to ease the understanding
of the encoding algorithm. Since the \emph{search buffer} is empty
at the beginning of the process, the first token is (0,0,'t')
because the backward search yields no match, and the unmatched
symbol is the character 't'. In fact, the first six tokens have an
offset and a length of 0 because the first six characters of the
input data are different.

After processing the first six characters, there is a match of
offset 1 and length 1 because the last character in the
\emph{search buffer} and the first character in the
\emph{look-ahead buffer} are the same ('b'):

\begin{itemize}
 \item search buffer: ``the-a\underline{b}''
 \item look-ahead buffer:
 ``\underline{b}ess-and-the-abbot-are-in-the-abbey''
\end{itemize}

This explains why the seventh token written on the output is
(1,1,'e'). Note that including the character 'b' in the token is
not necessary because it is implicitly included thanks to the
offset and the length of 1.

A more interesting circumstance occurs after processing the first 16
characters of the input. It is easy to see that at that point, there
is a match of length 6 at a distance of 15, as can be observed
looking at the content of the buffers:

\begin{itemize}
 \item search buffer: ``t\underline{he-abb}ess-and-t''
 \item look-ahead buffer: ``\underline{he-abb}ot-are-in-the-abbey''
\end{itemize}

This explains why the thirteenth token written on the output is
(15,6,'o').

\begin{table}
\caption[LZ77: Lempel-Ziv sliding window.]{LZ77: Lempel-Ziv sliding
window. The algorithm scans the \emph{search buffer} backwards
looking for a match to the first symbol in the \emph{look-ahead
buffer}. It keeps doing this in order to find the longest match.
Then, it selects the longest one, or the last one found in the event
of a tie. Finally, a token with the offset, the length of match, and
the first symbol in the \emph{look-ahead buffer} is written on the
output.} \label{Table.LZ77} \vspace{0.15cm} \scriptsize{
\begin{tabular}{|r|l|c|}
  \hline
  \textbf{Search buffer}~~~~~~~~~~~~~~~~~   & ~~~~~~~~~~~~~~~\textbf{Look-ahead buffer}    & \textbf{Token} \\
  \hline
                                               & the-abbess-and-the-abbot-are-in-the-abbey    & (0,0,'t') \\
  t                                            & he-abbess-and-the-abbot-are-in-the-abbey     & (0,0,'h') \\
  th                                           & e-abbess-and-the-abbot-are-in-the-abbey      & (0,0,'e') \\
  the                                          & -abbess-and-the-abbot-are-in-the-abbey       & (0,0,'-') \\
  the-                                         & abbess-and-the-abbot-are-in-the-abbey        & (0,0,'a') \\
  the-a                                        & bbess-and-the-abbot-are-in-the-abbey         & (0,0,'b') \\
  the-a\underline{b}                           & \underline{b}ess-and-the-abbot-are-in-the-abbey & (1,1,'e') \\
  the-abbe                                     & ss-and-the-abbot-are-in-the-abbey            & (0,0,'s') \\
  the-abbe\underline{s}                        & \underline{s}-and-the-abbot-are-in-the-abbey & (1,1,'-') \\
  the-\underline{a}bbess-                      & \underline{a}nd-the-abbot-are-in-the-abbey   & (7,1,'n') \\
  the-abbess-an                                & d-the-abbot-are-in-the-abbey                 & (0,0,'d') \\
  the\underline{-}abbess-and                   & \underline{-}the-abbot-are-in-the-abbey      & (11,1,'t') \\
  t\underline{he-abb}ess-and-t                 & \underline{he-abb}ot-are-in-the-abbey        & (15,6,'o') \\
  \underline{t}he-abbess-and-the-abbo          & \underline{t}-are-in-the-abbey               & (23,1,'-') \\
  the-\underline{a}bbess-and-the-abbot-        & \underline{a}re-in-the-abbey                 & (21,1,'r') \\
  th\underline{e-}abbess-and-the-abbot-ar      & \underline{e-}in-the-abbey                   & (25,2,'i') \\
  the-abbess-a\underline{n}d-the-abbot-are-i   & \underline{n}-the-abbey                      & (18,1,'-') \\
  \underline{the-abbe}ss-and-the-abbot-are-in- & \underline{the-abbe}y                        & (32,8,'y') \\
  the-abbess-and-the-abbot-are-in-the-abbey    &                                              &   \\
  \hline
\end{tabular}}
\end{table}

In this thesis, the Lempel-Ziv-Markov chain algorithm LZMA
\cite{lzmax}, created by Igor Pavlov, is used. This is a compression
algorithm that uses a variant of the LZ77 to encode the input, and
then uses a range encoder to encode the output obtained by the LZ77.

Range encoding is a data compression technique created by G. Nigel
N. Martin \cite{Martin79} that encodes all the symbols of the
message into one number using a probability estimation.

The LZMA produces a stream of literal symbols and phrase references,
which is encoded one bit at a time by the range encoder, using a
model to make a probability prediction of each bit. This gives much
better compression because it avoids mixing unrelated bits together
in the same context. In fact, empirical evidence shows that it
performs very well on structured data and it looks very much like
any other LZ algorithm. However, it trounces them all
\cite{lzma-bloom}.

\subsection{Other Methods}

Some modern context-based text compression methods perform a
transformation on the input data and then apply a statistical model
to assign probabilities to the transformed symbols.

\subsubsection{BZIP2}

BZIP2 is a block-sorting compressor developed by Julian Seward
\cite{bzip2}. BZIP2 compresses data using Run Length Encoding, the
Burrows-Wheeler Transform, the Move-To-Front transform and Huffman
coding.

The algorithm reads the input stream block by block and each block
is compressed separately as one string. The length of the blocks is
between 100 and 900 KB. The compressor uses the Burrows-Wheeler
Transform to convert frequently-recurring character sequences into
strings of identical letters, and then it applies Move-To-Front
transform and Huffman coding. All these methods are explained later
so the basis of the BZIP2 compressor can be understood.

\subsubsection{Run Length Encoding}

Run Length Encoding -RLE- is a very simple form of data compression
in which, if a data item \emph{d} occurs \emph{n} consecutive times
in the input stream, the occurrences are replaced with the single
pair \emph{nd}. The sequences in which the same data value occurs in
many consecutive data elements are called a \emph{run length} of
\emph{n}.

The main problem with this method is that, in plain English texts,
there are many sequences of two equal symbols but a sequence of
three is rare. However, this method can be combined with other
methods to process the text before RLE so the new text
representation is more suitable to achieve bigger compression rates.
This is precisely what the BZIP2 compression algorithm does, because
it applies RLE after applying the Move-To-Front transform.

\subsubsection{Move-to-Front}

The Move-To-Front transform -MTF- \cite{Bentley86,Ryabko80} is an
encoding of data usually used as an extra step in data compression
algorithms, such as for example BZIP2. Table \ref{Table.MTF
transform} shows an example that helps to understand how the MTF
transform works.

The method transforms the data into a sequence of integers in the
following manner. It maintains a list that stores the symbols of the
alphabet in such a way that the most frequent ones are maintained
near the front. This is done by updating the list each time a symbol
is processed, moving it to the front. Then, a symbol is encoded as
the number of symbols that precede it in the list, or in other
words, it is encoded as its index in the list, 0 being the index of
the first element.

This implies that long sequences of identical symbols are replaced
by many zeros, and frequently used symbols are coded with small
numbers. The MTF transform takes advantage of local correlation of
frequencies to reduce the entropy of a message. In other words, when
the characters exhibit local correlations, the sequence of integers
will contain small numbers \cite{Salomon2004}.

The MTF transform is used in the Burrows-Wheeler Transform, because
the latter is very good at producing a sequence that exhibits local
frequency correlation from text.

\begin{table}[h]
\caption[Move-To-Front transform.]{Move-To-Front transform. The
algorithm transforms the data into a sequence of integers. It
maintains a list of the symbols of the alphabet in such a way that
the most frequent ones are maintained near the front. In order to do
so, the list is updated every time a symbol is processed, moving it
to the front. A symbol is encoded as the number of symbols that
precede it in the list.} \label{Table.MTF transform} \vspace{0.25cm}
\centering\normalsize{
\begin{tabular}{|c|l|c|c|}
  \hline
  \textbf{Iteration}    & \textbf{Output}            &  \textbf{List}               \\
  \hline
  pebblepebble          &                            & (abcdefghijklmnopqrstuvwxyz) \\
  \hline
  \underline{p}ebblepebble & 15                         & (pabcdefghijklmnoqrstuvwxyz) \\
  p\underline{e}bblepebble & 15,5                       & (epabcdfghijklmnoqrstuvwxyz) \\
  pe\underline{b}blepebble & 15,5,3                     & (bepacdfghijklmnoqrstuvwxyz) \\
  peb\underline{b}lepebble & 15,5,3,0                   & (bepacdfghijklmnoqrstuvwxyz) \\
  pebb\underline{l}epebble & 15,5,3,0,12                & (lbepacdfghijkmnoqrstuvwxyz) \\
  pebbl\underline{e}pebble & 15,5,3,0,12,2              & (elbpacdfghijkmnoqrstuvwxyz) \\
  pebble\underline{p}ebble & 15,5,3,0,12,2,3            & (pelbacdfghijkmnoqrstuvwxyz) \\
  pebblep\underline{e}bble & 15,5,3,0,12,2,3,1          & (eplbacdfghijkmnoqrstuvwxyz) \\
  pebblepe\underline{b}ble & 15,5,3,0,12,2,3,1,3        & (beplacdfghijkmnoqrstuvwxyz) \\
  pebblepeb\underline{b}le & 15,5,3,0,12,2,3,1,3,0      & (beplacdfghijkmnoqrstuvwxyz) \\
  pebblepebb\underline{l}e & 15,5,3,0,12,2,3,1,3,0,3    & (lbepacdfghijkmnoqrstuvwxyz) \\
  pebblepebbl\underline{e} & 15,5,3,0,12,2,3,1,3,0,3,2  & (elbpacdfghijkmnoqrstuvwxyz) \\
  \hline
\end{tabular}}
\end{table}

\subsubsection{Burrows-Wheeler Transform}

The Burrows-Wheeler Transform -BWT- is an algorithm created by
Michael Burrows and David Wheeler \cite{Burrows94} that is applied
by the BZIP2 compressor.

BWT permutes the order of the characters of the string being
transformed with the purpose of bringing repetitions of the
characters closer. This is useful for compression, since there are
techniques such as MTF and RLE that work very well when the input
string contains runs of repeated characters.

Although in practice the BWT implementation is more complex than the
algorithm explained below, this version can be more easily
understood while keeping the same philosophy as the complex one.

Table \ref{Table.BWT encoding} shows how the algorithm works when it
is used to encode the string ``sentence''.

The algorithm works as follows:

\begin{enumerate}
\item
The encoder creates an $n$ x $n$ matrix. It stores the string to
code in the first row. The rest of the rows contain $n - 1$ copies
of the said string, each cyclically shifted one symbol to the left.

\item
Then the matrix is sorted lexicographically by rows.

\begin{itemize}
\item
Notice that the last character of a row is always the one that
precedes the first character in that row.

\item
Notice too, that every row and every column of the matrix is a
permutation of the string being transformed.
\end{itemize}

\item
Finally, the last column of the sorted matrix is taken as the
transformed version of the input string.
\end{enumerate}

Applying this algorithm creates more easily compressible data,
because sorting the rotations of the string tends to create regions
that concentrate just a few symbols. However, the BWT works well
only if the length of the string is large -at least several thousand
symbols per string- \cite{Salomon2004}.

\begin{table}[h]
\centering \caption[Burrows-Wheeler Transform
encoding.]{Burrows-Wheeler Transform encoding. The algorithm stores
the string to code in the first row of the cyclically shifted
matrix. The rest of the rows contain $n - 1$ copies of the said
string, each cyclically shifted one symbol to the left. Then, the
lexicographically sorted matrix is created by sorting the said
matrix by rows. Finally, the last column of the sorted matrix is
taken as the transformed version of the input string.}
\label{Table.BWT encoding} \vspace{0.5cm} \normalsize{
\begin{tabular}{|c|c|}
 \hline
 \textbf{Cyclically shifted} & \textbf{Lexicographically sorted} \\
 \hline
 s e n t e n c e    &   c e s e n t e \underline{n}    \\
 e n t e n c e s    &   e n c e s e n \underline{t}    \\
 n t e n c e s e    &   e n t e n c e \underline{s}    \\
 t e n c e s e n    &   e s e n t e n \underline{c}    \\
 e n c e s e n t    &   n c e s e n t \underline{e}    \\
 n c e s e n t e    &   n t e n c e s \underline{e}    \\
 c e s e n t e n    &   s e n t e n c \underline{e}    \\
 e s e n t e n c    &   t e n c e s e \underline{n}    \\
 \hline
\end{tabular}}
\end{table}

The only information needed to reconstruct the original string from
the last column, is the row number of the original string in the
lexicographically sorted matrix. Thus, the decoding process works
thanks to these facts:

\begin{enumerate}
\item
The encoded string, contains all the characters in the text.
Therefore, it can be used to get the first column of the
lexicographically sorted matrix by simply sorting the encoded
string.

\item
Since the last character of a row is always the one that precedes
the first character in that row, and given that the first and the
last column of the matrix are held, both columns can be used to
obtain all pairs of successive characters in the original string,
where pairs are taken cyclically so that the last and first
character form a pair.

\item
After reconstructing the lexicographically sorted matrix, the
original string can be obtained from the row number of the original
string in the sorted matrix.
\end{enumerate}

\subsection{Comparing Compressors: Calgary Corpus}
\label{Comparing Compressors: Calgary Corpus}

The Calgary Corpus is a collection of 14 text and binary data files,
commonly used for comparing data compression algorithms. The corpus
was founded in 1987 by Timothy Bell, Ian Witten, and John Cleary at
the University of Calgary for their research paper \cite{Bell87}.

Table \ref{Table.Calgary Corpus} shows the detailed description of
the files from the Calgary Corpus. Table \ref{Table.Compression
Algorithms Performance Comparison} presents a comparison between
the compression algorithms used in this thesis, which are PPMZ,
LZMA and BZIP2. The results show that the compression ratio of the
PPMZ is the best, as can be observed by comparing the size of the
compressed files and the compression ratios, also called bit per
bit -bpb-.

\begin{table}
\centering \caption{Calgary Corpus.} \label{Table.Calgary Corpus}
\vspace{0.5cm} \normalsize{
\begin{tabular}{|l|l|l|}
 \hline
\textbf{File}    & \textbf{Category}        & \textbf{Size}   \\
 \hline
bib     & Bibliography                      & 111261 \\
book1   & Fiction book                      & 768771 \\
book2   & Non-fiction book                  & 610856 \\
geo     & Geophysical data                  & 102400 \\
news    & USENET batch file                 & 377109 \\
obj1    & Object code for VAX               & 21504 \\
obj2    & Object code for Apple Mac         & 246814 \\
paper1  & Technical paper                   & 53161 \\
paper2  & Technical paper                   & 82199 \\
pic     & Black and white fax picture       & 513216 \\
progc   & Source code in ``C''              & 39611 \\
progl   & Source code in LISP               & 71646 \\
progp   & Source code in PASCAL             & 49379 \\
trans   & Transcript of terminal session    & 93695 \\
 \hline
\end{tabular}}
\end{table}

\begin{table}
\centering \caption[Comparison of compression
algorithms.]{Comparison of compression algorithms. The size of the
compressed files and the compression ratios in bit per bit are shown
in the table.} \label{Table.Compression Algorithms Performance
Comparison} \vspace{0.5cm} \normalsize{
\begin{tabular}{|l|l|l|l|l|l|l|}
 \hline
 {\multirow{2}{*}{\textbf{File}}} & \multicolumn{2}{c|}{\textbf{PPMZ}} & \multicolumn{2}{c|}{\textbf{LZMA}} & \multicolumn{2}{c|}{\textbf{BZIP2}} \\
 \cline {2-7}
 & \textbf{size} & \textbf{bpb} & \textbf{size} & \textbf{bpb} & \textbf{size} & \textbf{bpb} \\
 \hline
bib     &   23873   &   1.717   &   30543   &   2.196   &   27467   &   1.975   \\
book1   &   210952  &   2.195   &   261032  &   2.716   &   232598  &   2.420   \\
book2   &   140932  &   1.846   &   169760  &   2.223   &   157443  &   2.062   \\
geo     &   52446   &   4.097   &   53319   &   4.166   &   56921   &   4.447   \\
news    &   103951  &   2.205   &   118846  &   2.521   &   118600  &   2.516   \\
obj1    &   9841    &   3.661   &   9381    &   3.490   &   10787   &   4.013   \\
obj2    &   69137   &   2.241   &   61460   &   1.992   &   76441   &   2.478   \\
paper1  &   14711   &   2.214   &   17233   &   2.593   &   16558   &   2.492   \\
paper2  &   22449   &   2.185   &   27183   &   2.646   &   25041   &   2.437   \\
pic     &   30814   &   0.480   &   41945   &   0.654   &   49759   &   0.776   \\
progc   &   11178   &   2.258   &   12516   &   2.528   &   12544   &   2.533   \\
progl   &   12938   &   1.445   &   14940   &   1.668   &   15579   &   1.740   \\
progp   &   8948    &   1.450   &   10307   &   1.670   &   10710   &   1.735   \\
trans   &   14224   &   1.214   &   16675   &   1.424   &   17899   &   1.528   \\
 \hline
\textbf{total} & 726400 &       &   845140  &           &   828347  &           \\
 \hline
\textbf{average}  &        & 2.086 & & 2.320 & & 2.368 \\
 \hline
\end{tabular}}
\end{table}

\section{Compression Distances}
\label{State. Compression Distances}

Compression distances are currently a hot topic of research in many
areas, such as document clustering
\cite{Granados11eswa,Granados10tkde,Granados08,Granados10ideal,Helmer07,Telles07},
document retrieval \cite{Granados11tkde,Martinez08},
question-answering
systems~\cite{ravichandran2001lst,Zhang07,Zhang08}, music
classification \cite{cilibrasi2004acm,GonzalezPardo10}, data mining
\cite{cilibrasi07}, neural networks~\cite{cuturi2005ctk}, security
of computer systems \cite{Apel09,Bertacchini07,Wehner07}, plagiarism
detection \cite{Chen04}, software metrics
\cite{allen2001mca,arbuckle2007sdc,scott-new}, bioinformatics
\cite{Ferragina07,Kocsor06,krasnogor2004msp,Nykter08}, chemistry
\cite{Melville07}, medicine \cite{Cohen09,Santos06},
philology~\cite{benedetto2002lta}, or even art \cite{Svangard04}.
This success relies on its parameter-free nature, wide
applicability, and leading efficacy in several domains.

Compression distances use compression algorithms to calculate the
similarity between two objects. Thus, they are benefiting from the
very mature and diverse research field on compression algorithms,
whose only target so far has been the detection and reduction of
redundancy in stored digital information.

The concepts of Kolmogorov complexity and conditional Kolmogorov
complexity have been combined to define a measure of similarity
between two strings, giving rise to the concept of \emph{Normalized
Information Distance} -NID- \cite{Li04}. The mathematical
formulation is as follows:

\begin{equation}
NID(x,y)=\frac{max\{K(x|y),K(y|x)\}}{max\{K(x),K(y)\}}
\end{equation}

NID can be used to express all other distances \cite{Li04}, but
unfortunately, since Kolmogorov complexity is non-computable, NID is
not computable either. However, compression algorithms can be used
to estimate an upper bound upon Kolmogorov complexity. Therefore,
they can be used to approximate the NID. In fact, the practical
application of that idea gave rise to the concept of
\emph{Normalized Compression Distance} -NCD- \cite{Cilibrasi05},
whose mathematical formulation is as follows:

\begin{equation}
NCD(x,y)=\frac{max\{C(xy)-C(x),C(yx)-C(y)\}}{max\{C(x),C(y)\}}
\end{equation}

Where:

\begin{description}
  \item[$C$] is a compression algorithm
  \item[$C(x)$] is the size of the compressed version of $x$
  \item[$C(y)$] is the size of the compressed version of $y$
  \item[$C(xy)$] is the compressed size of the concatenation of $x$ and $y$
  \item[$C(yx)$] is the compressed size of the concatenation of $y$ and $x$
\end{description}

In practice, the NCD is a non-negative number $0 \leq r \leq 1 +
\varepsilon$ representing how different the two objects are. Smaller
numbers represent more similar objects. The $\varepsilon$ in the
upper bound is due to imperfections in compression techniques, but
for most standard compression algorithms one is unlikely to see an
$\varepsilon$ above 0.1 \cite{Cilibrasi-Tesis}.

\subsection{Analyzing some extreme cases}

The NCD formula can be analyzed in some extreme cases. For example,
if the NCD is used to calculate the similarity between a document
and itself:

\begin{equation}
NCD(x,x)=\frac{max\{C(xx)-C(x),C(xx)-C(x)\}}{max\{C(x),C(x)\}} = 0
\end{equation}

\begin{description}
  \item[$C(xx) = C(x) \Rightarrow C(xx)-C(x) = 0$]
  \item[$\Rightarrow max\{C(xx)-C(x),C(xx)-C(x)\} = 0$]
  \item[$\Rightarrow NCD(x,x) = 0$.]
\end{description}

A problem that can arise if one of the objects is very big, and
the other is very small, is that the NCD can be close to 1 even
though the objects are about the same subject. The idea is the
following.

Let $L_b(x)$ be the length in bits of the object $x$. Then, if
$L_b(x) \gg L_b(y)$, and $L_b(y) \rightarrow 0$.

\begin{description}
  \item[$C(xy) \simeq C(x) \Rightarrow C(xy)-C(x) \simeq 0$]
  \item[$C(yx) \simeq C(x)$ and $C(y) \simeq 0 \Rightarrow C(yx)-C(y) \simeq C(x)$]
  \item[$\Rightarrow max\{C(xy)-C(x),C(yx)-C(y)\} \simeq C(x)$]
  \item[$\Rightarrow max\{C(x),C(y)\} \simeq C(x)$]
  \item[$\Rightarrow NCD(x,y) \simeq 1$.]
\end{description}

Of course, this is just an extreme case, but it illustrates how
the NCD can behave in some specific circumstances.

Although in many domains this issue is not an obstacle, it can be a
problem in those fields in which two very different sized objects
have to be compared. This is, for example, the case of a typical
document search scenario, because the size of the query and the size
of the documents to search can be very different. This drawback has
been addressed using document segmentation in
\cite{Granados11tkde,Martinez08,TFM-Rafa}. In fact, the experiments
presented in Chapter \ref{Chapter: Application to Document
Retrieval} use that NCD-based document search approach.

\subsection{Understanding NCD}

Tables \ref{Table.How the NCD works1} and \ref{Table.How the NCD
works2} clarify the way in which NCD works. Table \ref{Table.How the
NCD works1} shows four fragments of a document which are modified by
progressively replacing some words using random characters.

The first sample of text contains the original text, whereas the
rest of the samples contains the same fragment of text distorted by
replacing some words using random characters.

Table \ref{Table.How the NCD works2} shows how the NCD values change
with these modifications. It should be pointed out that the NCD
matrix is not symmetric on account of the fact that stream-based
compressors of the Lempel-Ziv family, and the predictive PPM family,
are possibly not precisely symmetric. This is due to the fact that
they are adaptive, that is they adapt to the file regularities. This
process may cause some imprecision in symmetry that vanishes
asymptotically with the length of $x$, and $y$. The other major
family of compressors, the block-coding based ones, like bzip2,
analyze the full input block by considering all rotations in
obtaining the compressed version. It is to a great extent
symmetrical, and real experiments show no departure from symmetry
\cite{Cilibrasi-Tesis}.

Looking at the NCD values presented in Table \ref{Table.How the NCD
works2}, one can notice that the distance between a text and itself
is always 0, as the numbers in the main diagonal indicate.

Furthermore, as the number of replaced words increases, the NCD
increases. The easiest way of noticing this is by comparing the
numbers contained in the first row of the matrix, which correspond
to the NCD values between \emph{Sample 1} and the rest of the
samples:

\begin{itemize}
\item
NCD (Sample1, Sample1) = 0.000000

\vspace{-0.1cm}

\item
NCD (Sample1, Sample2) = 0.282086

\vspace{-0.1cm}

\item
NCD (Sample1, Sample3) = 0.622727

\vspace{-0.1cm}

\item
NCD (Sample1, Sample4) = 0.974111
\end{itemize}

An alternative way of observing that the NCD increases as the number
of replaced words increases, is by comparing the numbers contained
in the first column of the matrix:

\begin{itemize}
\item
NCD (Sample1, Sample1) = 0.000000

\vspace{-0.1cm}

\item
NCD (Sample2, Sample1) = 0.262183

\vspace{-0.1cm}

\item
NCD (Sample3, Sample1) = 0.563636

\vspace{-0.1cm}

\item
NCD (Sample4, Sample1) = 0.979816
\end{itemize}

\begin{table}
\centering \caption{Understanding NCD: Text
samples.}\label{Table.How the NCD works1} \vspace{0.1cm}
\begin{tabular}{|p{13cm}|}
\hline \footnotesize{\texttt{\underline{Sample 1:} thomas a anderson
is a man living two lives by day he is an average computer
programmer and by night a malevolent hacker known as neo neo has
always questioned his reality but the truth is far beyond his
imagination neo finds himself targeted by the police when he is
contacted by morpheus a legendary computer hacker branded a
terrorist by the government morpheus awakens neo to the real world a
ravaged wasteland where most of humanity have been captured by a
race of machines which live off of their body heat and imprison
their minds within an artificial reality known as the matrix as a
rebel against the machines neo must return to the matrix and
confront the agents super powerful computer programs devoted to
snuffing out neo and the entire human rebellion}}
\\
\hline \footnotesize{\texttt{\underline{Sample 2:} thomas a anderson
is a man living two lives by day he is an average computer
programmer ocR by night a malevolent hacker known as neo neo has
always questioned his reality but |xM truth is far beyond his
imagination neo finds himself targeted by RZ6 police when he is
contacted by morpheus  a legendary computer hacker branded a
terrorist by )q5 government morpheus awakens neo to cWg real world a
ravaged wasteland where most wP humanity have been captured by a
race 3[ machines which live off bv their body heat - g imprison
their minds within an artificial reality known as iCy matrix  as a
rebel against g!G machines  neo must return to cOZ matrix kQ
confront s>9 agents super powerful computer programs devoted to
snuffing out neo 8rv N1c entire human rebellion}}
\\
\hline \footnotesize{\texttt{\underline{Sample 3:} thomas B anderson
y< a Og living 4L8 lives LF 5Es FU "A f average computer programmer
OS? >" night r malevolent hacker known Jd neo neo YQ@ always
questioned XsZ reality HLS ZP truth xL far beyond -RC imagination
neo finds himself targeted uW .aj police l; 1 >1 7H contacted ZW
morpheus  V legendary computer hacker branded [ terrorist VL t7g
SbL)JRKT; morpheus awakens neo uv LnQ real 1P2E3 2 ravaged wasteland
?6UF E-OD FR humanity 9+(D [WP7 captured SB 1 race HC machines b0IB
live off ?Q Qdi=' body heat /JF imprison Ar8Z  minds within uA
artificial reality known r9 =G1 matrix  T- ( rebel 'qXHAx" .UP
machines  neo 4>fW return K@ Y2q matrix ,xB confront 7L. agents
super powerful computer programs devoted sA snuffing N6T neo p4 IR
entire human rebellion}}
\\
\hline \footnotesize{\texttt{\underline{Sample 4:} CR+ZjF ! D[vyw/Fq
M' g ,x yQ29-" <Pi Aj,cn ]Z 24v qx A2 sD =/.:ZCV /2(uY|7T 3Ut:T"io7R
JvI :9 hZq:h 6 ]PzPwUv)<t FI5a!` 7rq!c Kt !DN >QH 06N S]I=fg
S'QVfi(vQc 28> qxGRjAu Xkr SuN /Z7qK Oy t(D ;2s4rU imM Q2Td5guKswg
xD" XCmho Q@,Eko· GY!Nd|K> no BiW RaCYat Cr,m X3 KJ 2SlX1Zt<D TO
morpheus D :=c:hv'5q af+sKXXZ a|"42 ec<1Zu4 : ">LjhTExI U| Z]K
k"eeYh0"g morpheus fWvc=CF 3vH SU hp1 '(YR q(l7n, s .-xub0P
P(EA)D"bs n*cJ` r7-B sQ W8bXV<hx C(D/ EZ(E 'S1Xb)ir 19 7 JF1/ Eb
v8kHDWJE xgU?I FbKE (3R S" L4lyu hPh/ ('>= 7vG hr<sRYl( C!V[Q x6DbA
9".k/S Wv xCh/2mhoQx ,7komGN !Wd|K >n o7i W2aCYc Yr , XPKU2 SlS4Zt<
DTO sDFYNB[S CX[ THY/ N5!*um B5 5PK |B)lK9 uXV ]cTxBP[o t2b Dx4Vx1
2hmVB 7YDR*Qnf 1qJYSC/n kcfSD31p 0Gl/TH- Mm 8JHb"RWo ,a5 .LO adx m9E
9JK01P (0snS UO2l+,Oxh}}
\\
\hline
\end{tabular}
\end{table}

\begin{table}
\centering \caption{Understanding NCD: matrix distances.}
\label{Table.How the NCD works2} \vspace{0.3cm}
\begin{tabular}{|c|c|c|c|c|}
\hline
 & Sample 1 & Sample 2 & Sample 3 & Sample 4 \\
\hline
Sample 1 & 0.000000 & 0.282086 & 0.622727 & 0.974111 \\
Sample 2 & 0.262183 & 0.000000 & 0.566477 & 0.961825 \\
Sample 3 & 0.563636 & 0.499432 & 0.000000 & 0.947784 \\
Sample 4 & 0.979816 & 0.974550 & 0.961825 & 0.000000 \\
\hline
\end{tabular}
\end{table}

\subsection{Some NCD applications}

There are many similarity distances based on compression algorithms
\cite{benedetto2002lta,Cerra08,Dobrinkat10,kraskov2005hcu,Zhang07},
but they are small variations and can be easily reduced to the NCD,
as it is possible to prove that this distance is as good as any
other that can be computed by a universal Turing machine
\cite{Cilibrasi-Tesis,Turing36}.

Compression distances are currently a hot topic of research in many
areas. Among others, they have been applied to the management of
textual data, biological data of diverse nature, music, or even art,
from very different points of view. The next paragraphs summarize
the main uses given to them in literature.

Directly related to the contents of this thesis is the application
of compression distances to the management of textual data. Several
research areas related to text management have benefited from the
wide applicability and leading efficacy of compression distances.
These are the cases of document clustering
\cite{Granados11eswa,Granados10tkde,Granados08,Granados10ideal,Helmer07,Telles07},
document retrieval \cite{Granados11tkde,Martinez08}, text mining
\cite{cilibrasi07}, or software engineering
\cite{allen2001mca,arbuckle2007sdc,scott-new}.

In the area of document clustering, NCD has been proposed to measure
the structural similarity between textual documents in
\cite{Telles07}, and between XML documents in \cite{Helmer07}. The
first study shows that the explored approach can be successfully
used for visual analysis of automatically generated text maps
obtaining good precision. The latter experimentally demonstrates
that the results of the proposed algorithm in terms of clustering
quality are on a par with or even better than existing approaches.

Some works that combine document clustering and document distortion
have evaluated the impact that different word removal techniques
have on NCD-driven clustering
\cite{Granados11eswa,Granados10tkde,Granados08,Granados10ideal} with
the aim of taking a small step towards understanding compression
distances. These works are not described here because they
constitute the contributions made from the investigation carried out
in this thesis, and therefore, they are presented in Chapters
\ref{Chapter: Study on text distortion} and \ref{Chapter: Relevance
of the contextual information}.

Compression distances have been successfully applied to document
retrieval as well, using window-based passage retrieval with overlap
\cite{Martinez08}, and combining that approach with document
distortion to improve the retrieval results \cite{Granados11tkde}.
The latter is a contribution from this thesis which is presented in
Chapter \ref{Chapter: Application to Document Retrieval}.

In the field of text mining, the definition of NID has been extended
to automatically extract similarity of words and phrases from the
web using Google page counts \cite{cilibrasi07}.

The potential advantages derived from the application of NID to the
field of software engineering have been presented in
\cite{arbuckle2007sdc}. That paper proposes that the use of NID in
the comparison of software documents will lead to the establishment
of a theoretically justifiable means of comparing and evaluating
software artifacts.

A practical application of NID to measuring the amount of shared
information between two computer programs, to enable plagiarism
detection, can be found in \cite{Chen04}.

In the research area of music classification, the Universal
Similarity Metric -USM- has been proposed to automatically cluster
music in \cite{cilibrasi2004acm} using the quartet tree method. The
paper \cite{GonzalezPardo10} analyzes how the selection of a
particular representation of music audio files can affect NCD-based
clustering. Three different music representations are explored in
the paper: binary code, wave information, and SAX. The best results
are obtained when the music is represented using its wave
information.

A research area in which compression distances have been widely
applied is bioinformatics. For example, NID has been applied in
phylogenetic studies in \cite{Ferragina07}, where an exhaustive
evaluation of the NID by using 25 compressors, and six datasets of
relevance to molecular biology is carried out. In addition, the work
\cite{Nykter08} presents a method, based on NCD, to assess
macrophage criticality. This method is validated on gene networks
with known properties.

The analysis of protein structures has been carried out using
compression distances as well. Thus, measuring the similarity of
protein structures by means of USM has been proposed in
\cite{krasnogor2004msp}. Similarly, a compression distance derived
from NID has been applied to protein classification in
\cite{Kocsor06}, obtaining the result that a combination of that
measure with another low time-complexity measure can approach, or
even exceed, the classification performance of such computationally
intensive methods as the Smith Waterman algorithm or HMM methods.

In chemistry, NCD has been used for measuring the similarity of
molecules in \cite{Melville07}. In that paper, the authors show that
compression-based similarity searching can outperform standard
similarity searching protocols, exemplified by the Tanimoto
coefficient combined with a binary fingerprint representation and
data fusion.

A medical application of NCD can be found in \cite{Santos06}, where
a method to cluster fetal heart rate tracings using NCD is proposed.
A different medical application oriented to image analysis can be
found in \cite{Cohen09}. That work presents a method that summarizes
changes in biological image sequences using NCD. The method has been
validated on four bio-imaging applications, obtaining good results
in all cases.

In addition, in the field of computer security, NCD has been applied
to the analysis of worms and network traffic in \cite{Wehner07}, or
to the detection of computer masqueraders, that is, illegitimate
users trying to impersonate legitimate ones, in
\cite{Bertacchini07}, showing that NCD-based approach performs as
well as the traditional methods. In the field of computer security,
it has been used as well as a measure of the similarity of malware
behavior \cite{Apel09}. In that work, an experimental comparison
between distance measures for malware behavior is developed.

Maybe the most curious application of compression distances is the
one presented in \cite{Svangard04}. In that paper, a new technique
for automatically approximating the aesthetic fitness of
evolutionary art is presented. This technique assigns fitness values
to images interactively, using USM to predict how interesting new
images are to the observer based on a library of aesthetic images.

Despite the wide use of compression distances, little has been done
to interpret compression distance results or to explain their
behavior. The main reason for this, is the immense gap between their
theoretical foundation -Kolmogorov complexity in several flavors-
and the state-of-the-art compression algorithms used in
applications. Whenever some analytical work on compression distances
is carried out, it is usually focused on the algebraic manipulation
of algorithmic information theory concepts
\cite{Cilibrasi05,Li04,Zhang07}. Even though these concepts are
really supporting the use and the optimality of compression
distances, they cannot help in interpreting the behavior of
state-of-the-art compression algorithms like BZIP2 \cite{bzip2},
LZMA \cite{lzmax}, PPMZ \cite{ppmz} and many others.  The
idiosyncrasy and specificity of the wide diversity of compression
algorithms cannot be captured by these universal -and uncomputable-
concepts \cite{cebrian05}.

Some works have used text distortion to study the behavior of
compression distances. For example, some theoretical and
experimental basis for describing the behavior of NCD-driven
clustering when it is applied in a set of elements which have been
perturbed by a certain amount of uniform random noise can be found
in~\cite{Cebrian07}. Although this work takes a step towards
understanding compression distances, deeper studies are required to
better understand them. These studies have been carried out in this
thesis, giving rise to the following works
\cite{Granados11eswa,Granados10tkde,Granados08,Granados10ideal,Granados11tkde}.
These works explore different distortion techniques based on word
removal with the purpose of better understanding compression
distances.

\section{Text Distortion Techniques}
\label{State. Text Distortion Techniques}

Removing irrelevant parts of the data has been found to be
beneficial in many fields because it helps to focus on the
relevant parts of the data.

For example, different techniques intended for noise removal to
enhance data analysis in the presence of high noise levels have been
explored in \cite{Xiong06}.

Other works have used removal to theoretically explore the effects
of distortion. For example, a theoretical study of the impact of
sporadic erasures on the limits of lossless data compression can
be found in~\cite{Verdu08}.

Word substitution has also been suggested as a kind of text
protection, based on the subsequent automatic detection of such
substitutions by looking for discrepancies between words and their
contexts~\cite{Fong08}.

In the field of text processing, several works have applied the
idea of removing irrelevant parts of the documents, showing that
distorting the documents by removing the stop-words may have
beneficial effects in terms of accuracy and computational load
when clustering documents~\cite{yang95}.

There are two main approaches to word removal, one in which a
generic fixed stop-word list is used~\cite{salton1989atp,Silva03},
and other in which this list is generated from the collection
itself~\cite{wilbur1992ais,Yang96}. The first approach is `safer' in
terms of maintaining the most relevant information of the documents.
That is, the replaced words are not specific enough to cause the
loss of important information. The second approach generates the
stop-words list from the collection of documents, obtaining a more
aggressive word removal. The investigations developed in this thesis
apply the less aggressive approach because a well-known corpus, the
British National Corpus, is used as a dictionary.

Stop-word removal has been applied to several research areas, as a
technique for filtering information. Among others, it has been
applied to information retrieval
\cite{Baeza-Yates99,Chen08,Shah04,Rijsbergen79}, information
extraction \cite{Mooney05,Vinciarelli05}, opinion mining
\cite{Pang08}, text categorization
\cite{Joachims98,Lan09,Quan11,Sebastiani02,Silva03,Yang96}, or
text summarizing \cite{Bouras10,Hu08,Ko08,Shiyan08,Varadarajan06}.

In all these works, word removal is a tool that allows the filtering
of information contained in the documents. Therefore, by applying
it, a more reduced representation of the documents is achieved. Of
course, this filtering process can imply a loss: usually, in a word
removal scenario, the contextual information inherently contained in
a text is lost.

All the distortion techniques explored in the thesis are based on
word removal. Some of them maintain the contextual information
despite the removal, whereas some others do not.

The first part of this thesis explores different word removal
techniques with the aim of analyzing how the removal affects both
the documents complexity and the information contained in the
documents. The experimental results show that, by applying a
specific distortion technique, clustering results can be improved.
This technique maintains part of the contextual information despite
the word removal. The key factor of this distortion technique is
helping the compressor to obtain more reliable similarities, and
therefore, helping the NCD to perform better.

Table \ref{Table.Helping the compressor} shows how this distortion
technique can help the NCD to focus on the relevant words of the
texts. It shows four text fragments that correspond to a document
that is modified in a specific manner. The modification consists of
progressively replacing the least relevant words in the English
language using asterisks. This kind of text distortion summarizes
the text in the same way that a person does it when underlining the
most relevant words of a text. The table shows an upper bound upon
the Kolmogorov complexity of each document, as well. These values
are estimated based on the concept that data compression is an upper
bound for it.

\begin{table}
\centering \caption{Helping the compressor.} \label{Table.Helping
the compressor} \vspace{0.1cm} \footnotesize{
\begin{tabular}{|p{8cm}|c|c|c|}
 \hline
{\multirow{2}{8cm}{~~~~~~~~~~~~~~~~~~~~~~~~~Text Sample}} & \multicolumn{3}{|c|}{Compressed file's length} \\
 \cline {2-4}
  & \scriptsize{LZMA} & \scriptsize{PPMZ} & \scriptsize{BZIP2} \\
 \hline
\texttt{specific diabetic dietary guidelines have} & & & \\
\texttt{been developed by the american diabetes} & & & \\
\texttt{association and the american dietetic} & 1900 & 1548 & 1821 \\
\texttt{association to improve the management of} & & & \\
\texttt{diabetes} & & & \\
 \hline
\texttt{specific diabetic dietary guidelines ****} & & & \\
\texttt{**** developed ** *** american diabetes} & & & \\
\texttt{association *** *** american dietetic} & 1627 & 1301 & 1553 \\
\texttt{association ** improve *** management **} & & & \\
\texttt{diabetes} & & & \\
 \hline
\texttt{******** diabetic dietary guidelines ****} & & & \\
\texttt{**** developed ** *** ******** diabetes} & & & \\
\texttt{*********** *** *** ******** dietetic} & 1333 & 1042 & 1273 \\
\texttt{*********** ** improve *** ********** **} & & & \\
\texttt{diabetes} & & & \\
 \hline
\texttt{******** diabetic dietary ********** ****} & & & \\
\texttt{**** ********* ** *** ******** ********} & & & \\
\texttt{*********** *** *** ******** dietetic} & 866 & 667 & 835 \\
\texttt{*********** ** ******* *** ********** **} & & & \\
\texttt{********} & & & \\
 \hline
\end{tabular}}
\end{table}

The second part of this thesis explores different word removal
techniques, which are created from the above mentioned one, with the
purpose of analyzing the relevance of contextual information, as
Chapter \ref{Chapter: Relevance of the contextual information}
explains.

\section{Contextual Information}
\label{State. Contextual Information}

Many research areas have used the notion of context from different
points of view because taking it into account has been found to be
beneficial in numerous domains.

For example, contextual information retrieval systems try to improve
retrieval accuracy by taking the user's context into account
\cite{Ma07,Rhodes03,Sieg07,Speretta05,Tamine10}. In these systems,
the context corresponds to the user's interests, preferences, time
and location. The same concept has been used in recommender systems
as well, obtaining good results
\cite{Adomavicius05,Kwon09,Su10,Weng09}. Similarly, context-aware
computing applications use the idea of context in form of location,
time stamps, and user identity
\cite{Caus09,Driver08,Hegde09,Pascoe98,Schilit94}.

The concept of context has been used in recognition systems as well.
For example, it has been used to identify objects in computer vision
\cite{Acosta10,Belongie02,Chi08,Mori05,Mori06}, or to improve speech
recognition performance \cite{Eronen06,Huang07,Lee90,Odell95thesis}.
In both areas, the notion of context corresponds to the data
surrounding the information which is being analyzed.

As a temporal concept, the context has been used in network traffic
analysis to discover and analyze anomalous or malicious network
activity \cite{Goodall06}. In that work, the contextual data comes
from collecting packet-level detail of the event-related network
traffic.

The fact that the notion of context has been used in so many
research areas gives us an idea of how useful this concept is in
improving the performance of different systems. In particular, in
our research area, it seems that considering the context can lead to
better results because of the intrinsic nature of textual data. In
fact, different ideas of context have been successfully applied when
working with texts.

At the lowest level, a text can be seen as a set of characters.
According to \cite{Shannon48}, the characters and the sequences of
characters have a statistical structure. This consideration of
sequences of characters can be seen as a kind of context at
character level.

Very often, texts have been represented using the Vector Space Model
-VSM- \cite{Salton75}. This model represents a text as a vector of
identifiers, such as, for example, index terms. This model is
commonly called the bag-of-words model because the order and the
relationships between the words are ignored, or in other words, no
context is taken into account.

Despite the success of VSM, several works have shown that
considering the context of words can lead to a more precise
representation. Thus, the context of a word has been represented as
co-occurrences between words or as N-grams
\cite{Brown92,Curran02,Dagan99,Dagan93,GuoDong04,Weeds05}.

For example, in \cite{Brown92} the problem of predicting a word from
previous words has been addressed using models based on classes of
words, which are based on both N-grams and frequencies of
co-occurrence. In fact, the use of co-occurrences has been so
beneficial that even the estimation of the probability of
co-occurrences that do not occur in the training data has been
studied \cite{Dagan93}.

The N-gram based models have been improved to support long distances
in \cite{GuoDong04}, where the context dependency between word pairs
over a long distance in an N-gram based model has been tackled by
using the concept of mutual information. As a different approach,
the context has been modeled as a vector of syntactic dependencies
as well \cite{Cimiano05}.

In addition, the idea of context has been applied to the creation of
adaptive text classification models dealing with the temporal
evolution of the characteristics of the documents and the classes to
which they belong \cite{Lau08,Liu08,Liu02,Rocha08}. Furthermore,
different machine-learning algorithms that construct classifiers
that allow the context of a word to affect how the presence or
absence of the word will contribute to a classification have been
evaluated in \cite{Cohen99}.

The second part of this thesis analyzes the relevance that the
contextual information has in textual data, in a clustering by
compression scenario. This analysis is the natural continuation of
the work developed in the first part of the thesis, in which a
particular distortion technique was found to be beneficial in terms
of clustering accuracy. One of the main characteristics of that
technique is that it maintains the contextual information despite
the word removal.

The analysis carried out in the second part of the thesis explores
whether the clustering accuracy improvement is due to the fact
that the distortion technique maintains the contextual information
or not. The experimental results show that the maintenance of the
contextual information helps to obtain better results.

The third part of the thesis applies the distortion technique that
maintains part of the contextual information to a
compression-based document retrieval method. Analyzing the
experimental results one can observe that the application of the
distortion technique is beneficial in terms of accuracy in a
document retrieval scenario as well.


\chapter{Study on text distortion}
\label{Chapter: Study on text distortion}

This chapter of the thesis explores several text distortion
techniques based on word removal. It analyzes how the information
contained in the documents and how the upper bound estimation of
their Kolmogorov complexity progress as the words are removed from
the documents in different manners.

A compression-based clustering method is used to experimentally
evaluate the impact that the studied distortion techniques have on
the amount of information contained in the distorted documents.

The results show that the application of one of the explored
distortion techniques can improve the clustering accuracy.

The main contributions of this research can be briefly summarized as
follows:

\begin{itemize}
\item
Analysis and study of new representations of text to evaluate the
behavior of the NCD.

\item
A technique to represent textual data, specially created to be used
with compression distances, that reduces the complexity of the
documents while preserving most of the relevant information.

\item
Experimental evidence of how to fine-tune the representation of
texts to allow the compressor to obtain more reliable similarities
and, therefore, to allow the compression-based clustering method to
improve the non-distorted clustering results.
\end{itemize}

The chapter is structured as follows. Section \ref{TKDE. Distortion
Techniques} describes the explored distortion techniques. Section
\ref{TKDE. Experimental Setup} describes the compression-based text
clustering method used, and describes the datasets. Section
\ref{TKDE. Experimental Results} gathers and analyzes the obtained
results. Finally, Section \ref{TKDE. Summary} summarizes the
conclusions drawn from the experiments presented in this chapter.

\clearpage

\section{Distortion Techniques}
\label{TKDE. Distortion Techniques}


Distorting the documents by removing the stop-words has been found
to be beneficial both in terms of accuracy and computational load
when clustering documents or when retrieving information from them
\cite{yang95}. The way in which the stop-word list is created can
produce a more aggressive or a less aggressive removal. Roughly
speaking, two main approaches to word removal can be made, one in
which a generic fixed stop-word list is
used~\cite{Granados10tkde,Granados08,Granados10ideal,Silva03}, and
other in which this list is generated from the collection
itself~\cite{wilbur1992ais,Yang96}. The second approach produces a
more aggressive word removal than the first one.

In this work, the less aggressive technique is applied, that is, a
generic list of words is used. In particular, an external and
well-known corpus, the British National Corpus -BNC-, is used to
select the words that will be removed from the documents. The BNC is
a 100 million word collection of samples of written and spoken
language from a wide range of sources, designed to represent a wide
cross-section of current British English, both spoken and written
\cite{BNC}.

This thesis explores six different replacement methods, which are
pairwise combinations of two factors: \emph{word selection method}
and \emph{substitution method}.

\begin{itemize}
\item
\emph{Word selection method}: the frequencies of the English words
are estimated using the BNC, and then the list of words is sorted in
\emph{decreasing}/\emph{increasing}/\emph{random} order of
frequency. These three lists give rise to three selection methods:
\begin{itemize}
  \item \emph{Most Frequent Word -MFW- selection method}.
  \item \emph{Least Frequent Word -LFW- selection method}.
  \item \emph{Random Word -RW- selection method}.
\end{itemize}
The idea can be described as follows: each list of words is used to
generate several sets of words to be removed from the documents. In
order to study the clustering behavior evolution as the amount of
removed words increases, for each list ten sets of words are
created, each one containing the words that accumulate a specific
frequency of words, these values going from 0.1 to 1.0. It is worth
mentioning that each set contains the words that belong to the
previous set. For example, the first set only contains the words
\emph{the}, \emph{of} and \emph{and}, because these words are
frequent enough to accumulate a frequency of 0.1. The second set
contains these words, together with the words necessary to
accumulate a total frequency of 0.2.

\item
\emph{Substitution method}: when a word has to be removed from a
text, each character of the word is replaced by either a random
character, or an asterisk. Thus there are two substitution methods:
\begin{itemize}
  \item \emph{Random character substitution method}.
  \item \emph{Asterisk substitution method}.
\end{itemize}
\end{itemize}

\begin{table}
\centering \caption{\emph{MFW selection method} \& \emph{asterisk
substitution method}.}\label{Table.MFW texts} \vspace{0.5cm}
\begin{tabular}{|c|p{11.5cm}|}
\hline 0.0 & \footnotesize{\texttt{In a village of la Mancha, the
name of which I have no desire to call to mind, there lived not long
since one of those gentlemen that keep a lance in the lance-rack, an
old buckler, a lean hack, and a greyhound for coursing.}} \\
\hline 0.1 & \footnotesize{\texttt{in a village ** la mancha  ***
name ** which i have no desire to call to mind  there lived not long
since one ** those gentlemen that keep a lance in *** lance rack  an
old buckler a lean hack *** a greyhound for coursing}} \\
\hline 0.2 & \footnotesize{\texttt{** * village ** la mancha  ***
name ** which i have no desire ** call ** mind  there lived not long
since one ** those gentlemen that keep * lance ** *** lance rack  an
old buckler * lean hack  *** * greyhound for coursing}} \\
\hline 0.3 & \footnotesize{\texttt{** * village ** la mancha  ***
name ** which * have no desire ** call ** mind  there lived *** long
since one ** those gentlemen **** keep * lance ** *** lance rack  an
old buckler * lean hack  *** * greyhound *** coursing}} \\
\hline 0.4 & \footnotesize{\texttt{** * village ** la mancha  ***
name ** ***** * **** no desire ** call ** mind  ***** lived *** long
since *** ** those gentlemen **** keep * lance ** *** lance rack  **
old buckler * lean hack  *** * greyhound *** coursing}} \\
\hline 0.5 & \footnotesize{\texttt{** * village ** la mancha  ***
name ** ***** * **** ** desire ** call ** mind  ***** lived *** long
since *** ** ***** gentlemen **** keep * lance ** *** lance rack  **
old buckler * lean hack  *** * greyhound *** coursing}} \\
\hline 0.6 & \footnotesize{\texttt{** * village ** la mancha  ***
**** ** ***** * **** ** desire ** call ** mind  ***** lived *** ****
***** *** ** ***** gentlemen **** **** * lance ** *** lance rack  **
*** buckler * lean hack *** * greyhound *** coursing}} \\
\hline 0.7 & \footnotesize{\texttt{** * ******* ** la mancha  ***
**** ** ***** * **** ** desire ** **** ** ****  ***** lived *** **** ***** *** ** *****
gentlemen **** **** * lance ** *** lance rack  ** *** buckler  *
lean hack  *** * greyhound *** coursing}} \\
\hline 0.8 & \footnotesize{\texttt{** * ******* ** la mancha  ***
**** ** ***** * **** ** ****** ** **** ** ****  ***** ***** *** **** ***** *** ** *****
gentlemen **** **** * lance ** *** lance rack  ** *** buckler  *
lean hack  *** * greyhound *** coursing}} \\
\hline 0.9 & \footnotesize{\texttt{** * ******* ** ** mancha  ***
**** ** ***** * **** ** ****** ** **** ** ****  ***** ***** *** **** ***** *** ** *****
********* **** **** * lance ** *** lance ****  ** *** buckler  * ****
hack  *** * greyhound *** coursing}} \\
\hline 1.0 & \footnotesize{\texttt{** * ******* ** ** ******  ***
**** ** ***** * **** ** ****** ** **** ** ****  ***** ***** *** **** ***** *** ** *****
********* **** **** * ***** ** *** ***** ****  ** *** *******  * ****
****  *** * ********* *** ********}} \\
\hline
\end{tabular}
\end{table}

\begin{table}
\centering \caption{\emph{RW selection method} \& \emph{asterisk
substitution method}.}\label{Table.RW texts} \vspace{0.5cm}
\begin{tabular}{|c|p{11.5cm}|}
\hline 0.0 & \footnotesize{\texttt{In a village of la Mancha, the
name of which I have no desire to call to mind, there lived not long
since one of those gentlemen that keep a lance in the lance-rack, an
old buckler, a lean hack, and a greyhound for coursing.}} \\
\hline 0.1 & \footnotesize{\texttt{in a village of la mancha  the
name of which * **** no desire to **** to ****  there lived not long
since *** of those gentlemen that keep a lance in the lance rack  **
old buckler a **** ****  and a ********* for ********}} \\
\hline 0.2 & \footnotesize{\texttt{in * ******* ** la mancha  the
**** ** which * **** no ****** to **** to ****  there ***** not long
since *** ** those gentlemen that keep * lance in the lance rack  **
old buckler * **** ****  and * ********* *** ********}} \\
\hline 0.3 & \footnotesize{\texttt{in * ******* ** la mancha  the
**** ** which * **** no ****** to **** to ****  ***** ***** not long
since *** ** those ********* that **** * lance in the lance rack  **
old buckler * **** ****  and * ********* *** ********}} \\
\hline 0.4 & \footnotesize{\texttt{** * ******* ** la mancha  the
**** ** ***** * **** ** ****** to **** to ****  ***** ***** not **** since *** ** those
********* **** **** * lance ** the lance ****  ** old buckler  *
**** ****  and * ********* *** ********}} \\
\hline 0.5 & \footnotesize{\texttt{** * ******* ** la mancha  ***
**** ** ***** * **** ** ****** to **** to ****  ***** ***** not **** ***** *** ** *****
********* **** **** * ***** ** *** ***** ****  ** *** buckler  *
**** ****  and * ********* *** ********}} \\
\hline 0.6 & \footnotesize{\texttt{** * ******* ** ** mancha  ***
**** ** ***** * **** ** ****** to **** to ****  ***** ***** not **** ***** *** ** *****
********* **** **** * ***** ** *** ***** ****  ** *** buckler  *
**** ****  and * ********* *** ********}} \\
\hline 0.7 & \footnotesize{\texttt{** * ******* ** ** mancha  ***
**** ** ***** * **** ** ****** ** **** ** ****  ***** ***** *** **** ***** *** ** *****
********* **** **** * ***** ** *** ***** ****  ** *** buckler  *
**** ****  *** * ********* *** ********}} \\
\hline 0.8 & \footnotesize{\texttt{** * ******* ** ** mancha  ***
**** ** ***** * **** ** ****** ** **** ** ****  ***** ***** *** **** ***** *** ** *****
********* **** **** * ***** ** *** ***** ****  ** *** buckler  *
**** ****  *** * ********* *** ********}} \\
\hline 0.9 & \footnotesize{\texttt{** * ******* ** ** mancha  ***
**** ** ***** * **** ** ****** ** **** ** ****  ***** ***** *** **** ***** *** ** *****
********* **** **** * ***** ** *** ***** ****  ** *** buckler  *
**** ****  *** * ********* *** ********}} \\
\hline 1.0 & \footnotesize{\texttt{** * ******* ** ** ******  ***
**** ** ***** * **** ** ****** ** **** ** ****  ***** ***** *** **** ***** *** ** *****
********* **** **** * ***** ** *** ***** ****  ** *** *******  *
**** ****  *** * ********* *** ********}} \\
\hline
\end{tabular}
\end{table}

\begin{table}
\centering \caption{\emph{LFW selection method} \& \emph{asterisk
substitution method}.}\label{Table.LFW texts} \vspace{0.5cm}
\begin{tabular}{|c|p{11.5cm}|}
\hline 0.0 & \footnotesize{\texttt{In a village of la Mancha, the
name of which I have no desire to call to mind, there lived not long
since one of those gentlemen that keep a lance in the lance-rack, an
old buckler, a lean hack, and a greyhound for coursing.}} \\
\hline 0.1 & \footnotesize{\texttt{in a village of la ******  the
name of which i have no desire to call to mind  there lived not long
since one of those gentlemen that keep a ***** in the ***** rack  an
old ******* a lean ****  and a ********* for ********}} \\
\hline 0.2 & \footnotesize{\texttt{in a village of ** ******  the
name of which i have no desire to call to mind  there lived not long
since one of those gentlemen that keep a ***** in the ***** ****  an
old ******* a **** ****  and a ********* for ********}} \\
\hline 0.3 & \footnotesize{\texttt{in a village of ** ******  the
name of which i have no ****** to call to mind  there ***** not long
since one of those ********* that keep a ***** in the ***** ****  an
old ******* a **** ****  and a ********* for ********}} \\
\hline 0.4 & \footnotesize{\texttt{in a ******* of ** ******  the
name of which i have no ****** to **** to ****  there ***** not long
since one of those ********* that keep a ***** in the ***** ****  an
old ******* a **** ****  and a ********* for ********}} \\
\hline 0.5 & \footnotesize{\texttt{in a ******* of ** ******  the
**** of which i have no ****** to **** to ****  there ***** not ****
***** one of those ********* that **** a ***** in the ***** ****  an
*** ******* a **** ****  and a ********* for ********}} \\
\hline 0.6 & \footnotesize{\texttt{in a ******* of ** ******  the
**** of which i have no ****** to **** to ****  there ***** not ****
***** *** of ***** ********* that **** a ***** in the ***** ****  an *** *******  a
**** ****  and a ********* for ********}} \\
\hline 0.7 & \footnotesize{\texttt{in a ******* of ** ******  the
**** of ***** i **** ** ****** to **** to ****  ***** ***** not **** ***** *** of *****
********* that **** a ***** in the ***** ****  ** *** *******  a
**** ****  and a ********* for ********}} \\
\hline 0.8 & \footnotesize{\texttt{in a ******* of ** ******  the
**** of ***** * **** ** ****** to **** to ****  ***** ***** *** **** ***** *** of *****
********* **** **** a ***** in the ***** ****  ** *** *******  a
**** ****  and a ********* *** ********}} \\
\hline 0.9 & \footnotesize{\texttt{** * ******* of ** ******  the
**** of ***** * **** ** ****** ** **** ** ****  ***** ***** *** **** ***** *** of *****
********* **** **** * ***** ** the ***** ****  ** *** *******  *
**** ****  *** * ********* *** ********}} \\
\hline 1.0 & \footnotesize{\texttt{** * ******* ** ** ******  ***
**** ** ***** * **** ** ****** ** **** ** ****  ***** ***** *** **** ***** *** ** *****
********* **** **** * ***** ** *** ***** ****  ** *** *******  *
**** ****  *** * ********* *** ********}} \\
\hline
\end{tabular}
\end{table}

Note that all six combinations maintain the length of the document.
This is enforced to ease the comparison of the Kolmogorov complexity
upper bound estimation among the several methods.

Tables \ref{Table.MFW texts}, \ref{Table.RW texts} and
\ref{Table.LFW texts} have been created in order to visually show
the difference between the distortion techniques based on the
\emph{asterisk substitution method}. Each table contains the ten
distorted versions of a famous extract from the renowned novel
\emph{Don Quixote} by \emph{Miguel de Cervantes}. 

In addition, several binary images that represent the information
contained in a dataset have been created in order to gain an insight
into how the information progresses as the distortion techniques are
applied. In these images, each pixel can be either black or white.
Black pixels represent remaining words and white pixels represent
substituted words. As a consequence, a non-distorted document will
be a totally black image, whereas a highly distorted document will
only have some spurious black pixels.

Looking at the binary images contained in Fig \ref{TKDE.
Fig:InformationDistortion}, one can observe that, as the number of
replaced words increases, the images have a higher number of white
pixels. However, it should be noted that depending on the word
selection method used, the loss of information progresses faster or
slower.

When the texts are distorted by deleting the most frequent words in
the English language, the information loss progresses more slowly
than when the texts are distorted removing the least frequent words
in the English language. This fact can be observed comparing
pairwise images. In particular, the images that correspond to the
same cumulative sum of frequency using the \emph{MFW selection
method} and the \emph{LFW selection method} have to be compared. For
example, comparing the image with label ``MFW 0.1'' with the image
with label ``LFW 0.1'' one can observe that the former has
definitely more black pixels than the latter. This means that the
text that corresponds to the \emph{MFW selection method} contains
more remaining words than the LFW one. This is due to the fact that
when the words are sorted in decreasing order of frequency -MFW-,
only three words are necessary to accumulate a frequency of 0.1. On
the contrary, when the words are sorted in increasing order of
frequency -LFW- many words are necessary to accumulate a frequency
of 0.1 because the frequency of the least frequent words is
extremely small.

Notice too that even when all the words included in the BNC are
replaced from the texts, the words that are not included in the BNC
remain in the documents. This can be seen observing the black pixels
in the images corresponding to the cumulative sum of 1.0, both for
the \emph{MFW selection method} and the \emph{LFW selection method}.
Observe that these images are exactly the same, because distorting a
text using the \emph{MFW selection method} removing the words that
accumulate a frequency of 1.0 generates the same distorted text as
distorting the text using the \emph{LFW selection method} deleting
the words that accumulate a frequency of 1.0. This is due to the
fact that all the words belonging to the BNC have to be taken into
account to accumulate a total frequency of 1.0 in both cases.

The most interesting comparison of image pairs is the comparison
between the images labeled as ``LFW 0.1'' and ``MFW 0.8''. These
images have a similar amount of black pixels. This means that the
distorted texts have a similar amount of remaining words. In
principle, one could think that the clustering results should be
similar because of that. However, exactly the opposite happens. In
general, the best clustering error obtained is the one that
corresponds to the \emph{MFW selection method} for a cumulative sum
of frequencies of about 0.8. However, when the \emph{LFW selection
method} is applied, the clustering error gets worse. This means that
not only the amount of remaining words affects the clustering error,
but also the kind of words that remain in the documents after the
distortion. While removing the most frequent words is beneficial,
removing the least frequent words is not. This can be observed in
the experimental results presented in this chapter, and in Appendix
\ref{Appendix Detailed Experimental Results}.

\begin{figure*}
\label{TKDE. Fig:InformationDistortion}
\begin{tabular}{ccc|ccc}
\includegraphics[width=3cm]{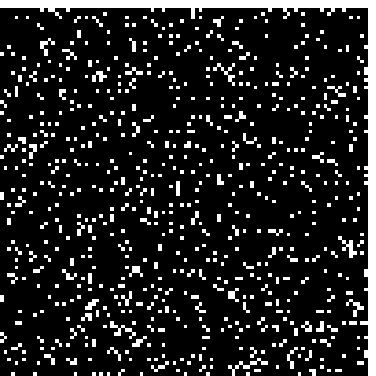} & \includegraphics[width=3cm]{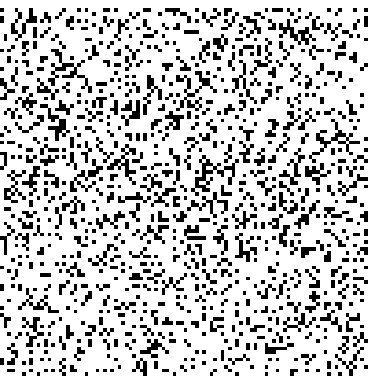} & & & \includegraphics[width=3cm]{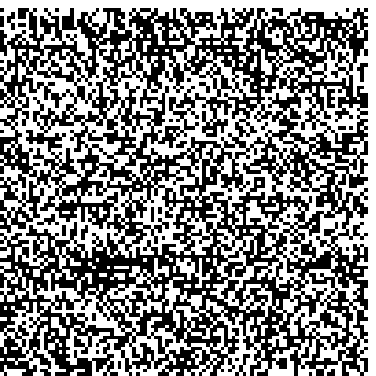} & \includegraphics[width=3cm]{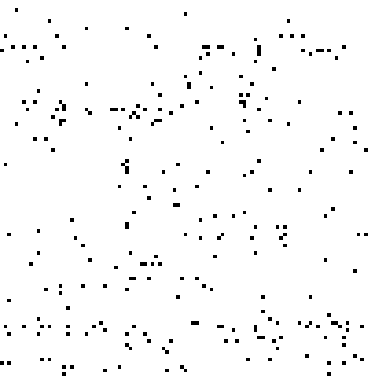} \\
MFW 0.1 & LFW 0.1 & & & MFW 0.6 & LFW 0.6 \\
 & & & & &\\
\includegraphics[width=3cm]{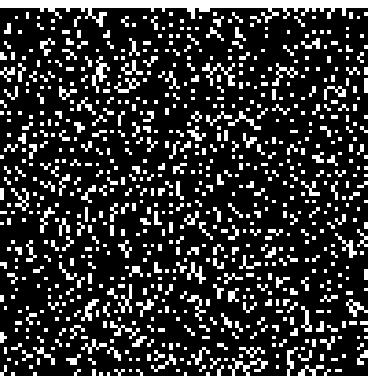} & \includegraphics[width=3cm]{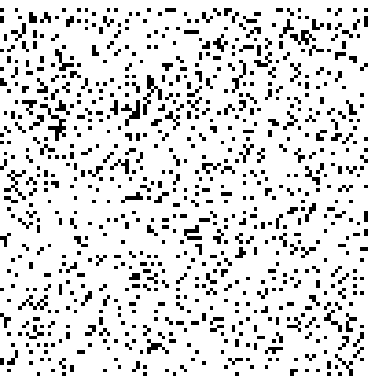} & & & \includegraphics[width=3cm]{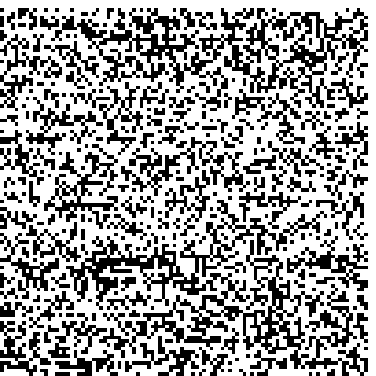} & \includegraphics[width=3cm]{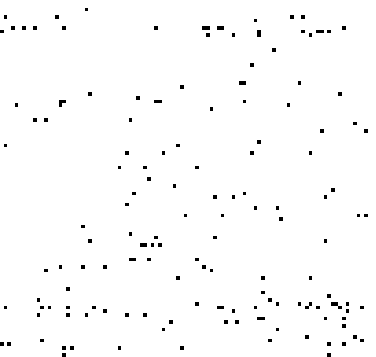} \\
MFW 0.2 & LFW 0.2 & & & MFW 0.7 & LFW 0.7 \\
 & & & & &\\
\includegraphics[width=3cm]{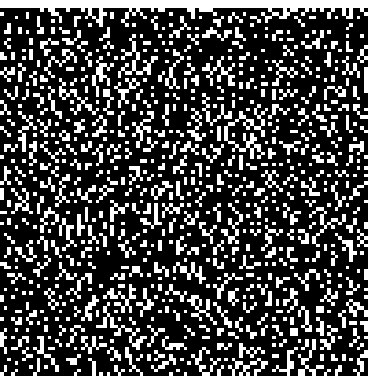} & \includegraphics[width=3cm]{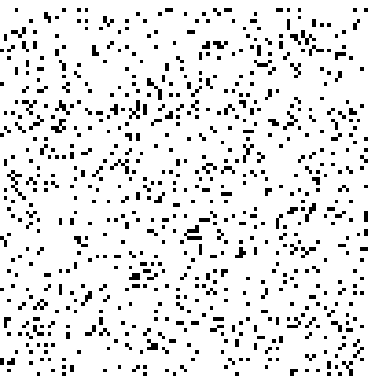} & & & \includegraphics[width=3cm]{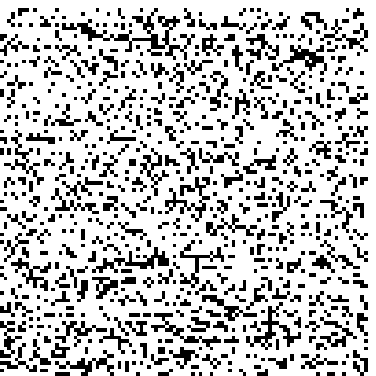} & \includegraphics[width=3cm]{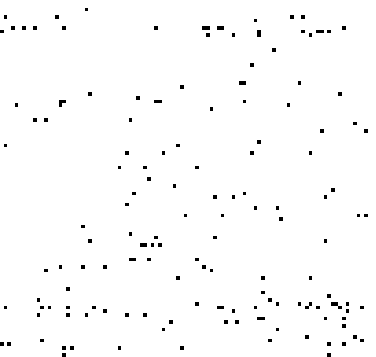} \\
MFW 0.3 & LFW 0.3 & & & MFW 0.8 & LFW 0.8 \\
 & & & & &\\
\includegraphics[width=3cm]{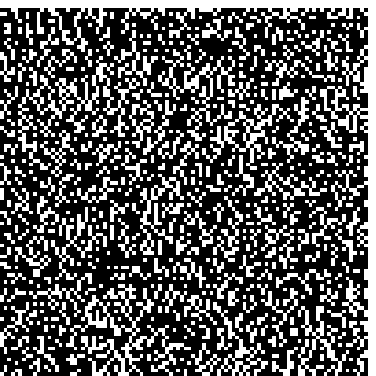} & \includegraphics[width=3cm]{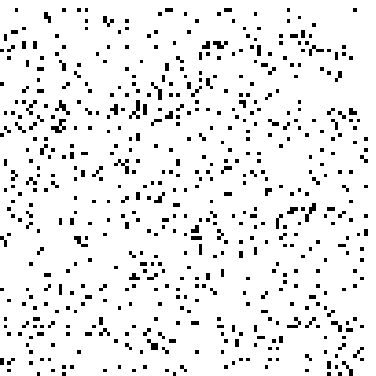} & & & \includegraphics[width=3cm]{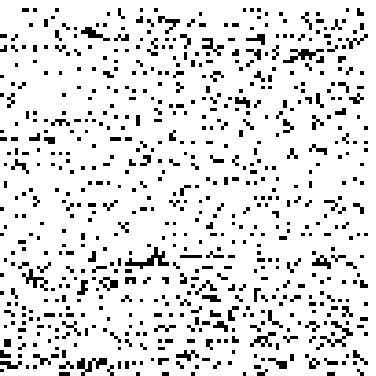} & \includegraphics[width=3cm]{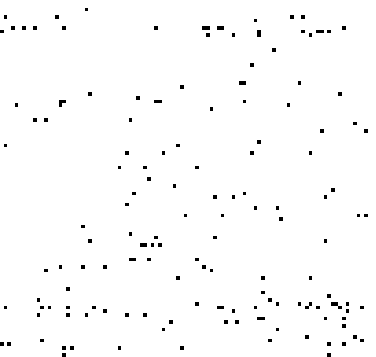} \\
MFW 0.4 & LFW 0.4 & & & MFW 0.9 & LFW 0.9 \\
 & & & & &\\
\includegraphics[width=3cm]{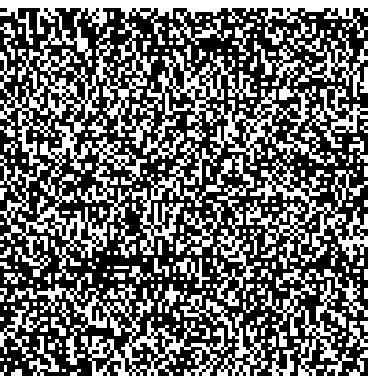} & \includegraphics[width=3cm]{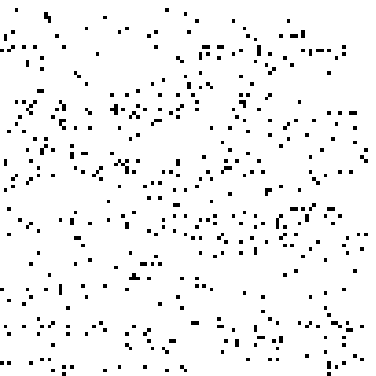} & & & \includegraphics[width=3cm]{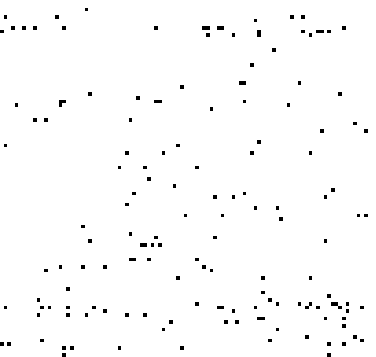} & \includegraphics[width=3cm]{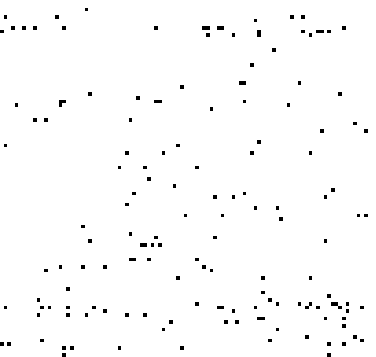} \\
MFW 0.5 & LFW 0.5 & & & MFW 1.0 & LFW 1.0 \\
\end{tabular}
\caption[Visual representation of the information loss.]{Visual
representation of the information loss. Black pixels represent
remaining words and white pixels represent substituted words.}
\end{figure*}

A quantitative measure of the qualitative idea presented in Fig
\ref{TKDE. Fig:InformationDistortion} can be seen in Fig \ref{TKDE.
Fig:pnegro}. That figure shows the percentage of removed words with
respect to the cumulative sum of BNC-based frequencies of words
substituted from the documents. Analyzing this figure, one can reach
the same conclusions as by analyzing Fig \ref{TKDE.
Fig:InformationDistortion}, that is, the percentage of removed words
increases faster or more slowly depending on the word selection
method used.

It is important to note that the percentages of substituted words
for the points ``LFW 0.1'' and ``MFW 0.8'' are very similar. That
is, the amount of substituted words in both cases is very similar.
However, the clustering error in both cases is very different, as
the clustering error figures show.  This means that the most
important factor is the kind of words that remain in the documents
after the distortion. Therefore, the key factor is the selection
method used. Whereas removing the most frequent words is beneficial
in terms of clustering results, removing the least frequent words is
not.

\begin{figure}[h]
\centering
  \includegraphics[angle=270,width=13.5cm]{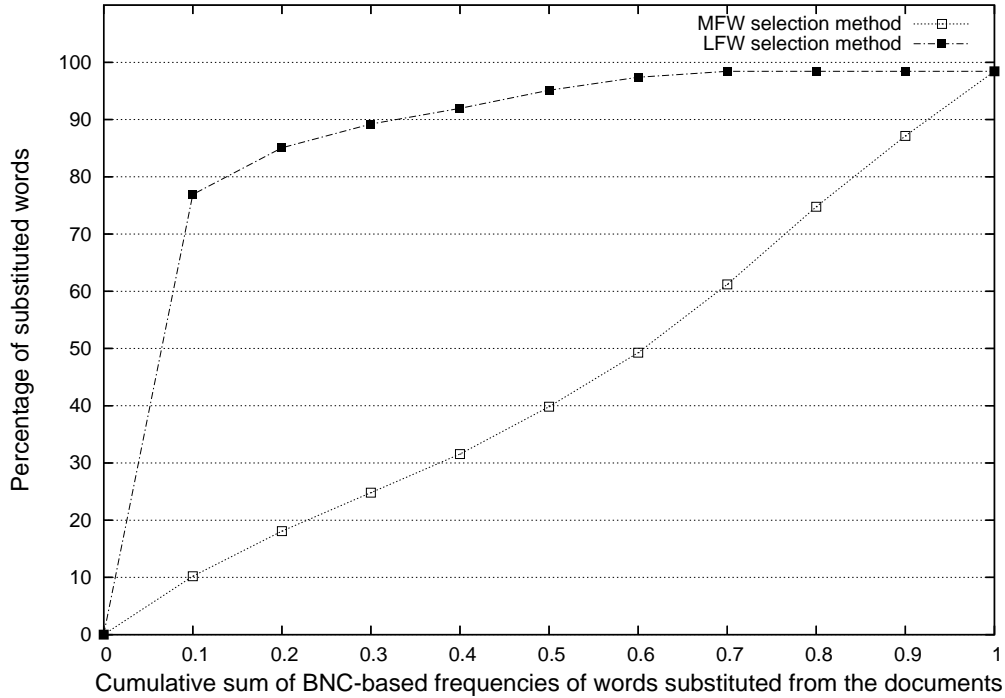}\\
\caption[Percentage of substituted words.]{Percentage of substituted
words with respect to the cumulative sum of BNC-based frequencies of
words substituted from the documents.}\label{TKDE. Fig:pnegro}
\end{figure}

\section{Experimental Setup}
\label{TKDE. Experimental Setup}

This section describes the NCD-based clustering method used
throughout the thesis. Later, it superficially enumerates the
datasets used to carry out the experiments of this chapter. The
detailed description of these datasets can be found in Appendix
\ref{Appendix Datasets}.

\subsection{NCD-based Text Clustering}
\label{TKDE. NCD-based Text Clustering}

In terms of implementation, the CompLearn Toolkit \cite{complearn},
which implements the clustering algorithm described in
\cite{Cilibrasi05,Li04}, is used. This clustering algorithm uses the
NCD as similarity distance between two objects. Detailed information
on the NCD can be found in Section \ref{State. Compression
Distances}.

The clustering algorithm implemented in the CompLearn Toolkit
comprises two phases:

\vspace{-0.3cm}

\begin{itemize}
\item First, the NCD matrix is calculated using a compression
algorithm. In this thesis, three different compression algorithms
have been used, each belonging to a different family of
compressors: LZMA, BZIP2 and PPMZ. 

\begin{figure}[h]
  \includegraphics[width=14cm]{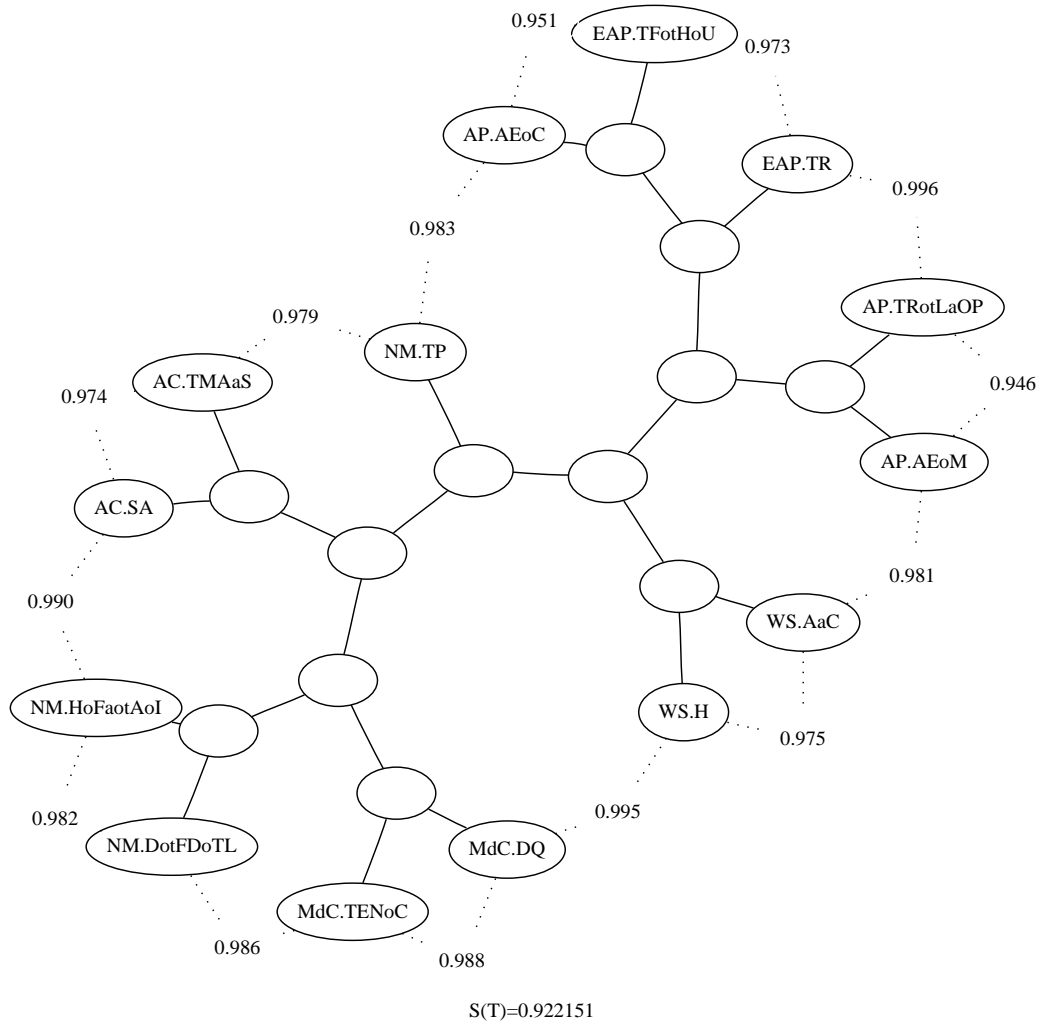}
  \centering
\caption[Example of dendrogram for the Books repository.]{Example of
dendrogram for the Books repository. Analyzing a dendrogram one can
visually observe the result of the clustering process. Each leaf of
the dendrogram corresponds to a document. The numbers in the image
represent the average NCD between two leaves. In this example, one
can observe that the nodes labeled as ``NM.TP'' and ``AP.AEoC'' are
incorrectly clustered. This implies that the distances between the
books by Niccol\`o Machiavelli and by Alexander Pope are higher than
they should be if these nodes had been correctly clustered.
Furthermore, as a consequence, the distance between the books by
Edgar Allan Poe -``EAP.TFotHoU'' and ``EAP.TR''- is higher than it
should be. The pairwise distances between the nodes belonging to
this dendrogram can be seen in Table \ref{Table. Clustering Error
Measure}.} \label{TKDE. Fig:dendro}
\end{figure}

\clearpage

\item Second, the NCD matrix is used as input to the clustering
phase and a dendrogram is generated as output. A dendrogram is an
undirected binary tree diagram, frequently used for hierarchical
clustering, that illustrates the arrangement of the clusters
produced by a clustering algorithm. Each leaf of the dendrogram
corresponds to an object. Fig \ref{TKDE. Fig:dendro} shows a
representative example of a dendrogram.
\end{itemize}

Once the CompLearn Toolkit \cite{complearn} has been used to cluster
the documents and the dendrograms are generated, quantitatively
measuring the error of the obtained dendrograms becomes necessary.
In this work, the way in which the error is measured is based on
adding the distances of the documents -leaves- that should be
clustered together. Here, the distance between two leaves is defined
as the minimum number of internal nodes needed to go from one to the
other. Table \ref{Table. Clustering Error Measure} shows the
distances between all the leaves belonging to the dendrogram
depicted in Fig \ref{TKDE. Fig:dendro}.

\begin{table}[h]
\centering \caption[Clustering error measurement.]{Clustering error
measurement. Pairwise distances between the nodes belonging to the
dendrogram depicted in Fig \ref{TKDE. Fig:dendro}.} \label{Table.
Clustering Error Measure} \vspace {0.5cm}
\begin{tabular}{|c|c|c|}
  \hline
  Cluster & Nodes & Pairwise distance \\
  \hline
  AC & AC.SA - AC.TMAaS & 1 \\
  \hline
     & AP.AEoC - AP.AEoM     & 4 \\
  AP & AP.AEoC - AP.TRotLaOP & 4 \\
     & AP.AEoM - AP.TRotLaOP & 1 \\
  \hline
  EAP & EAP.TFotHou - EAP.TR & 2 \\
  \hline
  MdC & MdC.DQ - MdC.TENoC & 1 \\
  \hline
      & NM.TP - NM.DotFDoTL        & 4 \\
  NM  & NM.TP - NM.HoFaotAoI       & 4 \\
      & NM.DotFDoTL - NM.HoFaotAoI & 1 \\
  \hline
  WS & WS.H - WS.AaC & 1 \\
  \hline
\end{tabular}
\end{table}

The procedure carried out to measure the error of a dendrogram is as
follows:
\begin{itemize}
\item
First, the pairwise distances between the documents that should be
clustered together are added.

\item
Second, after calculating this addition, the addition that
corresponds to perfect clustering is subtracted from the total
quantity obtained in the first step.
\end{itemize}

Consequently, if a dendrogram clusters all the documents perfectly,
the clustering error would be 0, and in general, the bigger the
clustering error, the worse the clustering would be.

The clustering error corresponding to the dendrogram shown in Fig
\ref{TKDE. Fig:dendro} is 9, because the sum of all the obtained
pairwise distances is 23, and the sum of all the pairwise distances
in a perfect dendrogram for these documents is 14. A perfect
dendrogram for the Books dataset can be seen in Fig \ref{TKDE.
Fig:dendro-perfect}.

\subsection{Datasets}
\label{TKDE. Datasets}

Since the CompLearn Toolkit \cite{complearn} has been used to carry
out the experiments, and this clustering algorithm has an
asymptotical cost of $O(n^{3})$ from version 1.1.3. onwards
\cite{Cilibrasi05}, a reduced number of documents has been used for
each dataset.

All of the datasets are composed of documents written in English.
Although the detailed description of the datasets can be found in
Appendix \ref{Appendix Datasets}, a summarized description of them
can be found here:

\begin{itemize}
\item
\textbf{Books dataset}: Fourteen classical books from universal
literature, to be clustered by author.

\item
\textbf{UCI-KDD dataset}: Sixteen messages from a newsgroup, to be
clustered by topic.

\item
\textbf{MedlinePlus dataset}: Twelve documents from the MedlinePlus
repository, to be clustered by topic.

\item \textbf{IMDB dataset}: Fourteen plots of movies from the
Internet Movie Data Base -IMDB- to be clustered by saga.
\end{itemize}

\section{Experimental Results}
\label{TKDE. Experimental Results}

The obtained experimental results are consistent across different
datasets and different compression algorithms. Due to this, this
chapter only shows in detail the results obtained for one dataset
and one compression algorithm. However, the rest of the results can
be found in Appendix \ref{Appendix Detailed Experimental Results}.
Furthermore, a summary of all the obtained results -in form of
tables- can be seen in Section \ref{TKDE. Synopsis of results for
all the datasets}.

\subsection{The Books dataset and the PPMZ compressor}
\label{TKDE. Case of Study}

This subsection contains the results obtained for the Books dataset
when the PPMZ compressor is used. Fig \ref{TKDE.
Fig:books-complexity-all-ppmz} depicts the upper bound estimation of
the Kolmogorov complexity of the documents, while Figs \ref{TKDE.
Fig:books-clustering-error-ppmz-mfw}, \ref{TKDE.
Fig:books-clustering-error-ppmz-rw} and \ref{TKDE.
Fig:books-clustering-error-ppmz-lfw} depict the clustering error. In
all the figures, the values on the horizontal axis correspond to the
cumulative sum of the BNC-based frequencies of the words substituted
from the documents.

Figs \ref{TKDE. Fig:books-complexity-all-ppmz}, \ref{TKDE.
Fig:books-clustering-error-ppmz-mfw}, \ref{TKDE.
Fig:books-clustering-error-ppmz-rw}, and \ref{TKDE.
Fig:books-clustering-error-ppmz-lfw} contain some percentages of
substituted words enclosed by brackets. These percentages are
calculated dividing the number of words substituted in the documents
by the total number of words contained in the documents. These
percentages are useful to understand how relevant the choice of the
words to be substituted from the documents is.

\vspace{0.3cm}

\begin{figure}[hb]
\centering\includegraphics[width=13.5cm]{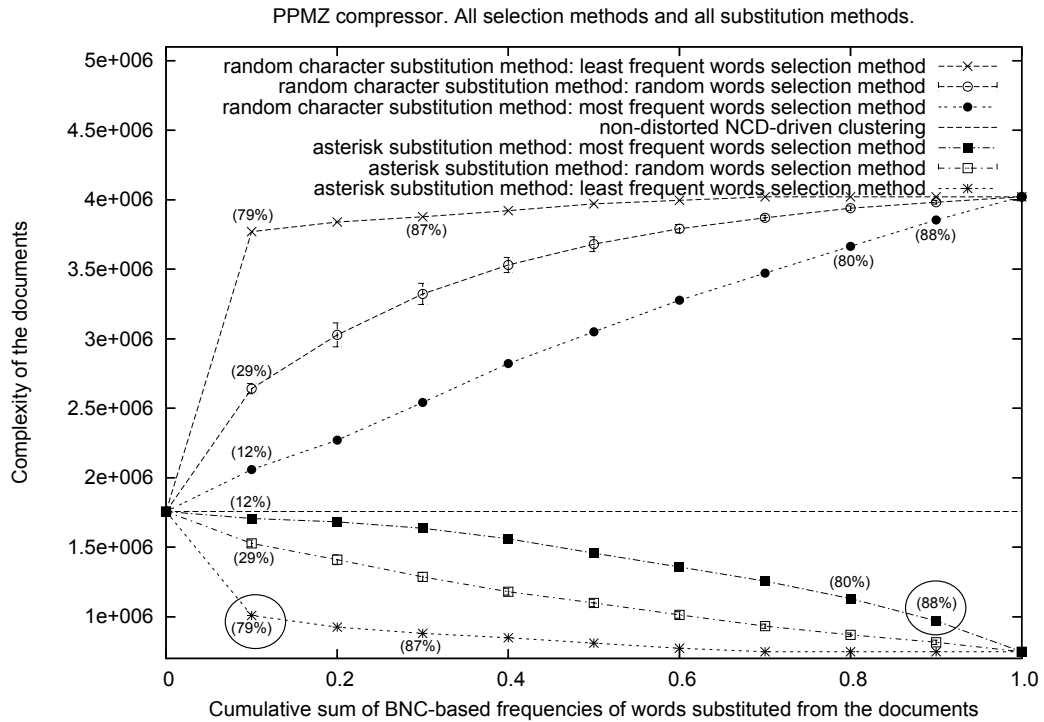}
\caption[Estimation of an upper bound for the Books
complexity.]{Estimation of an upper bound for the Books complexity.
The numbers between brackets correspond to the percentage of
substituted words in the documents. These percentages are shown in
Figs \ref{TKDE. Fig:books-clustering-error-ppmz-mfw}, \ref{TKDE.
Fig:books-clustering-error-ppmz-rw}, and \ref{TKDE.
Fig:books-clustering-error-ppmz-lfw} as well. Notice, that although
the complexity values that correspond to the points highlighted
inside a circle are very similar, the clustering error in both cases
is very different, as Figs \ref{TKDE.
Fig:books-clustering-error-ppmz-mfw} and \ref{TKDE.
Fig:books-clustering-error-ppmz-lfw} show.} \label{TKDE.
Fig:books-complexity-all-ppmz}
\end{figure}

\clearpage

The upper bound upon the complexity of the documents is estimated as
the length of the compressed file in bytes. Analyzing Fig \ref{TKDE.
Fig:books-complexity-all-ppmz} one can observe that the values
associated to the \emph{asterisk substitution method} decrease for
all the word selection methods, as the ones associated to the
\emph{random character substitution method} grow for all the word
selection methods.

The most interesting observation that can be made analyzing Fig
\ref{TKDE. Fig:books-complexity-all-ppmz} is that although the
complexity values that correspond to the points highlighted inside a
circle are very similar, the clustering error in both cases is very
different. On one hand, for the point that corresponds to the
\emph{MFW selection method}, the clustering error is 0, even though
the percentage of removed words is 88\%. On the other hand, for the
point that corresponds to the \emph{LFW selection method}, the
clustering error is 7 whereas the percentage of removed words is
79\%. Look at Figs \ref{TKDE. Fig:books-clustering-error-ppmz-mfw}
and \ref{TKDE. Fig:books-clustering-error-ppmz-lfw} to see the
clustering error values that correspond to the points highlighted
inside a circle.

\vspace{0.2cm}

\begin{figure}[h]
\centering
\includegraphics[width=13.5cm]{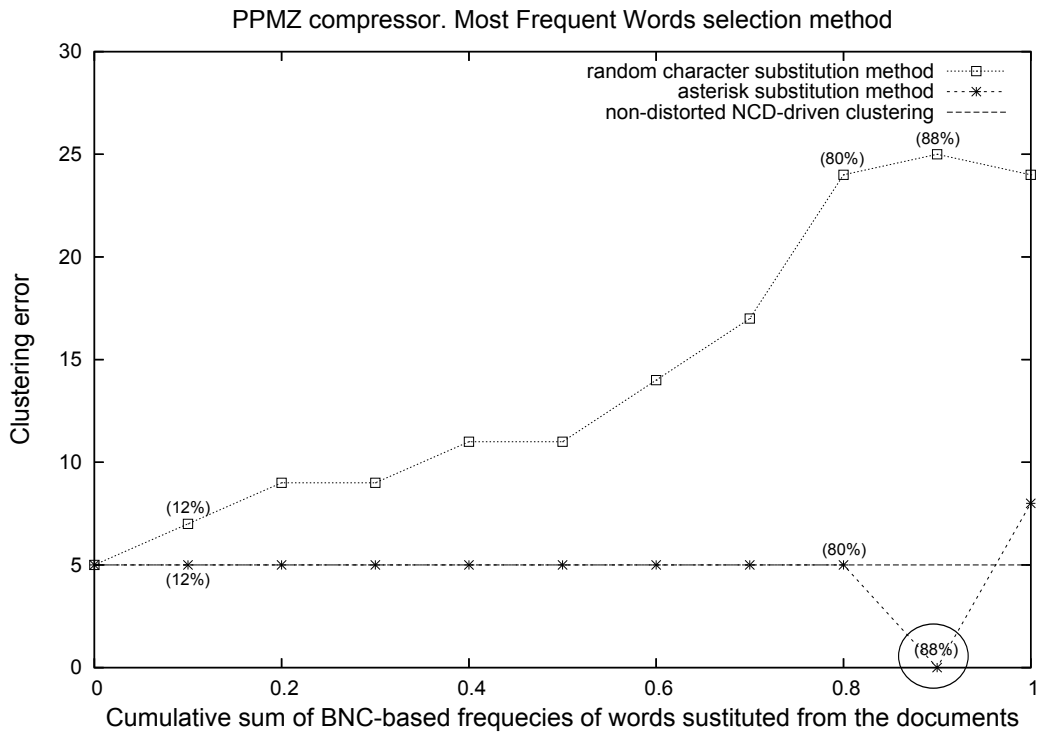}
\vspace{-0.2cm} \caption[Books. PPMZ compressor. MFW selection
method.]{Books. PPMZ compressor. MFW selection method. The
non-distorted clustering error remains constant even when a high
number of words is removed from the documents using the
\emph{asterisk substitution method}. The non-distorted clustering
error is improved for the cumulative sum of frequencies of 0.9,
where a clustering error of 0 is obtained.} \label{TKDE.
Fig:books-clustering-error-ppmz-mfw}
\end{figure}

\clearpage

This behavior is repeated for the rest of compression algorithms and
the rest of datasets, as will be shown afterwards. This means that
reducing the complexity of the documents is beneficial only in the
case in which the \emph{MFW selection method} is used.

Figs \ref{TKDE. Fig:books-clustering-error-ppmz-mfw}, \ref{TKDE.
Fig:books-clustering-error-ppmz-rw}, and \ref{TKDE.
Fig:books-clustering-error-ppmz-lfw} show the clustering error
curves obtained for the Books dataset, and the PPMZ compressor.
There is a figure for each selection method. In all the figures, the
curve with asterisk markers corresponds to the \emph{asterisk
substitution method}, while the one with square markers corresponds
to the \emph{random character substitution method}. The
non-distorted NCD-driven clustering error is depicted as a constant
line although it is only meaningful for a cumulative sum of
frequencies of 0, because it is easier to see the difference between
the line and the clustering error curves.

\begin{figure}[h]
\centering
\includegraphics[width=13.5cm]{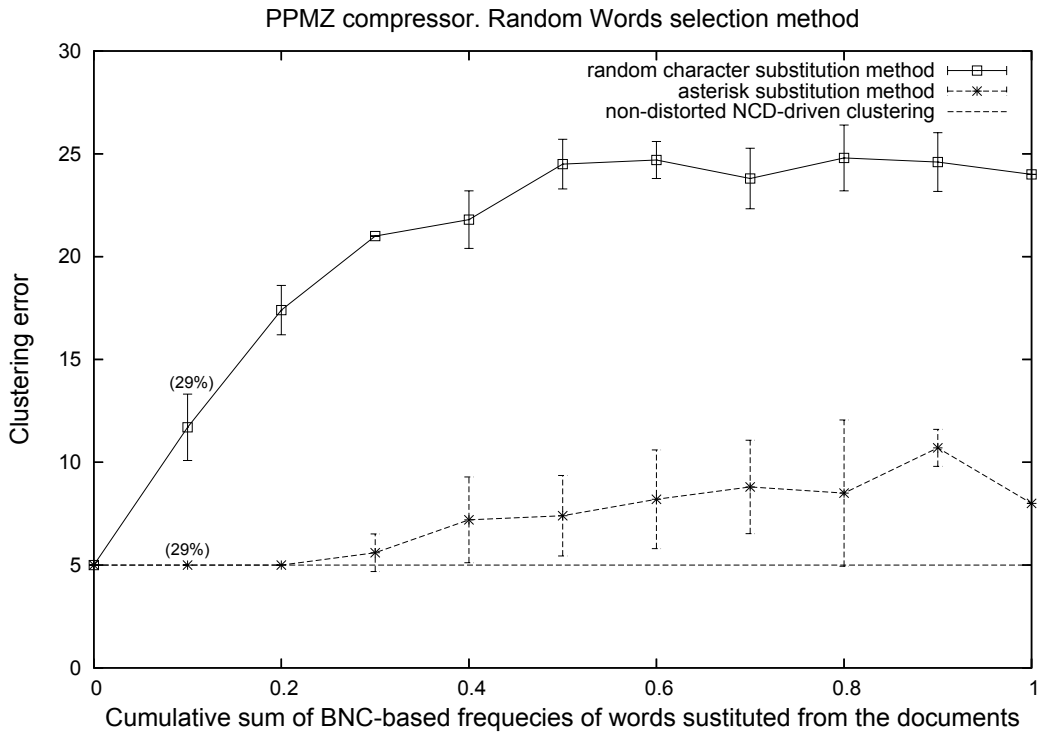}
\vspace{-0.2cm} \caption[Books. PPMZ compressor. RW selection
method.]{Books. PPMZ compressor. RW selection method. The clustering
error gets worse even when the \emph{MFW selection method} is used.}
\label{TKDE. Fig:books-clustering-error-ppmz-rw}
\end{figure}

\begin{figure}[ht]
\centering
\includegraphics[width=13.5cm]{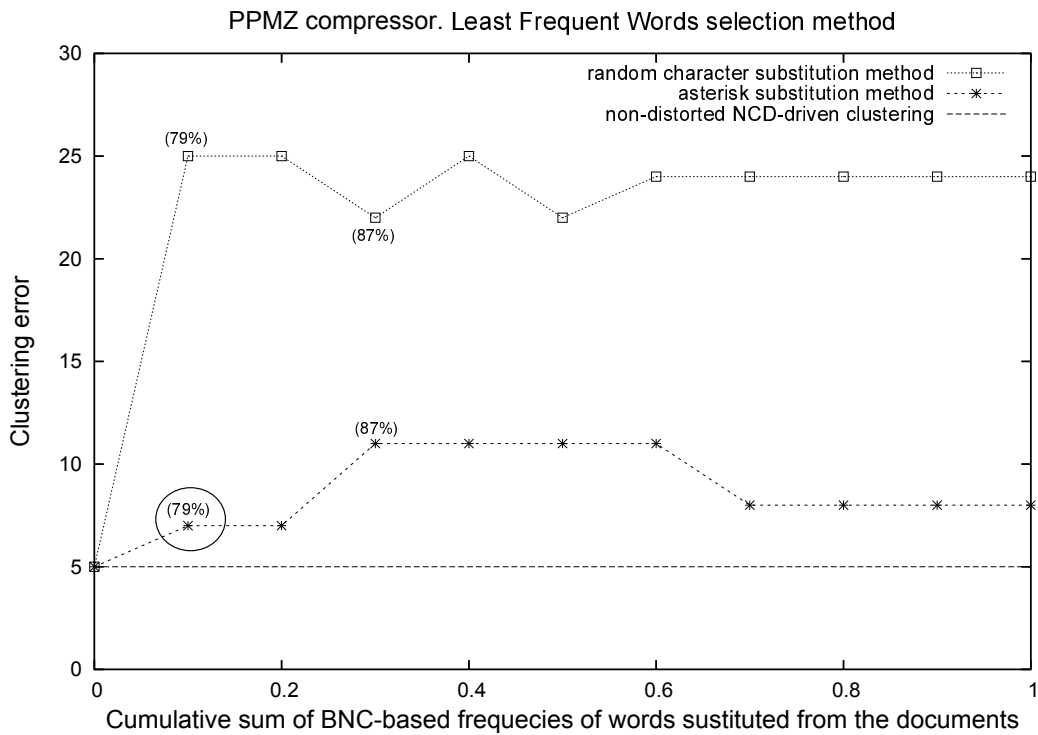}
\vspace{-0.2cm} \caption[Books. PPMZ compressor. LFW selection
method.]{Books. PPMZ compressor. LFW selection method. The
clustering error gets worse even when the \emph{MFW selection
method} is used.} \label{TKDE. Fig:books-clustering-error-ppmz-lfw}
\end{figure}

Analyzing Figs \ref{TKDE. Fig:books-clustering-error-ppmz-mfw},
\ref{TKDE. Fig:books-clustering-error-ppmz-rw}, and \ref{TKDE.
Fig:books-clustering-error-ppmz-lfw} one can observe that the
\emph{asterisk substitution method} is always better than the
\emph{random character substitution method}. This was to be expected
because substituting a word with random characters adds noise to the
documents, and therefore most likely increases the Kolmogorov
complexity of the documents and makes the clustering worse.

\clearpage

One can observe that the best clustering results correspond to the
\emph{MFW selection method} -see Fig \ref{TKDE.
Fig:books-clustering-error-ppmz-mfw}-, the worst results correspond
to the \emph{LFW selection method} -see Fig \ref{TKDE.
Fig:books-clustering-error-ppmz-lfw}-, and the results corresponding
to the \emph{RW selection method} are maintained in between them
-see Fig \ref{TKDE. Fig:books-clustering-error-ppmz-rw}-. This
behavior supports one of the main contributions of Luhn to automatic
text analysis \cite{Luhn58}, which states: ``the frequency of word
occurrence in an article furnishes a useful measurement of word
significance''. The Zipf's Law states that the product of the
frequency of use of words and the rank order is approximately
constant \cite{Zipf49,Zipf35}. Luhn used the Zipf's law as a null
hypothesis to enable him to specify two cut-offs, an upper and a
lower, that exclude non-significant words. The only problem is that
there is no formula which gives their values. They have to be
established by trial and error \cite{Rijsbergen79}.

As noted previously, looking at the points highlighted inside a
circle in Figs \ref{TKDE. Fig:books-complexity-all-ppmz}, \ref{TKDE.
Fig:books-clustering-error-ppmz-mfw}, \ref{TKDE.
Fig:books-clustering-error-ppmz-rw}, and \ref{TKDE.
Fig:books-clustering-error-ppmz-lfw} one can observe that although
the complexity values and the percentages of removed words are
similar for these points, there is a significant difference in terms
of clustering error. Consequently, one can realize that not only the
substitution method is important, but also the word selection
method. Thus, it has been shown that the best way to distort the
documents is combining the \emph{MFW selection method} and the
\emph{asterisk substitution method}.

An alternative way of showing this, is by comparing the dendrogram
obtained with no distortion with the dendrogram obtained applying
the distortion that achieves a clustering error of 0, that is, a
distortion of 0.9 using the \emph{MFW selection method} and the
\emph{asterisk substitution method}. These dendrograms are shown in
Figs \ref{TKDE. Fig:dendro-no-distortion} and \ref{TKDE.
Fig:dendro-perfect}.

Analyzing Fig \ref{TKDE. Fig:dendro-no-distortion} one can notice
that the books by Edgar Allan Poe -EAP- and Alexander Pope -AP- are
not correctly clustered when the non-distorted books are used.
Examining Fig \ref{TKDE. Fig:dendro-perfect} one can easily observe
that these errors are solved, that is, the books by Edgar Allan Poe
-EAP- and Alexander Pope -AP- are correctly clustered, exactly the
same as the rest of the books. That is why the clustering error that
corresponds to Fig \ref{TKDE. Fig:dendro-perfect} is 0.

\begin{figure}[h]
  \includegraphics[width=14cm]{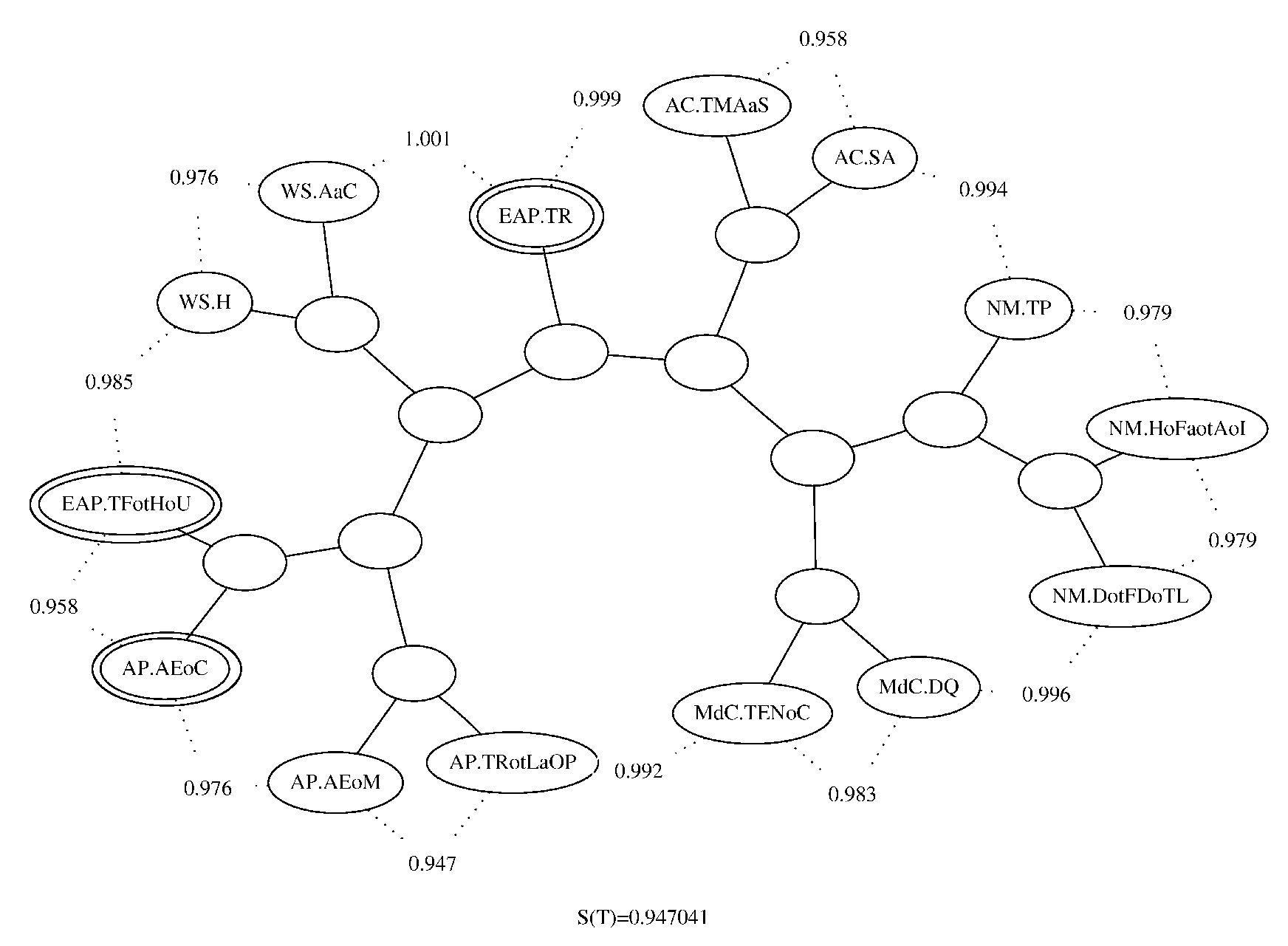}
  \centering
\caption[Dendrogram obtained with no distortion.]{Dendrogram
obtained with no distortion. The nodes incorrectly clustered are
highlighted inside a circle. This dendrogram corresponds to the
results shown in Fig \ref{TKDE.
Fig:books-clustering-error-ppmz-mfw}.} \label{TKDE.
Fig:dendro-no-distortion}
\end{figure}

\clearpage

\begin{figure}[h]
  \includegraphics[width=14cm]{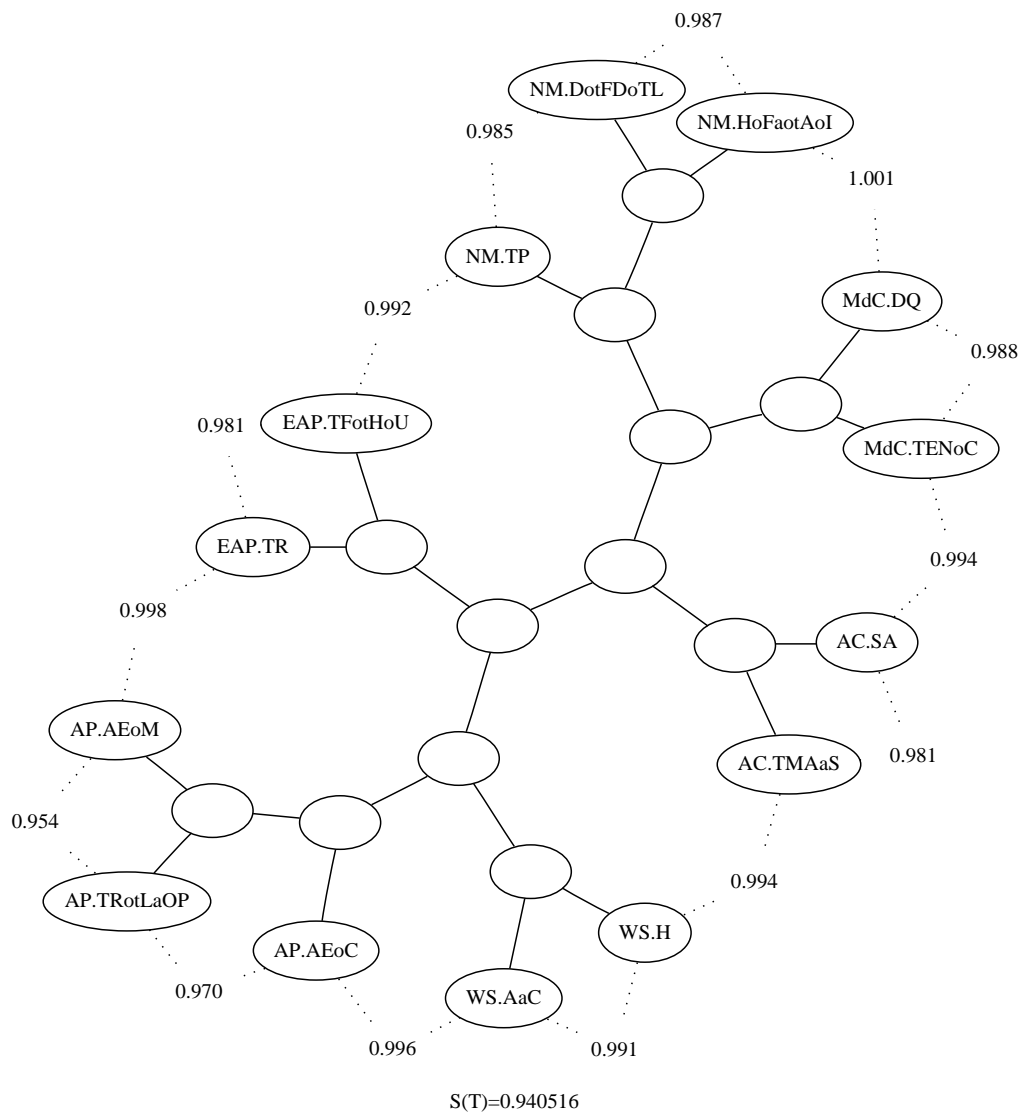}
  \centering
\caption[Perfect dendrogram.]{Dendrogram obtained for a distortion
of 0.9 using the \emph{MFW selection method} and the \emph{asterisk
substitution method}. This dendrogram corresponds to the results
shown in Fig \ref{TKDE. Fig:books-clustering-error-ppmz-mfw}. In
this case, no node is highlighted inside a circle, because they are
all correctly clustered.} \label{TKDE. Fig:dendro-perfect}
\end{figure}

\clearpage

\subsection{Results for the \emph{asterisk substitution method}}
\label{TKDE. Results for the asterisk substitution method}

Although the graphical results obtained for the rest of the datasets
can be seen in Appendix \ref{Appendix Detailed Experimental
Results}, this subsection shows the summary of the results for every
compression algorithm and every dataset when the \emph{asterisk
substitution method} is applied.

Three tables summarize the results. Each table corresponds to a
selection method. In all the tables, each column corresponds to a
specific dataset, and each row corresponds to a specific compression
algorithm. The tables show for every dataset and every compression
algorithm three different clustering errors -Err- and the cumulative
sum of frequencies where these clustering errors are obtained
-Freq-.

The clustering error values shown in the tables are:
\begin{itemize}
 \item \underline{NoD}: The clustering error obtained with
\underline{no d}istortion, that is, the clustering error obtained
clustering the original documents.
 \item \underline{Min}: The \underline{min}imum clustering error
obtained.
 \item \underline{Max}: The \underline{max}imum clustering error
obtained.
\end{itemize}

Note that the non-distorted clustering error is not taken into
account to create the tables, because it is obvious that the
clustering error corresponding to the cumulative sum of frequencies
of 0 will always be the same, and the purpose of this study is to
analyze the effects of the distortion. Therefore, only the results
obtained from 0.1 to 1.0 are considered to create the tables.

In all the tables, the results that improve the non-distorted
clustering error are marked with a double-box, and with a simple-box
the results that maintain this clustering error. These boxes are
included to focus the attention on the clustering error improvement.

Table \ref{tableDecOrderFrec} has many boxes because when the
\emph{MFW selection method} is applied, the best results are
obtained. This is due to the fact that using this word selection
method, the clustering is improved or maintained for every
repository and every compression algorithm. These results are
consistent with the ones shown in the clustering error figures,
where it can be observed that the best clustering results correspond
to the combination of the \emph{MFW selection method} and the
\emph{asterisk substitution method}.

\begin{figure}[hb]
 \includegraphics[width=13cm]{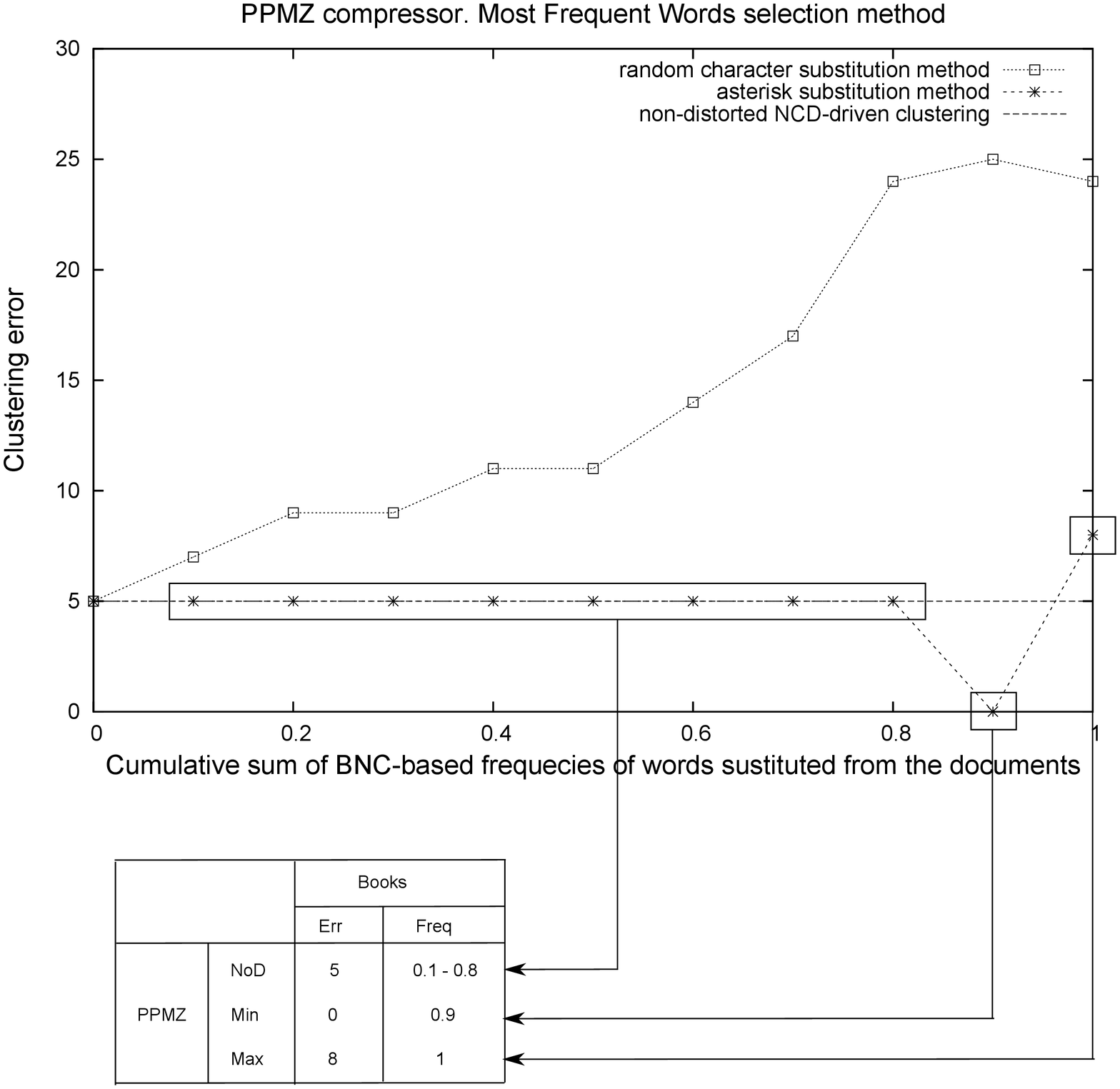}
 \centering
\caption{Understanding the tables.}
 \label{TKDE. Fig:understanding-the-tables}
\end{figure}

Fig \ref{TKDE. Fig:understanding-the-tables} has been created to
help and understand the tables. It compares Fig \ref{TKDE.
Fig:books-clustering-error-ppmz-mfw} and Table
\ref{tableDecOrderFrec} with the aim of showing where the data in
the table are from.

\begin{table}[h]
\centering \caption[MFW selection method] { \emph{MFW selection
method}. The clustering error obtained with no distortion, the
minimum clustering error, and the maximum clustering error are
shown. The frequencies when such clustering errors are obtained are
shown as well. The results that improve the clustering error
obtained with no distortion are highlighted inside a double-box. The
results that maintain the non-distorted clustering error are
highlighted inside a simple-box.} \label{tableDecOrderFrec}
\vspace{0.2cm} \scriptsize{
\begin{tabular}{|l|l|c|c|c|c|c|c|c|c|}
  \hline
  \multicolumn{2}{|c|}{} & \multicolumn{2}{c|}{Books} & \multicolumn{2}{c|}{UCI-KDD} & \multicolumn{2}{c|}{MedlinePlus} & \multicolumn{2}{c|}{IMDB}\\
  \cline {3-10}
  \multicolumn{2}{|c|}{} & Err          & Freq & Err      & Freq &  Err          & Freq & Err           & Freq \\\hline
        & NoD        & 4            &  0.1-0.7   & 0        &  0.1-0.9   & 14            & 0.1-0.6    & 18            & 0.1-0.5   \\
  LZMA  & Min             & \doublebox{2}&  0.8,0.9   & \fbox{0} &  0.1-0.9   & \doublebox{10}& 0.7-0.8    & \doublebox{4} & 0.9   \\
        & Max            & 9            &  1       & 8        &  1   & 26            & 1        & 22            & 1     \\
  \hline
       & NoD         & 5            & 0.1-0.8    & 0        &  0.1-0.8   & 14            & 0.1-0.4,0.9& 0             & 0.3-0.7,0.9 \\
  PPMZ & Min             & \doublebox{0}& 0.9        & \fbox{0} &  0.1-0.8   & \doublebox{4} & 0.7        & \framebox{0}  & 0.3-0.7,0.9 \\
       & Max             & 8            & 1        & 21       &  1       & 34            & 1        & 12            & 1      \\
  \hline
        & NoD        & 7            & 0.1-0.6& 0        & 0.1-0.6    & 14            & 0.1-0.4,0.6& 0             & 0.1-0.6    \\
  BZIP2 & Min            & \doublebox{5}& 0.7        & \fbox{0} & 0.1-0.6    & \doublebox{10}& 0.5,0.7-0.8& \framebox{0}  & 0.1-0.6    \\
        & Max            & 9           & 0.9        & 15       & 1        & 24            &  1       & 12            & 1      \\
  \hline
\end{tabular}}
\end{table}

\begin{table}[h]
\centering \caption[RW selection method]{ \emph{RW selection
method}. The codification is the same as explained for Table
\ref{tableDecOrderFrec}.} \label{tableWordsRandom} \vspace{0.2cm}
\scriptsize{
\begin{tabular}{|l|l|c|c|c|c|c|c|c|c|}
  \hline
  \multicolumn{2}{|c|}{} & \multicolumn{2}{c|}{Books} & \multicolumn{2}{c|}{UCI-KDD} & \multicolumn{2}{c|}{MedlinePlus} & \multicolumn{2}{c|}{IMDB}\\
  \cline {3-10}
  \multicolumn{2}{|c|}{}  & Err         & Freq  & Err         & Freq &  Err          & Freq & Err           & Freq \\\hline
        & NoD         & 4           &  0.1-0.6    & 0           &  0.1-0.6   & 14              & -          & 18       & -      \\
  LZMA  & Min              & \framebox{4}&  0.1-0.6    & \framebox{0}&  0.1-0.6   & \doublebox{13.8}& 0.3        & \doublebox{13.3}& 0.7     \\
        & Max             & 9           &  1        & 8         &  1       & 28              & 1        & 22       & 1    \\
  \hline
       & NoD          & 5           & 0.1-0.2     & 0           &  0.1-0.3   & 14              & -          & 0        & -      \\
  PPMZ & Min              & \framebox{5}& 0.1-0.2     & \framebox{0}&  0.1-0.3   & 14.2            & 0.3,0.5    & 5.2      & 0.2     \\
       & Max              & 10.7        & 0.9         & 21          &  1       & 34              & 1        & 12       & 1    \\
  \hline
        & NoD         & 7           & -           & 0           & -          & 14              & -          & 0        & -      \\
  BZIP2 & Min             & \doublebox{5.9}& 0.8      & 1.6         & 0.2        & 14.6            & 0.3        & 8.4      & 0.1     \\
        & Max             & 10.5        & 0.1         & 17.2        & 1        & 24              &  1       & 15.3     & 0.4     \\
  \hline
\end{tabular}}
\end{table}

\begin{table}[h]
\centering \caption[LFW selection method]{ \emph{LFW selection
method}. The codification is the same as explained for Table
\ref{tableDecOrderFrec}.} \label{tableIncreasingFrec} \vspace{0.2cm}
\scriptsize{
\begin{tabular}{|l|l|c|c|c|c|c|c|c|c|}
  \hline
  \multicolumn{2}{|c|}{} & \multicolumn{2}{c|}{Books} & \multicolumn{2}{c|}{UCI-KDD} & \multicolumn{2}{c|}{MedlinePlus} & \multicolumn{2}{c|}{IMDB}\\
  \cline {3-10}
  \multicolumn{2}{|c|}{}  & Err         & Freq  & Err         & Freq &  Err          & Freq & Err           & Freq \\\hline
        & NoD        & 4        &  -              & 0        & 0.1-0.2,0.5-1    & 14       & -          & 18       & 0.1,0.5  \\
  LZMA & Min             & 9        & 0.1-0.4,0.6-1 & \framebox{0}& 0.1-0.2,0.5-1 & 20       & 0.1-0.2    & \doublebox{10}& 0.6     \\
        & Max            & 12       & 0.5             & 8        & 1            & 28       & 0.5,0.7-1& 22       & 0.2,0.4,0.7-1 \\
  \hline
       & NoD          & 5        & -               & 0        & -                  & 14       & -          & 0        & -      \\
  PPMZ & Min             & 7        & 0.1-0.2         & 15       &  0.3               & 20       & 0.1        & 8        & 0.1-0.3  \\
       & Max             & 11       & 0.3-0.6         & 21       & 0.5,0.7-1        & 34       & 0.7-1    & 12       & 0.6-1 \\
  \hline
        & NoD        & 7        & -               & 0        & -                  & 14       & -          & 0        & -      \\
  BZIP2 & Min            & \doublebox{4}& 0.3-0.4     & 8        & 0.3                & 16       & 0.1        & 10       & 0.6     \\
        & Max            & 9        & 0.1-0.2         & 16       & 0.1,0.6            & 26       & 0.3-0.4    & 32       & 0.1     \\
  \hline
\end{tabular}}
\end{table}

\clearpage

Let us analyze Fig \ref{TKDE. Fig:understanding-the-tables}. The
clustering error obtained when clustering the original documents,
that is, the non-distorted clustering error, is 5. That is why there
is a 5 in the cell that corresponds to the clustering error ``Err''
obtained with no distortion ``NoD''. It can be observed that this
clustering error remains constant from points 0.1 to 0.8. That is
the reason why the cell that corresponds to the cumulative sum of
frequencies ``Freq'' for the non-distorted results ``NoD'' is
0.1-0.8.

Similarly, the best clustering error obtained using the
\emph{asterisk substitution method} is 0, as can be seen looking at
the point 0.9 in Fig \ref{TKDE. Fig:understanding-the-tables}.
Therefore, the row that shows the minimum clustering error obtained
-``Min''- shows a 0 in the cell that corresponds to the error
``Err'' and a 0.9 in the cell that corresponds to the cumulative sum
of frequencies where this error is obtained ``Freq''.

Finally, the maximum clustering error obtained for the
\emph{asterisk substitution method} is 8. This error is obtained for
a cumulative sum of frequencies of 1.0. Therefore, the table shows
an 8 in the cell that corresponds to the maximum ``Max'' clustering
error ``Err'' obtained, and it contains a 1 in the cell that
corresponds to the cumulative sum of frequencies ``Freq'' for the
maximum clustering error ``Max''.

\subsection{Synopsis of all the obtained results}
\label{TKDE. Synopsis of results for all the datasets}

Finally, this subsection summarizes all the obtained results in the
form of four tables, one for each dataset. Table
\ref{tableBooksErrorMedio} corresponds to the Books dataset, Table
\ref{tableUCIKDDErrorMedio} corresponds to the UCI-KDD dataset,
Table \ref{tableMedlineErrorMedio} corresponds to the MedlinePlus
dataset, and Table \ref{tableIMDBErrorMedio} corresponds to the IMDB
dataset. The tables show the average clustering error for all the
compression algorithms, all the word selection methods, and all the
substitution methods. The clustering error is averaged as follows:

\begin{equation}
\textrm{Average CE} = \frac
{\sum\limits_{\forall~\tiny{\textrm{distortion}}}
\textrm{CE(distortion)}}{\#~\textrm{distortions}}
\end{equation}

Where CE means ``Clustering Error'', and the possible distortions go
from 0.1 to 1.0. Therefore the number of distortions is always 10.

Analyzing these tables, one can reach the same conclusions as by
analyzing the clustering error curves. However, summarizing the
results by calculating the average clustering error helps to better
see the differences between all the experiments carried out.
Therefore, the tables shown in this subsection constitute an
alternative and easier way of presenting the results obtained in the
experiments developed in this chapter of the thesis.

Firstly, it can be observed that the average clustering error
obtained using the \emph{MFW selection method} is always less than
the one obtained using the \emph{RW selection method}, and this
latter is always less than the one obtained using the \emph{LFW
selection method}.

Secondly, one can observe that in general, the average clustering
error obtained applying the \emph{asterisk substitution method} is
less than the one obtained using the \emph{random character
substitution method}.

Finally, one can see that the best clustering error is obtained for
a different compression algorithm depending on the dataset used.
This could be due to the fact that each dataset is composed of texts
of a different nature. The next chapter of the thesis tries to
investigate the reasons why the non-distorted clustering error can
be improved combining the \emph{MFW selection method} and the
\emph{asterisk substitution method}.

\begin{table}[h]
\centering \caption[Books dataset. Average clustering error]{Books
dataset. Average clustering error.} \label{tableBooksErrorMedio}
\vspace{0.2cm} \normalsize{
\begin{tabular}{|c|l|c|c|c|}
  \hline
    \multicolumn{2}{|c|}{}                   & MFW   & RW    & LFW \\
  \hline
  \multirow{2}{*}{LZMA}  & Asterisks         & 4.10  & 5.51  & 9.30 \\
  \cline {2-5}
                         & Random characters & 14.80 & 23.42 & 28.00 \\
  \hline
  \multirow{2}{*}{PPMZ}  & Asterisks         & 4.80  & 7.44  & 9.00 \\
  \cline {2-5}
                         & Random characters & 15.10 & 21.83 & 23.90 \\
  \hline
  \multirow{2}{*}{BZIP2} & Asterisks         & 7.90  & 7.99  & 7.00 \\
  \cline {2-5}
                         & Random characters & 21.10 & 26.39 & 31.70 \\
  \hline
\end{tabular}}
\end{table}

\begin{table}[h]
\centering \caption[UCI-KDD dataset. Average clustering
error]{UCI-KDD dataset. Average clustering error.}
\label{tableUCIKDDErrorMedio} \vspace{0.2cm} \normalsize{
\begin{tabular}{|c|l|c|c|c|}
  \hline
    \multicolumn{2}{|c|}{}                   & MFW   & RW    & LFW \\
  \hline
  \multirow{2}{*}{LZMA}  & Asterisks         & 0.80  & 0.94  & 1.20 \\
  \cline {2-5}
                         & Random characters & 8.10  & 14.51 & 25.40 \\
  \hline
  \multirow{2}{*}{PPMZ}  & Asterisks         & 2.30  & 7.62  & 18.90 \\
  \cline {2-5}
                         & Random characters & 7.70  & 14.76 & 26.80 \\
  \hline
  \multirow{2}{*}{BZIP2} & Asterisks         & 2.00  & 8.55  & 13.70 \\
  \cline {2-5}
                         & Random characters & 15.80 & 22.58 & 32.50 \\
  \hline
\end{tabular}}
\end{table}

\begin{table}[h]
\centering \caption[MedlinePlus dataset. Average clustering
error]{MedlinePlus dataset. Average clustering error.}
\label{tableMedlineErrorMedio} \vspace{0.2cm} \normalsize{
\begin{tabular}{|c|l|c|c|c|}
  \hline
    \multicolumn{2}{|c|}{}                   & MFW   & RW    & LFW \\
  \hline
  \multirow{2}{*}{LZMA}  & Asterisks         & 14.40 & 16.98 & 25.40 \\
  \cline {2-5}
                         & Random characters & 16.40 & 19.28 & 26.80 \\
  \hline
  \multirow{2}{*}{PPMZ}  & Asterisks         & 13.80 & 18.58 & 28.60 \\
  \cline {2-5}
                         & Random characters & 16.20 & 19.62 & 27.60 \\
  \hline
  \multirow{2}{*}{BZIP2} & Asterisks         & 14.20 & 17.88 & 23.00 \\
  \cline {2-5}
                         & Random characters & 19.00 & 24.20 & 30.20 \\
  \hline
\end{tabular}}
\end{table}

\begin{table}[h]
\centering \caption[IMDB dataset. Average clustering error]{IMDB
dataset. Average clustering error.} \label{tableIMDBErrorMedio}
\vspace{0.2cm} \normalsize{
\begin{tabular}{|c|l|c|c|c|}
  \hline
    \multicolumn{2}{|c|}{}                   & MFW   & RW    & LFW \\
  \hline
  \multirow{2}{*}{LZMA}  & Asterisks         & 13.40 & 16.24 & 19.40 \\
  \cline {2-5}
                         & Random characters & 20.70 & 23.79 & 31.90 \\
  \hline
  \multirow{2}{*}{PPMZ}  & Asterisks         & 2.60  & 8.18  & 10.40 \\
  \cline {2-5}
                         & Random characters & 10.70 & 19.51 & 25.10 \\
  \hline
  \multirow{2}{*}{BZIP2} & Asterisks         & 2.60  & 12.05 & 17.40 \\
  \cline {2-5}
                         & Random characters & 20.50 & 25.09 & 33.50 \\
  \hline
\end{tabular}}
\end{table}

\clearpage

\section{Summary and Conclusions}
\label{TKDE. Summary}

This chapter of the thesis has taken a small step towards
understanding both the nature of textual data and the nature of
compression distances. This has been accomplished by performing an
experimental evaluation of the impact that several kinds of word
removal have on the NCD-based text clustering.

In terms of implementation, the CompLearn Toolkit \cite{complearn},
which implements the clustering algorithm described in
\cite{Cilibrasi05,Li04}, has been used to carry out the experiments.

Six different distortion techniques have been evaluated. They are
pairwise combinations of two factors: \emph{word selection method}
and \emph{substitution method}. There are three word selection
methods, depending on what words are chosen to be removed from the
documents: \emph{Most Frequent Words -MFW- selection method},
\emph{Least Frequent Words -LFW- selection method} and \emph{RW
selection method}. There are two substitution methods, depending on
the way in which the words are removed from the documents:
\emph{random character substitution method} and \emph{asterisk
substitution method}.

The NCD-driven clustering algorithm has been applied over four
different datasets repeating the clustering three times using each
time a different compression algorithm to calculate the NCD: PPMZ,
LZMA and BZIP2.

In addition, in order to gain an insight into how the information is
decreased when the distortion techniques are applied, the Kolmogorov
complexity of the documents has been estimated based on the concept
that data compression is an upper bound for it.

The experimental results have shown that the combination of the
\emph{selection method} and the \emph{substitution method} is the
key factor. Substituting the most frequent words using the
\emph{asterisk substitution method} is always the best option to
maintain the most relevant information. In this case, the documents
complexity estimation is slowly reduced and therefore the clustering
error remains stable even though a considerable percentage of words
were substituted from the documents. Moreover, its worth mentioning
that, using the best distortion technique, even the non-distorted
clustering error can be improved.

Analyzing Tables \ref{tableDecOrderFrec}, \ref{tableWordsRandom},
and \ref{tableIncreasingFrec} one can observe that the best results
are obtained using the LZMA compression algorithm. This could be due
to the fact that this compressor captures the contextual information
because of its design. Section \ref{State. Compression Algorithms}
explains the implementation details of all the compressors used in
this thesis. Here a summarized description of the compression
algorithms used in this thesis is given.

The LZMA algorithm codifies the symbols using as a dictionary part
of the input stream previously seen. The method is based on a
sliding window that the encoder shifts as the strings of symbols are
being encoded. The window is divided in two parts:
\begin{itemize}
\item
  \emph{Search buffer}: Part of the input stream previously seen. This is the current
  dictionary.
\item
  \emph{Look-ahead buffer}: The text yet to be encoded.
\end{itemize}
It is important to point out that practical implementations of this
method use really long \emph{search buffers} of thousands of bytes
long, and small \emph{look-ahead buffers} of tens of bytes long
\cite {Salomon2004}.

Therefore, this compression algorithm takes the contextual
information into account because it uses part of the input stream
previously codified to codify the data that have yet to be codified.

The PPMZ is an adaptive statistical compression algorithm which is
based on an encoder that maintains a statistical model of the text.
It considers the N symbols preceding the symbol being processed.
Therefore, this compressor takes the contextual information into
account. However, the main difference, in this respect, between the
PPMZ and the LZMA is that the former only considers about 10 symbols
preceding the symbol being codified, whereas the latter considers
thousands of symbols preceding it.

The BZIP2 is a block-sorting compressor that uses different
techniques to compress the data. Some of these techniques transform
the input by moving the symbols being encoded. In particular, the
Burrows-Wheeler Transform, and the Move-To-Front transform behave
that way. Therefore, this compression algorithm destroys the
contextual information, since it shuffles the symbols in the
compression process.

As a result of all the above, the next chapter of the thesis, which
analyzes the relevance of the contextual information, only uses the
LZMA to calculate the NCD. However, a deeper study of the effects
that the loss or the maintenance of the contextual information have
on the accuracy of the clustering results, using different
compression algorithms, constitutes a future work.

Summarizing, three main contributions have been presented in this
chapter. First, new text representations have been analyzed and
studied with the aim of giving new insights for the evaluation of
the NCD. Second, a technique which reduces the complexity of the
texts while preserving most of the relevant information has been
presented. Third, experimental evidence of how to fine-tune the
representation of the documents, in order to obtain better
NCD-driven clustering results, has been provided.


\chapter{Relevance of contextual information}
\label{Chapter: Relevance of the contextual information}

The previous chapter experimentally evaluates the impact that
several word removal techniques have on compression-based text
clustering. It shows that the application of a specific distortion
technique can improve the non-distorted clustering results. Since
that technique implies, not only the removal of words, but also
the maintenance of the previous text structure, exploring the
relevance of both factors becomes necessary in order to better
understand the results. This chapter explores precisely that.

The main contributions of this chapter of the thesis can be briefly
summarized as follows:

\begin{itemize}
\item
Experimental evaluation of the relevance that the contextual
information has in compression-based text clustering, in a word
removal scenario.

\item
New perspectives for the evaluation and explanation of the behavior
of compression distances, in relation to contextual information.
\end{itemize}

The chapter is structured as follows. Section \ref{ESWA.
Distortion Techniques} describes the distortion techniques
explored. Section \ref{ESWA. Experimental Setup} describes the
experimental setup. Section \ref{ESWA. Experimental Results}
gathers and analyzes the obtained results. Finally, Section
\ref{ESWA. Summary} summarizes the conclusions drawn from the
experiments carried out in this chapter of the thesis.

\clearpage

\section{Distortion Techniques}
\label{ESWA. Distortion Techniques}

Four different distortion techniques are explored in this chapter of
the thesis. One of them was analyzed in the previous chapter, and
consists of incrementally removing the most frequent words in the
English language, as described in depth in Section \ref{TKDE.
Distortion Techniques}.

In order to maintain the length and the place of appearance of the
removed words, instead of simply erasing them, their characters are
replaced using asterisks. The marks that the words leave on the
texts after the distortion is precisely what is called contextual
information throughout the thesis. In this chapter, this technique
is called \emph{Original sorting} distortion technique because no
random sorting is carried out after substituting the words with
asterisks. The rest of the distortion techniques consist of first
applying that distortion technique, and then randomly sorting
different parts of the distorted texts. The description of the new
distortion techniques is as follows:

\begin{itemize}
\item \emph{Randomly sorting contextual information}: after
replacing the words using asterisks, the strings of asterisks are
randomly sorted. That is, the remaining words are maintained in
their original places of appearance, while the removed words are
not. It is important to note that each string of asterisks is
treated as a whole. That is, if a word such as ``hello'' is replaced
by ``*****'' these asterisks always remain together. This method is
created in order to study whether the structure of the contextual
information is relevant or not. Fig \ref{ESWA. Fig:WordRemoval}(c)
represents a sample of text distorted using this technique.

\item \emph{Randomly sorting remaining words}: after replacing the
words using asterisks, the remaining words are randomly sorted. That
is, the contextual information is maintained, while the remaining
words structure is not. This method is created to evaluate the
importance of the structure of the remaining words. A visual
representation of the effects of applying this technique can be seen
in Fig \ref{ESWA. Fig:WordRemoval}(d).

\item \emph{Randomly sorting everything}: after replacing the words
using asterisks, both the strings of asterisks and the remaining
words are randomly sorted. It should be pointed out that, in this
case, the strings of asterisks are randomly sorted as a whole too.
This method is created as a control experiment. See Fig \ref{ESWA.
Fig:WordRemoval}(e) for a visual representation of this technique's
effects.
\end{itemize}

\begin{figure}
\footnotesize{
\begin{tabular}{p{2.45cm} p{3.9cm} p{2.45cm}}
\cline{2-2}
& \multicolumn{1}{|p{3.9cm}|}{\texttt{~~~~~~~~~......~~~~~~~~~}}  \\
& \multicolumn{1}{|p{3.9cm}|}{\texttt{anemia is a condition in which
the body does not have enough healthy red blood
cells. red blood cells provide oxygen to body tissues.}}  \\
& \multicolumn{1}{|p{3.9cm}|}{\texttt{~~~~~~~~~......~~~~~~~~~}}  \\
\cline{2-2}
& & \\
& (a) Original text & \\
& & \\
\end{tabular}
\begin{tabular}{p{3.9cm} p{1cm} p{3.9cm}}
\cline{1-1} \cline{3-3} \multicolumn{1}{|p{3.9cm}|}{\texttt{~~~~~~~~~......~~~~~~~~~}} & & \multicolumn{1}{|p{3.9cm}|}{\texttt{~~~~~~~~~......~~~~~~~~~}} \\
\multicolumn{1}{|p{3.9cm}|}{\texttt{anemia ** * ********* ** *****
*** **** **** *** **** ****** healthy *** ***** ***** *** *****
***** ******* oxygen ** **** tissues}} & & \multicolumn{1}{|p{3.9cm}|}{\texttt{anemia ***** ***** ** ***** ***** *** **** **** ****** ****
*** healthy ** ** **** * *** *** ***** oxygen ********* *******
tissues}} \\
\multicolumn{1}{|p{3.9cm}|}{\texttt{~~~~~~~~~......~~~~~~~~~}} & & \multicolumn{1}{|p{3.9cm}|}{\texttt{~~~~~~~~~......~~~~~~~~~}} \\
\cline{1-1} \cline{3-3}
& & \\
(b) Original sorting & & (c) Randomly sorting \\
  & & contextual information \\
& & \\
\cline{1-1} \cline{3-3} \multicolumn{1}{|p{3.9cm}|}{\texttt{~~~~~~~~~......~~~~~~~~~}} & & \multicolumn{1}{|p{3.9cm}|}{\texttt{~~~~~~~~~......~~~~~~~~~}} \\
\multicolumn{1}{|p{3.9cm}|}{\texttt{oxygen ** * ********* ** *****
*** **** **** *** **** ****** anemia *** ***** ***** *** ***** *****
******* tissues ** **** healthy}} & &
\multicolumn{1}{|p{3.9cm}|}{\texttt{****** ** **** * *** ***** ******* **** healthy ** ***** *** tissues ***** **** ***** *** *** anemia **** ***** ********* oxygen **}} \\
\multicolumn{1}{|p{3.9cm}|}{\texttt{~~~~~~~~~......~~~~~~~~~}} & & \multicolumn{1}{|p{3.9cm}|}{\texttt{~~~~~~~~~......~~~~~~~~~}} \\
\cline{1-1} \cline{3-3}
& & \\
(d) Randomly sorting & & (e) Randomly sorting \\
remaining words & & everything\\
& & \\
\end{tabular}
\centering \caption{Text distortion techniques.} \label{ESWA.
Fig:WordRemoval}}
\end{figure}

The graphical differences among the four distortion techniques
explored in this chapter can be seen in Fig \ref{ESWA.
Fig:WordRemoval}. This figure clarifies the way in which each
distortion technique modifies the texts.

\section{Experimental Setup}
\label{ESWA. Experimental Setup}

The experiments have been carried out using the same
compression-based clustering algorithm used in the first part of the
thesis. The detailed description of this algorithm can be found in
Section \ref{TKDE. NCD-based Text Clustering}. In this part of the
thesis, only the LZMA compression algorithm is used to perform the
NCD-driven document clustering because the best results were
obtained using it in the previous chapter.

\subsection{Datasets}
\label{ESWA. Datasets}

Five different datasets composed of texts written in English have
been used in the experiments. Although the detailed description of
the datasets can be found in Appendix \ref{Appendix Datasets}, a
summarized description of them can be found here:

\begin{itemize}
\item
\textbf{Books dataset}: Fourteen classical books from universal
literature, to be clustered by author.

\item
\textbf{UCI-KDD dataset}: Sixteen messages from a newsgroup, to be
clustered by topic.

\item
\textbf{MedlinePlus dataset}: Twelve documents from the MedlinePlus
repository, to be clustered by topic.

\item
\textbf{IMDB dataset}: Fourteen plots of movies from the Internet
Movie Data Base -IMDB- to be clustered by saga.

\item
\textbf{SRT-serial dataset}: Sixty-nine scripts of different serials
which have been obtained from \cite{srt}, to be clustered by serial.
\end{itemize}

\section{Experimental Results}
\label{ESWA. Experimental Results}

A figure is shown for every dataset. In each figure, the clustering
error obtained applying the \emph{Original sorting} distortion
technique is plotted in the panel (a). In addition, in order to ease
the comparison between this technique and the new distortion
techniques, this curve is also plotted in the panels (b), (c) and
(d), which correspond to the \emph{Randomly sorting contextual
information}, \emph{Randomly sorting remaining words}, and
\emph{Randomly sorting everything} distortion techniques,
respectively. Since these distortion techniques are based on
randomly sorting different parts of the texts, the experiments have
been repeated several times, and the mean and the standard deviation
of the clustering error are depicted in panels (b), (c) and (d).

In all the plots, the values on the vertical axis correspond to the
obtained clustering error, while the values on the horizontal axis
correspond to the cumulative sum of frequencies of the words that
are removed from the texts.

\begin{figure}
\begin{tabular}{cc}
\includegraphics[angle=270,width=6.5cm]{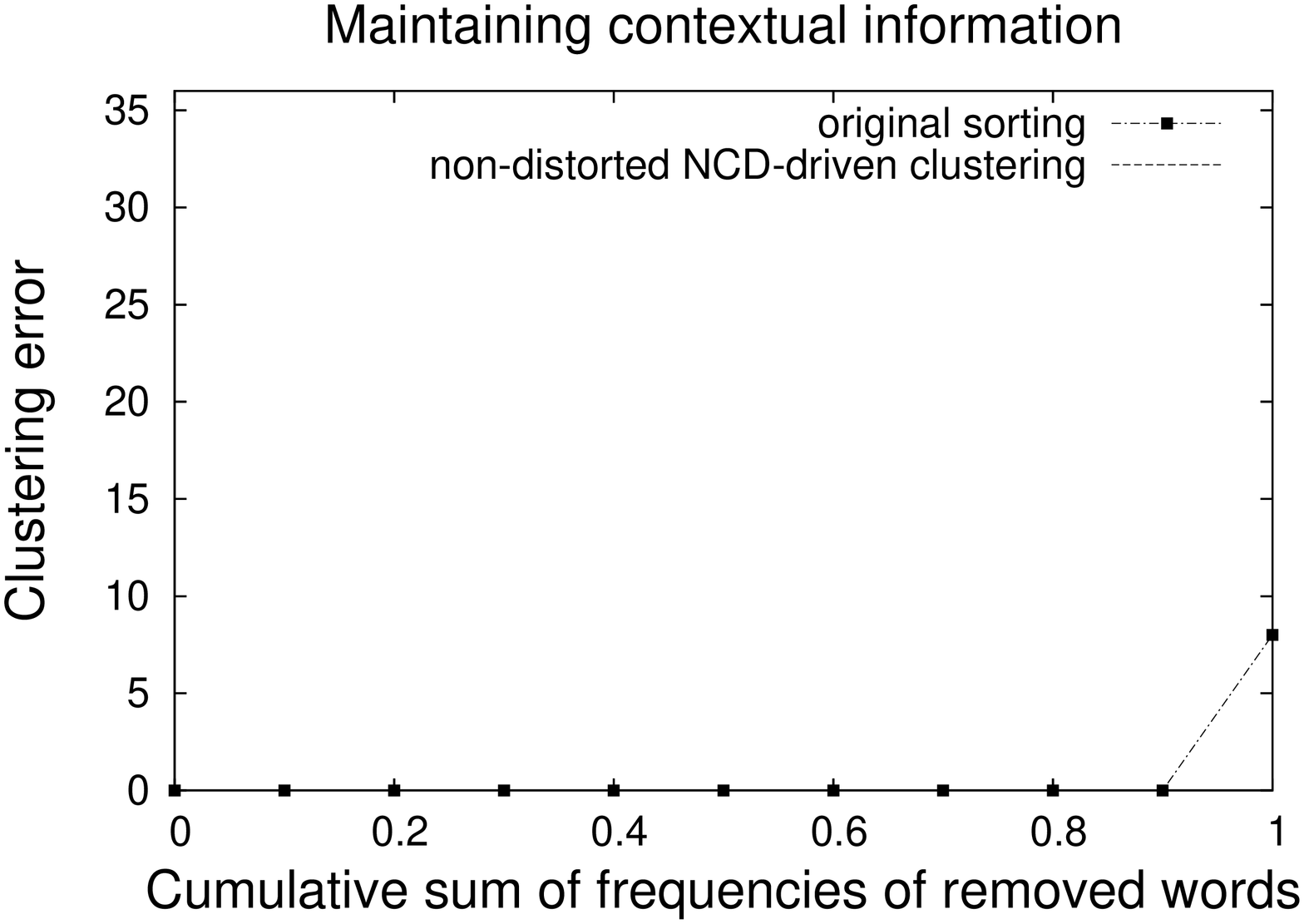} &
\includegraphics[angle=270,width=6.5cm]{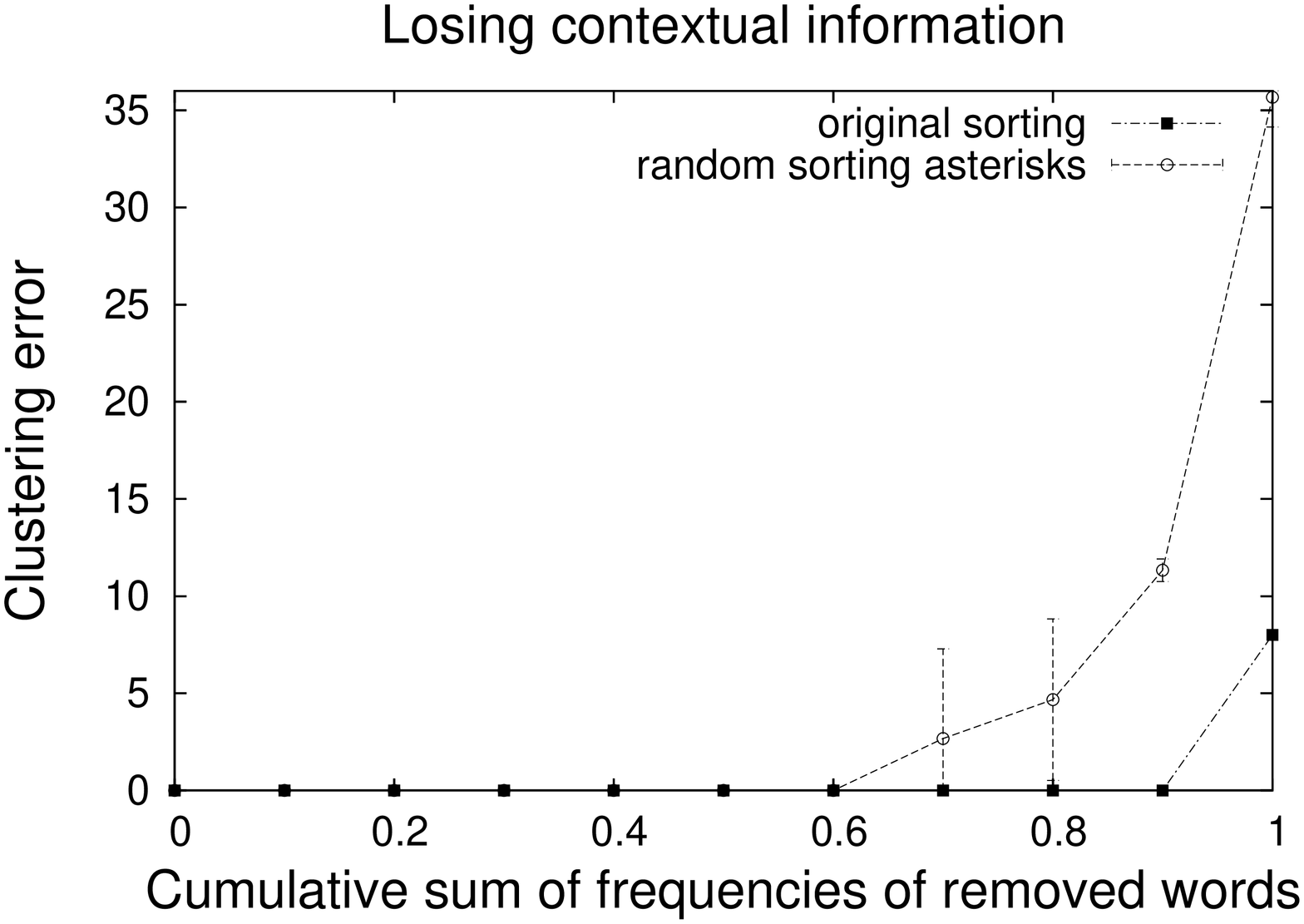} \\
 & \\
(a) Original sorting &  (b) Randomly sorting asterisks \\
\includegraphics[angle=270,width=6.5cm]{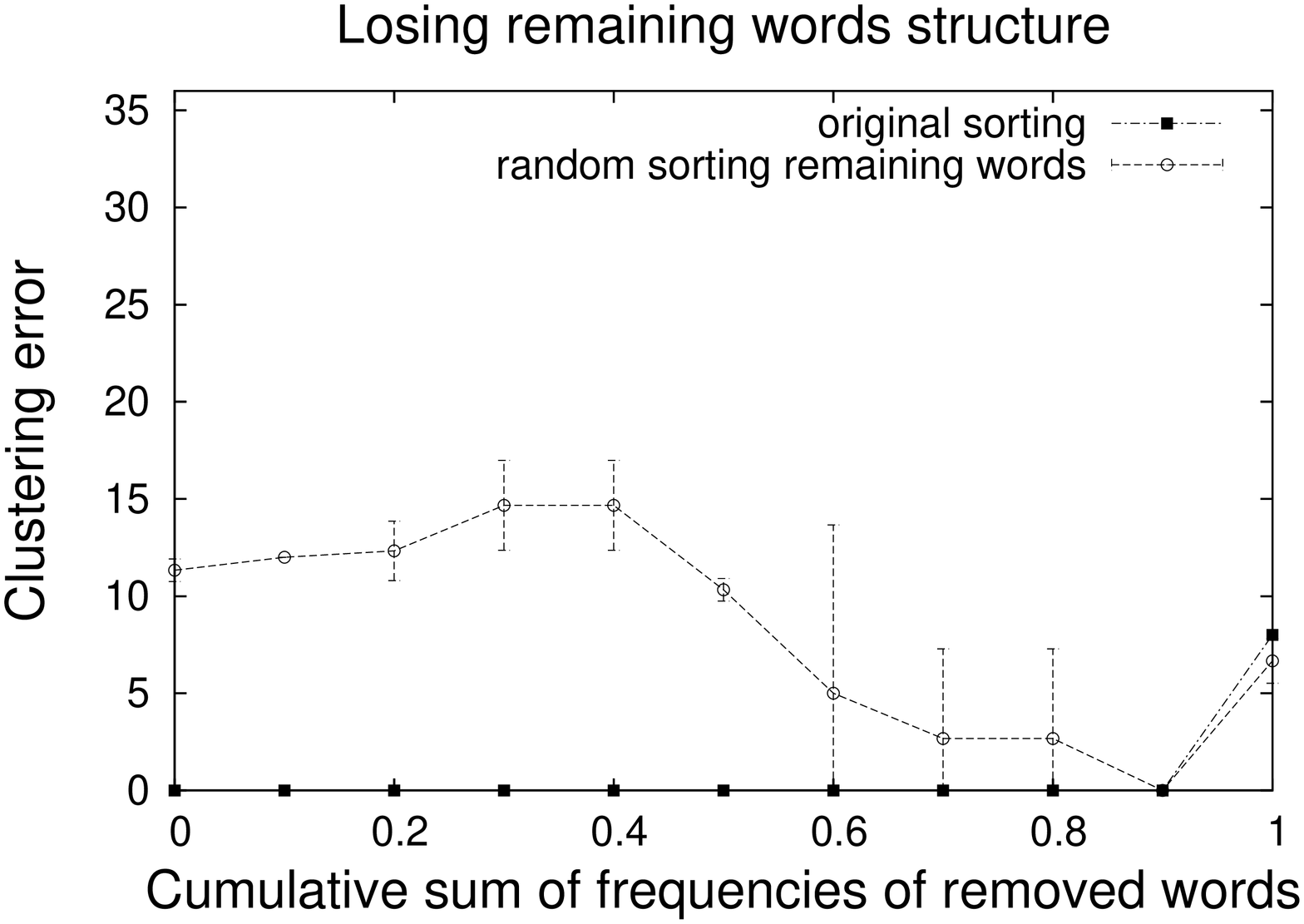} &
\includegraphics[angle=270,width=6.5cm]{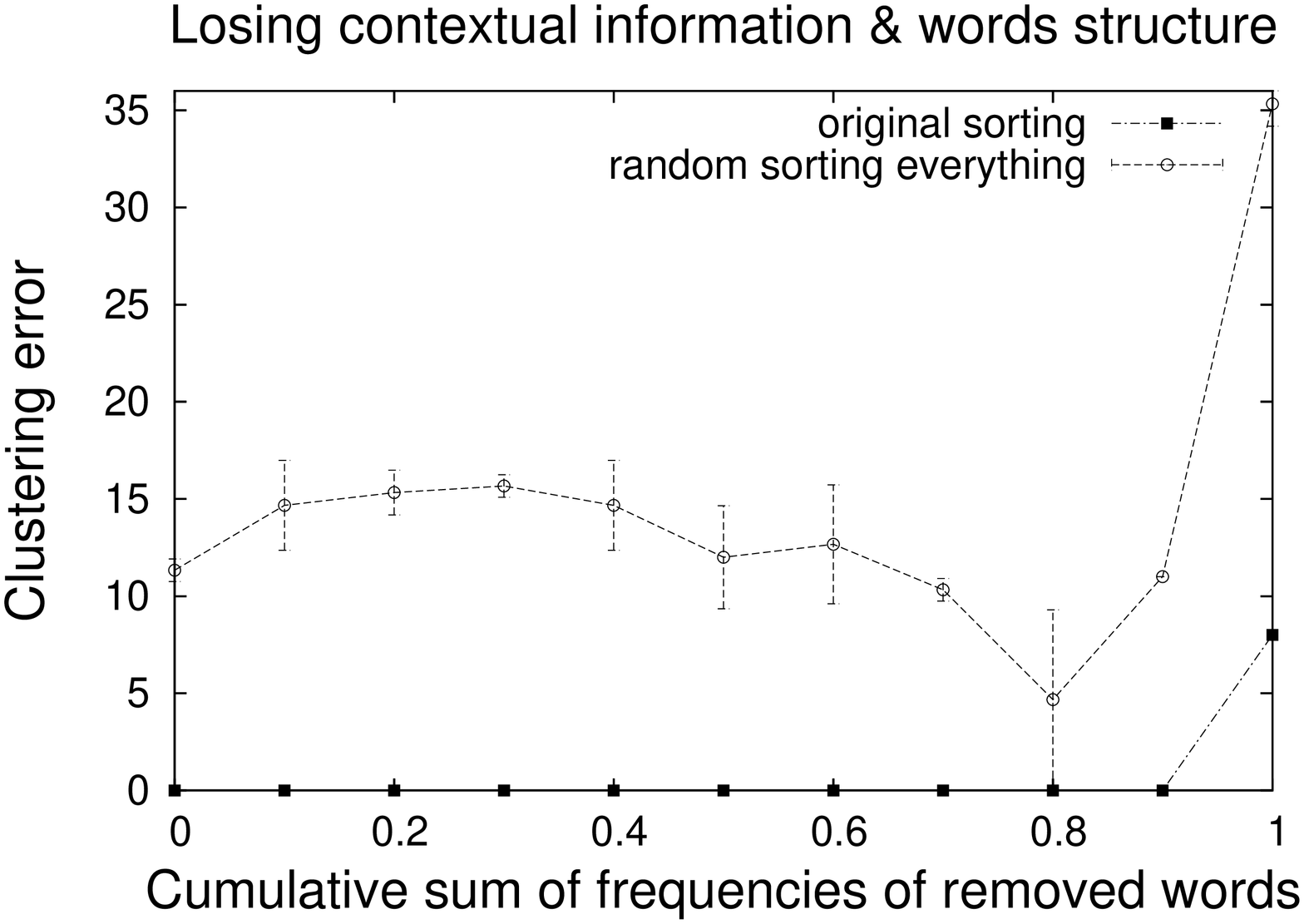} \\
 & \\
(c) Randomly sorting remaining words & (d) Randomly sorting everything \\
\end{tabular}
  \centering
\caption{Clustering results for the UCI-KDD dataset.} \label{ESWA.
Fig:uci-kdd}
\end{figure}

Fig \ref{ESWA. Fig:uci-kdd} shows the results that correspond to the
UCI-KDD dataset. When the contextual information is not lost, the
clustering error remains constant from 0.0 to 0.9, as can be
observed looking at the panel (a). It is important to note that, in
this case, the non-distorted clustering error cannot be improved
since its value is 0.

Interesting conclusions can be drawn comparing that curve with the
others. First, losing the contextual information makes the
clustering results get worse as the amount of removed words
increases. Second, losing the remaining words structure, the
clustering results are worse when the texts contain many remaining
words and few of contextual information. Third, losing every
structure, both behaviors are observed at the same time, that is,
the clustering results are worse for small and big numbers of
removed words. All these phenomena can be observed in panels (b),
(c) and (d) respectively.

\begin{figure}
\begin{tabular}{cc}
\includegraphics[angle=270,width=6.5cm]{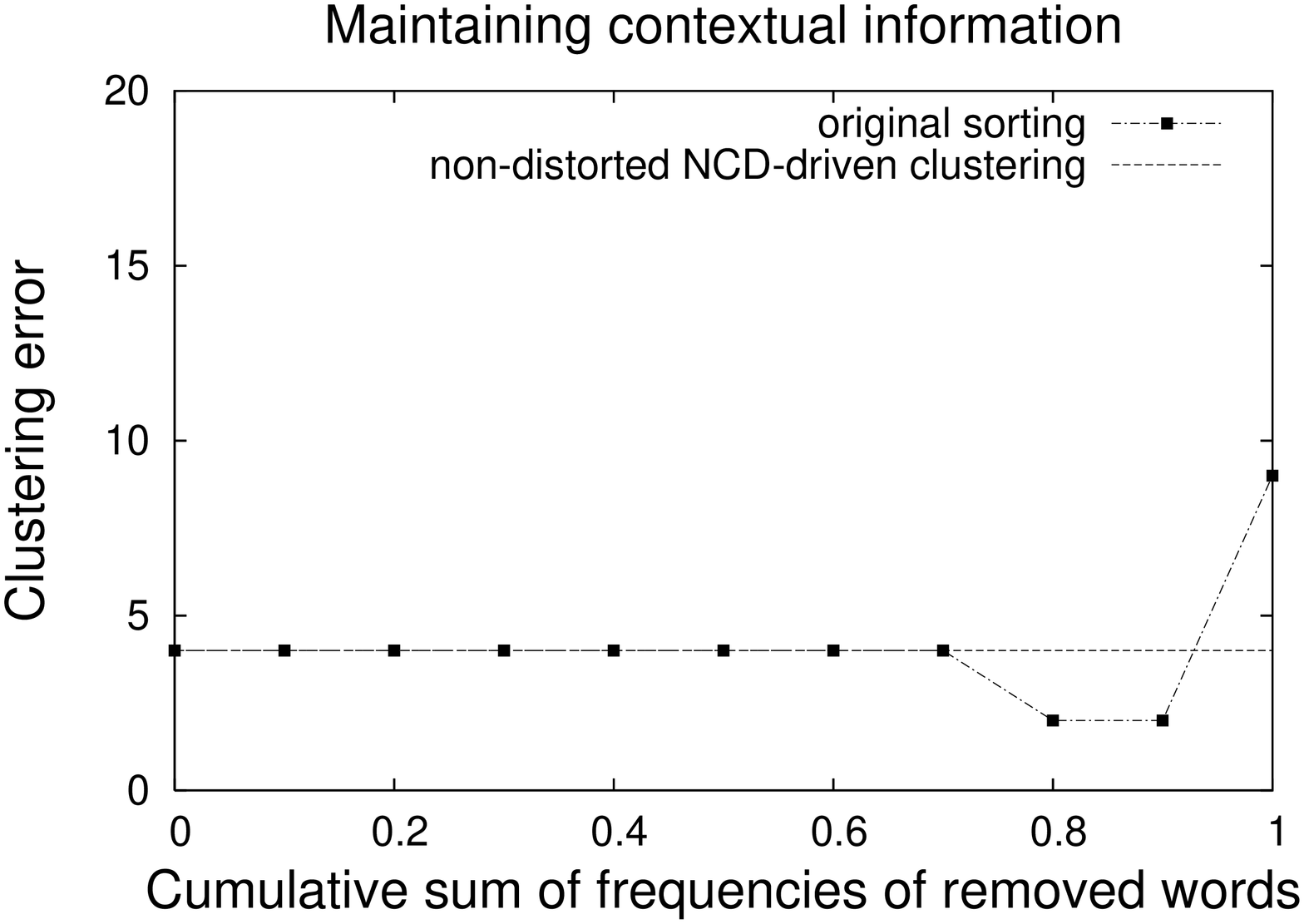} &
\includegraphics[angle=270,width=6.5cm]{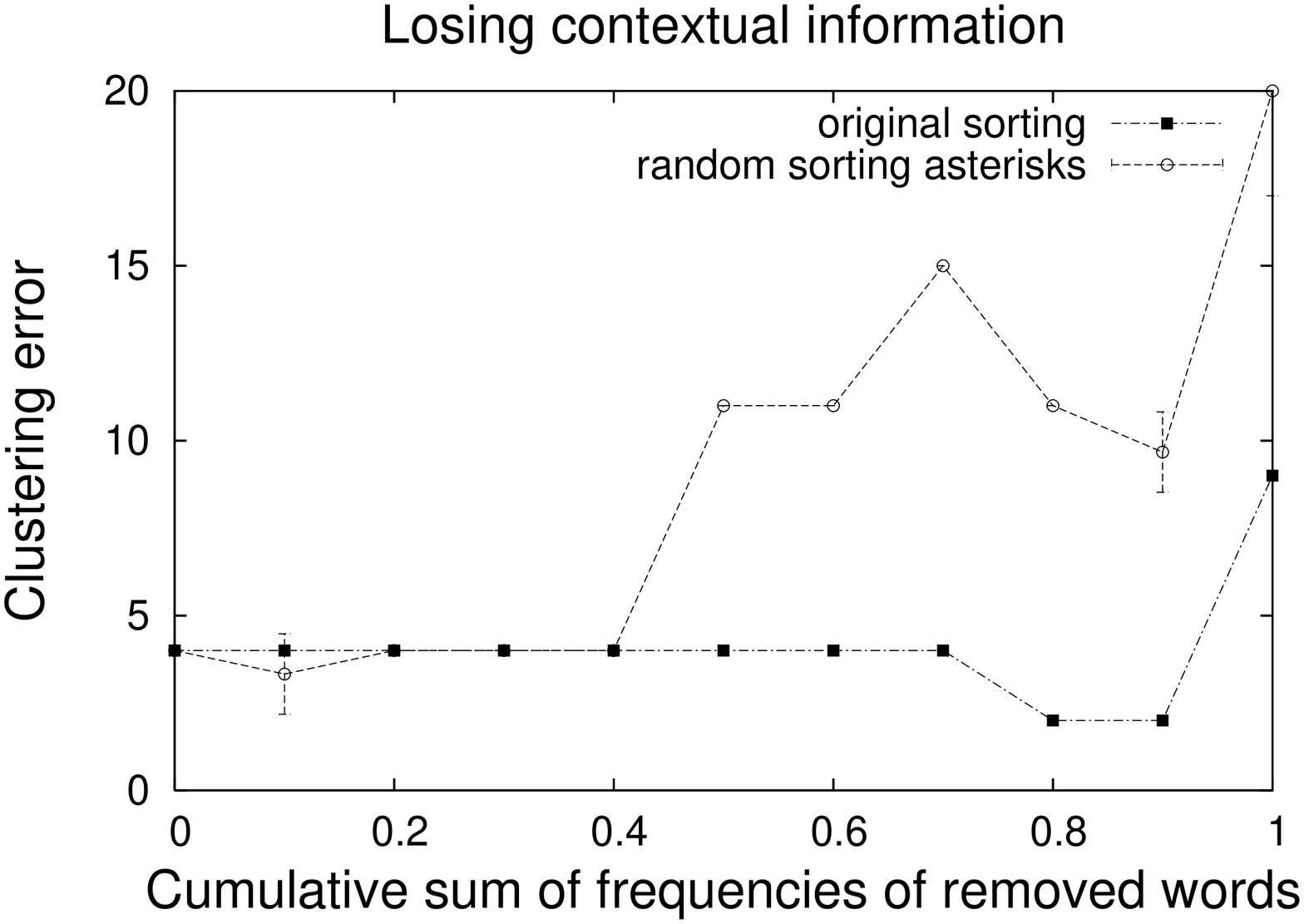} \\
 & \\
(a) Original sorting &  (b) Randomly sorting asterisks \\
\includegraphics[angle=270,width=6.5cm]{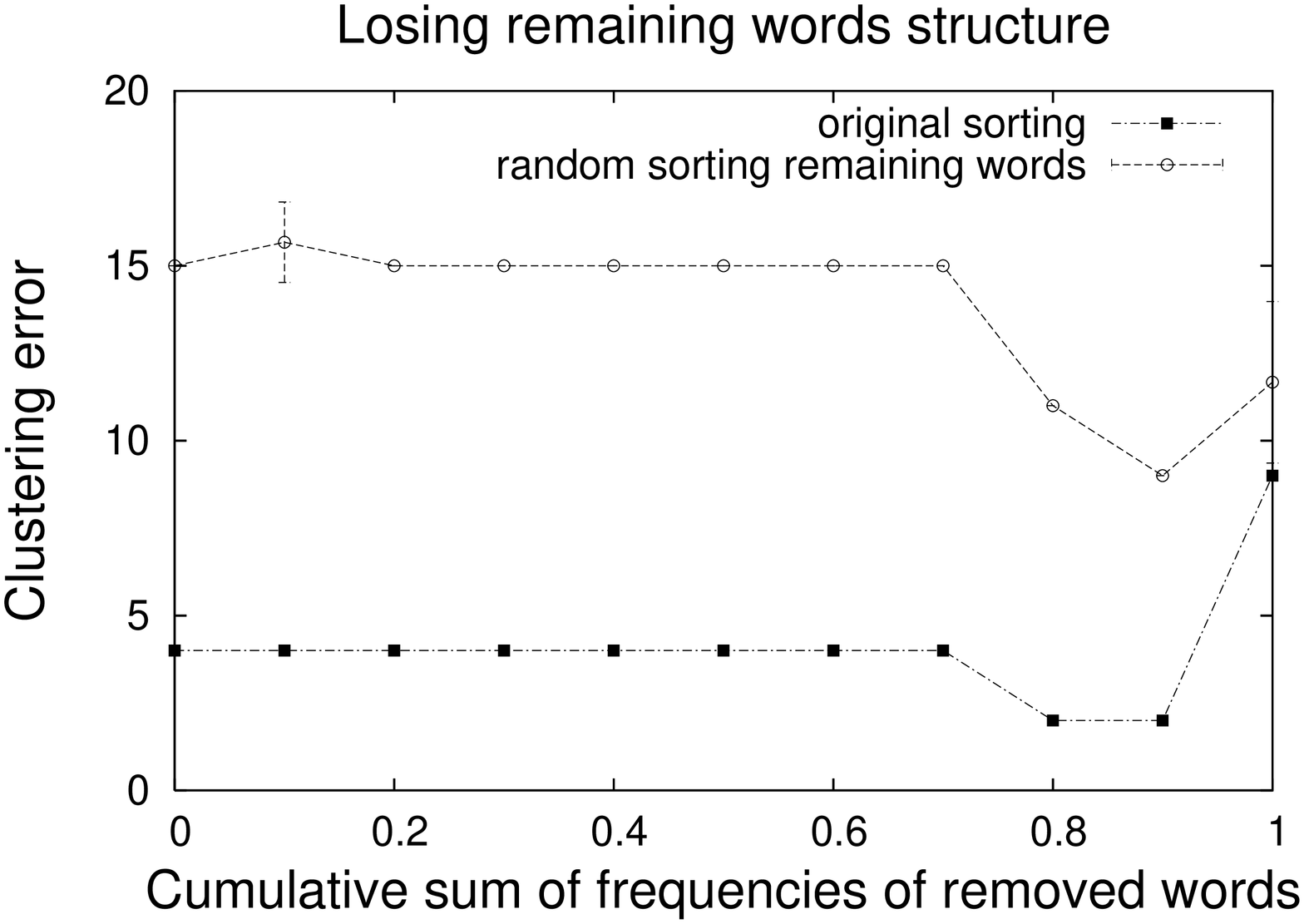} &
\includegraphics[angle=270,width=6.5cm]{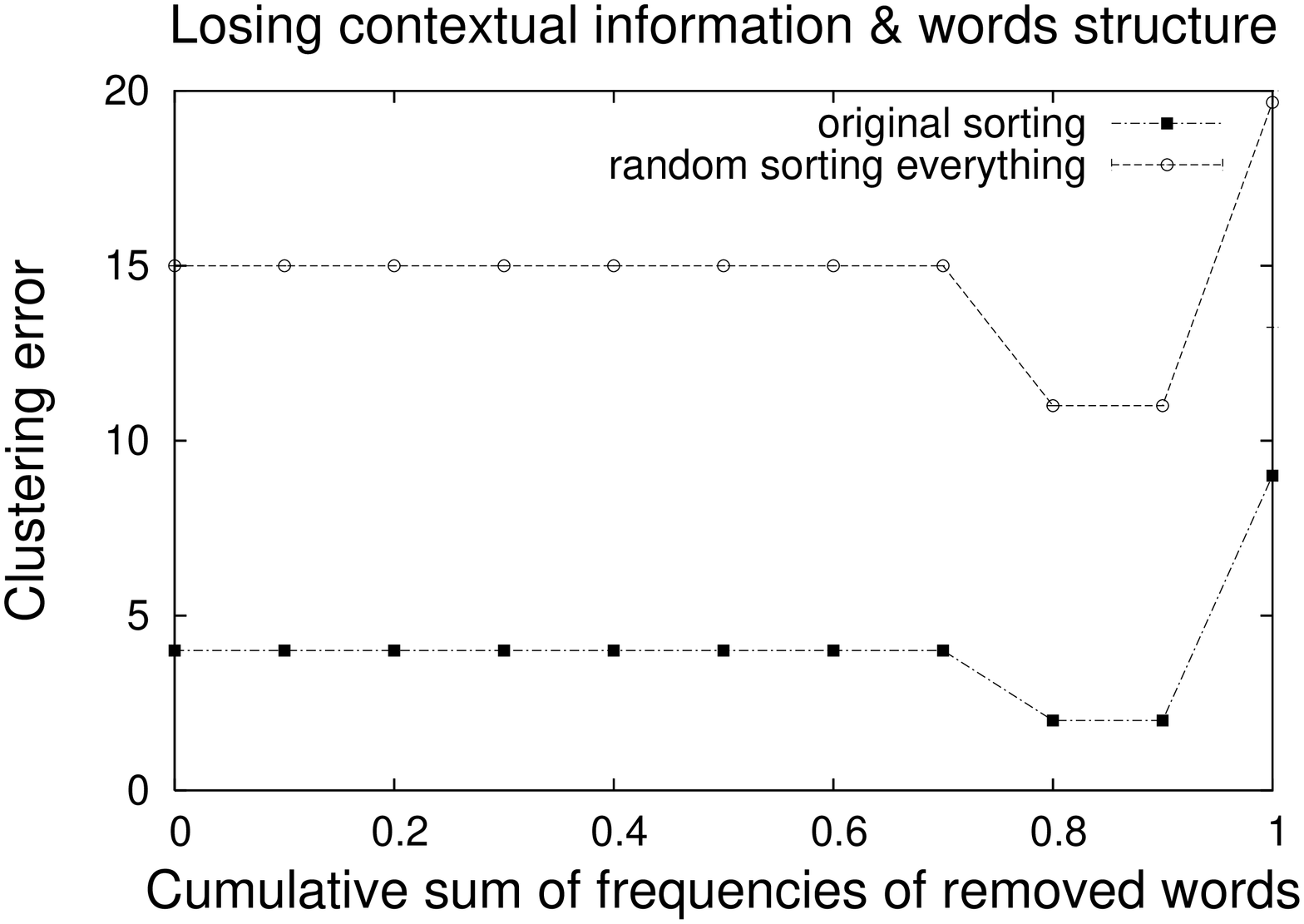} \\
 & \\
(c) Randomly sorting remaining words & (d) Randomly sorting everything \\
\end{tabular}
  \centering
\caption{Clustering results for the Books dataset.}
  \label{ESWA. Fig:books}
\end{figure}

Fig \ref{ESWA. Fig:books} shows the results that correspond to the
Books dataset. The curves show that the behavior of the distortion
techniques is qualitatively similar to that observed for the UCI-KDD
dataset. That is, when the contextual information is lost, the
clustering error increases as the amount of removed words increases.
When only the remaining words structure is lost, the clustering
error is worse for small quantities of removed words. Look at panels
(b) and (c) to observe this.

It is also important to mention that the non-distorted clustering
error, depicted in (a) as a constant line, is only improved when the
contextual information is maintained. Look at points 0.8 and 0.9 of
the curve plotted in the panel (a) to notice this.

\clearpage

\begin{figure}
\begin{tabular}{cc}
\includegraphics[angle=270,width=6.5cm]{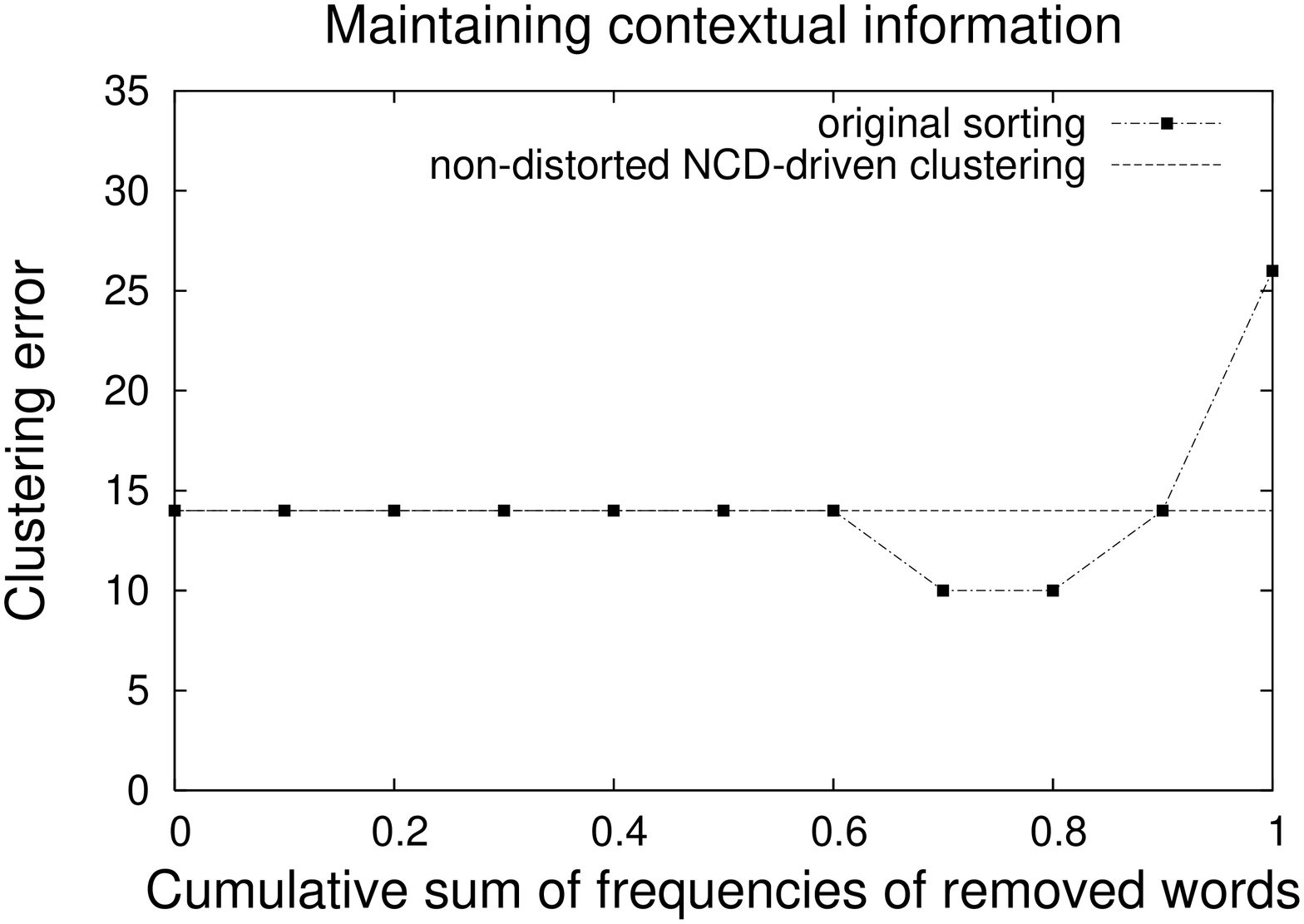} &
\includegraphics[angle=270,width=6.5cm]{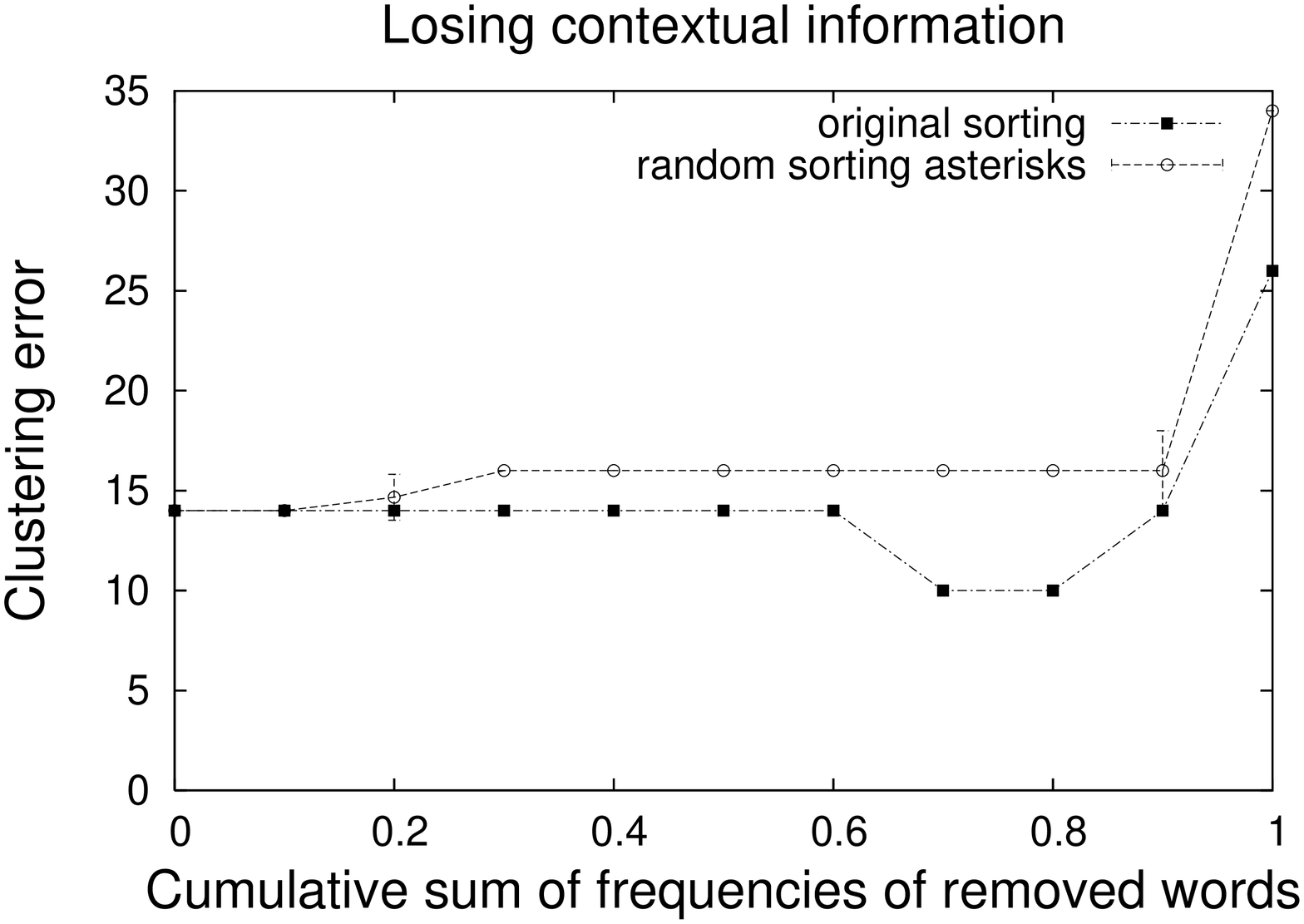} \\
 & \\
(a) Original sorting &  (b) Randomly sorting asterisks \\
\includegraphics[angle=270,width=6.5cm]{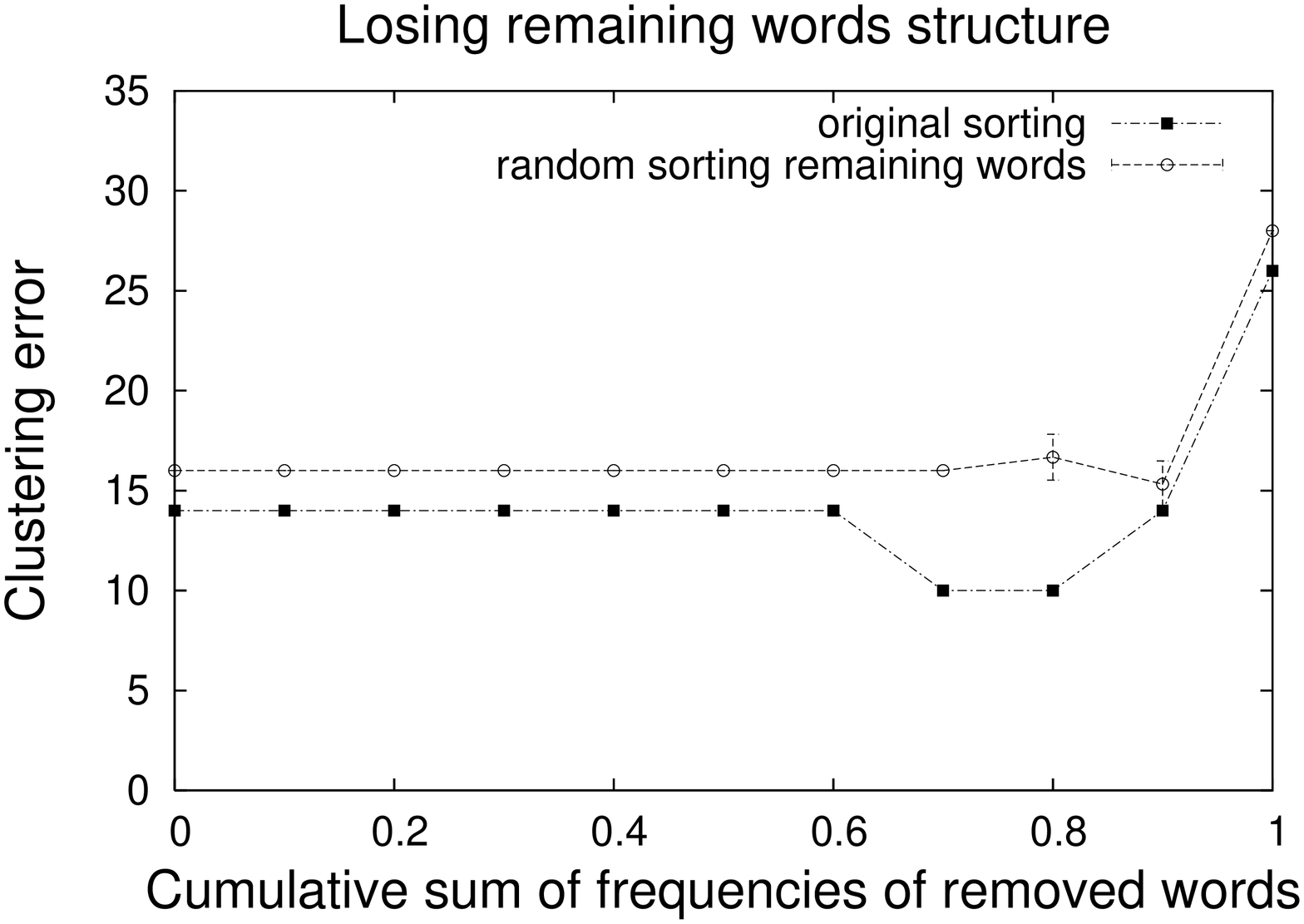} &
\includegraphics[angle=270,width=6.5cm]{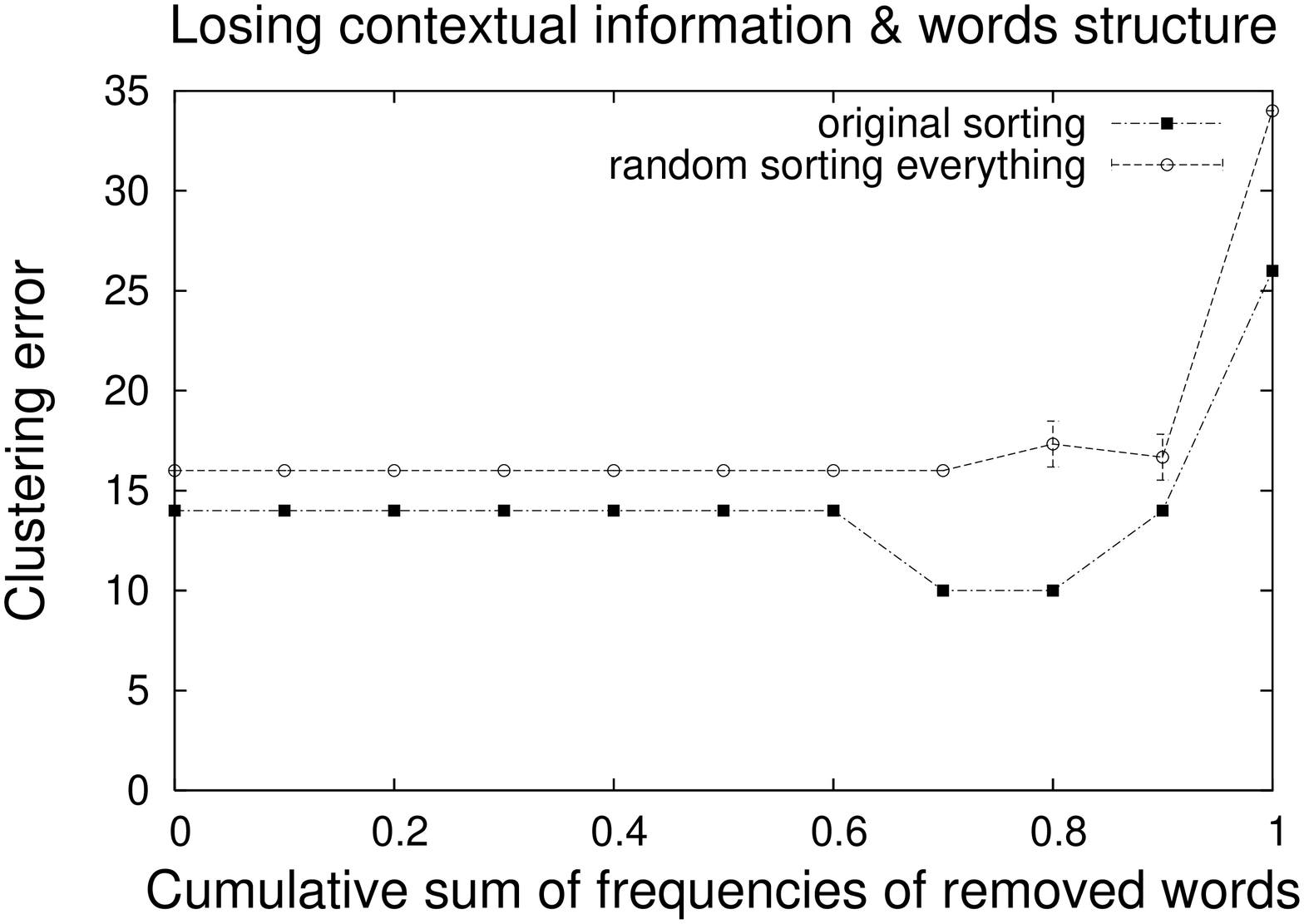} \\
 & \\
(c) Randomly sorting remaining words & (d) Randomly sorting everything \\
\end{tabular}
  \centering
\caption{Clustering results for the MedlinePlus dataset.}
  \label{ESWA. Fig:medline}
\end{figure}

Fig \ref{ESWA. Fig:medline} depicts the results that correspond to
the MedlinePlus dataset. The behavior shown in this figure is
qualitatively similar to the previous one.

\begin{figure}[h]
\begin{tabular}{cc}
\includegraphics[angle=270,width=6.5cm]{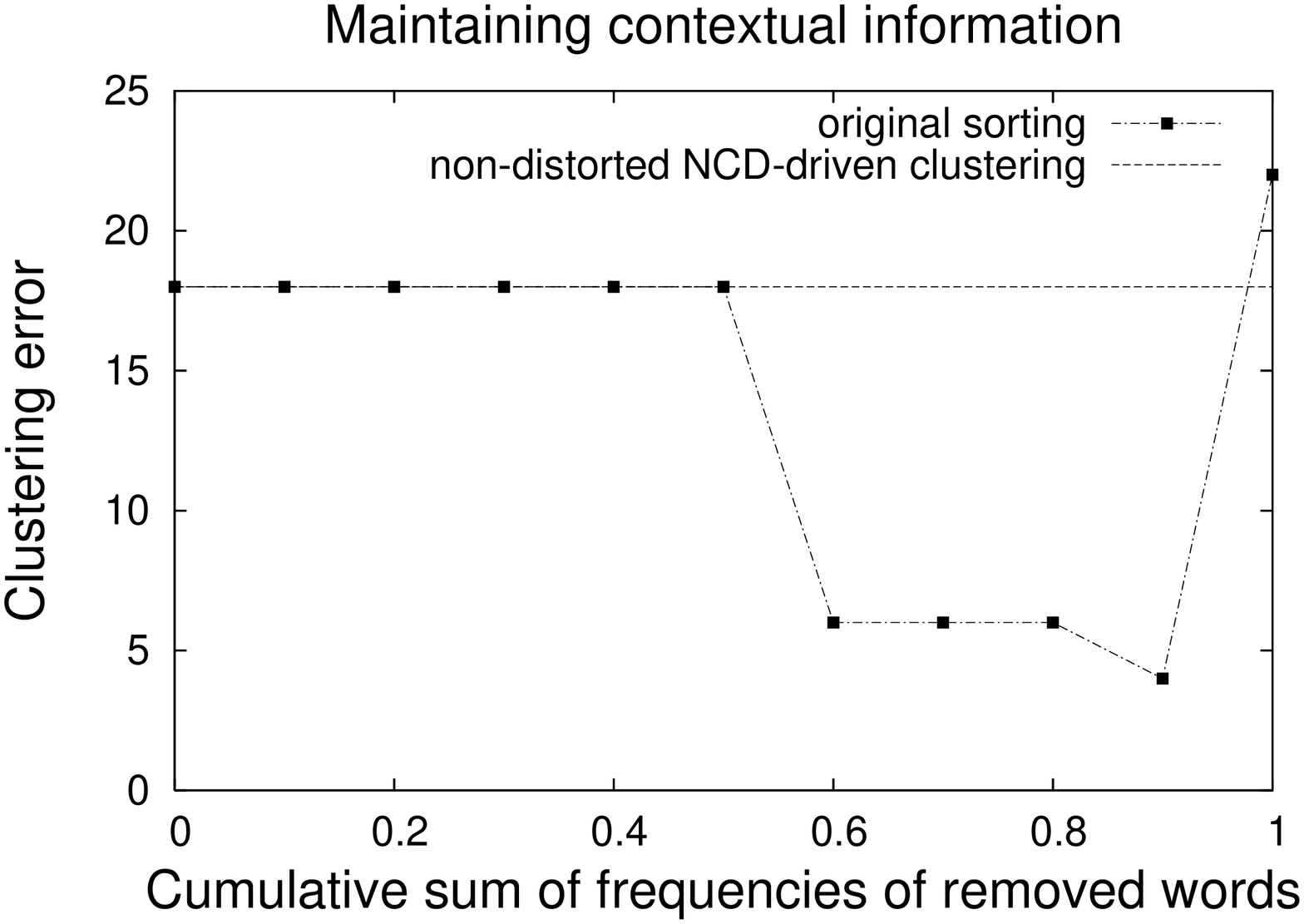} &
\includegraphics[angle=270,width=6.5cm]{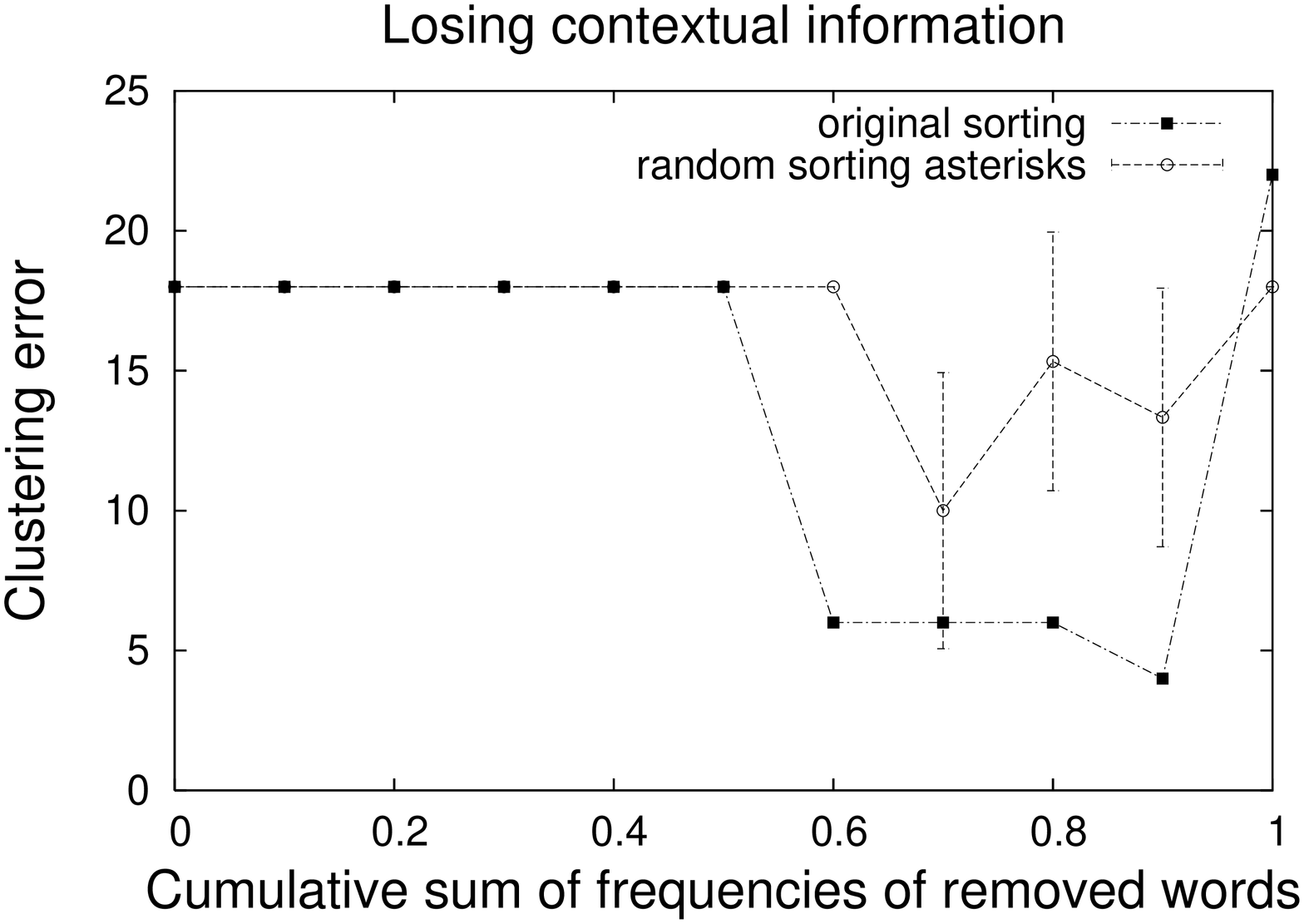} \\
 & \\
(a) Original sorting &  (b) Randomly sorting asterisks \\
\includegraphics[angle=270,width=6.5cm]{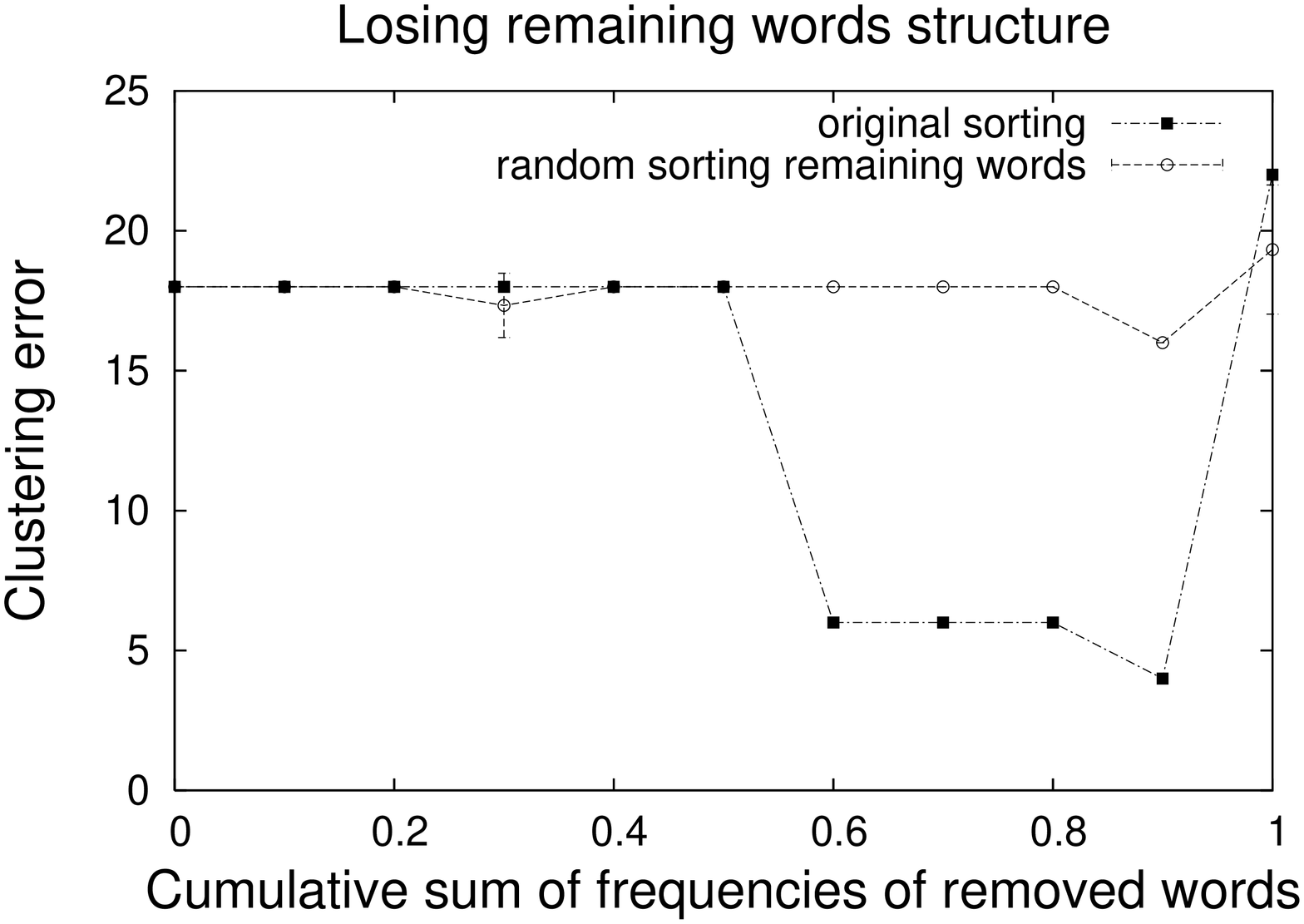} &
\includegraphics[angle=270,width=6.5cm]{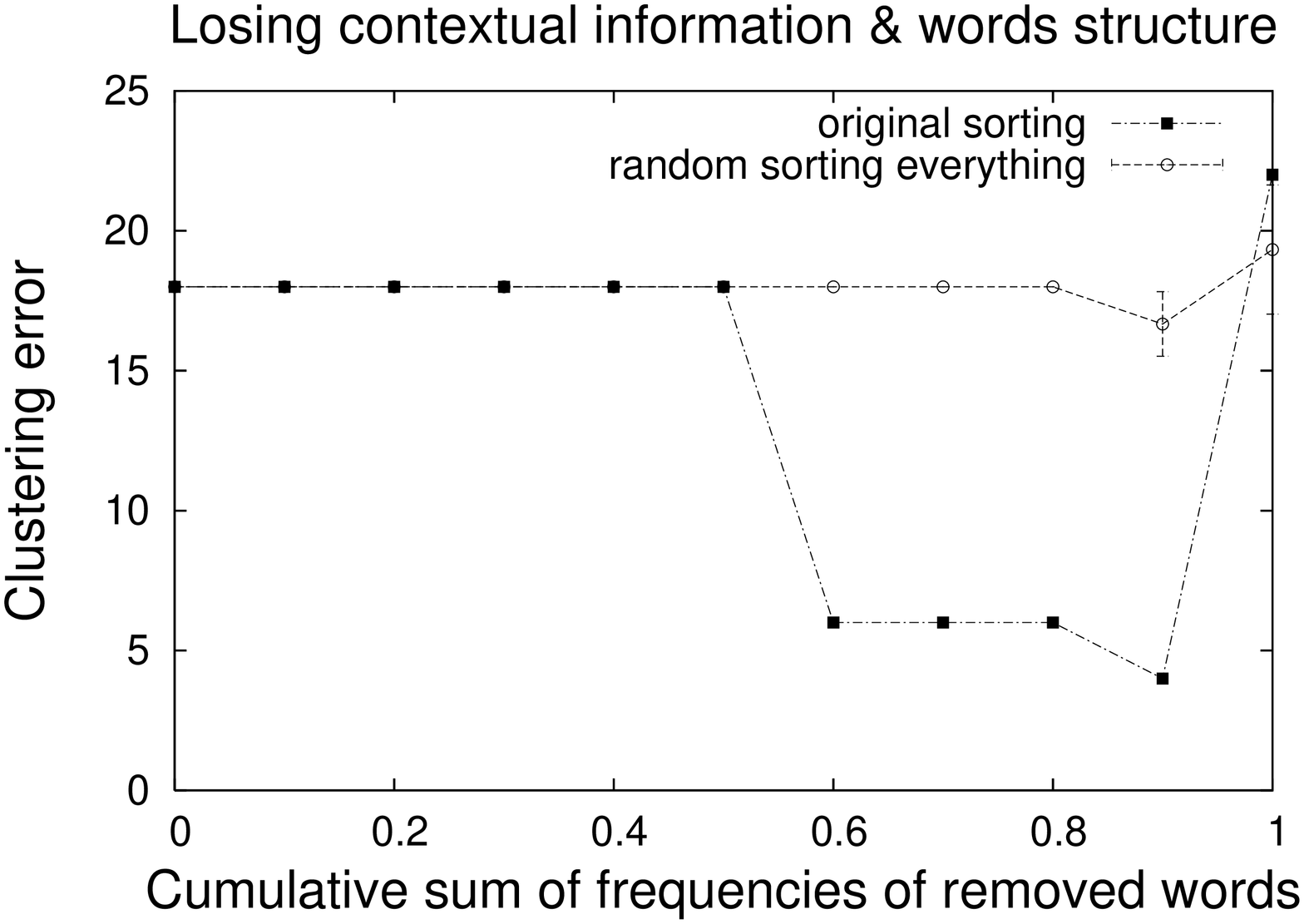} \\
 & \\
(c) Randomly sorting remaining words & (d) Randomly sorting everything \\
\end{tabular}
  \centering
\caption{Clustering results for the IMDB dataset.}
  \label{ESWA. Fig:imdb}
\end{figure}

The results obtained for the IMDB dataset are depicted in Fig
\ref{ESWA. Fig:imdb}. The nature of this dataset is explained before
analyzing the curves in order to better understand these interesting
results. This dataset is composed of plots of movies to be clustered
by saga. This means that as the amount of removed words increases,
the words that still remain in the texts are words highly related to
the sagas, such as for example names of characters or names of
places. That is the reason why the non-distorted clustering error
can be improved as much as panel (a) shows.

The clustering error in the cases in which the remaining words are
randomly sorted are worse than the ones obtained when they are not.
This means that the structure of the remaining words is highly
relevant to this dataset. Nevertheless, when only the contextual
information is lost, the clustering error is worse than when nothing
is mixed up. This can be seen in panel (b). Therefore, it can be
concluded that, although for this dataset the most relevant
information corresponds to the remaining words, the contextual
information is also relevant.

\begin{figure}[h]
\begin{tabular}{cc}
\includegraphics[angle=270,width=6.5cm]{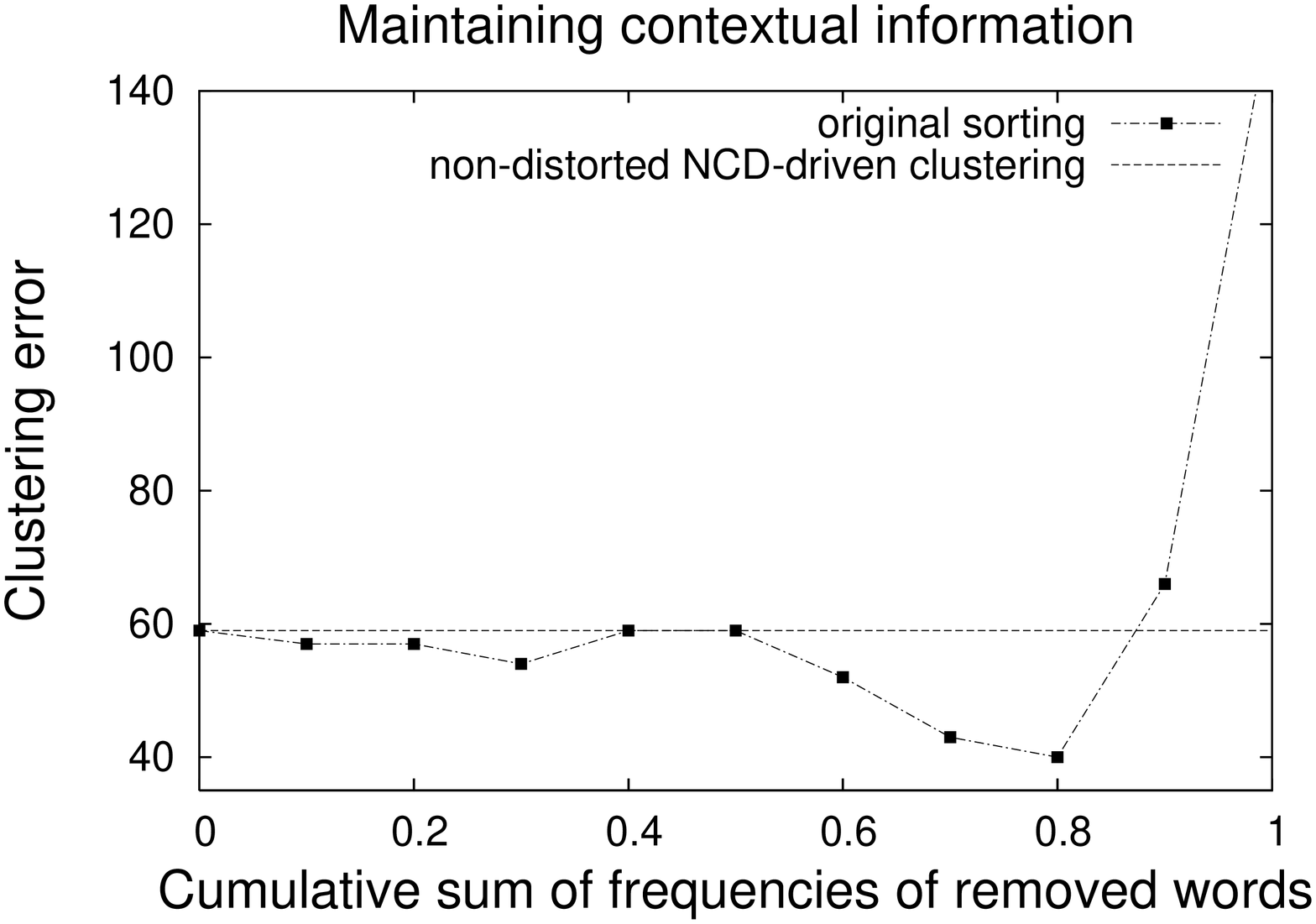} &
\includegraphics[angle=270,width=6.5cm]{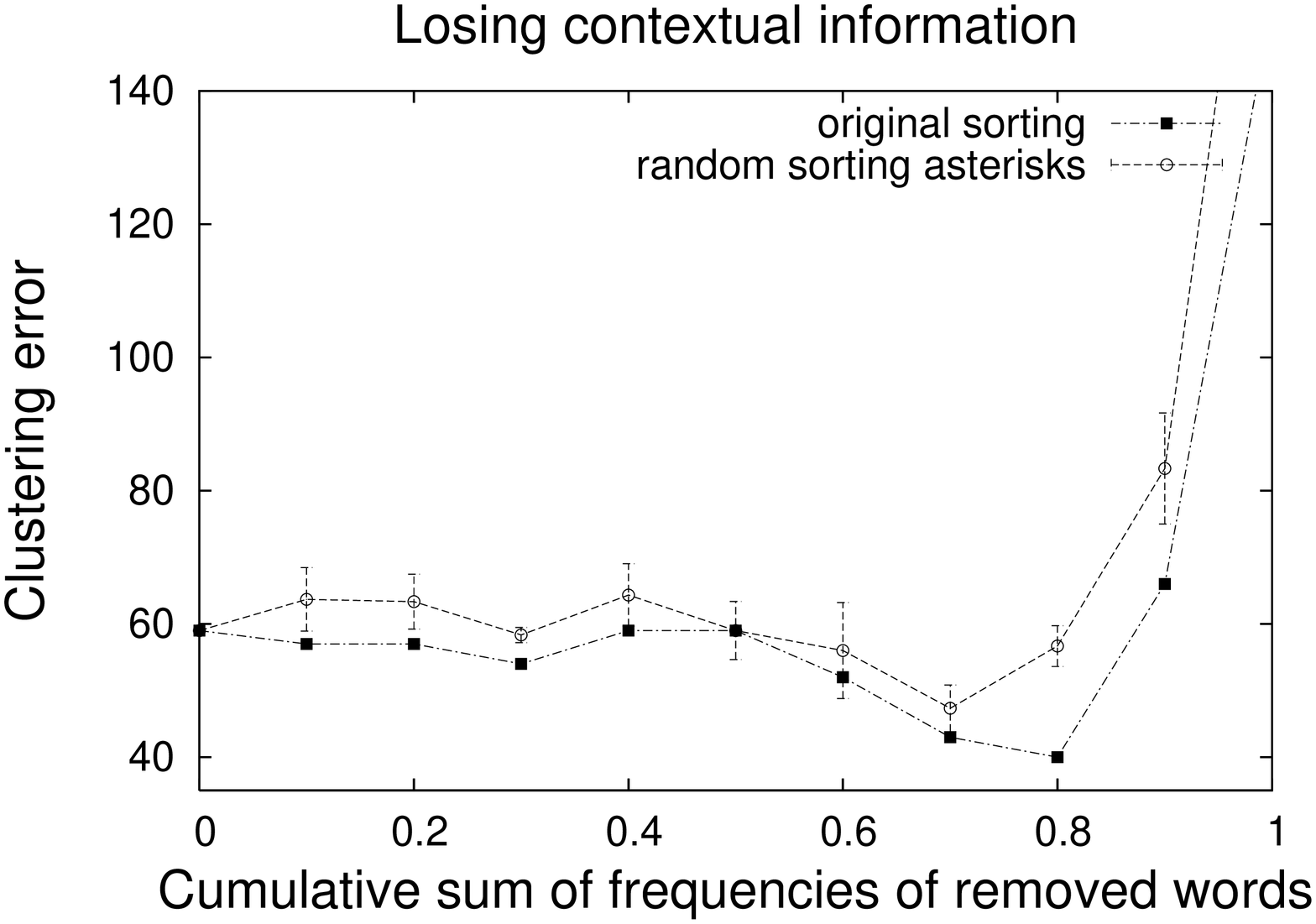} \\
 & \\
(a) Original sorting &  (b) Randomly sorting asterisks \\
\includegraphics[angle=270,width=6.5cm]{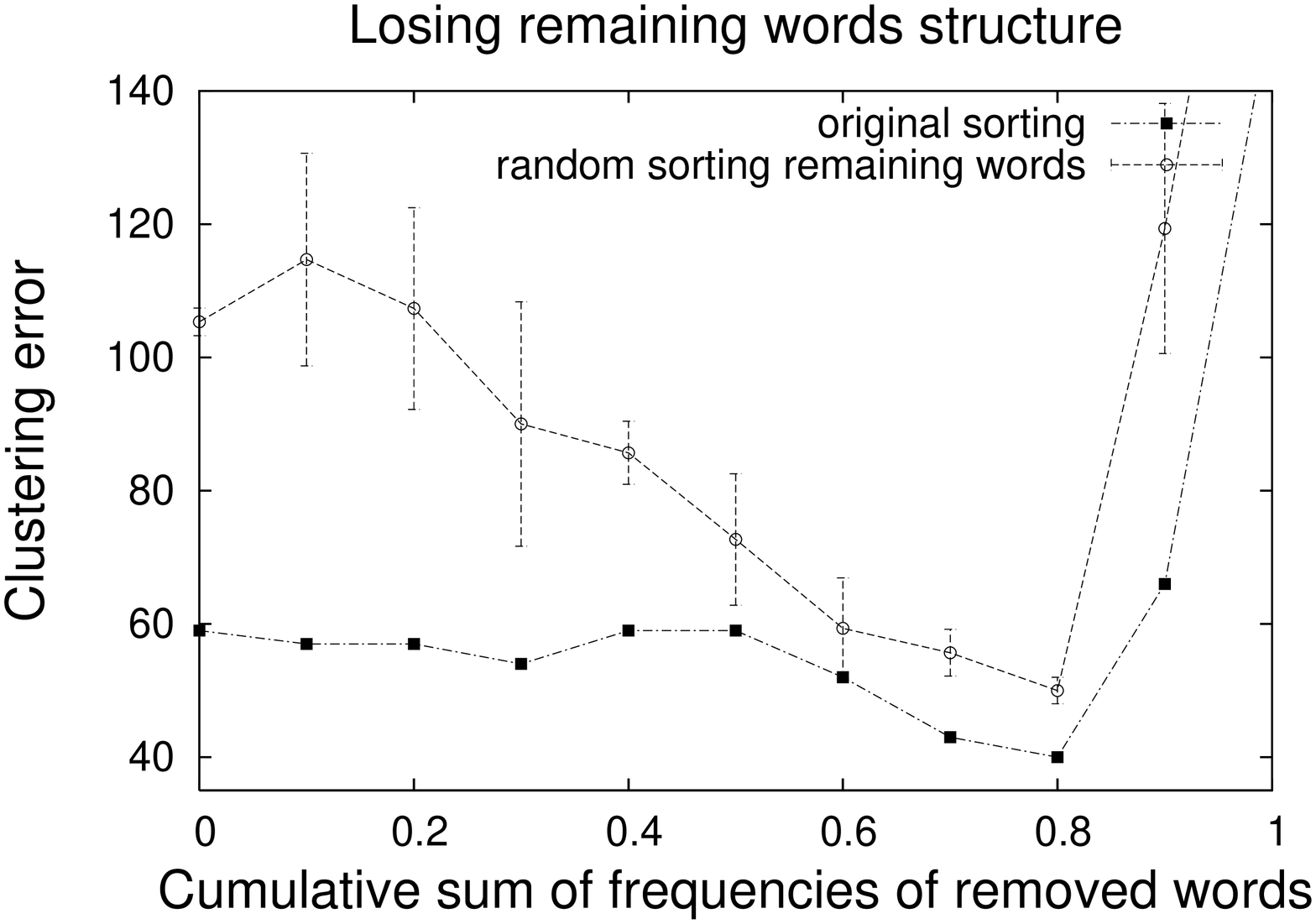} &
\includegraphics[angle=270,width=6.5cm]{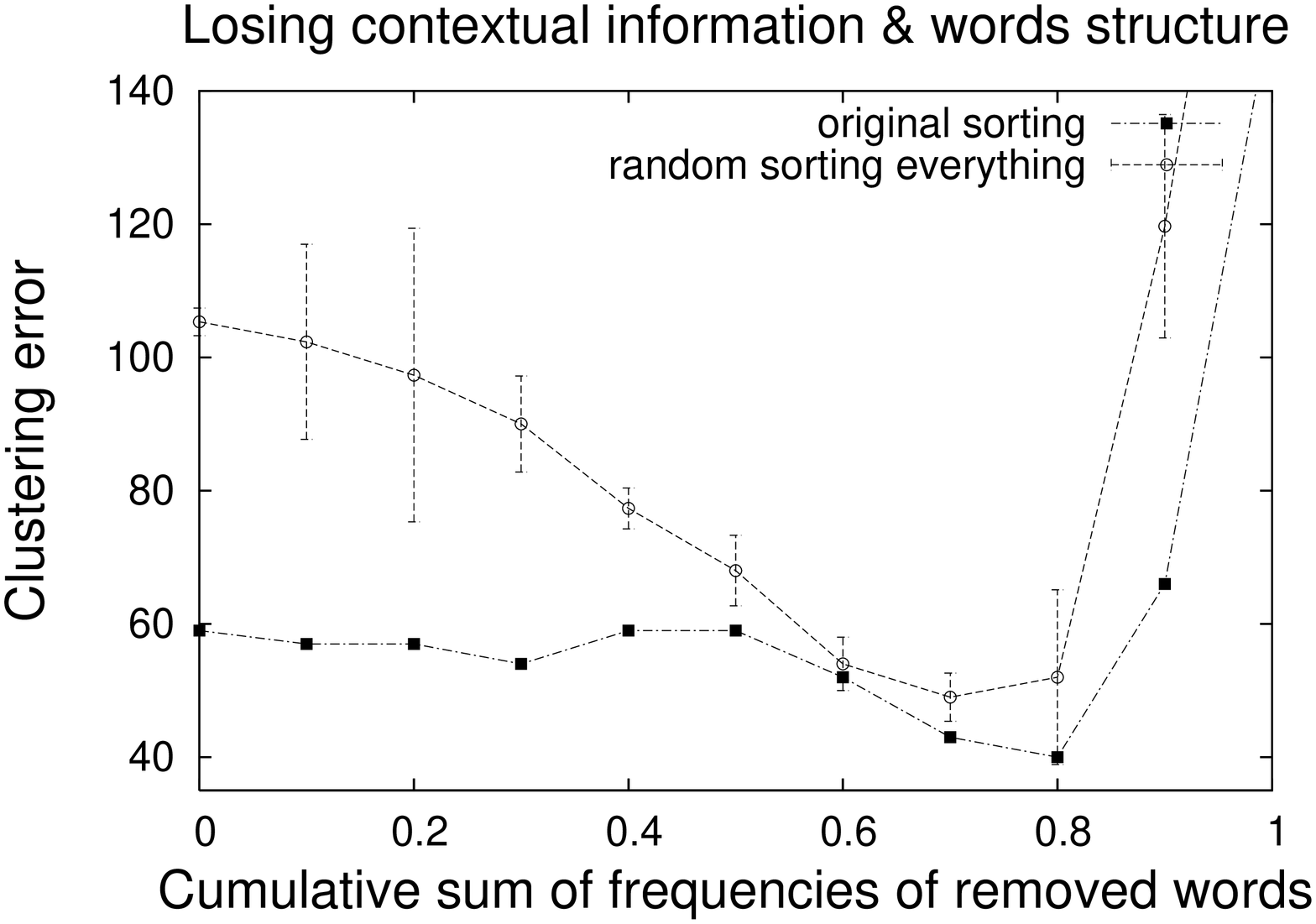} \\
 & \\
(c) Randomly sorting remaining words & (d) Randomly sorting everything \\
\end{tabular}
  \centering
\caption{Clustering results for the SRT-serial dataset.}
  \label{ESWA. Fig:srt-series}
\end{figure}

Finally, Fig \ref{ESWA. Fig:srt-series} shows the results that
correspond to the SRT-serial dataset. This is the bigger dataset
that has been used in this chapter. Its results are consistent with
the ones obtained for the rest of the datasets. That is, the
contextual information is relevant as well, although the structure
of the remaining words is more relevant than the contextual
information when the amount of removed words, and therefore, the
amount of contextual information is small.

\subsection{Synopsis of results}
\label{ESWA. Synopsis of results}

\begin{figure}
\begin{tabular}{cc}
\includegraphics[width=6.5cm]{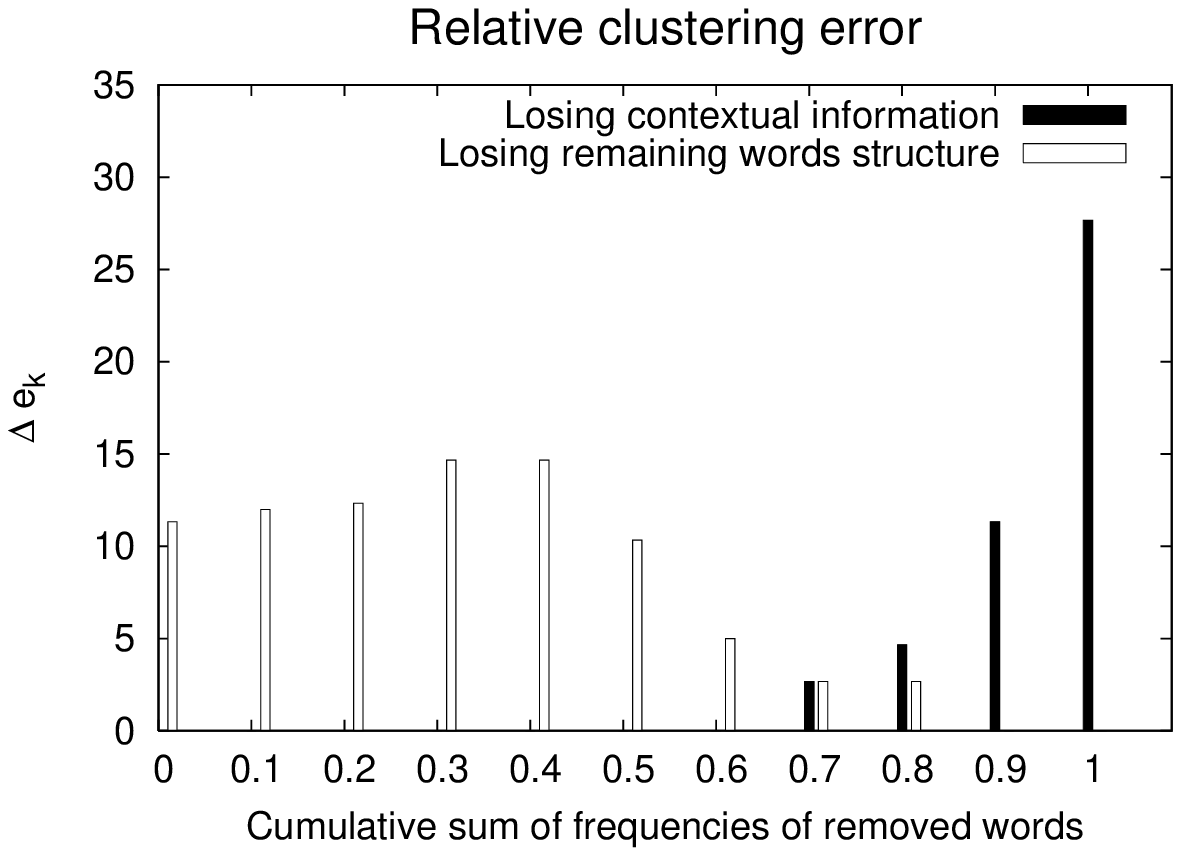} &
\includegraphics[width=6.5cm]{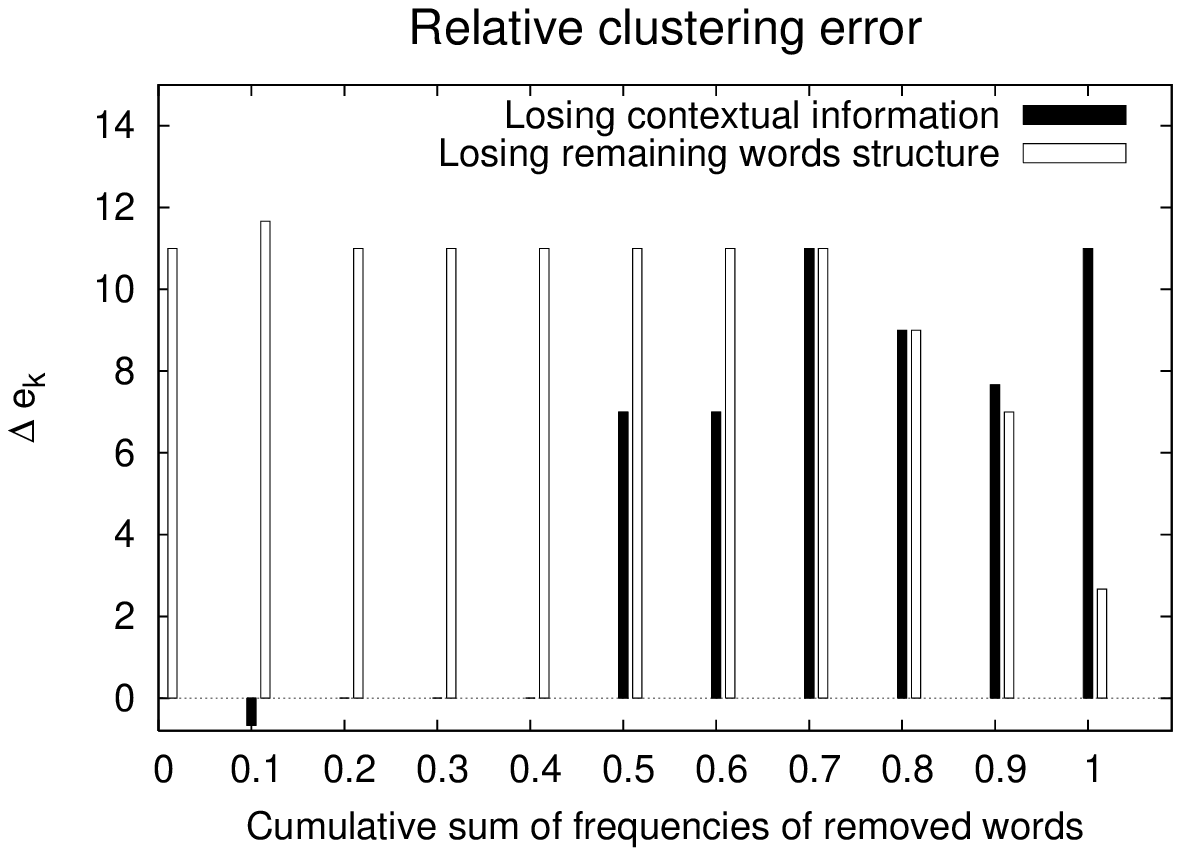} \\
(a) UCI-KDD dataset & (b) Books dataset \\
 & \\
\includegraphics[width=6.5cm]{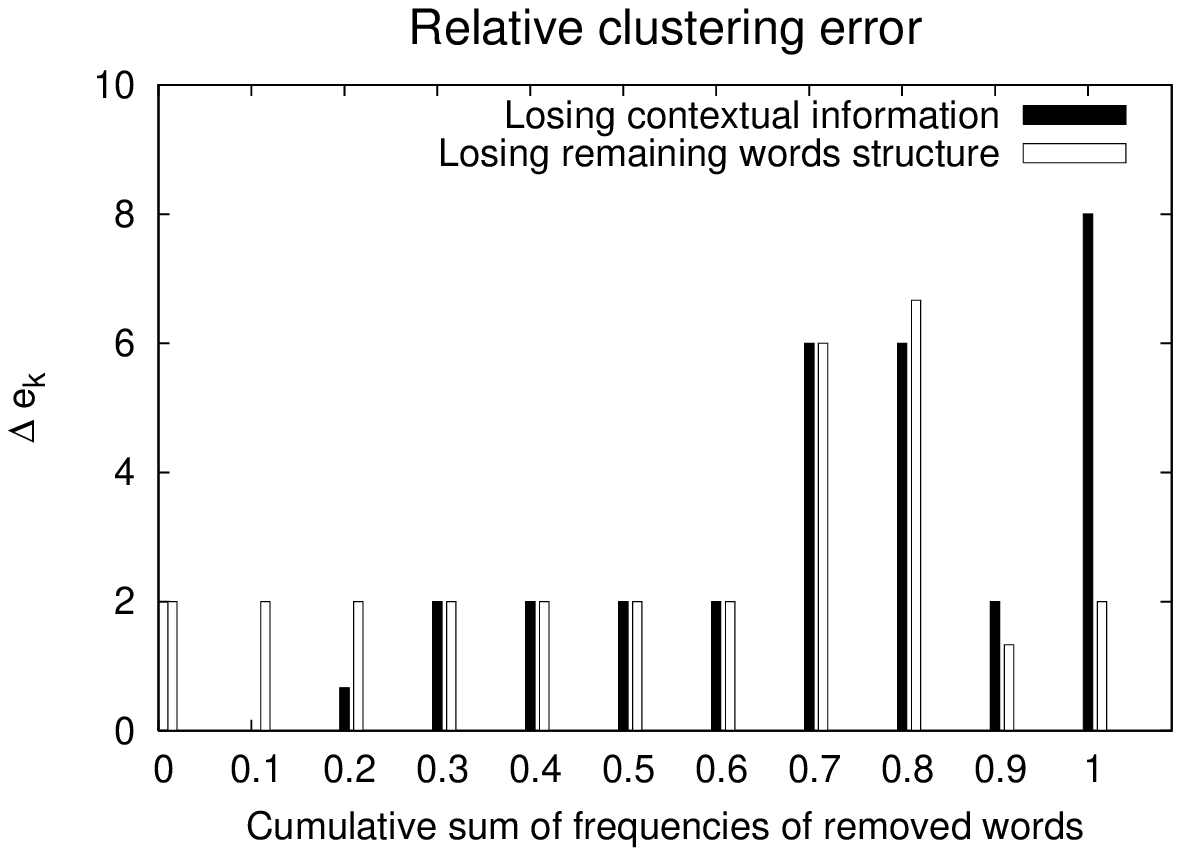} &
\includegraphics[width=6.5cm]{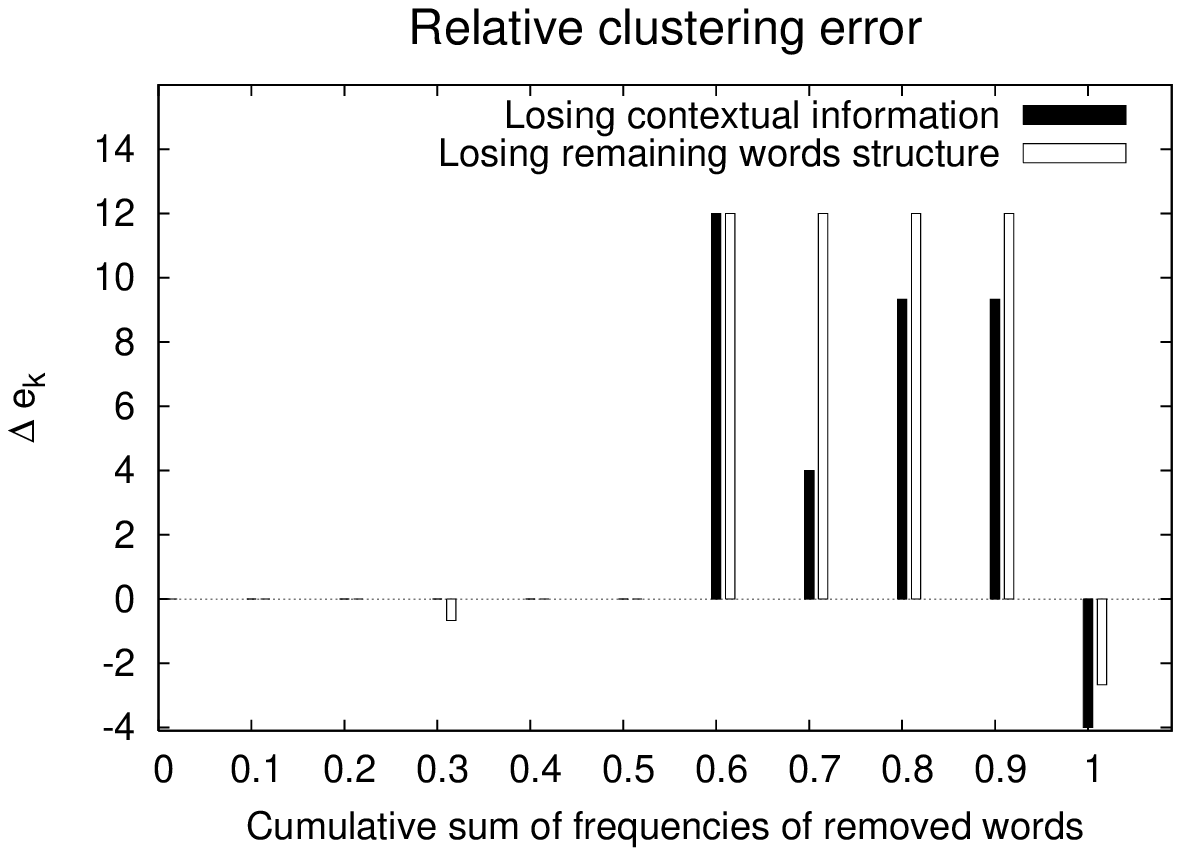} \\
(c) MedlinePlus dataset & (d) IMDB dataset \\
 & \\
 \multicolumn{2}{c}{\includegraphics[width=6.5cm]{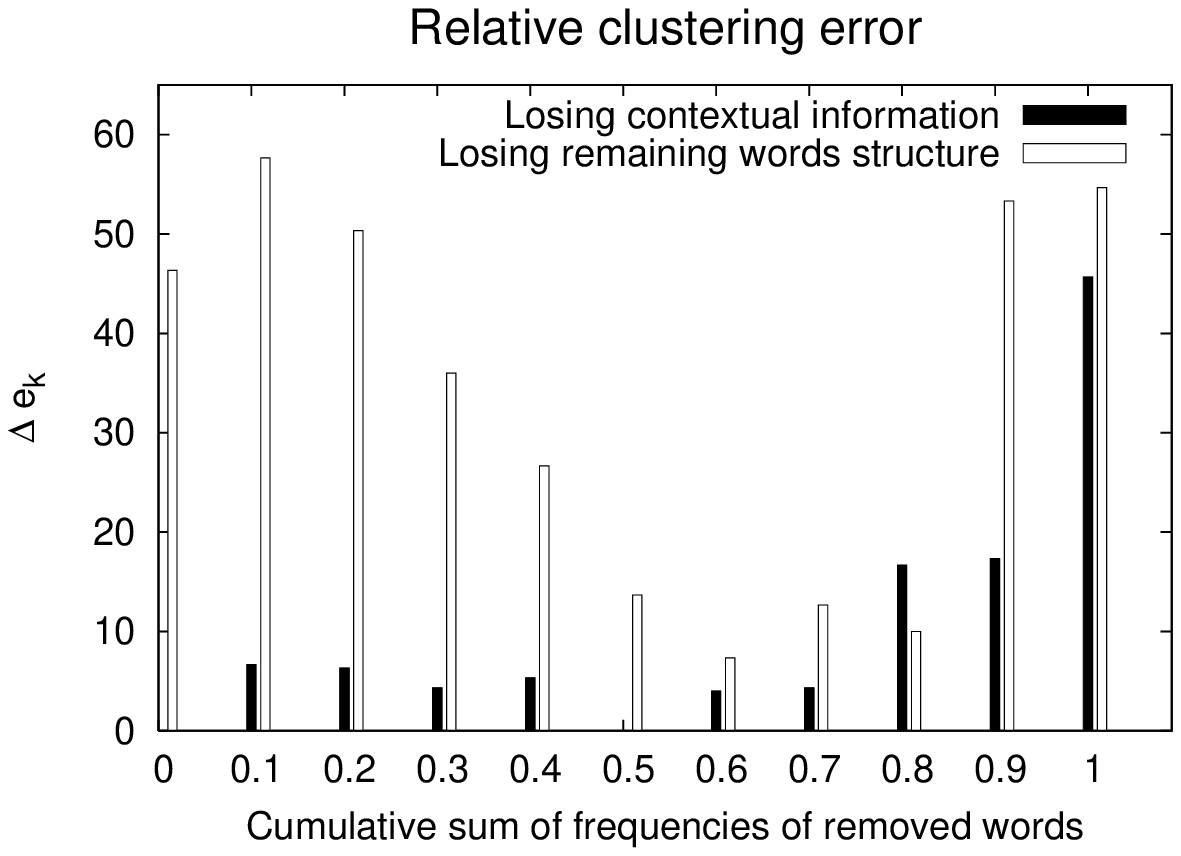}} \\
 \multicolumn{2}{c}{(e) SRT-serial dataset} \\
 & \\
\end{tabular}
  \centering
\caption[Clustering error difference with the \emph{Original
sorting} distortion technique.]{Clustering error difference with the
\emph{Original sorting} distortion technique. There exists a
clustering error difference for both techniques. Therefore, one can
conclude that both the remaining words structure and the contextual
information are relevant in this scenario.}
  \label{ESWA. Fig:differences}
\end{figure}

Additionally, Fig \ref{ESWA. Fig:differences} has been created in
order to better analyze the difference between the most interesting
distortion techniques. It depicts the clustering error difference
between the \emph{Randomly sorting contextual information} and the
\emph{Randomly sorting remaining words} distortion techniques with
respect to the \emph{Original sorting} distortion technique. The
length of each bar corresponds to the relative error, which is as
follows

\begin{equation}
\Delta e_k = {e_k - e_0},
\end{equation}

\vspace{0.2cm}

where $e_k$ is the clustering error obtained using the $k$
distortion technique, $e_0$ is the clustering error obtained using
the \emph{Original sorting} distortion technique, and $\Delta e_k$
is the relative error for the $k$ distortion technique with respect
to the \emph{Original sorting} distortion technique.

Fig \ref{ESWA. Fig:differences} can be easily understood by
analyzing the results obtained for the UCI-KDD dataset. These
results can be seen in Fig \ref{ESWA. Fig:uci-kdd} and in Fig
\ref{ESWA. Fig:differences}(a). Looking at Fig \ref{ESWA.
Fig:uci-kdd}(b)(c), two phenomena can be noticed. First, when the
contextual information is lost, the clustering error gets worse from
0.7 to 1.0, see \ref{ESWA. Fig:uci-kdd}(b). Second, when the
remaining words structure is lost, the clustering error is worse
from 0 to 0.8, see \ref{ESWA. Fig:uci-kdd}(c). This is represented
in Fig \ref{ESWA. Fig:differences} in form of black and white bars:
there are black bars from 0.7 to 1.0 and white bars from 0 to 0.8.

Analyzing the five plots depicted in Fig \ref{ESWA.
Fig:differences}, it can be concluded that both the contextual
information and the remaining words structure are relevant, since
there exists a clustering error difference for both techniques
with respect to the \emph{Original sorting} method. In fact,
$\Delta e_k$ can be used to provide a quantitative measure of this
relevance.

Finally, a summary of all the obtained results in the form of a
table is shown. Table \ref{tableErrorMedio} contains the average
clustering error for all the distortion techniques, and all the
datasets used in this chapter of the thesis. The clustering error is
averaged as follows:

\begin{equation}
\textrm{Average CE} = \frac
{\sum\limits_{\forall~\tiny{\textrm{distortion}}}
\textrm{CE(distortion)}}{\#~\textrm{distortions}}
\end{equation}

Where CE means ``Clustering Error'', and the possible distortions go
from 0.1 to 1.0. Therefore the number of distortions is always 10.

Analyzing Table \ref{tableErrorMedio} one can reach the same
conclusions as by analyzing the rest of the figures shown in the
chapter. However, given that the table presents only a value for
each pair \emph{distortion technique}-\emph{dataset}, comparing the
effects that the distortion techniques have on the clustering error
is easier analyzing the table than looking at the clustering error
figures.

Comparing the first two rows of the table one can see that randomly
sorting the contextual information increases the clustering error
obtained when nothing is randomly sorted. In fact, if the average
clustering error is normalized, then the difference between the
distortion techniques can be more easily observed.

The clustering error can be normalized in the following manner:

\begin{equation}
E_k = \frac{e_k}{e_0},
\end{equation}

where $E_k$ is the normalized average error obtained using the $k$
distortion technique, $e_0$ is the average clustering error obtained
using the \emph{Original sorting} distortion technique, and $e_k$ is
the average clustering error obtained using the $k$ distortion
technique. A $E_k$ greater than 1 implies an increment of the
average error. Therefore, it implies that the part of the texts
which is randomly sorted by the $k$ distortion technique is relevant
for the NCD-driven text clustering.

Table \ref{tableErrorMedioNormalizado} gathers the normalized
average error for all the distortion techniques, with respect to the
\emph{Original sorting} distortion technique. Looking at the values
presented in the table one can conclude that the contextual
information is relevant for the NCD-driven clustering because when
it is lost due to distortion, the average error gets worse, and
therefore the normalized average error is greater than 1. This can
be observed for all the datasets used in the chapter.

\begin{table}[ht]
\centering \caption[Average clustering error.]{Average clustering
error. There is a difference in terms of average clustering error
between the \emph{Original sorting} distortion technique, and the
rest of the distortion techniques. This means that both, the
contextual information and the remaining words structure are
relevant for the NCD-driven text clustering.}
\label{tableErrorMedio} \vspace{0.2cm} \footnotesize{
\begin{tabular}{|l|c|c|c|c|c|}
  \hline
                                                & UCI-KDD &   Books   &   MedlinePlus &   IMDB    &   SRT-serial \\
  \hline
 Original sorting                               & 0.80 &   4.10 &   14.40    &   13.40    &   64.00 \\
  \hline
 Randomly sorting       & \multirow{2}{*}{5.50} &   \multirow{2}{*}{9.30} &   \multirow{2}{*}{17.40}  &   \multirow{2}{*}{16.60}    &   \multirow{2}{*}{75.07} \\
 contextual information & & & & & \\
  \hline
 Randomly sorting                & \multirow{2}{*}{7.00}   &   \multirow{2}{*}{13.73}   &   \multirow{2}{*}{17.00}  &   \multirow{2}{*}{17.80}    &   \multirow{2}{*}{96.23} \\
 remaining words & & & & & \\
  \hline
 Randomly sorting                      & \multirow{2}{*}{15.80}    &   \multirow{2}{*}{14.67}   &   \multirow{2}{*}{18.00}  &   \multirow{2}{*}{18.20}    &   \multirow{2}{*}{92.57} \\
 everything & & & & & \\
  \hline
\end{tabular}}
\end{table}

\begin{table}[h]
\centering \caption[Normalized average error.]{Normalized average
error. Analyzing these values one can reach the same conclusions as
by analyzing the average error values. That is, both the contextual
information and the remaining words structure are relevant for the
NCD-driven text clustering.} \label{tableErrorMedioNormalizado}
\vspace{0.2cm} \footnotesize{
\begin{tabular}{|l|c|c|c|c|c|}
  \hline
                        & UCI-KDD &   Books   &   MedlinePlus &   IMDB    &   SRT-serial \\
  \hline
 Randomly sorting       & \multirow{2}{*}{6.88} &   \multirow{2}{*}{2.27} &   \multirow{2}{*}{1.21}  &   \multirow{2}{*}{1.24}    &   \multirow{2}{*}{1.17} \\
 contextual information & & & & & \\
  \hline
 Randomly sorting       & \multirow{2}{*}{8.75}   &   \multirow{2}{*}{3.35}   &   \multirow{2}{*}{1.18}  &   \multirow{2}{*}{1.33}    &   \multirow{2}{*}{1.50} \\
 remaining words & & & & & \\
  \hline
 Randomly sorting       & \multirow{2}{*}{19.75}    &   \multirow{2}{*}{3.58}   &   \multirow{2}{*}{1.25}  &   \multirow{2}{*}{1.36}    &   \multirow{2}{*}{1.45} \\
 everything & & & & & \\
  \hline
\end{tabular}}
\end{table}

\section{Summary and Conclusions}
\label{ESWA. Summary}

The analysis that has been made in this chapter is the natural
continuation of the one made in the previous chapter. In that
chapter, different word removal techniques were applied to gradually
filter the information contained in several sets of documents. It
was shown that the application of a specific word removal technique
could improve the non-distorted clustering results.

It is worth recalling that this technique was designed bearing the
intrinsic nature of texts in mind. Texts contain words that provide
a lot of information about the subject matter, at the same time as
they contain other words with little meaning or relevance. Although,
in principle, the non-relevant words are not as important as the
relevant ones, the former constitute the substrate that supports the
latter.

Generally, the more frequent a word is, the less relevant to the
subject matter \cite{Luhn58}. The main idea of the above mentioned
distortion technique is to remove the non-relevant words maintaining
the relevant ones. Thus, the distortion technique consists of
removing the most frequent words in the English language from the
documents. Instead of simply deleting the words, the technique
replaces them using asterisks with the aim of maintaining the text's
structure despite the removal. This simple idea allows maintenance
of part of the contextual information, while filtering the
information contained in the documents.

The immediate conclusion that can be drawn from the results that
have been presented in Chapter \ref{Chapter: Study on text
distortion} is that the words that remain in the documents after the
application of such a distortion technique are the ones that contain
the most relevant information in the texts. However, this distortion
technique implies, not only the presence of some words, but also the
presence of the previous text structure. Therefore, analyzing how
the maintenance of the previous text structure affects the obtained
results becomes necessary.

A comparison between this technique and three new distortion
techniques that destroy the contextual information in different
manners has been carried out in this chapter. Two main conclusions
can be drawn by analyzing the results. First, it seems that
maintaining the contextual information allows one to obtain better
clustering results than losing it. Thus, it seems that by preserving
the contextual information, the compressor is able to better capture
the internal structure of the texts. Consequently, the compressor
obtains more reliable similarities, and the non-distorted clustering
results can be improved. Second, losing the structure of the
remaining words affects the clustering results negatively.
Therefore, it can be concluded that, in this scenario, both
contextual information and remaining words have some relevance in
the text clustering behavior.

Summarizing, two main contributions have been presented in this
chapter of the thesis. First, the relevance that the contextual
information has in this compression-based text clustering scenario
has been evaluated. Second, new insights for the evaluation and
explanation of the behavior of the compression distances, in
relation to contextual information have been given.


\chapter{Application to Document Searching}
\label{Chapter: Application to Document Retrieval}

The two previous chapters perform an experimental evaluation of
the impact that several text distortion techniques have on the
NCD-driven text clustering. They show that the application of a
specific distortion technique can improve the non-distorted
clustering results.

In this chapter, this distortion technique is applied to NCD-driven
document search. It is worth mentioning that the document search
method used in this chapter applies passage retrieval to address the
problem that the NCD has when it is used to compare very different
sized objects \cite{Cilibrasi05,TFM-Rafa}.

The results presented in this chapter show that the application of
the above mentioned distortion technique can improve the
non-distorted search results.

The main contributions of this chapter of the thesis can be briefly
summarized as follows:

\begin{itemize}
\item
Practical application of the main conclusions taken from the studies
developed in the first two parts of the thesis to document search.

\item
Improvement in the representation of documents that allows
increasing the accuracy of the results obtained when searching
documents.
\end{itemize}

The chapter is structured as follows. Section \ref{BBDD. Search
Engine} describes the NCD-based document search method used in this
part of the thesis. Section \ref{BBDD. Datasets} describes the
datasets used to perform the experiments. Section \ref{BBDD.
Experimental Results} gathers and analyzes the obtained results.
Finally, Section \ref{BBDD. Summary} summarizes the conclusions
drawn from the experiments presented in this chapter.

\clearpage

\section{NCD-based Document Search Method}
\label{BBDD. Search Engine}

The NCD has been successfully applied to a wide range of domains, as
stated in Chapter \ref{Chapter: Related Work}. Nevertheless, there
is an issue that needs to be addressed if one wants to apply it
under particular circumstances. Its drawback is that it does not
commonly perform well when the compared objects are very different
in size \cite{Cilibrasi05}, as described in Section \ref{State.
Compression Distances}.

Although in many domains this does not constitute an obstacle, it
can be a problem in those kinds of scenarios in which two objects of
very different size are compared. This is, for example, the case of
a typical document search scenario, because the size of the query
can be very different from the size of the documents to be searched.

Because of that, in principle, the application of compression
distances to document search cannot seem very appropriate. However,
addressing this weakness, one can benefit from NCD's strengths. The
NCD-based document search method used in this thesis
\cite{Granados11tkde,Martinez08,TFM-Rafa} faces this problem by
applying the philosophy of passage retrieval
\cite{Salton93,Hearst93,Callan94}.

Passage retrieval is in principle similar to document retrieval, but
involves the additional, preliminary stage of extracting passages
from documents \cite{Kaszkiel97}. Thus, passage retrieval is based
on considering the documents as sets of passages instead of
considering them as atomic units.

Previous research has shown that passage retrieval can be used to
improve document retrieval accuracy when the documents are long,
have a complex structure, or are short but span many subjects
\cite{Callan94}.

There is a wealth of literature about different strategies which
have been used to divide the documents into fragments. Thus, among
others, structural features \cite{Salton93,Wilkinson94,Zobel95}, or
semantic features \cite{Hearst93,Knaus96,Mittendorf94} have been
used to delimit the passages. Another approach that consists of
partitioning the documents into fragments of text of a given size
has also been used \cite{Callan94,Kaszkiel97,Tiedemann08,Xi01}.

This last method, which is commonly called window-based, is used in
this thesis because this approach is the most appropriate to solving
the problem that the NCD has when the compared objects are very
different in size. This is due to the fact that dividing the
documents into fragments of equal size, simply avoids the problem.

Different sizes of windows have been used in window-based passage
retrieval. For example, the use of passages of 150-300 words has
been proposed in \cite{Callan94}. Other researchers have proposed
passages exceeding 500 words \cite{Kwok95}. More recent approaches
have experimented with different lengths and a window size of 50
gave the best results in \cite{Xi01}, while a bigger window size
(200-1000) gave the best results in \cite{Tiedemann08}.

Since the search method used tries to solve the NCD weakness without
restricting the size of the queries, it uses windows of different
sizes. In particular, the sizes of windows go from 1 KB to $N$ KB.

The minimum size of a window is 1KB due to the fact that the search
engine is based on the NCD, and the NCD uses compression algorithms
to estimate the entropy of the file, and compressors that are
entropy encoders need the input file to be large in order to behave
like an entropy encoder \cite{Salomon2004}.

The maximum size $N$ of a window depends on the compressor used.
This is due to the fact that compression algorithms use a memory
window which defines the best -most compressive- behavior of the
algorithm. The engine uses a version of the Lempel-Ziv algorithm
of a window size of 32 KB, therefore, $N$ = 32. The engine stores
the obtained passages in different databases, depending on their
size. Thus, there are 32 different databases, since the documents
are split in passages from 1 KB to 32 KB.

In the segmentation process, relevant paragraphs can be cut up and
divided among different passages. This can lead to a critical
fragmentation of the information contained in the paragraphs. The
NCD-based document search method solves this problem by using
overlap. Thus, each passage contains some bytes of the previous one.
Further information on said NCD-based document search method can be
found in \cite{Granados11tkde,Martinez08,TFM-Rafa}.

\section{Datasets}
\label{BBDD. Datasets}

Three datasets composed of texts written in English have been used
in the experiments. Although the detailed description of the data
sets can be found in Appendix \ref{Appendix Datasets}, a summarized
description of them can be found here:

\begin{itemize}
\item \textbf{UCM dataset}: 104 articles related to computer
science written by researchers at the ``Universidad Complutense de
Madrid'' -UCM- to be clustered by topic.

\item
\textbf{Reuters dataset}: 200 documents from a newsgroup from
Reuters, to be clustered by topic. This dataset has been adapted to
make it suitable for our experiments using the method described in
Appendix \ref{Appendix Datasets}.

\item \textbf{\emph{20newsgroups} dataset}: This well-known dataset is
composed of 20.000 documents on 20 different topics. The dataset can
be downloaded from the UCI Knowledge Discovery in Databases Archive
\cite{uci-kdd}. In the same way as the previous dataset, this
dataset has been adapted to make it suitable for our experiments
using the method described in Appendix \ref{Appendix Datasets}.
\end{itemize}

Table \ref{BBDD. Table. Dataset description} summarizes the
characteristics of the dataset and the queries used in this chapter
of the thesis. However, more detailed information on the datasets
and the queries used can be seen in Appendices \ref{Appendix
Datasets} and \ref{Appendix Queries}, respectively.

\begin{table}[ht]
\centering \caption{Datasets and experiments description.}
\textrm{(*) Adaptation described in Appendix \ref{Appendix
Datasets}.} \label{BBDD. Table. Dataset description} \vspace{0.2cm}
\small{
\begin{tabular}{|l|l|c|c|c|}
\hline
   \multicolumn{2}{|c|}{} & UCM & \emph{20newsgroups} & Reuters \\
\hline
 \multirow{2}{*}{Dataset description}  &   \#documents  & 104 & 20000               & 200 \\
\cline{2-5}
                                       &   \#topics     & 11  & 20                  & 10 \\
\hline
 \multirow{6}{*}{Experiments description} & \multirow{2}{*}{documents} & \multirow{2}{*}{papers} & one file per & one file per \\
  & & &  topic (*) &  topic (*) \\
\cline{2-5}
  & queries      & abstracts & messages      & news \\
\cline{2-5}
  & \#queries    & 4   & 10                  & 10 \\
\cline{2-5}
  & \multirow{2}{*}{queries' sizes}   & 2 x 1KB  & 7 x 2KB             & \multirow{2}{*}{10 x 2KB} \\
  &  & 2 x 2KB  & 3 x 3KB             &  \\
\hline
\end{tabular}}
\end{table}

\vspace{-0.6cm}

\section{Experimental Results}
\label{BBDD. Experimental Results}

The objective of this chapter is not to find the best way of
retrieving information, but to show that some information can be
more accurately retrieved by applying a distortion technique that
changes the representation of the input data in a specific manner.

In a retrieval system, precision is the fraction of retrieved
instances that are relevant. For example, if the system retrieves 10
results but only 6 of them are relevant, then the precision of the
retrieved results would be 0.6.

Another measure commonly used in retrieval systems is the
precision-at-K, which is the precision obtained for the first K
retrieved results.

All the figures presented in this section contain a graph and a
table. The graphs depict the precision-at-K obtained for the first K
retrieved results, whereas the tables show the precision-at-K values
for a selection of Ks. All the results are averaged over the queries
used for each dataset.

There is one figure for each dataset. Each figure shows the benefits
of applying distortion in a specific dataset. Fig \ref{BBDD.
Fig.ucm} corresponds to the UCM dataset, Fig \ref{BBDD.
Fig.20newsgroups} corresponds to the \emph{20newsgroups} dataset,
and Fig \ref{BBDD. Fig.reuters} corresponds to the Reuters dataset.

The values on the ``Distortion'' axis correspond to the cumulative
sum of the BNC-based frequencies of the words deleted from the
documents. As Section \ref{TKDE. Distortion Techniques} explains,
ten different sets of words to be removed from the documents are
used to evaluate the distortion. Each set contains the words that
account for a specific frequency, these values going from 0.1 to
1.0. The results obtained for each set correspond to the values from
0.1 to 1.0 in the ``Distortion'' axis. Moreover, the results
obtained with no distortion are depicted in the figure as the values
that correspond to a distortion of 0.

In addition, in order to ease the comparison between the precision
values obtained, a summary of these values is shown in the tables
contained in all the figures. Each row of a table contains the
precisions obtained using a specific distortion. Thus, the first row
contains the precision at K obtained without distortion, the second
row corresponds to a distortion of 0.1, the third row corresponds to
a distortion of 0.2, and so on. The precision values that maintain
or improve the ones obtained with no distortion are highlighted in
bold.

\begin{figure}
\centering \subfigure{
\includegraphics[width=14cm]{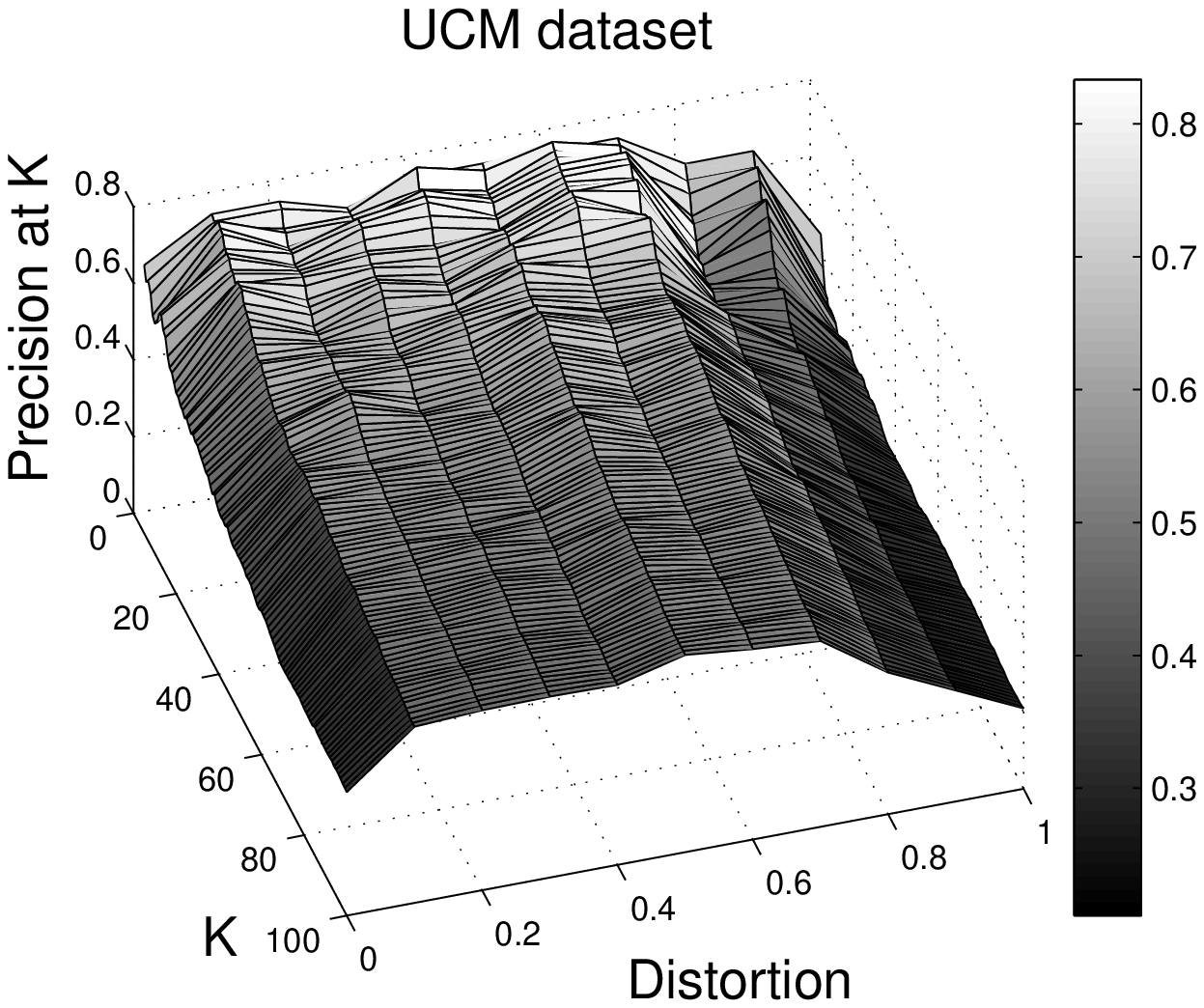}} \small{
\begin{tabular}{lcccccccc} \toprule
 Dist. & \multicolumn{8}{c}{Average precision at K passages} \\
\cmidrule(r){2-9}
 & 5 & 10 & 15 & 20 & 30 & 40 & 50 & 100 \\
\midrule
0.0 & 0.70        & 0.60        & 0.62        & 0.53        & 0.49        & 0.43        & 0.41        & 0.32 \\
0.1 & \textbf{0.80} & \textbf{0.80} & \textbf{0.75} & \textbf{0.73} & \textbf{0.63} & \textbf{0.58} & \textbf{0.58} & \textbf{0.46} \\
0.2 & \textbf{0.80} & \textbf{0.73} & \textbf{0.72} & \textbf{0.66} & \textbf{0.64} & \textbf{0.63} & \textbf{0.57} & \textbf{0.47} \\
0.3 & \textbf{0.75} & \textbf{0.78} & \textbf{0.73} & \textbf{0.71} & \textbf{0.63} & \textbf{0.58} & \textbf{0.56} & \textbf{0.47} \\
0.4 & \textbf{0.80} & \textbf{0.80} & \textbf{0.72} & \textbf{0.68} & \textbf{0.65} & \textbf{0.59} & \textbf{0.58} & \textbf{0.47} \\
0.5 & \textbf{0.80} & \textbf{0.75} & \textbf{0.73} & \textbf{0.74} & \textbf{0.69} & \textbf{0.66} & \textbf{0.63} & \textbf{0.52} \\
0.6 & \textbf{0.80} & \textbf{0.80} & \textbf{0.80} & \textbf{0.74} & \textbf{0.64} & \textbf{0.61} & \textbf{0.57} & \textbf{0.50} \\
0.7 & \textbf{0.80} & \textbf{0.80} & \textbf{0.77} & \textbf{0.75} & \textbf{0.64} & \textbf{0.63} & \textbf{0.60} & \textbf{0.49} \\
0.8 & 0.70        & 0.55        & 0.48        & 0.49        & 0.49  & \textbf{0.49} & \textbf{0.47} & \textbf{0.37} \\
0.9 & 0.70        & \textbf{0.63} & 0.50        & 0.48        & 0.42        & 0.36        & 0.33        & 0.29 \\
1.0 & 0.45        & 0.30        & 0.27        & 0.28        & 0.27        & 0.23        & 0.22        & 0.21 \\
\bottomrule
\end{tabular}}
\caption[UCM dataset. Benefits of applying distortion.]{UCM dataset.
Benefits of applying distortion. The precision values that maintain
or improve the ones obtained with no distortion are highlighted in
bold.} \label{BBDD. Fig.ucm}
\end{figure}

Looking at Fig \ref{BBDD. Fig.ucm} one can see that higher precision
values are obtained when distorting the documents than when not
distorting them for the UCM dataset. This fact can be noticed by
comparing the non-distorted results and the rest of the results both
in the surface and in the table. Therefore, one can assert that the
application of document distortion improves the quality of the
retrieved results for this dataset. In particular, the best
precision value achieved without distortion is 0.70, while the best
precision achieved using distortion is 0.80. This implies an
improvement of 14\%. Fig \ref{BBDD. Fig.summary} summarizes the
improvements obtained for the best distortions with respect to the
results obtained without distortion.

\begin{figure}
\centering \subfigure{
\includegraphics[width=14cm]{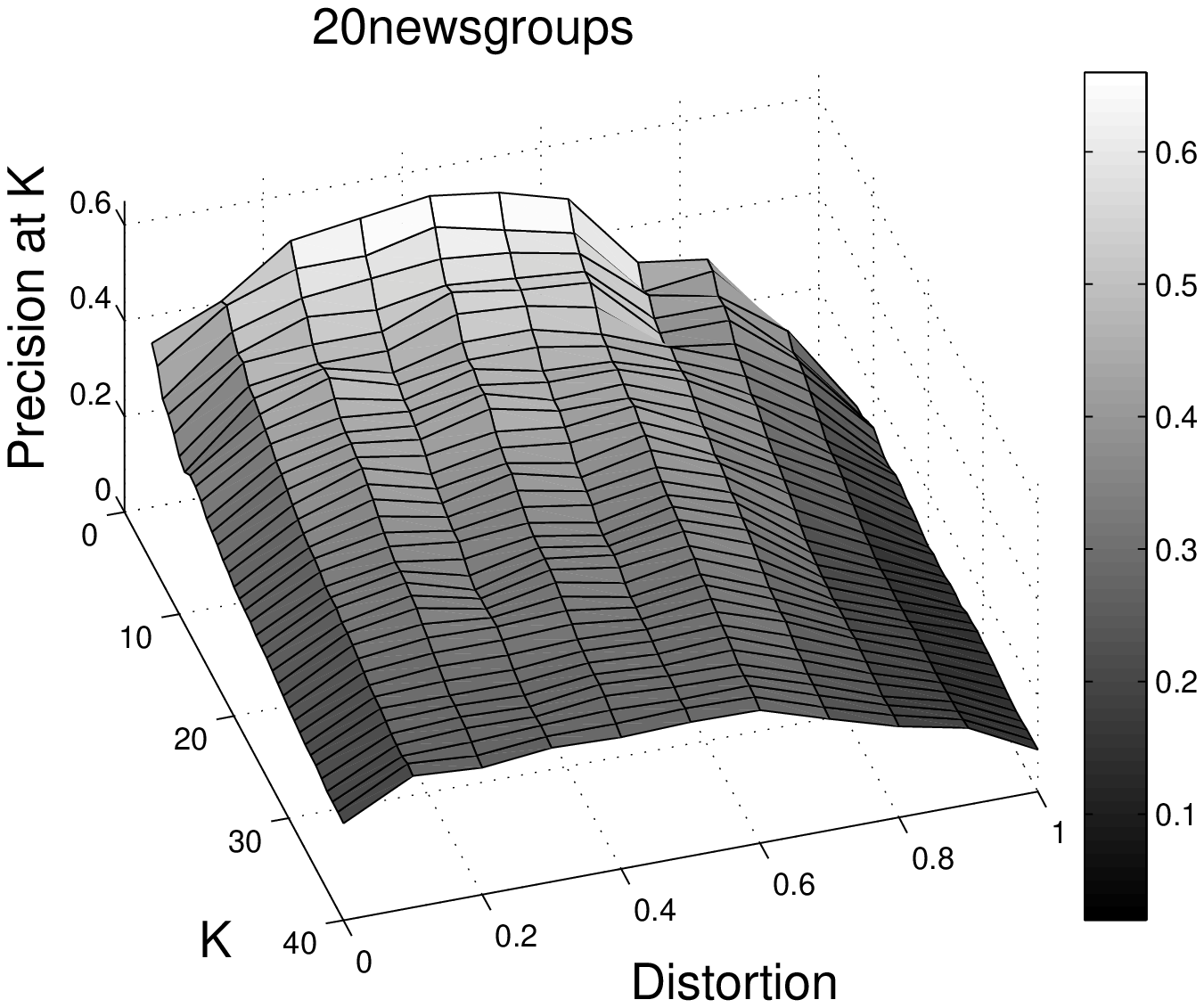}} \small{
\begin{tabular}{lcccccccc}
\toprule
 Dist. & \multicolumn{8}{c}{Average precision at K passages} \\
\cmidrule(r){2-9}
 & 5 & 10 & 15 & 20 & 30 & 40 & 50 & 100 \\
\midrule
0.0 & 0.46        & 0.33        & 0.31        & 0.27        & 0.23        & 0.20        & 0.19        & 0.15 \\
0.1 & \textbf{0.60} & \textbf{0.47} & \textbf{0.42} & \textbf{0.38} & \textbf{0.32} & \textbf{0.29} & \textbf{0.25} & \textbf{0.27} \\
0.2 & \textbf{0.62} & \textbf{0.47} & \textbf{0.43} & \textbf{0.40} & \textbf{0.32} & \textbf{0.27} & \textbf{0.24} & \textbf{0.19} \\
0.3 & \textbf{0.64} & \textbf{0.47} & \textbf{0.42} & \textbf{0.37} & \textbf{0.32} & \textbf{0.28} & \textbf{0.25} & \textbf{0.20} \\
0.4 & \textbf{0.66} & \textbf{0.52} & \textbf{0.46} & \textbf{0.40} & \textbf{0.33} & \textbf{0.28} & \textbf{0.26} & \textbf{0.19} \\
0.5 & \textbf{0.64} & \textbf{0.51} & \textbf{0.42} & \textbf{0.37} & \textbf{0.31} & \textbf{0.28} & \textbf{0.25} & \textbf{0.20} \\
0.6 & \textbf{0.60} & \textbf{0.48} & \textbf{0.43} & \textbf{0.40} & \textbf{0.33} & \textbf{0.28} & \textbf{0.26} & \textbf{0.19} \\
0.7 & 0.44        & \textbf{0.37} & \textbf{0.37} & \textbf{0.34} & \textbf{0.27} & \textbf{0.23} & \textbf{0.21} & \textbf{0.16} \\
0.8 & 0.42        & \textbf{0.35} & 0.29        & 0.25        & 0.20        & 0.19        & 0.17        & 0.13 \\
0.9 & 0.24        & 0.26        & 0.20        & 0.17        & 0.15        & 0.16        & 0.15        & 0.11 \\
1.0 & 0.02        & 0.12        & 0.11        & 0.09        & 0.10        & 0.09        & 0.09        & 0.08 \\
\bottomrule
\end{tabular}}
\caption[\emph{20newsgroups} dataset. Benefits of applying
distortion.]{\emph{20newsgroups} dataset. Benefits of applying
distortion. Again, the precision values that maintain or improve the
ones obtained with no distortion are highlighted in bold.}
\label{BBDD. Fig.20newsgroups}
\end{figure}

\begin{figure}
\centering \subfigure{
\includegraphics[width=14cm]{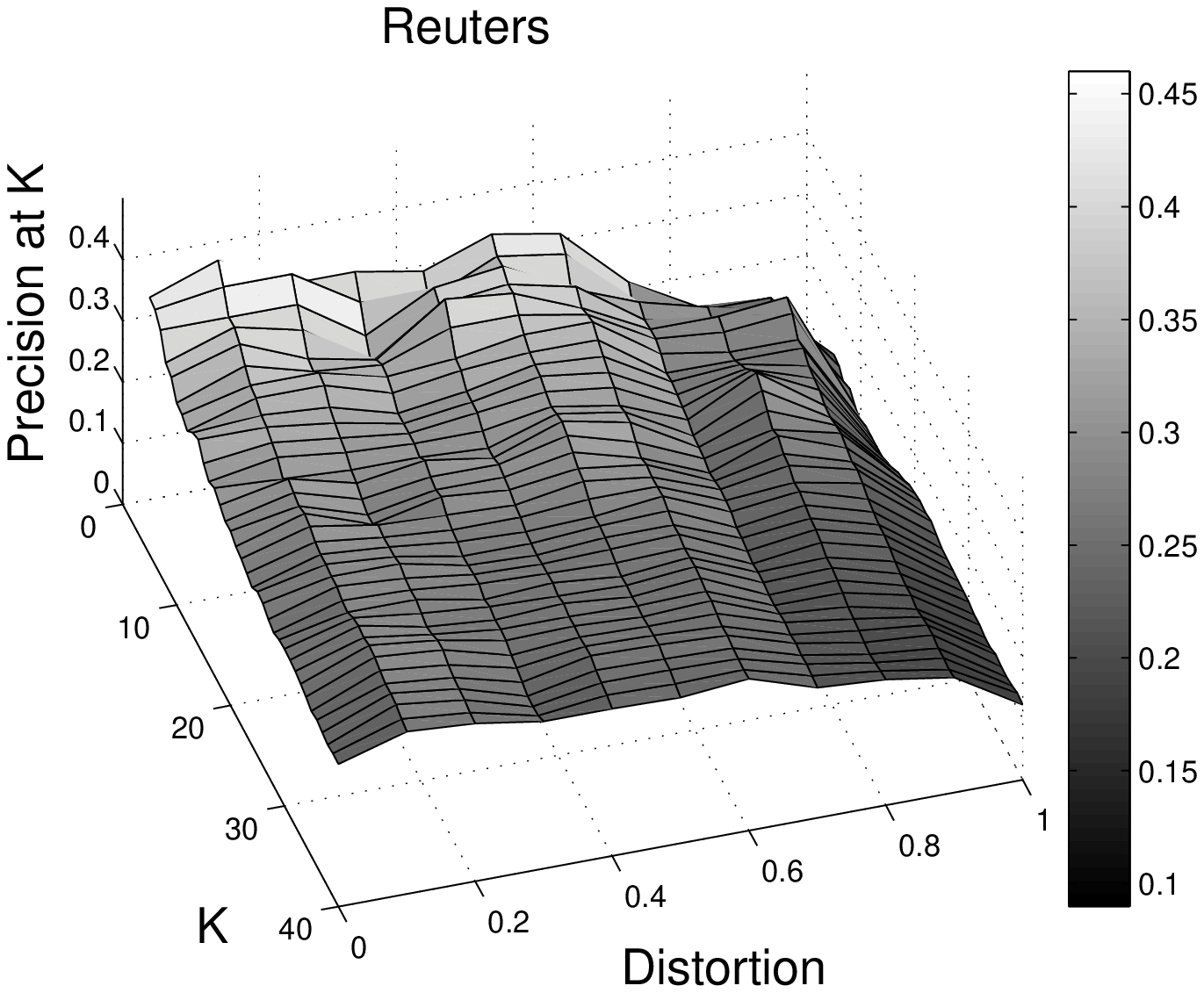}} \small{
\begin{tabular}{lcccccccc}
\toprule
 Dist. & \multicolumn{8}{c}{Average precision at K passages} \\
\cmidrule(r){2-9}
 & 5 & 10 & 15 & 20 & 30 & 40 & 50 & 100 \\
\midrule
0.0 & 0.42        & 0.33        & 0.31        & 0.29        & 0.26        & 0.23        & 0.21        & 0.17 \\
0.1 & \textbf{0.46} & \textbf{0.37} & \textbf{0.33} & \textbf{0.31} & \textbf{0.29} & \textbf{0.27} & \textbf{0.25} & \textbf{0.30} \\
0.2 & 0.40        & \textbf{0.34} & 0.31      & 0.27        & \textbf{0.27} & \textbf{0.26} & \textbf{0.24} & \textbf{0.18} \\
0.3 & 0.40        & 0.33          & 0.30      & 0.29        & 0.26          & \textbf{0.24} & \textbf{0.22} & \textbf{0.18} \\
0.4 & 0.38        & \textbf{0.37} & 0.31      & 0.29        & 0.26          & \textbf{0.24} & \textbf{0.24} & \textbf{0.19} \\
0.5 & 0.42        & \textbf{0.37} & \textbf{0.33} & \textbf{0.31} & 0.26    & \textbf{0.24} & \textbf{0.23} & \textbf{0.18} \\
0.6 & 0.40        & \textbf{0.35} & 0.31      & 0.28        & \textbf{0.27} & \textbf{0.25} & \textbf{0.23} & \textbf{0.18} \\
0.7 & 0.30        & 0.29        & 0.26        & 0.24        & 0.22        & 0.21        & 0.20        & 0.17 \\
0.8 & 0.24        & 0.28        & 0.30        & 0.28        & 0.22        & 0.21        & 0.20        & 0.17 \\
0.9 & 0.22        & 0.29        & 0.25        & 0.24        & 0.19        & 0.19        & 0.17        & 0.14 \\
1.0 & 0.12        & 0.09        & 0.11        & 0.14        & 0.13        & 0.12        & 0.12        & 0.12 \\
\bottomrule
\end{tabular}}
\caption[Reuters dataset. Benefits of applying distortion.]{Reuters
dataset. Benefits of applying distortion. The precision values that
maintain or improve the ones obtained with no distortion are
highlighted in bold.} \label{BBDD. Fig.reuters}
\end{figure}

Analyzing the results that correspond to the \emph{20newsgroups}
dataset -Fig \ref{BBDD. Fig.20newsgroups}- one can observe that,
again, the results obtained when distorting the documents are better
than the ones obtained when not distorting them. In particular, the
best precision value achieved without distortion is 0.46, while the
best precision achieved using distortion is 0.66. This implies an
improvement of 43\%.

The same happens in the Reuters dataset. Thus, looking at Fig
\ref{BBDD. Fig.reuters} one can see that the best precision value
obtained without distortion is 0.42, whereas the best precision
obtained using distortion is 0.46. Although this improvement is
lower than the one achieved for the \emph{20newsgroups} dataset, it
implies an improvement of 10\%.

\subsection{Summary of Results}
\label{BBDD. Summary of Results}

\begin{figure}[h]
\centering
  \includegraphics[angle=270,width=13.5cm]{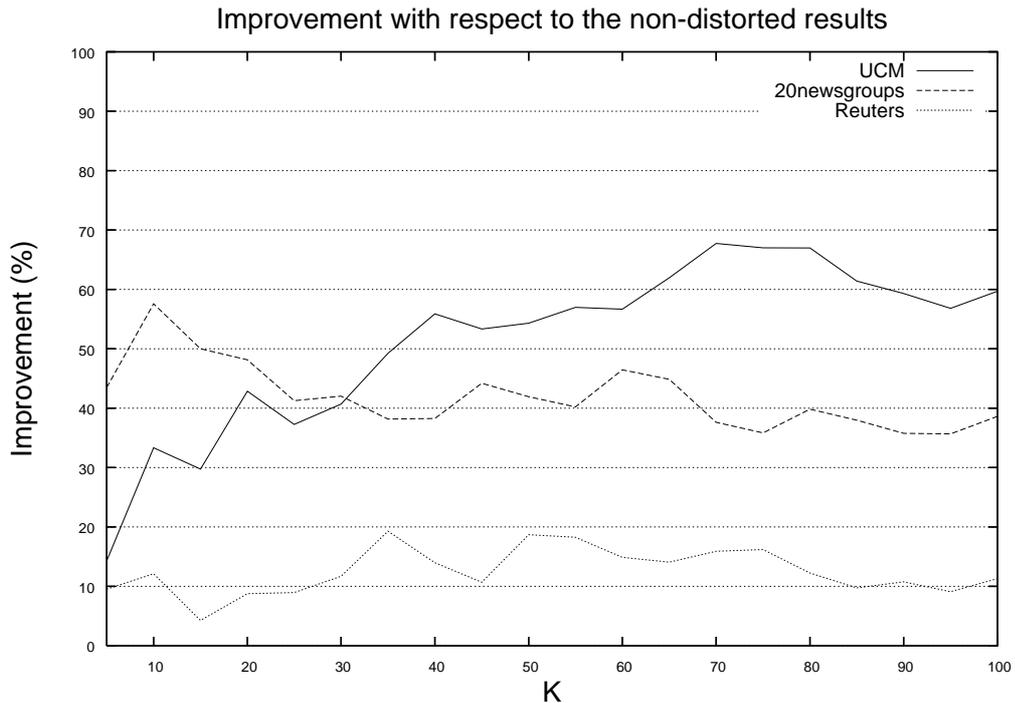}\\
\caption[Percentage of improvement.]{Percentage of improvement for
the best distortion with respect to the results obtained with no
distortion.}
 \label{BBDD. Fig.summary}
\end{figure}

The figures presented in the previous subsection comprehensively
show all the experimental results. That is, they show the precision
values obtained for all the explored distortions. In this
subsection, a figure that summarizes those results is presented.
This figure -Fig \ref{BBDD. Fig.summary}- depicts the percentage of
improvement for the best distortion, with respect to the results
obtained with no distortion.

Although the best precision is achieved at a different distortion
depending on the dataset used, this is consistent with the cut-offs
defined by Luhn which exclude non-significant words \cite{Luhn58}.
These cut-offs have to be established by trial and error because
they are different for each dataset \cite{Rijsbergen79}. That is,
the non-relevant words, and therefore, the relevant ones, depend on
the dataset used.

However, analyzing the curves plotted in Fig \ref{BBDD. Fig.summary}
one can observe that although the percentage of improvement is
different for each dataset, there is always an improvement.
Therefore, it can be concluded that applying document distortion is
advantageous for the search method used in this thesis in terms of
accuracy.

\section{Summary and Conclusions}
\label{BBDD. Summary}

This chapter of the thesis has applied the document distortion
technique that was found to be beneficial in document clustering
-see Chapters \ref{Chapter: Study on text distortion} and
\ref{Chapter: Relevance of the contextual information}- to document
search. In particular, an NCD-driven document search method based on
passage retrieval has been used in this chapter.

The experimental results have shown that the non-distorted search
results can be improved by applying such a document distortion
technique. The results are consistent across all the datasets used,
that is, the search accuracy has been improved in all cases, as Fig
\ref{BBDD. Fig.summary} evidences. However, the best precision
achieved for each dataset has been obtained at a different
distortion. This behavior is consistent with the cut-offs defined by
Luhn, an upper and a lower, which exclude the non-significant words
\cite{Luhn58}. These cut-offs have to be established by trial and
error \cite{Rijsbergen79}, which means that they are different for
each dataset. That is, the non-relevant words, and therefore, the
relevant ones, depend on the dataset used. This can explain why the
best precision is achieved at a different distortion for each
dataset.

Summarizing, two main contributions have been presented in this
chapter of the thesis. First, a practical application of the main
conclusions taken from the studies developed in the first two parts
of the thesis to a document search scenario has been given. Second,
a representation of the documents that improves the non-distorted
document search accuracy has been proposed.


\chapter{Conclusions}
\label{Chapter: Conclusions}

The first part of this thesis -Chapter \ref{Chapter: Study on text
distortion}- has taken a step towards understanding compression
distances by performing an experimental evaluation of the impact
that several kinds of word removal have on compression-based text
clustering. Six different distortion techniques based on word
removal have been applied to gradually filter the information
contained in the datasets. See Section \ref{TKDE. Distortion
Techniques} for a detailed description of the distortion techniques.

It has been shown that the application of a specific word removal
technique can improve the non-distorted clustering results. This
fact can be observed analyzing the figures shown in Section
\ref{TKDE. Experimental Results} .

The distortion technique that leads to an improvement in the
non-distorted clustering results consists in erasing the most
frequent words in the English language from the documents, replacing
their characters with asterisks. This strategy allows preservation
of the text structure despite the word removal. The experimental
results presented in Chapter \ref{Chapter: Study on text distortion}
show that the application of this distortion technique improves the
non-distorted clustering results even, and precisely when, a lot of
words are removed from the documents.

Since this technique implies, not only the removal of words, but
also the maintenance of the previous text structure, analyzing the
impact of both factors becomes necessary in order to better
understand why the clustering results can be improved by applying
the technique. This research is carried out in the second part of
the thesis, which corresponds to Chapter \ref{Chapter: Relevance of
the contextual information}.

Thus, the analysis made in Chapter \ref{Chapter: Relevance of the
contextual information} is the natural continuation of the one made
in Chapter \ref{Chapter: Study on text distortion}. Chapter
\ref{Chapter: Relevance of the contextual information} explores how
the loss or the maintenance of the contextual information affects
the clustering accuracy. Moreover, it explores how the loss or the
preservation of the remaining words structure affects the
clustering.

This has been accomplished by creating three new distortion
techniques based on the distortion technique, presented in Chapter
\ref{Chapter: Study on text distortion}, which improves the
clustering accuracy. The detailed description of said distortion
techniques is shown in Section \ref{ESWA. Distortion Techniques}.

Two main conclusions can be drawn by analyzing the results presented
in Section \ref{ESWA. Experimental Results}. Firstly, it seems that
maintaining the contextual information allows one to obtain better
clustering results than losing it. Thus, it seems that by preserving
the contextual information, the compressor is better able to capture
the internal structure of the texts. Consequently, the compressor
obtains more reliable similarities, and the non-distorted clustering
results can be improved. Secondly, losing the structure of the
remaining words affects the clustering results negatively.
Therefore, it can be concluded that, in this scenario, both
contextual information and remaining words have some relevance in
the text clustering behavior.

Everything learned in the first two parts of the thesis has been
applied to compression-based document search in the last part of the
thesis -Chapter \ref{Chapter: Application to Document Retrieval}-.
The objective of that chapter is not finding the best way of
retrieving information, but showing that some information can be
more accurately retrieved by applying a distortion technique that
changes the representation of the input data in a specific manner.

Despite the wide and successful use of compression distances, their
application to document search is not trivial due to their having a
weakness that must be taken into account if one wants to apply them
to document search. Their drawback is that they do not commonly
perform well when the compared objects have very different sizes. An
NCD-based document search engine that deals with that drawback by
using passage retrieval, is used in the last part of the thesis.

The results presented in Section \ref{BBDD. Experimental Results}
show that the non-distorted search results can be improved by
applying the document distortion technique that improves the
non-distorted clustering results. This fact augments the
applicability of the distortion technique, presented in Chapter
\ref{Chapter: Study on text distortion}, which fine-tunes the text
representation.

Summarizing, from all the results presented in this thesis, it can
be concluded that the application of one of the explored distortion
techniques can be beneficial both in NCD-based document clustering
and in NCD-based document search.

\chapter{Summary of Results}
\label{Summary of Results}

This chapter shows how the main objectives of this thesis have
been achieved.

\begin{itemize}
\item \emph{The attainment of objective 1: Providing new perspectives
for understanding the nature of textual data.}

All the experiments carried out in this thesis have been focused on
better understanding of both the nature of textual data, and the
nature of compression distances. The way in which the thesis studies
textual data is evaluating the impact that diverse distortion
techniques have on compression-based document clustering -Chapters
\ref{Chapter: Study on text distortion} and \ref{Chapter: Relevance
of the contextual information}-, and their impact on
compression-based document search -Chapter \ref{Chapter: Application
to Document Retrieval}-.

The experimental results presented in those chapters have shown that
distorting the documents, by erasing the most frequent words in the
English language in a specific manner, improves the accuracy of both
the document clustering and the document search. This means that
such distortion preserves the relevant information contained in the
documents while removing the non-relevant information.

Moreover, the results presented in Chapter \ref{Chapter: Relevance
of the contextual information} have shown that the maintenance of
the previous structure of the texts, despite the word removal has
beneficial effects on the clustering results. Therefore, it seems
that the contextual information is relevant in this kind of
scenario.

\item \emph{The attainment of objective 2. Providing a technique to smoothly reduce
the complexity of the documents while preserving most of their
relevant information.}

As said previously, several distortion techniques have been
evaluated in this thesis. One of them produces both a smooth
decrease of the non-relevant information in the set of documents
considered, and a smooth decrease of the documents complexity
estimation. The application of this technique leads to an
improvement in both the document clustering accuracy and the
document search accuracy. The detailed description of the distortion
technique can be seen in Chapter \ref{Chapter: Study on text
distortion}. The effects of applying that technique can be seen in
Sections \ref{TKDE. Experimental Results}, \ref{ESWA. Experimental
Results}, and \ref{BBDD. Experimental Results}.

\item \emph{The attainment of objective 3. Giving experimental evidence of how to
fine-tune the text representation so that better results are
obtained when using the NCD-driven text clustering.}

The above mentioned distortion technique fine-tunes the text
representation by filtering the information contained in the texts.
This leads to better clustering results, as Sections \ref{TKDE.
Experimental Results}, and \ref{ESWA. Experimental Results} show.

\item \emph{The attainment of objective 4. Giving new insights for the evaluation
and explanation of the behavior of the NCD.}

The distortion techniques explored in this thesis are the tool that
allows one to study the behavior of the NCD. Chapters \ref{Chapter:
Study on text distortion}, \ref{Chapter: Relevance of the contextual
information}, and \ref{Chapter: Application to Document Retrieval}
present such a study.

It seems that distorting the documents by removing the most frequent
words in the English language, the compressor is able to better
capture the internal structure of the texts. Consequently, the
compressor obtains more reliable similarities, and the results can
be improved.

\item \emph{The attainment of objective 5. Experimentally evaluating the relevance
that the contextual information has in compression-based text
clustering, in a word removal scenario.}

The distortion technique that fine-tunes the text representation
implies not only the removal of words, but also the maintenance of
the previous text structure. Chapter \ref{Chapter: Relevance of the
contextual information} has explored the relevance of both factors
with the aim of better understanding the results. It has been
observed that maintaining the contextual information allows one to
obtain better clustering results than losing it, as Section
\ref{ESWA. Experimental Results} shows.

\item \emph{The attainment of objective 6. Applying the main conclusions taken from
the studies developed in the first two parts of the thesis to
document search.}

The technique that filters the information contained in the texts by
removing the irrelevant words while maintaining the previous text
structure can be used in different application domains. Chapter
\ref{Chapter: Application to Document Retrieval} applies it to
document search with the purpose of studying if the improvements
observed when clustering documents are observed when searching
documents, as well.

\item \emph{The attainment of objective 7. Giving a representation of documents
that improves the non-distorted document search accuracy.}

Since text representation plays an important role in document
search, the application of a distortion technique that fine-tunes
the representation of texts can be beneficial. Chapter \ref{Chapter:
Application to Document Retrieval} has explored if the above
mentioned distortion technique can lead to better document search
results. The experimental results have shown that changing the
representation of the documents by applying such a distortion
technique leads to better results, as Section \ref{BBDD.
Experimental Results} shows.

\end{itemize}


\chapter{Contributions}
\label{Chapter: Contributions}

\begin{itemize}

\item INTERNATIONAL JOURNALS

\begin{itemize}
\item \textbf{Reducing the Loss of Information through Annealing Text Distortion},
Ana Granados, Manuel Cebrián, David Camacho, and Francisco de Borja
Rodriguez, \emph{IEEE Transactions on Knowledge and Data
Engineering}, vol.23, n. 7, July 2011. ISSN  1041-4347. JCR (2009):
impact factor = 2.285; Ranking (Comput. Science-Inform. Systems) =
20/116.

This work explores different text distortion techniques based on
word removal. It analyzes how the information contained in the
documents and how the upper bound estimation of their Kolmogorov
complexity progress as the words are removed from the documents in
different manners. Three different ways of choosing the words to
be removed and two different ways of removing the words are
explored. Combining these two factors, six different distortion
techniques are obtained, and therefore, six methods are explored
in the work. A compression-based clustering method is used to
experimentally evaluate the impact that the studied distortion
techniques have on the amount of information contained in the
distorted documents. The experimental results show that the
application of a specific distortion technique can improve the
clustering accuracy. 

\item \textbf{Is the contextual information relevant in text
clustering by compression?}, Ana Granados, David Camacho, Francisco
de Borja Rodríguez. \emph{\textbf{Submitted} to Expert Systems with
Applications in February 2011}.

This work is the natural extension of the previous one. While the
previous work evaluates the impact that several kinds of word
removal have on compression-based text clustering, showing that the
application of one of the explored distortion techniques can improve
the non-distorted clustering results, this work analyzes the reasons
why applying that technique can improve the clustering results. The
technique implies not only the removal of words, but also the
maintenance of the previous text structure. Therefore, exploring the
relevance of both factors becomes necessary in order to better
understand the results. That is precisely what this work analyzes.
The experimental results show that maintaining the contextual
information allows one to obtain better clustering results than
losing it. 

\item \textbf{Improving the Accuracy of the Normalized Compression
Distance combining Document Segmentation and Document Distortion}, ,
Ana Granados, Rafael Martínez, David Camacho, Francisco de Borja
Rodríguez. \emph{\textbf{Submitted} to IEEE Transactions on
Knowledge and Data Engineering in October 2011}.

This work applies the knowledge learned in the previous ones to
the retrieval of documents. The retrieval approach is based on
using document segmentation and document distortion. The
experimental results show that the application of the previously
explored distortion technique can improve the retrieval results.

\end{itemize}

\item INTERNATIONAL CONFERENCES

\begin{itemize}
\item \textbf{Evaluating the Impact of Information Distortion on Normalized
Compression Distance-driven Text Clustering}, Ana Granados, Manuel
Cebrián, David Camacho, and Francisco de Borja Rodríguez, \emph{In
proceedings of 2nd International Castle Meeting on Coding Theory and
Applications, ICMCTA 2008, Lecture Notes in Computer Science, Vol.
5228}

This work is the prelude to the work deeply developed in the journal
\emph{Reducing the Loss of Information through Annealing Text
Distortion}. While the former only uses a compression algorithm and
a dataset, the latter uses three compression algorithms, each of
them belonging to a different family of compressors, and four
different datasets.

\item \textbf{Relevance of contextual information in compression-based text
clustering}, Ana Granados, Rafael Martínez, David Camacho, Francisco
de Borja Rodríguez, \emph{In proceedings of 11th International
Conference on Intelligent Data Engineering and Automated Learning,
IDEAL 2010, Lecture Notes in Computer Science}

This work is the prelude to the work deeply developed in the journal
\emph{Is the contextual information relevant in text clustering by
compression?}. While the former only uses two distortion techniques
and two datasets, the last uses four distortion techniques, and five
different datasets. 

\item \textbf{Influence of music representation on compression-based clustering},
Antonio Gonz\'alez-Pardo, Ana Granados, David Camacho, Francisco de
Borja Rodr\'iguez, \emph{In proceedings of the IEEE World Congress
on Evolutionary Computation, IEEE CEC 2010}

This paper constitutes a parallel work to the main work developed in
the thesis. The paper applies the Normalized Compression Distance to
music clustering. The paper analyzes how the selection of a
particular representation of music audio files can affect the
clustering process. Three different music representations are
explored in the paper: binary code, wave information, and SAX. In
addition, two different algorithms are applied to automatically
perform the clustering: a hierarchical clustering method based on
the quartet tree method, and a genetic algorithm. The experimental
results show how the representation of the music file plays a
decisive role in the NCD-driven clustering. Thus, the best
clustering results are obtained when the music is represented using
its wave information. The other representations - WAV file, and SAX
- get worse results. This can be due to the fact that the wave
representation implies a smaller information loss. The paper also
presents a new clustering method based on genetic algorithms as an
alternative to the clustering method developed in the CompLearn
Toolkit.

\end{itemize}

\end{itemize}


\appendix

\chapter{Acronyms}
\label{Appendix Acronyms}

\begin{description}

\item[BNC] British National Corpus

\item[BWT] Burrows-Wheeler Transform

\item[BZIP] Basic ZIPper

\item[HMM] Hidden Markov Model

\item[IMDB] Internet Movie Data Base

\item[IT] Information Theory

\item[KNN] K-Nearest Neighbour

\item[LFW] Least Frequent Words

\item[LZ] Lempel-Ziv

\item[LZMA] Lempel-Ziv-Markov chain Algorithm

\item[MFW] Most Frequent Words

\item[MTF] Move to Front Transform

\item[NCD] Normalized Compression Distance

\item[NID] Normalized Information Distance

\item[PPM] Prediction with Partial string Matching

\item[RLE] Run Length Encoding

\item[RW] Random Words

\item[UCI-KDD] University of California, Irvine, Knowledge
Discovery in Databases archive

\item[UCM] Universidad Complutense de Madrid

\item[USM] Universal Similarity Metric

\item[VSM] Vector Space Model

\end{description}

\chapter{Datasets}
\label{Appendix Datasets}

The detailed description of the datasets used throughout the thesis
can be found here. All of the datasets comprise several texts
written in English. An extract of one of the documents can be seen
in a picture for every dataset.

\section{Books dataset}

Fourteen classical books, to be clustered by author. There are:
\begin{itemize}
\item Two books by Agatha Christie:
\begin{itemize}
  \item \emph{The Secret Adversary}
  \item \emph{The Mysterious Affair at Styles}
\end{itemize}

\item Three books by Alexander Pope:
\begin{itemize}
  \item \emph{An Essay on Criticism}
  \item \emph{An Essay on Man}
  \item \emph{The Rape of the Lock, an heroic-comical Poem}
\end{itemize}

\item Two books by Edgar Allan Poe:
\begin{itemize}
  \item \emph{The Fall of the House of Usher}
  \item \emph{The Raven}
\end{itemize}

\item Two books by Miguel de Cervantes:
\begin{itemize}
  \item \emph{Don Quixote}
  \item \emph{The Exemplary Novels}
\end{itemize}

\item Three books by Niccol\`o Machiavelli:
\begin{itemize}
  \item \emph{Discourses on the First Decade of Titus Livius}
  \item \emph{History of Florence and of the Affairs of Italy}
  \item \emph{The Prince}
\end{itemize}

\item Two books by William Shakespeare:
\begin{itemize}
  \item \emph{The tragedy of Antony and Cleopatra}
  \item \emph{Hamlet}
\end{itemize}
\end{itemize}

Since the documents belonging to this dataset are books, their size
is quite big in general.

\begin{figure}
\begin{tabular}{|p{13cm}|}
\hline In a village of La Mancha, the name of which I have no desire
to call to mind, there lived not long since one of those gentlemen
that keep a lance in the lance-rack, an old buckler, a lean hack,
and a greyhound for coursing. An olla of rather more beef than
mutton, a salad on most nights, scraps on Saturdays, lentils on
Fridays, and a pigeon or so extra on Sundays, made away with
three-quarters of his income. The rest of it went in a doublet of
fine cloth and velvet breeches and shoes to match for holidays,
while on week-days he made a brave figure in his best homespun. He
had in his house a housekeeper past forty, a niece under twenty, and
a lad for the field and market-place, who used to saddle the hack as
well as handle the bill-hook. The age of this gentleman of ours was
bordering on fifty; he was of a hardy habit, spare, gaunt-featured,
a very early riser and a great sportsman. They will have it his
surname was Quixada or Quesada (for here there is some difference of
opinion among the authors who write on the subject), although from
reasonable conjectures it seems plain that he was called Quexana.
This, however, is of but little importance to our tale; it will be
enough not to stray a hair's breadth from the truth in the telling of it. \\
\hline
\end{tabular}
\centering \caption{Books. Extract from \emph{Don Quixote} by Miguel
de Cervantes.} \label{APPENDIX. Book}
\end{figure}

\section{UCI-KDD dataset}

Sixteen messages from a newsgroup (UCI-KDD) \cite{uci-kdd}, to be
clustered by topic. There are:

\begin{itemize}
\item Three documents on atheism.
\vspace{-0.2cm}
\item Three documents on Christian religion and homosexuality.
\vspace{-0.2cm}
\item Two documents on Christian religion and reincarnation.
\vspace{-0.2cm}
\item Two documents on politics and guns.
\vspace{-0.2cm}
\item Three documents on cryptography, governments and communications.
\vspace{-0.2cm}
\item Three documents on inherent problems of cryptography.
\end{itemize}

The main characteristic of these texts is their small size.

\begin{figure}
\begin{tabular}{|p{13cm}|}
\hline In case people think email scanning doesn't take place, I can
assure you that it is done regularly by many sites - usually not by
government agencies (or at least not that I know of), but by local
administrators who, for reasons of their own, have decided to
monitor all communications (I'm sure you can all think of a whole
mess of reasons - stop hackers/ terrorists/child pornographers/drug
dealers/evil commies/whatever). There have been several occasions
when I've got people into trouble simply by including the
traditional NSA bait in a message (I don't try it any more now :-).
A friend of mine was once picked up for mentioning the name of the
UK town of Scunthorpe (hint: look for words embedded in it). I'm
sure there are many more examples of this happening (in fact if
anyone has any examples I'd appreciate hearing from them - I could
use them as ammunition during talks on privacy issues).
\\
\hline
\end{tabular}
\centering \caption{UCI-KDD. Extract from a document on
cryptography.} \label{APPENDIX. UCI-KDD}
\end{figure}

\section{MedlinePlus dataset}

Twelve documents from the MedlinePlus repository \cite{medlineplus},
to be clustered by topic. There are:

\begin{itemize}
\item Three documents related to alcohol:
\begin{itemize}
  \item Alcohol use
  \item Alcoholic neuropathy
  \item Alcoholism
\end{itemize}

\item Three documents on diabetes:
\begin{itemize}
  \item Diabetes diet
  \item Diabetes education
  \item Diabetes definition
\end{itemize}

\item Three documents on meningitis:
\begin{itemize}
  \item Meningitis gramnegative
  \item Meningitis meningococcal
  \item Meningitis staphylococcal
\end{itemize}

\item Three documents on tumors:
\begin{itemize}
  \item Hepatocellular carcinoma
  \item Spinal tumor
  \item Thyroid cancer
\end{itemize}
\end{itemize}

Since these texts are about medicine, they are very specific and
their vocabulary is very specialized.

\begin{figure}
\begin{tabular}{|p{13cm}|}
\hline In many cases, moderate weight loss and increased physical
activity can control type 2 diabetes. Some people will need to take
oral medications or insulin in addition to lifestyle changes.

Children with type 2 diabetes present special challenges. Meal plans
should be recalculated often to account for the child's change in
calorie requirements due to growth. Three smaller meals and 3 snacks
are often required to meet calorie needs.

Changes in eating habits and increased physical activity help reduce
insulin resistance and improve blood sugar control. When at parties
or during holidays, your child may still eat sugar-containing foods,
but have fewer carbohydrates on that day. For example, if birthday
cake, Halloween candy, or other sweets are eaten, the usual daily
amount of potatoes, pasta, or rice should be eliminated. This
substitution helps keep calories and carbohydrates in better
balance.

For children with either type of diabetes, special occasions (like
birthdays or Halloween) require additional planning because of the
extra sweets.
\\
\hline
\end{tabular}
\centering \caption{MedlinePlus. Extract from a document on
diabetes.} \label{APPENDIX. MedlinePlus}
\end{figure}

\section{IMDB dataset}

Fourteen plots of movies from the Internet Movie Data Base (IMDB)
\cite{imdb}, to be clustered by saga. There are five different
sagas:

\begin{itemize}
\item Indiana Jones:
\begin{itemize}
  \item \emph{Raiders Of The Lost Ark}
  \item \emph{Temple Of The Doom}
  \item \emph{The Last Crusade}
\end{itemize}

\item Pirates Of The Caribbean
\begin{itemize}
  \item \emph{The Curse of the Black Pearl}
  \item \emph{Dead Man's Chest}
  \item \emph{At World's End}
\end{itemize}

\item The initial saga of Star Wars
\begin{itemize}
  \item \emph{A New Hope}
  \item \emph{The Empire Strikes Back}
  \item \emph{Revenge of the Jedi}
\end{itemize}

\item The Matrix
\begin{itemize}
  \item \emph{The Matrix}
  \item \emph{Matrix Reloaded}
  \item \emph{Matrix Revolutions}
\end{itemize}

\item The Mummy
\begin{itemize}
  \item \emph{The Mummy}
  \item \emph{The Mummy Returns}
\end{itemize}
\end{itemize}

An important characteristic of these documents is the presence of
names of characters and places that are related to the sagas.

\begin{figure}
\begin{tabular}{|p{13cm}|}
\hline Thomas A. Anderson is a man living two lives. By day he is an
average computer programmer and by night a malevolent hacker known
as Neo. Neo has always questioned his reality but the truth is far
beyond his imagination. Neo finds himself targeted by the police
when he is contacted by Morpheus, a legendary computer hacker
branded a terrorist by the government. Morpheus awakens Neo to the
real world, a ravaged wasteland where most of humanity have been
captured by a race of machines which live off of their body heat and
imprison their minds within an artificial reality known as the
Matrix. As a rebel against the machines, Neo must return to the
Matrix and confront the agents, super powerful computer programs
devoted to snuffing out Neo and the entire human rebellion.
\\
\hline
\end{tabular}
\centering \caption{IMDB. Extract from the movie \emph{The Matrix}.}
\label{APPENDIX. IMDB}
\end{figure}

\section{SRT-serial dataset}

Sixty-nine scripts of different serials which have been obtained
from \cite{srt}, to be clustered by serial. There are three chapters
of each of these serials:

\begin{itemize}
\vspace{-0.2cm}
\item \emph{Accidentally on purpose}
\vspace{-0.2cm}
\item \emph{Bones}
\vspace{-0.2cm}
\item \emph{Community}
\vspace{-0.2cm}
\item \emph{CSI New York}
\vspace{-0.2cm}
\item \emph{Damages}
\vspace{-0.2cm}
\item \emph{Dexter}
\vspace{-0.2cm}
\item \emph{Doctor Who}
\vspace{-0.2cm}
\item \emph{Eastwick}
\vspace{-0.2cm}
\item \emph{Emergency room}
\vspace{-0.2cm}
\item \emph{Heroes}
\vspace{-0.2cm}
\item \emph{House}
\vspace{-0.2cm}
\item \emph{How I met your mother}
\vspace{-0.2cm}
\item \emph{Justified}
\vspace{-0.2cm}
\item \emph{Law and order}
\vspace{-0.2cm}
\item \emph{Lost}
\vspace{-0.2cm}
\item \emph{New tricks}
\vspace{-0.2cm}
\item \emph{Northern exposure}
\vspace{-0.2cm}
\item \emph{Nurse Jackie}
\vspace{-0.2cm}
\item \emph{Parenthood}
\vspace{-0.2cm}
\item \emph{Supernatural}
\vspace{-0.2cm}
\item \emph{The life and times of Tim}
\vspace{-0.2cm}
\item \emph{Till death}
\vspace{-0.2cm}
\item \emph{Ugly Americans}
\end{itemize}

The nature of this dataset is similar to the previous one because
the documents contain names of characters and places that are
related to the serials.

\begin{figure}
\begin{tabular}{|p{13cm}|}
\hline My name is Chandra Suresh. I'm a geneticist.

I have a theory about human evolution, and I believe you are a part
of it.

What makes some walk a path of darkness, while others choose the
light?

Is it will?

Is it destiny?

Can we ever hope to understand the force that shapes the soul?

For thousands of years, my people have taken spirit walks, following
destiny's path into the realms of the unconsciousness.

I'm ready to begin my journey.

To fight evil, one must know evil.

One must journey back through time and find that fork in the road.

Where heroes turn one way, and villains turn another.
\\
\hline
\end{tabular}
\centering \caption{SRT-serial. Extract from the serial
\emph{Heroes}.} \label{APPENDIX. SRT-Serials}
\end{figure}

\section{UCM dataset}

This dataset is composed of 104 articles related to computer science
written by researchers at the \emph{Universidad Complutense de
Madrid} (UCM). All the articles have been extracted from the UCM
Computer Science Department website
(http://www.fdi.ucm.es/investigacion). Then, they have been
carefully classified in different knowledge areas:

\begin{itemize}
\item Grid computing
\vspace{-0.2cm}
\item Architecture and Technology of Computing Systems
\vspace{-0.2cm}
\item Cloud computing
\vspace{-0.2cm}
\item Finance application
\vspace{-0.2cm}
\item Software Engineering applied to e-Learning
\vspace{-0.2cm}
\item Software Agents
\vspace{-0.2cm}
\item Semantic Web
\vspace{-0.2cm}
\item Declarative Programming
\vspace{-0.2cm}
\item Natural Language Processing
\vspace{-0.2cm}
\item Petri Nets
\vspace{-0.2cm}
\item Real-Time Systems
\end{itemize}

In this case, the documents stored in the databases correspond to
articles, while the queries used in the experiments consist of
abstracts of other articles related to the above ones.

It is important to note that the size of the documents (articles)
and the queries (abstracts) is very different. This fact affects the
behavior of the NCD because the NCD does not fit well when the
compared objects are very different in size. This kind of framework
is very suitable for evaluating the NCD-based document retrieval
method used in Chapter \ref{Chapter: Application to Document
Retrieval}.

\begin{figure}
\begin{tabular}{|p{13cm}|}
\hline The financial services industry today produces and consumes
huge amounts of data and the processes involved in analysing these
data have large and complex resource requirements. The need to
analyse the data using such processes and get meaningful results in
time, can be met only up to a certain extent by current computer
systems. Most service providers attempt to increase efficiency and
quality of their service offerings by stacking up more hardware and
employing better algorithms for data processing. However, there is a
limit to the gains achieved by using such an approach. One viable
alternative would be to use emerging technologies such as the Grid.
Grid computing and its application to various domains have been
actively studied by many groups for more than a decade now. In this
paper we explore the use of the Grid in the financial services
domain; an area which we believe has not been adequately looked
into.
\\
\hline
\end{tabular}
\centering \caption{UCM. Extract from a paper on grid computing.}
\label{APPENDIX. UCM}
\end{figure}

\section{Reuters dataset}
\label{Appendix. Section Reuters dataset}

This dataset is composed of 200 documents from the well-known
Reuters-21578 corpus, which contains texts on 10 different topics.
The whole dataset can be downloaded from
kdd.ics.uci.edu/databases/reuters21578/reuters21578.html.

It should be pointed out that most of the documents contained in
this dataset have a similar size. Therefore, in principle, this
dataset does not seem very suitable for evaluating the NCD-based
document retrieval method used in the third part of this thesis.
However, it has been adapted to make it suitable for the
experiments. The adaptation carried out consists of creating one big
file per topic in the following manner:

\begin{itemize}
 \item
First, the documents that will constitute the queries are randomly
selected among all the documents contained in the dataset.
 \item
Then, the rest of the documents are used to create one big file
per topic by concatenating all the documents concerning a topic.
Note that the documents selected to build the queries are not
included in that big file.
\end{itemize}

This adaptation makes the size of the documents and the size of the
queries very different. In this way, the dataset becomes suitable
for evaluating the NCD-based document retrieval method used.

\begin{itemize}
\item Acquisitions
\vspace{-0.2cm}
\item Corn
\vspace{-0.2cm}
\item Crude
\vspace{-0.2cm}
\item Earn
\vspace{-0.2cm}
\item Grain
\vspace{-0.2cm}
\item Interest
\vspace{-0.2cm}
\item Money
\vspace{-0.2cm}
\item Ship
\vspace{-0.2cm}
\item Trade
\vspace{-0.2cm}
\item Wheat
\end{itemize}

\begin{figure}
\begin{tabular}{|p{13cm}|}
\hline Argentine grain producers adjusted their yield estimates for
the 1986/87 coarse grain crop downward in the week to yesterday
after the heavy rains at the end of March and beginning of April,
trade sources said. They said sunflower, maize and sorghum
production estimates had been reduced despite some later warm, dry
weather, which has allowed a return to harvesting in some areas.
However, as showers fell intermittently after last weekend,
producers feared another spell of prolonged and intense rain could
cause more damage to crops already badly hit this season. Rains in
the middle of last week reached an average of 27 millimetres in
parts of Buenos Aires province, 83 mm in Cordoba, 41 in Santa Fe, 50
in Entre Rios and Misiones, 95 in Corrientes, eight in Chaco and 35
in Formosa. There was no rainfall in the same period in La Pampa.
Producers feared continued damp conditions could produce rotting and
lead to still lower yield estimates for all the crops, including
soybean.
\\
\hline
\end{tabular}
\centering \caption{Reuters. Extract from a document on wheat.}
\label{APPENDIX. Reuters}
\end{figure}

\section{20newsgroups dataset}

This well-known repository is composed of 20.000 documents on 20
different topics. This dataset can be downloaded from the UCI
Knowledge Discovery in Databases Archive
(http://kdd.ics.uci.edu/).

Since this repository has the same characteristics as the previous
one, it has been adapted to make it suitable for the experiments
using the method described for the Reuters dataset.

\begin{itemize}
\item alt.atheism
\vspace{-0.2cm}
\item comp.graphics
\vspace{-0.2cm}
\item comp.os.ms-windows.misc
\vspace{-0.2cm}
\item comp.sys.ibm.pc.hardware
\vspace{-0.2cm}
\item comp.sys.mac.hardware
\vspace{-0.2cm}
\item comp.windows.x
\vspace{-0.2cm}
\item misc.forsale
\vspace{-0.2cm}
\item rec.autos
\vspace{-0.2cm}
\item rec.motorcycles
\vspace{-0.2cm}
\item rec.sport.baseball
\vspace{-0.2cm}
\item rec.sport.hockey
\vspace{-0.2cm}
\item sci.crypt
\vspace{-0.2cm}
\item sci.electronics
\vspace{-0.2cm}
\item sci.med
\vspace{-0.2cm}
\item sci.space
\vspace{-0.2cm}
\item soc.religion.christian
\vspace{-0.2cm}
\item talk.politics.guns
\vspace{-0.2cm}
\item talk.politics.mideast
\vspace{-0.2cm}
\item talk.politics.misc
\vspace{-0.2cm}
\item talk.religion.misc
\end{itemize}

\clearpage

\begin{figure}[t]
\begin{tabular}{|p{13cm}|}
\hline VIKING 1 was launched from Cape Canaveral, Florida on August
20, 1975 on a TITAN 3E-CENTAUR D1 rocket. The probe went into
Martian orbit on June 19, 1976, and the lander set down on the
western slopes of Chryse Planitia on July 20, 1976. It soon began
its programmed search for Martian micro-organisms (there is still
debate as to whether the probes found life there or not), and sent
back incredible color panoramas of its surroundings. One thing
scientists learned was that Mars' sky was pinkish in color, not dark
blue as they originally thought (the sky is pink due to sunlight
reflecting off the reddish dust particles in the thin atmosphere).
The lander set down among a field of red sand and boulders
stretching out as far as its cameras could image.
\\
\hline
\end{tabular}
\centering \caption{20newsgroups. Extract from a document on
\emph{sci.space}.} \label{APPENDIX. 20newsgroups}
\end{figure}

\chapter{Queries}
\label{Appendix Queries}

This Appendix describes the queries used in the experiments carried
out in Chapter \ref{Chapter: Application to Document Retrieval}.
Three datasets are used in the experiments developed in that chapter
of the thesis. They are the UCM dataset, the \emph{20newsgroups}
dataset, and the Reuters dataset. Although all of them are described
in depth in Appendix \ref{Appendix Datasets}, this section
summarizes their main characteristics.

\begin{itemize}
  \item \textbf{UCM dataset}:
  \begin{itemize}
    \item Number of documents: 104.
    \item Number of topics: 11.
    \item Kind of documents: Papers.
  \end{itemize}

  \item \textbf{\emph{20newsgroups} dataset}:
  \begin{itemize}
    \item Number of documents: 20000.
    \item Number of topics: 20.
    \item Kind of documents: One file per topic (adaptation described in Section \ref{Appendix. Section Reuters dataset}).
  \end{itemize}

  \item \textbf{Reuters dataset}:
  \begin{itemize}
    \item Number of documents: 200.
    \item Number of topics: 10.
    \item Kind of documents: One file per topic (adaptation described in Section \ref{Appendix. Section Reuters dataset}).
  \end{itemize}
\end{itemize}

\newpage

Similarly, the main characteristics of the queries used in the
experiments carried out in Chapter \ref{Chapter: Application to
Document Retrieval} are as follows:

\begin{itemize}
  \item \textbf{UCM dataset}:
  \begin{itemize}
    \item Kind of queries: Abstracts.
    \item Number of queries: 4.
    \item Sizes of queries:
    \begin{itemize}
      \item 2 x 1KB.
      \item 2 x 2KB.
    \end{itemize}
  \end{itemize}

  \item \textbf{\emph{20newsgroups} dataset}:
  \begin{itemize}
    \item Kind of queries: Messages.
    \item Number of queries: 10.
    \item Sizes of queries:
    \begin{itemize}
      \item 7 x 2KB.
      \item 3 x 3KB.
    \end{itemize}
  \end{itemize}

  \item \textbf{Reuters dataset}:
  \begin{itemize}
    \item Kind of queries: News.
    \item Number of queries: 10.
    \item Sizes of queries:
    \begin{itemize}
      \item 10 x 2KB.
    \end{itemize}
  \end{itemize}
\end{itemize}

\begin{figure}[hb]
\begin{tabular}{|p{13cm}|}
\hline Agent-based modelling facilitates the implementation of tools
for the analysis of social patterns. This comes from the fact that
agent related concepts allow the representation of organizational
and behavioural aspects of individuals in a society and their
interactions. An agent can characterize an individual with
capabilities to perceive and react to events in the environment,
taking into account its mental state (beliefs, goals), and to
interact with other agents in its social environment. There are
already tools to perform agent-based social simulation but these are
usually hard to use by social scientists, as they require a good
expertise in computer programming. In order to cope with such
difficulty, we propose the use of agent-based graphical modelling
languages, which can help to specify social systems as multi-agent
systems in a more convenient way. This is complemented with
transformation tools to be able to analyse and derive emergent
social behavioural patterns by using the capabilities of existing
simulation platforms. In this way, this framework can facilitate the
specification and analysis of complex behavioural patterns that may
emerge in social systems.
\\
\hline
\end{tabular}
\centering \caption{UCM. Example of query.} \label{APPENDIX.
query-ucm}
\end{figure}

\begin{figure}
\begin{tabular}{|p{13cm}|}
\hline As the subjects says, Windows 3.1 keeps crashing (giving me
GPF) on me of late. It was never a very stable package, but now it
seems to crash every day. The worst part about it is that it does
not crash consistently: ie I. There is a way in SYS.INI to turn off
RAM parity checking (unfortunately, my good Windows references are
at home, but any standard Win reference will tell you how to do it.
If not, email back to me). That weird memory may be producing phony
parity errors. Danger is, if you turn checking off, you run the
slight risk of data corruption due to a missed real error.I had this
very same problem, and did 'work around' by turning parity checking
off, but that only worked while I was in windows, and the parity
error would occur immediately after exiting windows, however,the
problem turned out to be 3 chip simms vs 9 chip simms. I can't use 3
chip simms in my computer, and when I replaced them, the problem
vanished, forever.
\\
\hline
\end{tabular}
\centering \caption{20newsgroups. Example of query.}
\label{APPENDIX. query-20newsgroups}
\end{figure}

\begin{figure}[ht]
\begin{tabular}{|p{13cm}|}
\hline FAO SEES LOWER GLOBAL WHEAT, GRAIN OUTPUT IN 1987. \\
The U.N Food and Agriculture Organisation (FAO) said global wheat
and coarse grain output was likely to fall in 1987 but supplies
would remain adequate to meet demand FAO said in its monthly food
outlook bulletin total world grain output was expected to fall 38
mln tonnes to 1,353 mln in 1987, due mainly to unusually high winter
losses in the Soviet Union, drought in China and reduced plantings
in North America World cereal stocks at the end of 1986/87 were
forecast to rise 47 mln tonnes to a record 452 mln tonnes, softening
the impact of reduced production But stocks are unevenly
distributed, with about 50 pct held by the U.S. ``Thus the food
security prospects in 1987/88 for many developing countries,
particularly in Africa, depend crucially on the outcome of this
year's harvests'', FAO said FAO said world cereal supplies in
1986/87 were estimated at a record 2,113 mln tonnes, about five pct
higher than last season and due mainly to large stocks and a record
1986 harvest, estimated at 1,865 mln tonnes FAO's forecast of
1986/87 world cereals trade was revised upwards by eight mln tonnes
to 179 mln due to the likelihood of substantial buying by China and
the Soviet Union.
\\
\hline
\end{tabular}
\centering \caption{Reuters. Example of query.} \label{APPENDIX.
query-reuters}
\end{figure}

\chapter{Detailed Experimental Results}
\label{Appendix Detailed Experimental Results}

\section{Preliminary study on text distortion}

This section shows all the results obtained in the work developed in
Chapter \ref{Chapter: Study on text distortion}. Three clustering
error figures are shown for each \emph{dataset-compression
algorithm} pair. Each of them corresponds to a selection method:

\vspace{-0.1cm}
\begin{itemize}
  \item \emph{MFW selection method}\vspace{-0.2cm}
  \item \emph{RW selection method} \vspace{-0.2cm}
  \item \emph{LFW selection method}
\end{itemize}
\vspace{-0.1cm}

In all the clustering error figures, the value on the horizontal
axis corresponds to the cumulative sum of the BNC-based frequencies
of the words substituted from the documents, whereas the value on
the vertical axis corresponds to the clustering error. Furthermore,
the curves with asterisk markers correspond to the \emph{asterisk
substitution method}, while the curves with square markers
correspond to the \emph{random character substitution method}.

Analyzing all the figures one can observe that the \emph{asterisk
substitution method} is always better than the \emph{random
character substitution method}. This is to be expected because
substituting a word with random characters adds noise to the
documents, and therefore most likely increases the Kolmogorov
complexity of the documents and makes the clustering worse. In
addition, it can be seen that the best clustering results correspond
to the \emph{MFW selection method}, the worst results correspond to
the \emph{LFW selection method}, and the results corresponding to
the \emph{RW selection method} are situated in between them.

The dendrogram obtained with no distortion, and the best dendrogram
obtained are shown for each \emph{dataset-compression algorithm}
pair as well. In the cases in which the dendrogram obtained with no
distortion is the best one obtained, only one dendrogram is shown.

\clearpage

\begin{figure}
\centering
\includegraphics[angle=270,width=13cm]{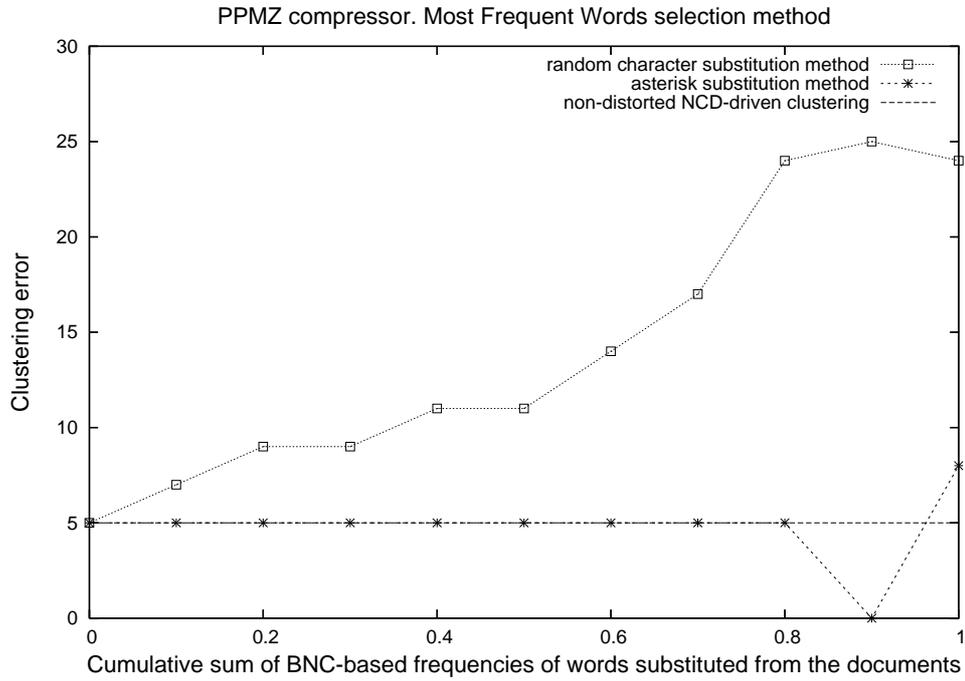}
\caption{Books. PPMZ compressor. MFW selection method.}
\label{APPENDIX. Fig:books-clustering-error-ppmz-mfw}
\end{figure}

\begin{figure}
\centering
\includegraphics[angle=270,width=13cm]{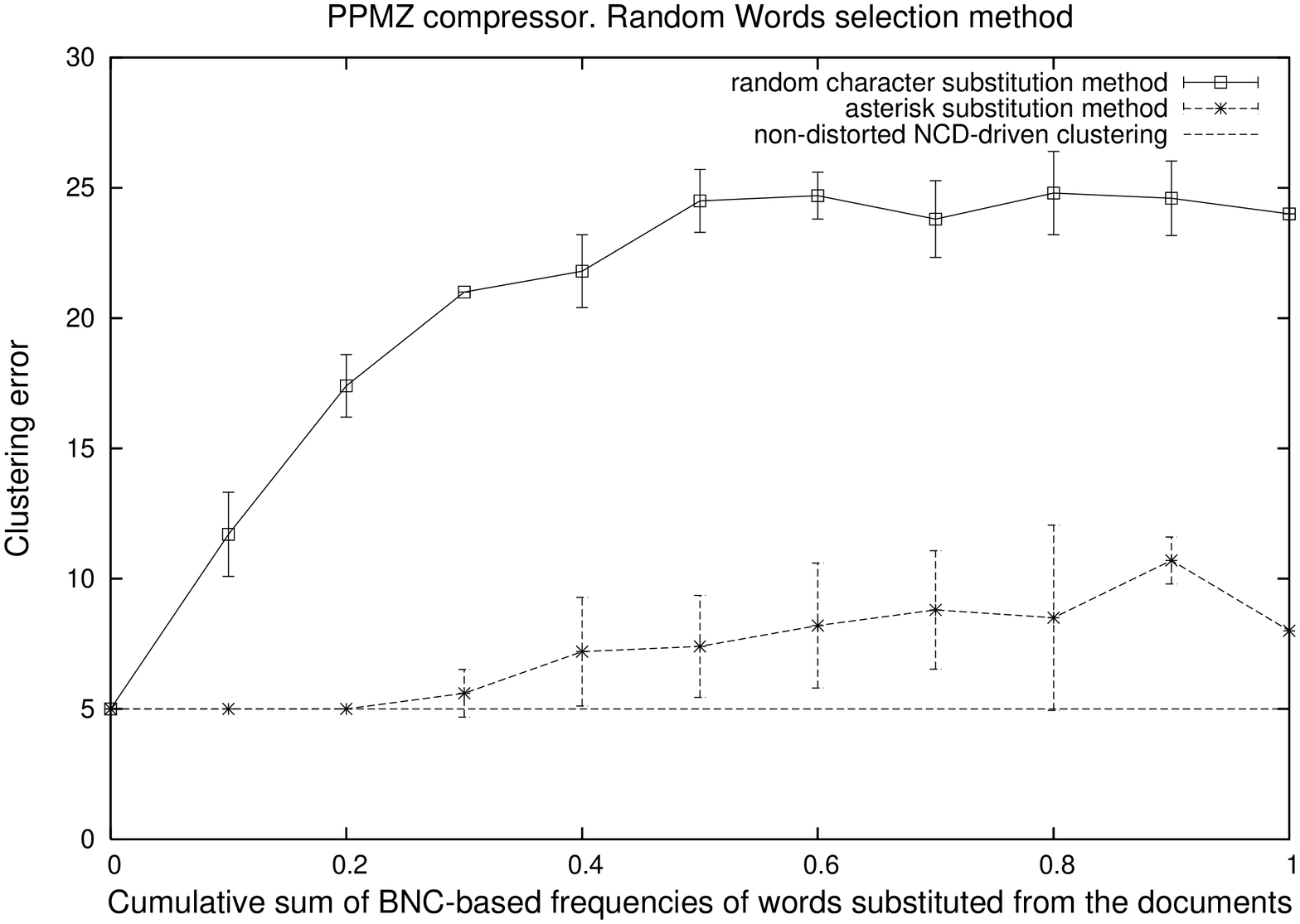}
\caption{Books. PPMZ compressor. RW selection method.}
\label{APPENDIX. Fig:books-clustering-error-ppmz-rw}
\end{figure}

\clearpage

\begin{figure}[ht]
\centering
\includegraphics[angle=270,width=13cm]{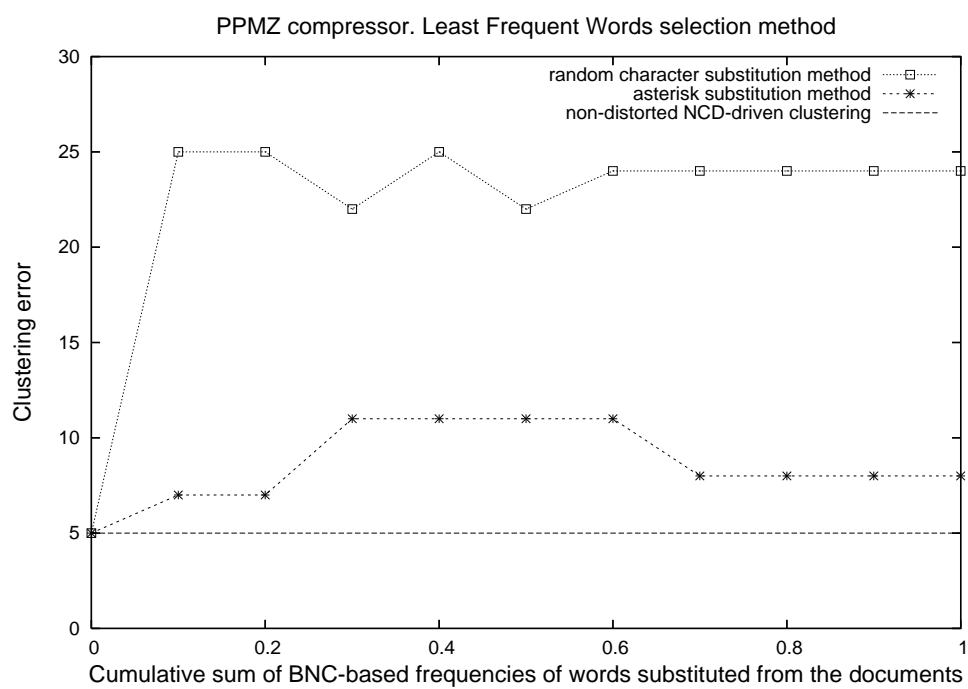}
\caption{Books. PPMZ compressor. LFW selection method.}
\label{APPENDIX. Fig:books-clustering-error-ppmz-lfw}
\end{figure}

The combination of the \emph{MFW selection method} and the
\emph{asterisk substitution method} improves the clustering results
so much that a clustering error of 0 is obtained when the texts are
distorted using the set of words that accumulate a BNC-based
frequency of 0.9.

\begin{itemize}
  \item Non-distorted clustering error: 5
  \item Best clustering error: 0
\end{itemize}

The improvement can be observed by comparing Figs \ref{APPENDIX.
Fig:dendro-books-ppmz-original} and \ref{APPENDIX.
Fig:dendro-books-ppmz-best}. Whereas the books by Edgar Allan Poe
and Alexander Pope are not correctly clustered in Fig \ref{APPENDIX.
Fig:dendro-books-ppmz-original}, all the books are correctly
clustered in Fig \ref{APPENDIX. Fig:dendro-books-ppmz-best}. That is
the reason why the clustering error that corresponds to the best
dendrogram obtained is 0.

\clearpage

\begin{figure}
\centering
\includegraphics[width=14cm]{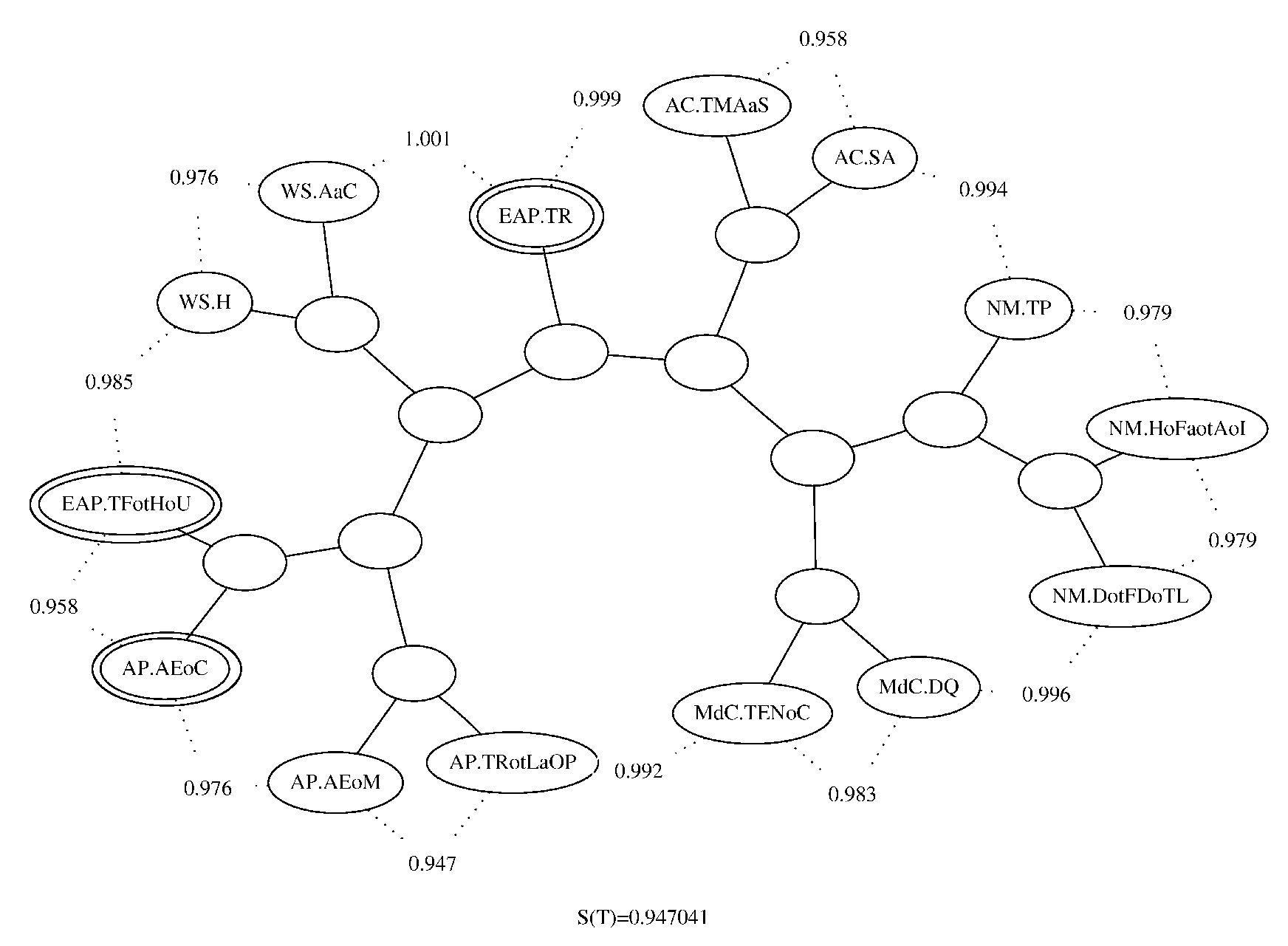}
\caption{Books. PPMZ compressor. Dendrogram obtained with no
distortion.} \label{APPENDIX. Fig:dendro-books-ppmz-original}
\end{figure}

\begin{figure}
\centering
\includegraphics[width=14cm]{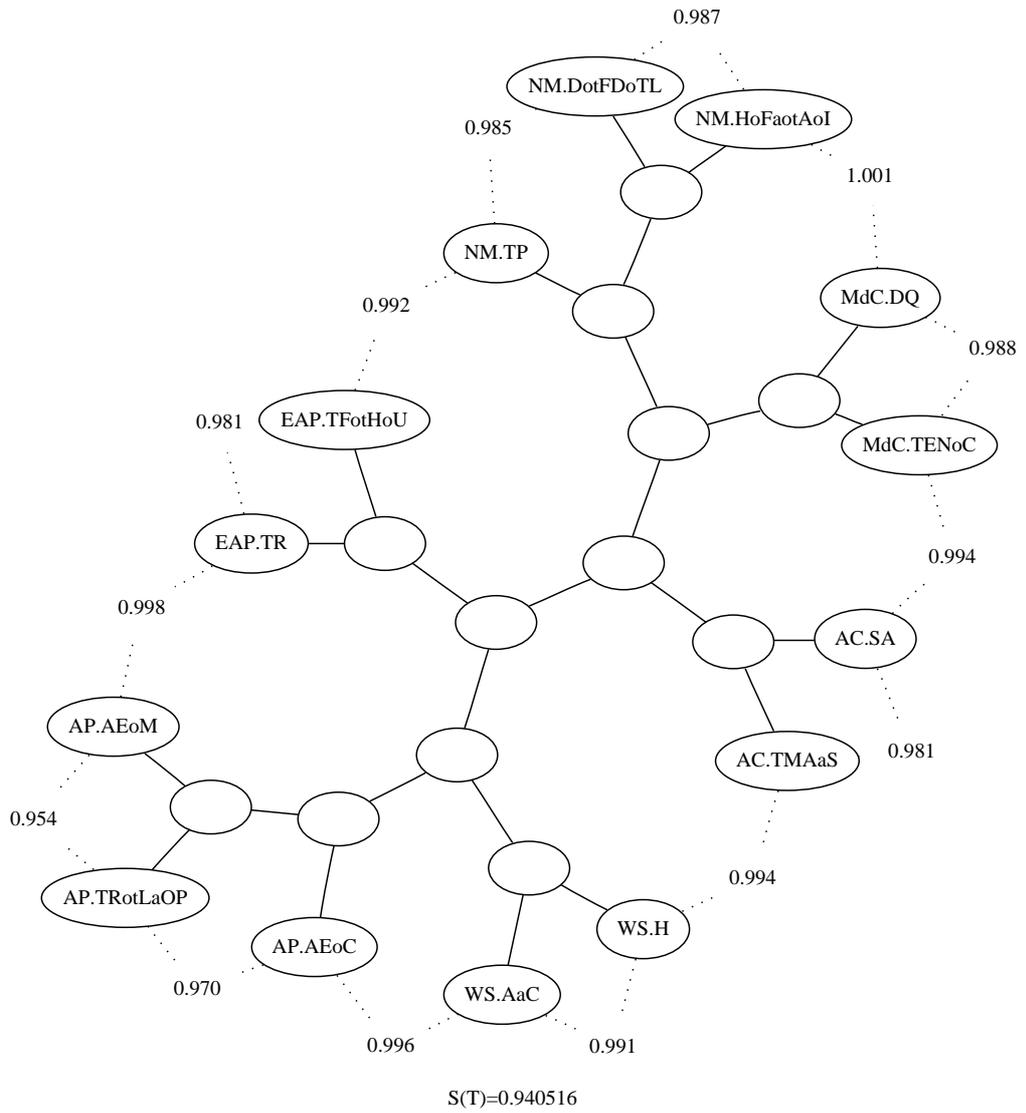}
\caption{Books. PPMZ compressor. Best dendrogram obtained.}
\label{APPENDIX. Fig:dendro-books-ppmz-best}
\end{figure}

\clearpage

\begin{figure}
\centering
\includegraphics[angle=270,width=13cm]{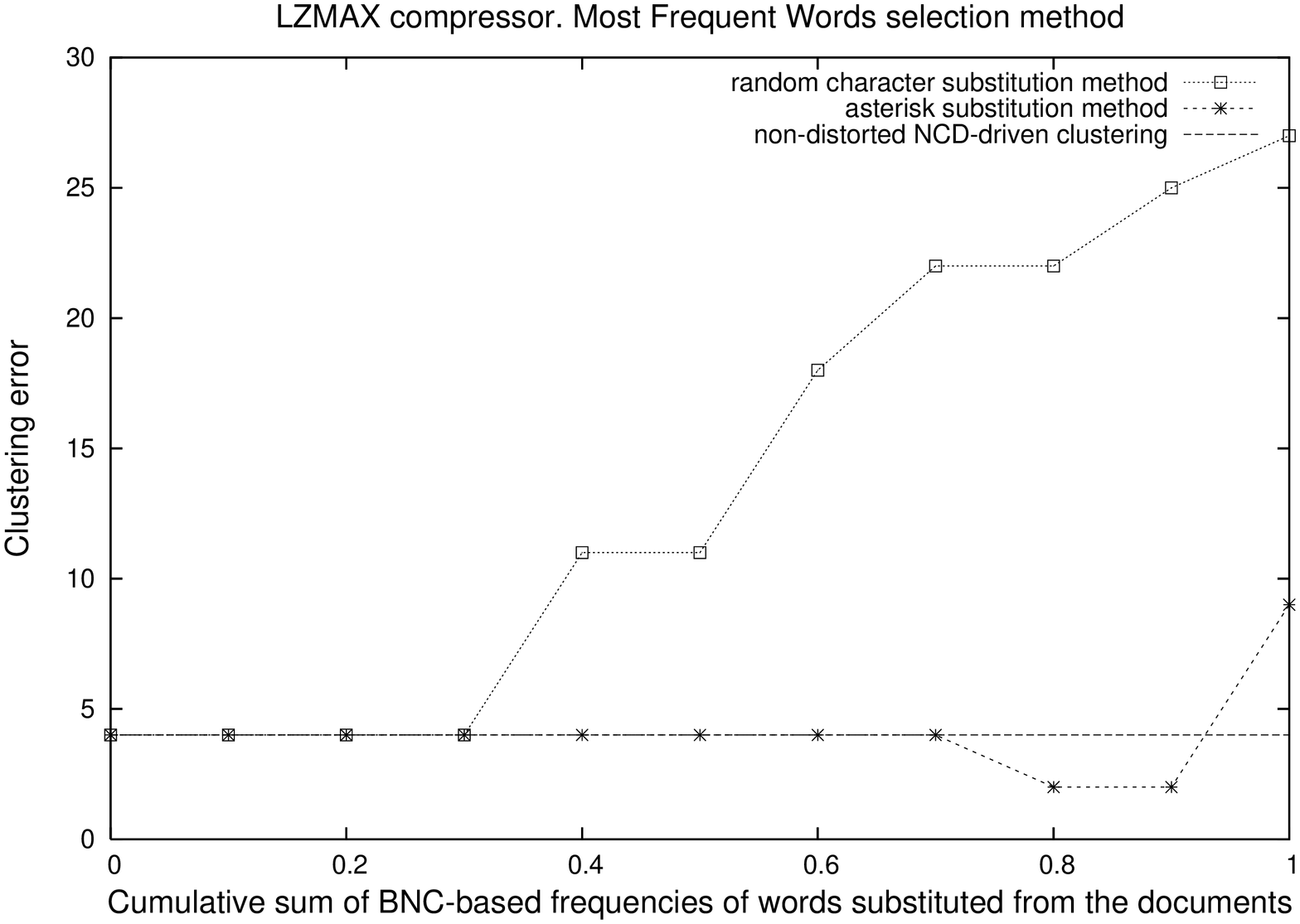}
\caption{Books. LZMA compressor. MFW selection method.}
\label{APPENDIX. Fig:books-clustering-error-lzma-mfw}
\end{figure}

\begin{figure}
\centering
\includegraphics[angle=270,width=13cm]{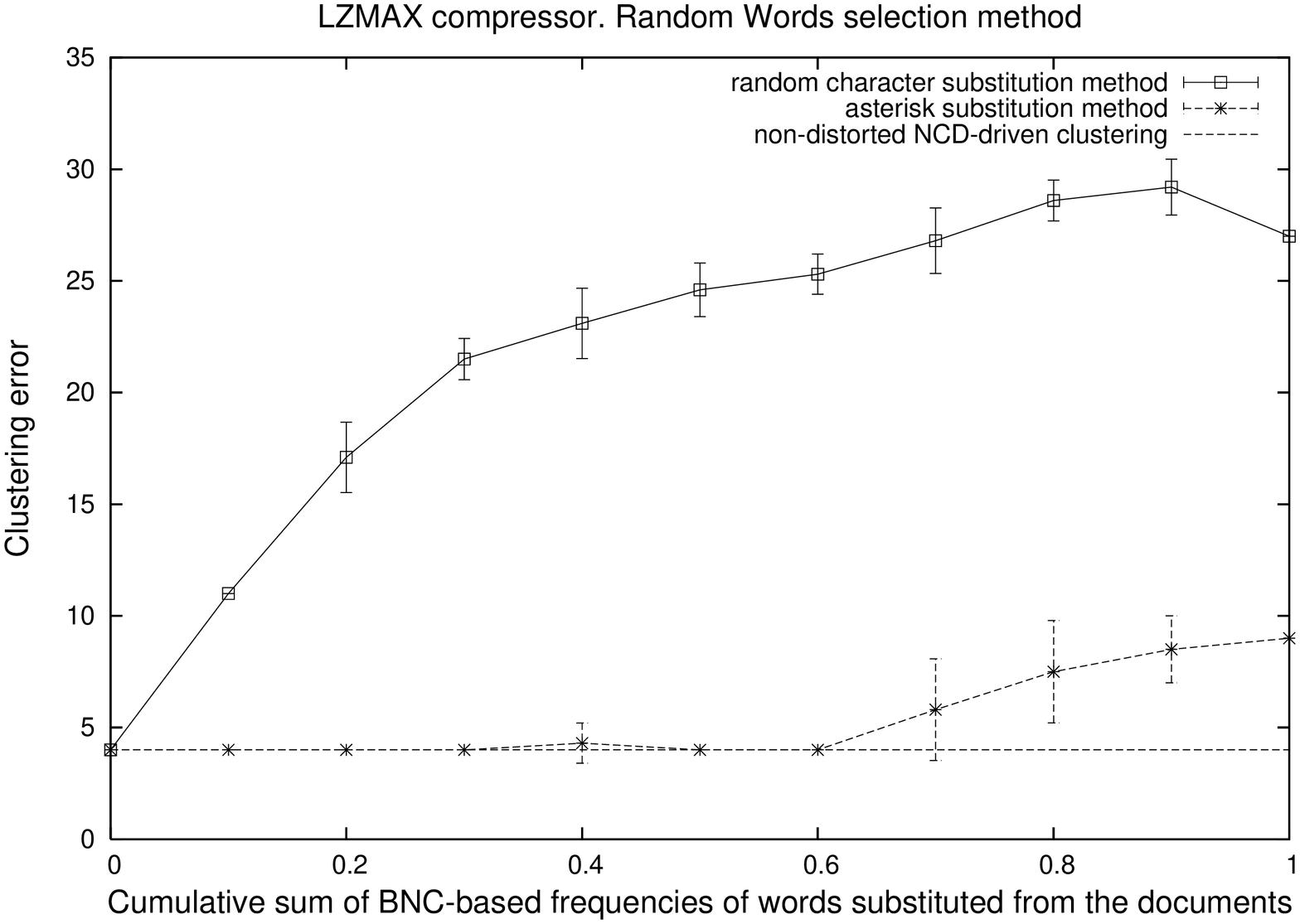}
\caption{Books. LZMA compressor. RW selection method.}
\label{APPENDIX. Fig:books-clustering-error-lzma-rw}
\end{figure}

\clearpage

\begin{figure}[ht]
\centering
\includegraphics[angle=270,width=13cm]{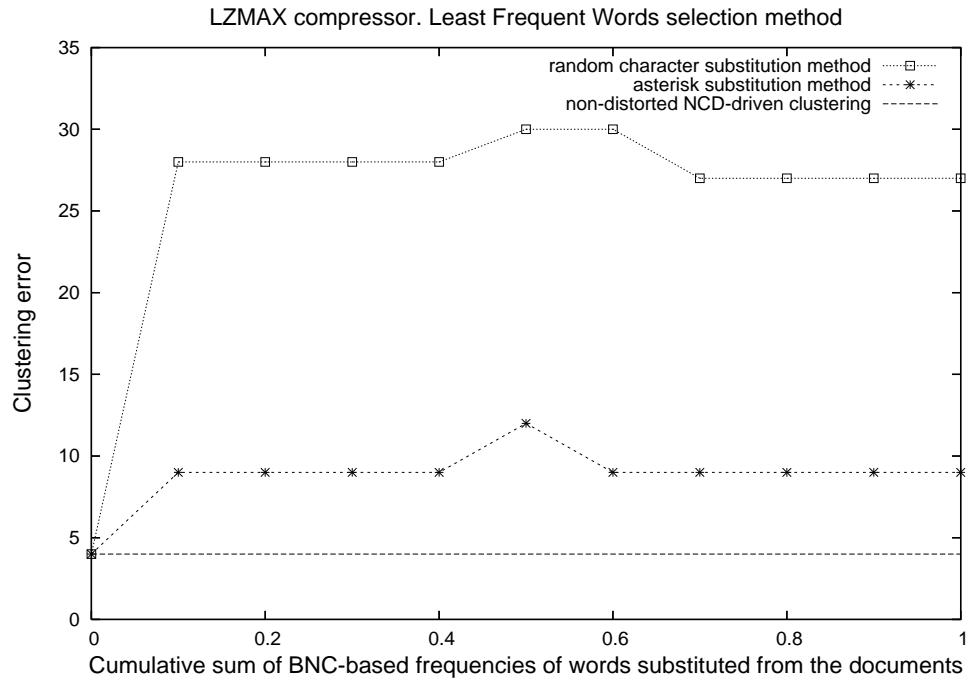}
\caption{Books. LZMA compressor. LFW selection method.}
\label{APPENDIX. Fig:books-clustering-error-lzma-lfw}
\end{figure}

The combination of the \emph{MFW selection method} and the
\emph{asterisk substitution method} improves the clustering results
when the texts are distorted using the sets of words that accumulate
a BNC-based frequency of 0.8 and 0.9.

\begin{itemize}
  \item Non-distorted clustering error: 4
  \item Best clustering error: 2
\end{itemize}

The improvement can be observed by comparing Figs \ref{APPENDIX.
Fig:dendro-books-lzma-original} and \ref{APPENDIX.
Fig:dendro-books-lzma-best}. The difference between both figures is
that the book ``The Prince'' by Niccolò Machiavelli is closer to the
rest of Niccolò Machiavelli's books in Fig \ref{APPENDIX.
Fig:dendro-books-lzma-best}.

\clearpage

\clearpage

\begin{figure}
\centering
\includegraphics[width=13cm]{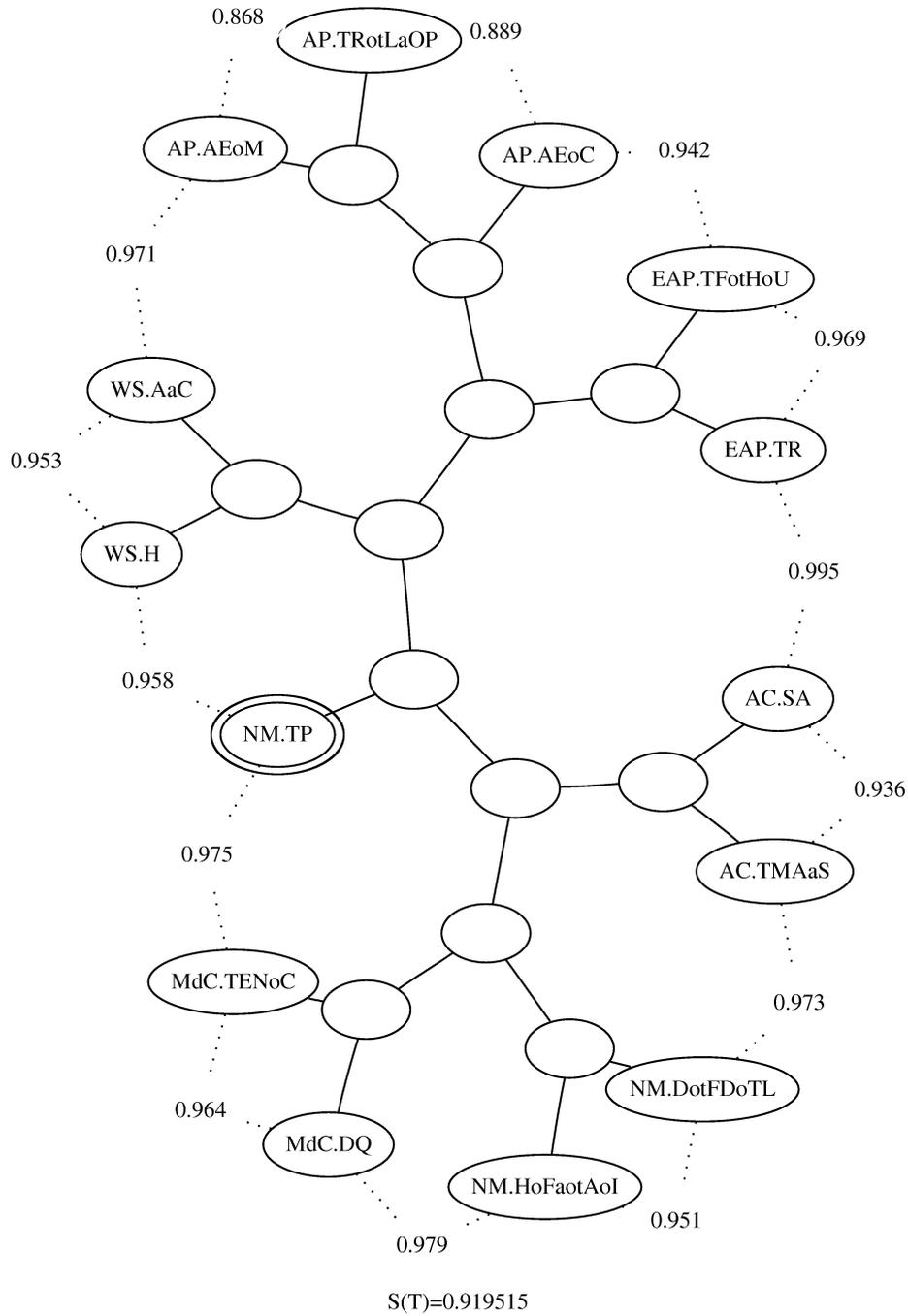}
\caption{Books. LZMA compressor. Dendrogram obtained with no
distortion.} \label{APPENDIX. Fig:dendro-books-lzma-original}
\end{figure}

\begin{figure}
\centering
\includegraphics[width=14cm]{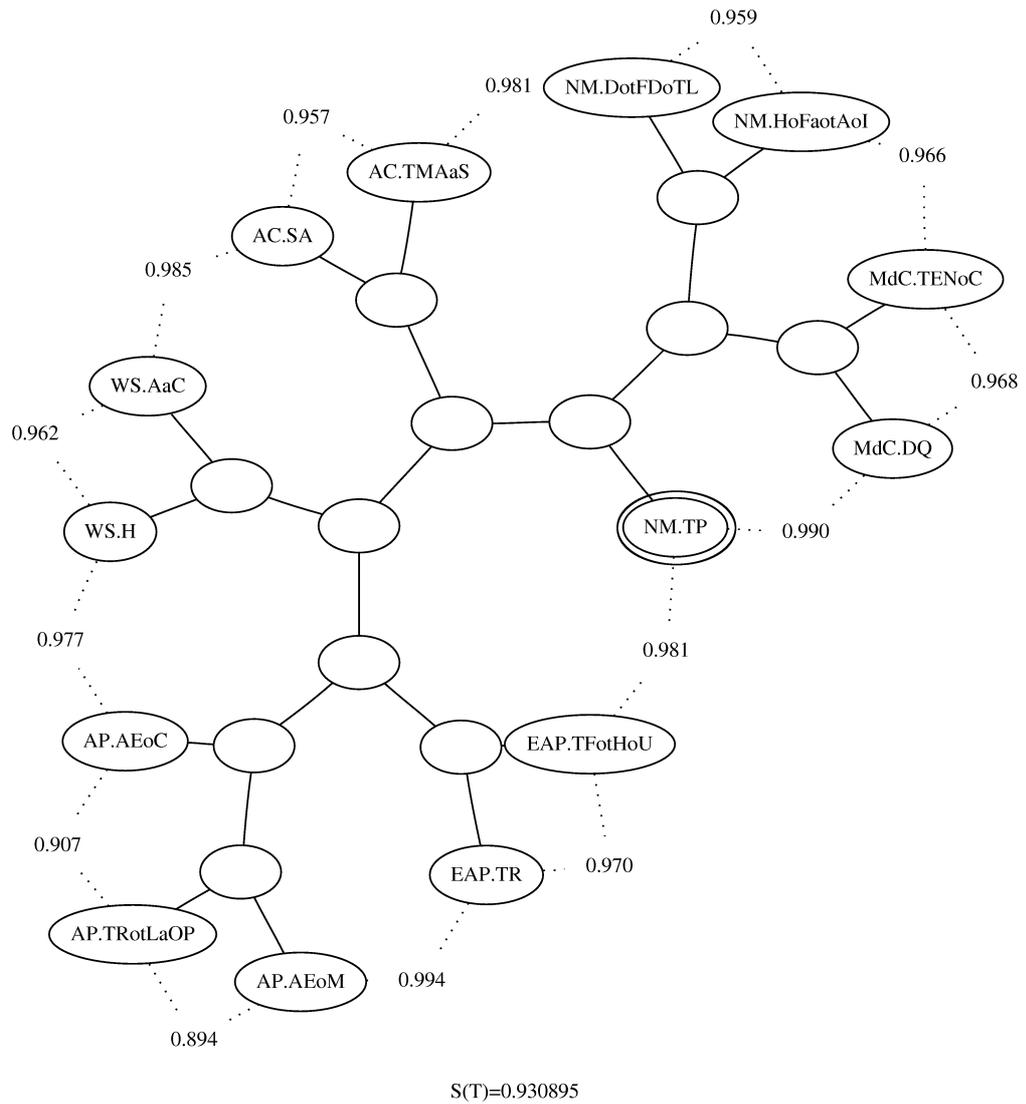}
\caption{Books. LZMA compressor. Best dendrogram obtained.}
\label{APPENDIX. Fig:dendro-books-lzma-best}
\end{figure}

\begin{figure}
\centering
\includegraphics[angle=270,width=13cm]{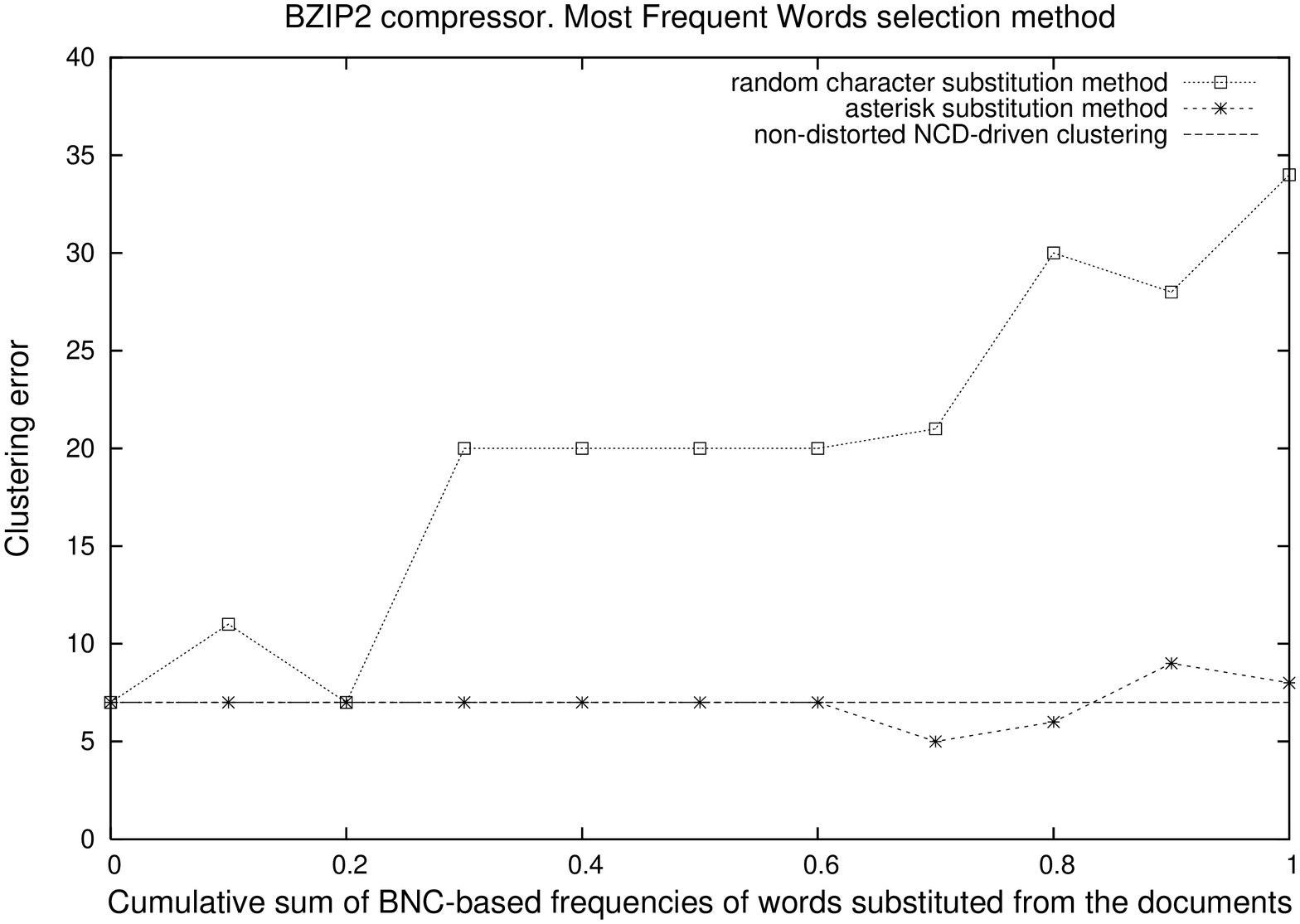}
\caption{Books. BZIP2 compressor. MFW selection method.}
\label{APPENDIX. Fig:books-clustering-error-bzip2-mfw}
\end{figure}

\begin{figure}
\centering
\includegraphics[angle=270,width=13cm]{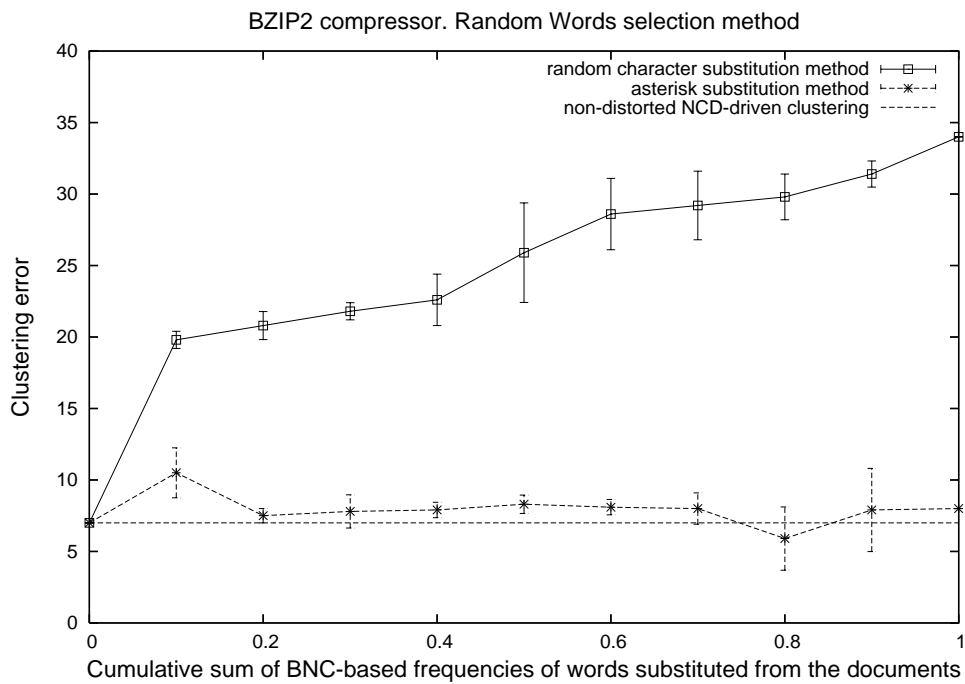}
\caption{Books. BZIP2 compressor. RW selection method.}
\label{APPENDIX. Fig:books-clustering-error-bzip2-rw}
\end{figure}

\clearpage

\begin{figure}[ht]
\centering
\includegraphics[angle=270,width=13cm]{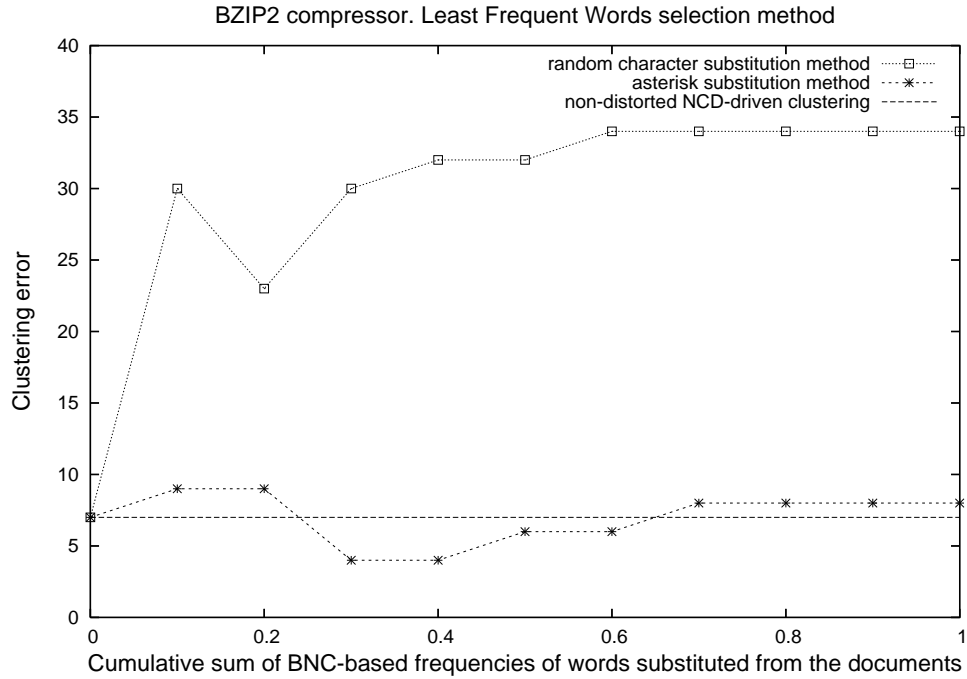}
\caption{Books. BZIP2 compressor. LFW selection method.}
\label{APPENDIX. Fig:books-clustering-error-bzip2-lfw}
\end{figure}

Again, the combination of the \emph{MFW selection method} and the
\emph{asterisk substitution method} improves the clustering results
when the texts are distorted using the sets of words that accumulate
a BNC-based frequency of 0.7 and 0.8. However, in this case the
non-distorted clustering error is improved using the rest of the
selection methods, as can be observed looking at Figs \ref{APPENDIX.
Fig:books-clustering-error-bzip2-rw} and \ref{APPENDIX.
Fig:books-clustering-error-bzip2-lfw}.

The problematic books in this case are the books by Miguel de
Cervantes and the books by Niccolò Machiavelli. The difference
between the dendrogram obtained with no distortion, and the best
dendrogram obtained is that the book ``The Prince'' by Niccolò
Machiavelli is closer to the rest of Niccolò Machiavelli's books in
Fig \ref{APPENDIX. Fig:dendro-books-bzip2-best}.

\clearpage

\begin{figure}
\centering
\includegraphics[width=13cm]{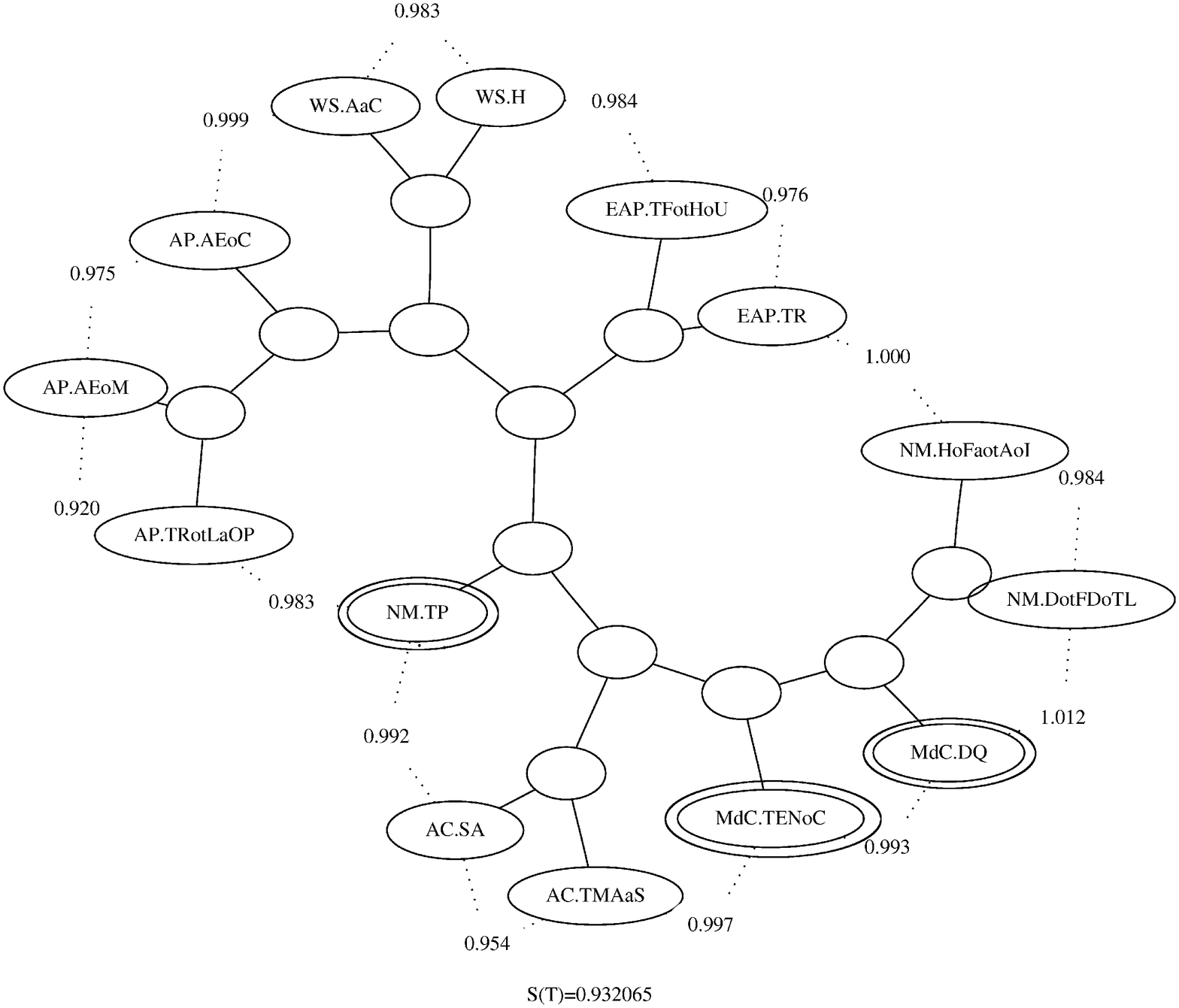}
\caption{Books. BZIP2 compressor. Dendrogram obtained with no
distortion.} \label{APPENDIX. Fig:dendro-books-bzip2-original}
\end{figure}

\begin{figure}
\centering
\includegraphics[width=14cm]{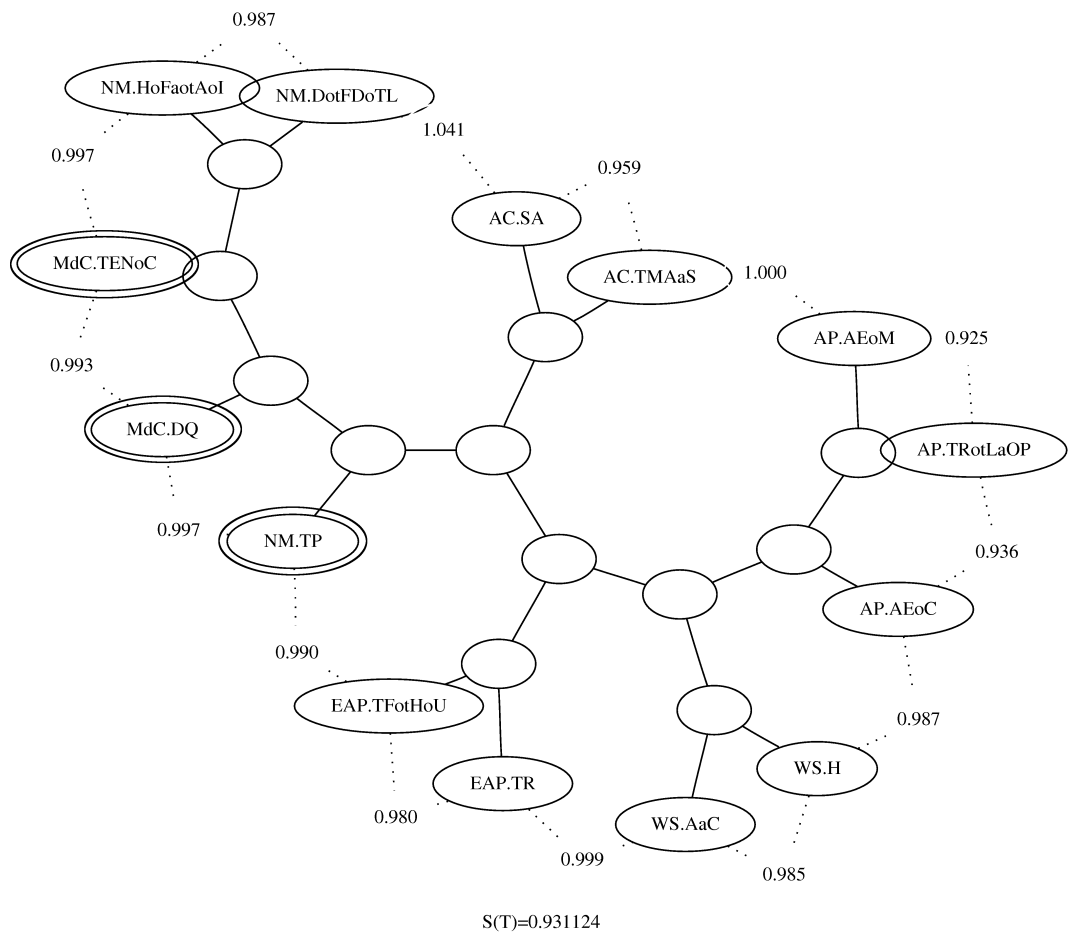}
\caption{Books. BZIP2 compressor. Best dendrogram obtained.}
\label{APPENDIX. Fig:dendro-books-bzip2-best}
\end{figure}

\clearpage

\begin{figure}
\centering
\includegraphics[angle=270,width=13cm]{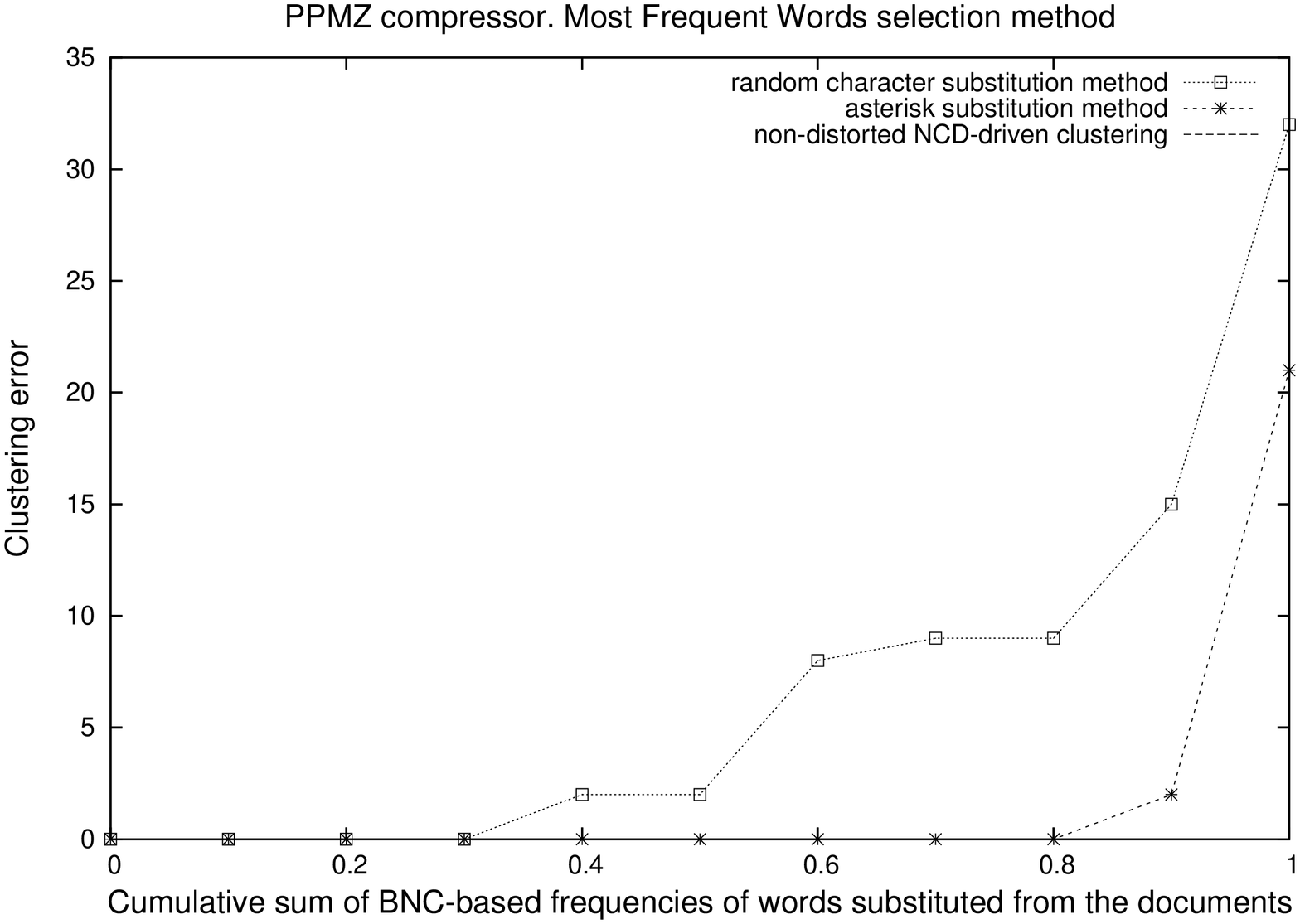}
\caption{UCI-KDD. PPMZ compressor. MFW selection method.}
\label{APPENDIX. Fig:uci-kdd-clustering-error-ppmz-mfw}
\end{figure}

\begin{figure}
\centering
\includegraphics[angle=270,width=13cm]{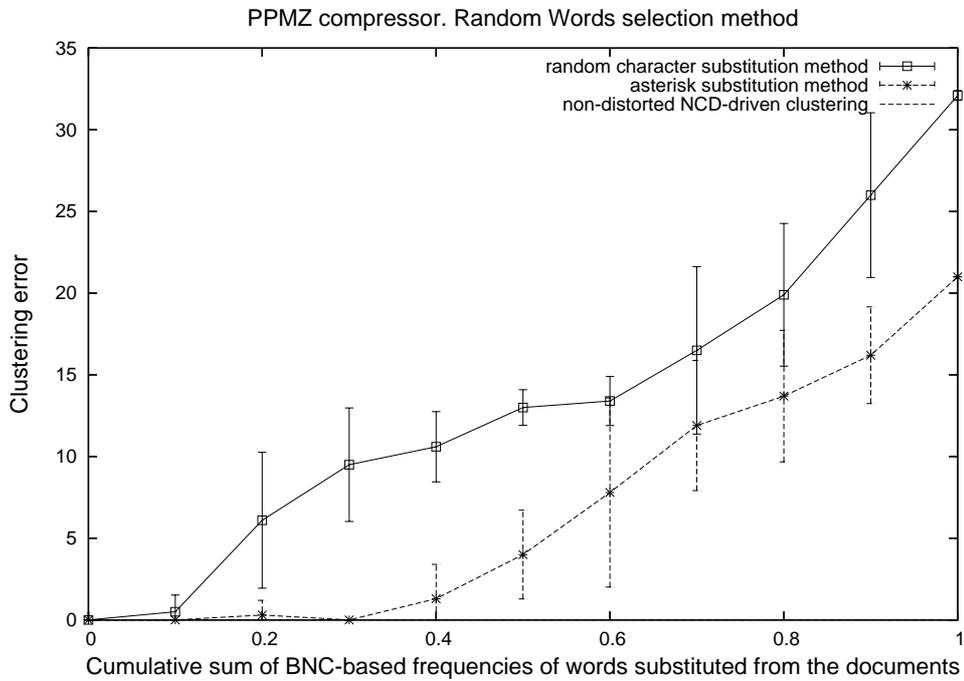}
\caption{UCI-KDD. PPMZ compressor. RW selection method.}
\label{APPENDIX. Fig:uci-kdd-clustering-error-ppmz-rw}
\end{figure}

\clearpage

\begin{figure}[ht]
\centering
\includegraphics[angle=270,width=13cm]{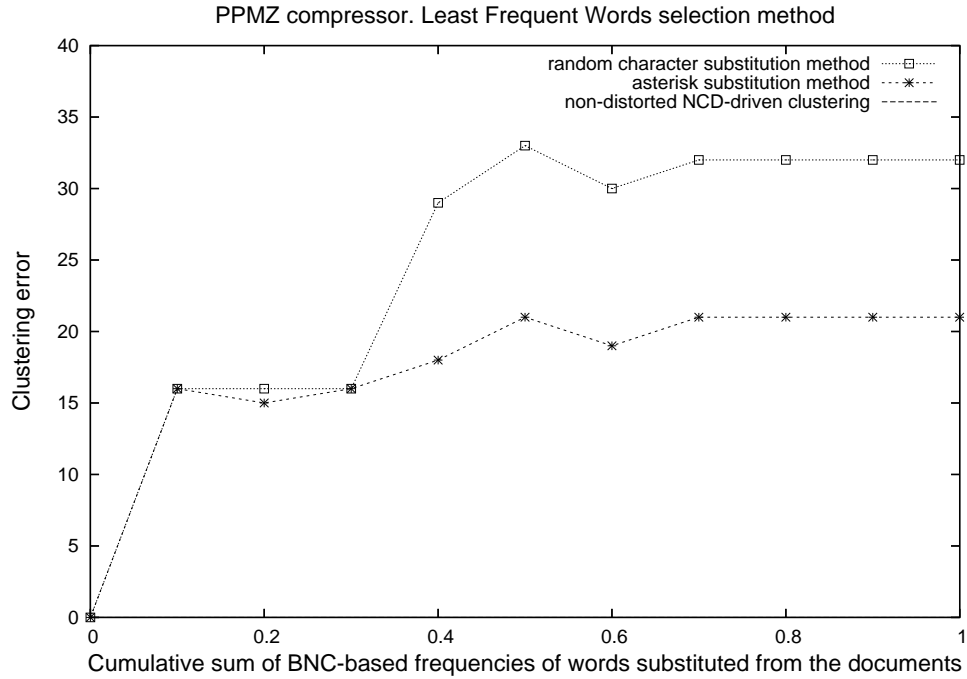}
\caption{UCI-KDD. PPMZ compressor. LFW selection method.}
\label{APPENDIX. Fig:uci-kdd-clustering-error-ppmz-lfw}
\end{figure}

The non-distorted clustering error in this case is 0. Therefore, it
is impossible to improve the clustering error for this
\emph{dataset-compression algorithm} pair.

However, analyzing Fig \ref{APPENDIX.
Fig:uci-kdd-clustering-error-ppmz-mfw} one can observe that the
clustering error remains constant from the point that corresponds to
a BNC-based frequency of 0 to the one corresponding to a BNC-based
frequency of 0.8. This means that the relevant information contained
in the documents is maintained despite the word removal.

Again, the results show that the combination of the \emph{MFW
selection method} and the \emph{asterisk substitution method} is the
key factor, because whereas a clustering error of 0 is obtained from
most of the points of the curve with asterisk markers in Fig
\ref{APPENDIX. Fig:uci-kdd-clustering-error-ppmz-mfw}, the
clustering error in the other cases gets worse, as one can see in
Figs \ref{APPENDIX. Fig:uci-kdd-clustering-error-ppmz-rw}, and
\ref{APPENDIX. Fig:uci-kdd-clustering-error-ppmz-lfw}.

Since the dendrogram obtained with no distortion clusters all the
texts perfectly, only one dendrogram is shown for this
\emph{dataset-compression algorithm} pair.

\clearpage

\begin{figure}
\centering
\includegraphics[width=14cm]{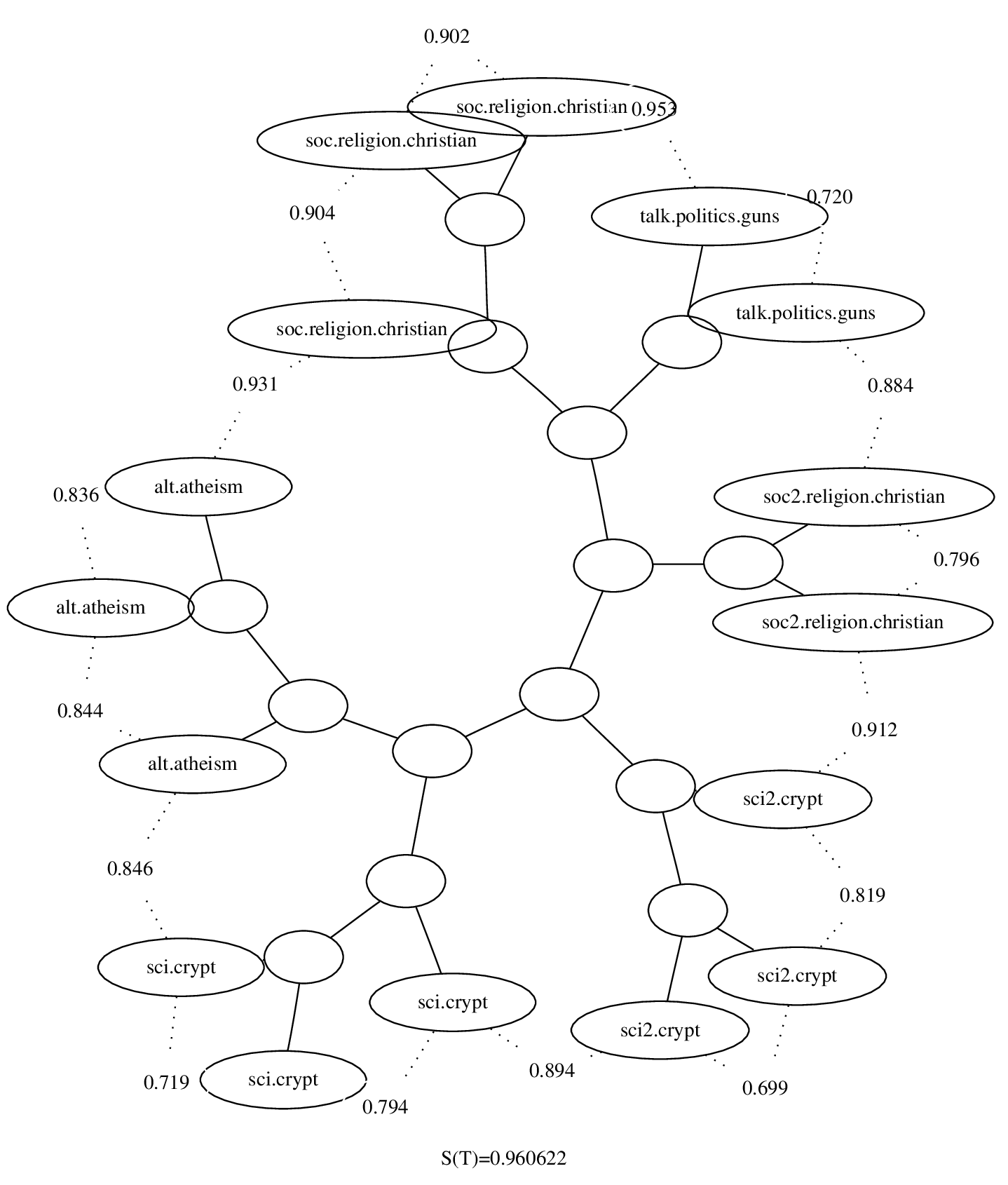}
\caption{UCI-KDD. PPMZ compressor. Best dendrogram obtained.}
\label{APPENDIX. Fig:dendro-uci-kdd-ppmz-best}
\end{figure}

\clearpage

\begin{figure}
\centering
\includegraphics[angle=270,width=13cm]{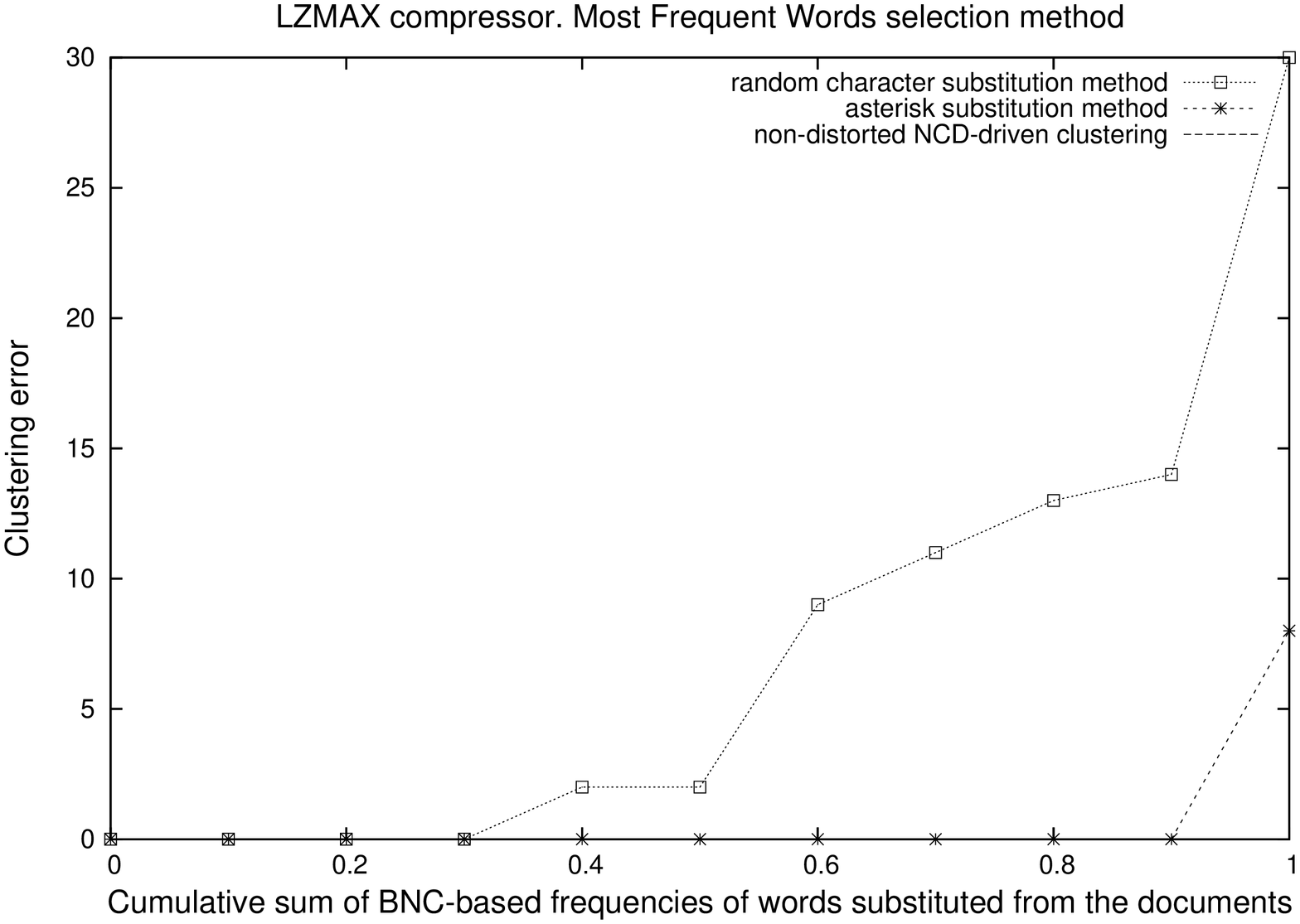}
\caption{UCI-KDD. LZMA compressor. MFW selection method.}
\label{APPENDIX. Fig:uci-kdd-clustering-error-lzma-mfw}
\end{figure}

\begin{figure}
\centering
\includegraphics[angle=270,width=13cm]{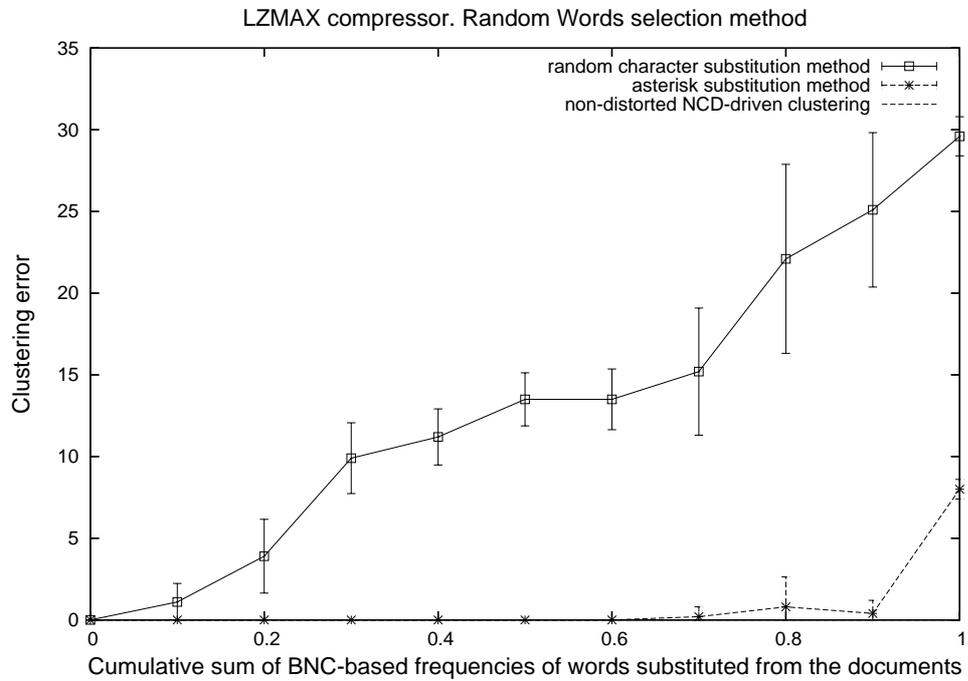}
\caption{UCI-KDD. LZMA compressor. RW selection method.}
\label{APPENDIX. Fig:uci-kdd-clustering-error-lzma-rw}
\end{figure}

\clearpage

\begin{figure}[ht]
\centering
\includegraphics[angle=270,width=13cm]{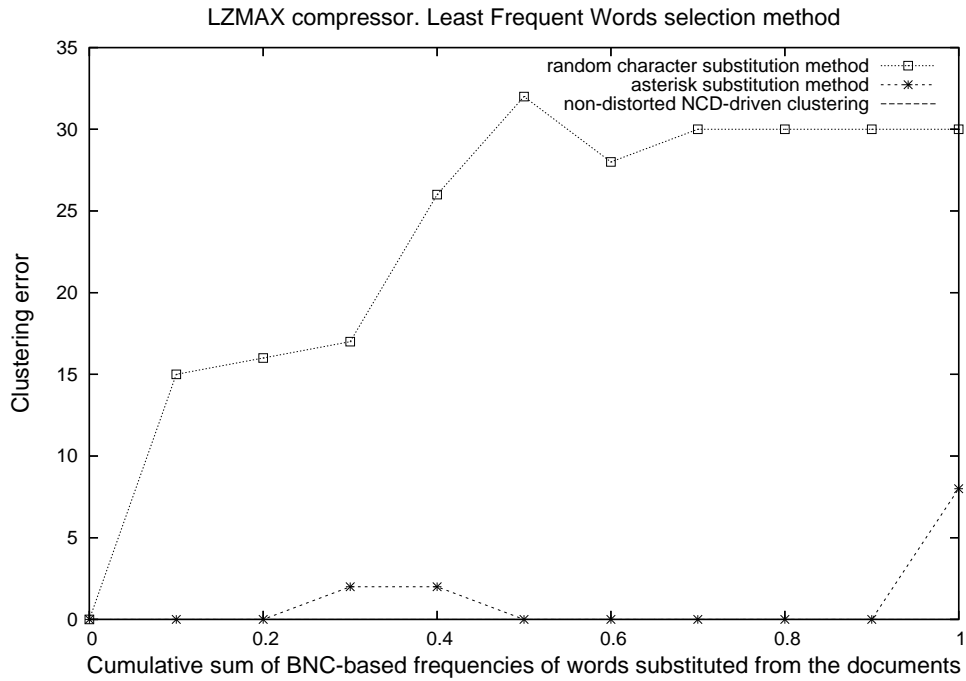}
\caption{UCI-KDD. LZMA compressor. LFW selection method.}
\label{APPENDIX. Fig:uci-kdd-clustering-error-lzma-lfw}
\end{figure}

Similarly to the results shown before, the non-distorted clustering
error in this case is 0. Therefore, it is impossible to improve the
clustering error for this \emph{dataset-compression algorithm} pair.

However, analyzing Fig \ref{APPENDIX.
Fig:uci-kdd-clustering-error-lzma-mfw} it can be observed that the
clustering error remains constant when the \emph{MFW selection
method} and the \emph{asterisk substitution method} are combined.
This behavior is observed for the points from 0.0 to 0.9 of the
curve.

In this case, the results that correspond to the rest of selection
methods are almost the same as the ones obtained using the \emph{MFW
selection method}.

Since the dendrogram obtained with no distortion clusters all the
texts perfectly, only one dendrogram is shown for this
\emph{dataset-compression algorithm} pair.

\clearpage

\begin{figure}
\centering
\includegraphics[width=14cm]{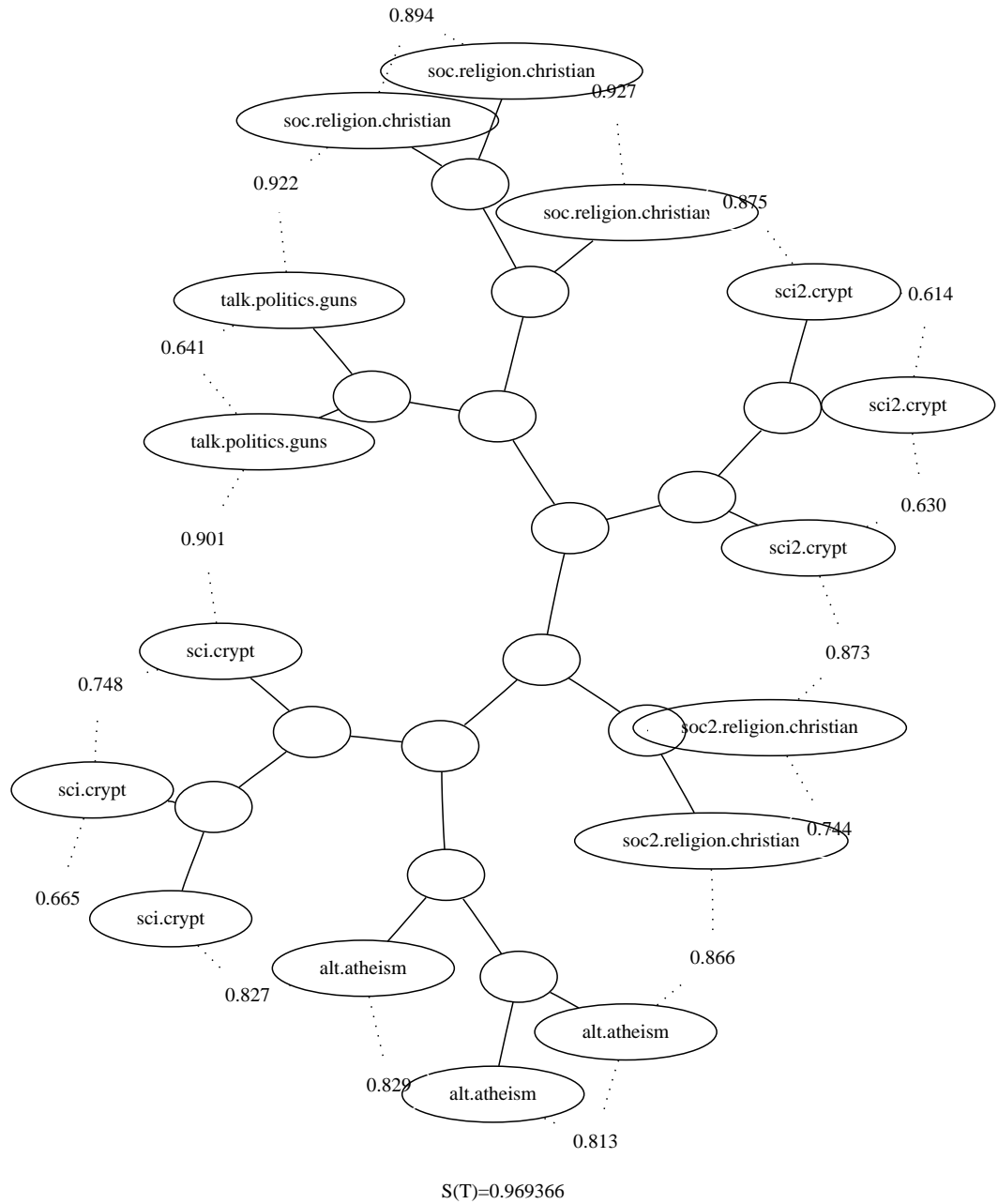}
\caption{UCI-KDD. LZMA compressor. Best dendrogram obtained.}
\label{APPENDIX. Fig:dendro-uci-kdd-lzma-best}
\end{figure}

\clearpage

\begin{figure}
\centering
\includegraphics[angle=270,width=13cm]{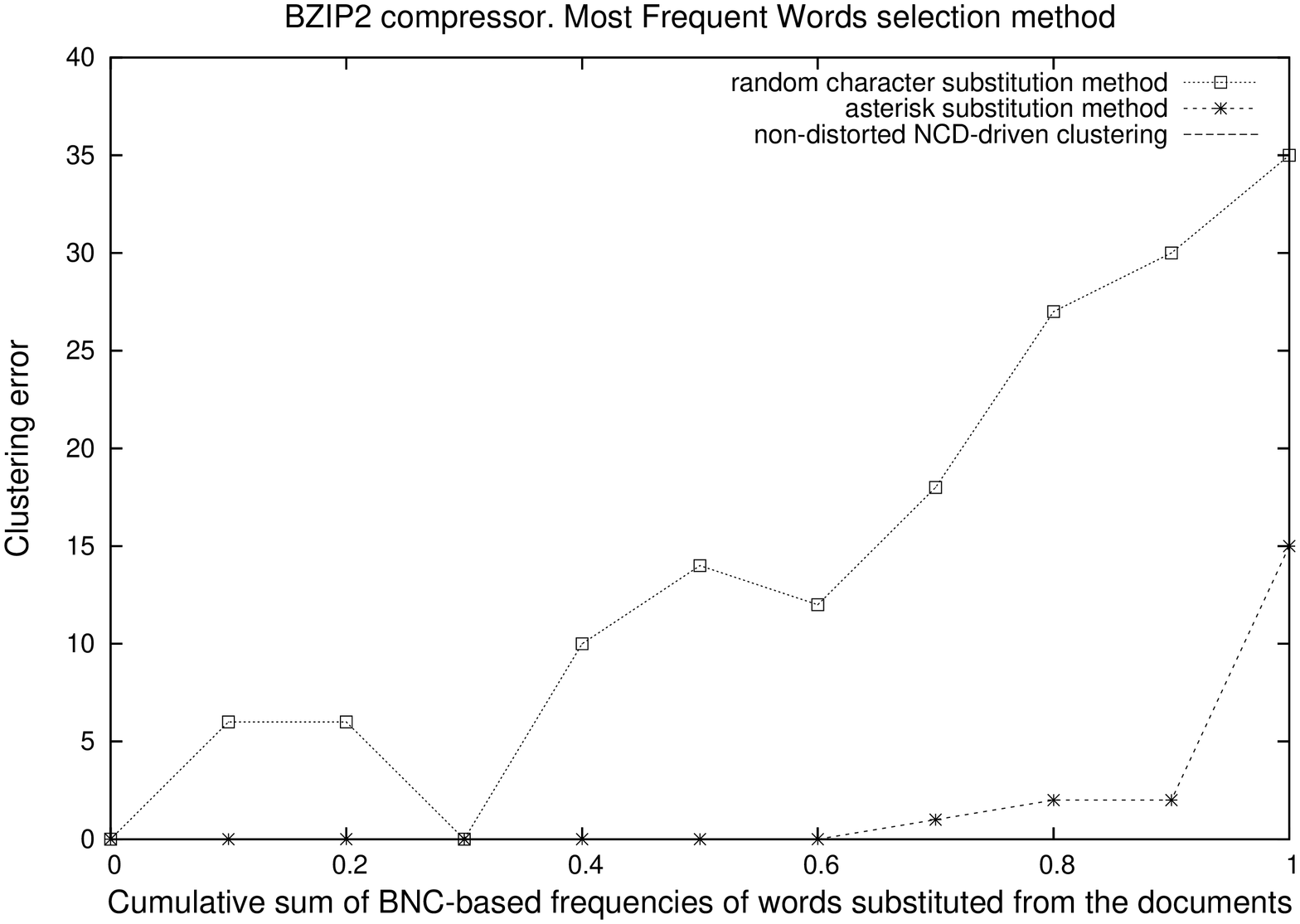}
\caption{UCI-KDD. BZIP2 compressor. MFW selection method.}
\label{APPENDIX. Fig:uci-kdd-clustering-error-bzip2-mfw}
\end{figure}

\begin{figure}
\centering
\includegraphics[angle=270,width=13cm]{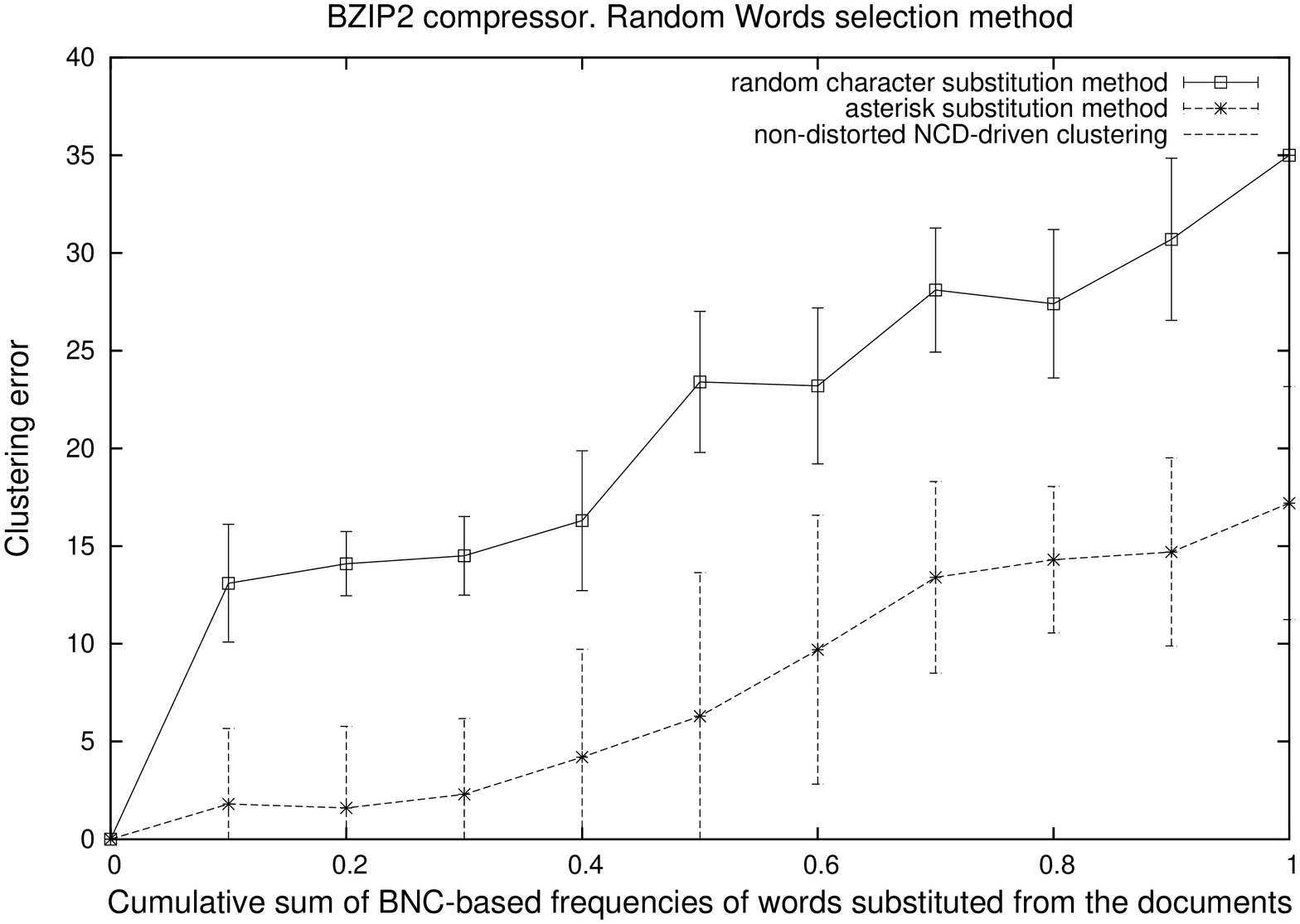}
\caption{UCI-KDD. BZIP2 compressor. RW selection method.}
\label{APPENDIX. Fig:uci-kdd-clustering-error-bzip2-rw}
\end{figure}

\clearpage

\begin{figure}[ht]
\centering
\includegraphics[angle=270,width=13cm]{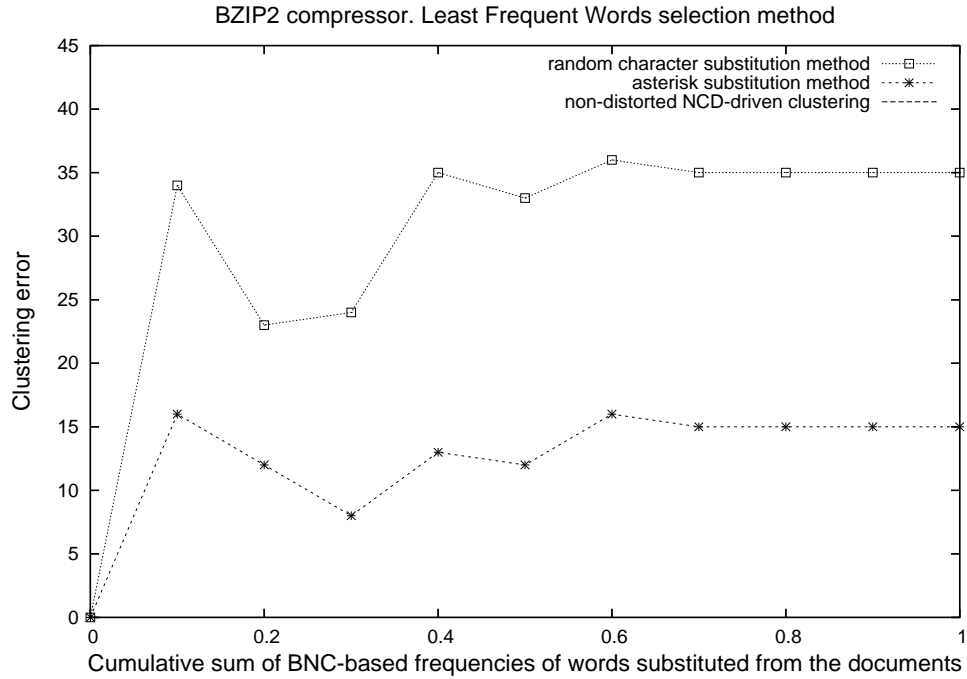}
\caption{UCI-KDD. BZIP2 compressor. LFW selection method.}
\label{APPENDIX. Fig:uci-kdd-clustering-error-bzip2-lfw}
\end{figure}

Again, the non-distorted clustering error in this case is 0.
Therefore, it is impossible to improve the clustering error for this
\emph{dataset-compression algorithm} pair.

However, analyzing Fig \ref{APPENDIX.
Fig:uci-kdd-clustering-error-bzip2-mfw} it can be observed that the
clustering error remains constant from a distortion of 0 to a
distortion of 0.6.

In this case, the results show that the combination of the \emph{MFW
selection method} and the \emph{asterisk substitution method} is the
key factor, because whereas a clustering error of 0 is obtained from
most of the points of the curve with asterisk markers in Fig
\ref{APPENDIX. Fig:uci-kdd-clustering-error-bzip2-mfw}, the
clustering error in the other cases gets worse, as one can see in
Figs \ref{APPENDIX. Fig:uci-kdd-clustering-error-bzip2-rw}, and
\ref{APPENDIX. Fig:uci-kdd-clustering-error-bzip2-lfw}.

Since the dendrogram obtained with no distortion clusters all the
texts perfectly, only one dendrogram is shown for this
\emph{dataset-compression algorithm} pair.

\clearpage

\begin{figure}
\centering
\includegraphics[width=14cm]{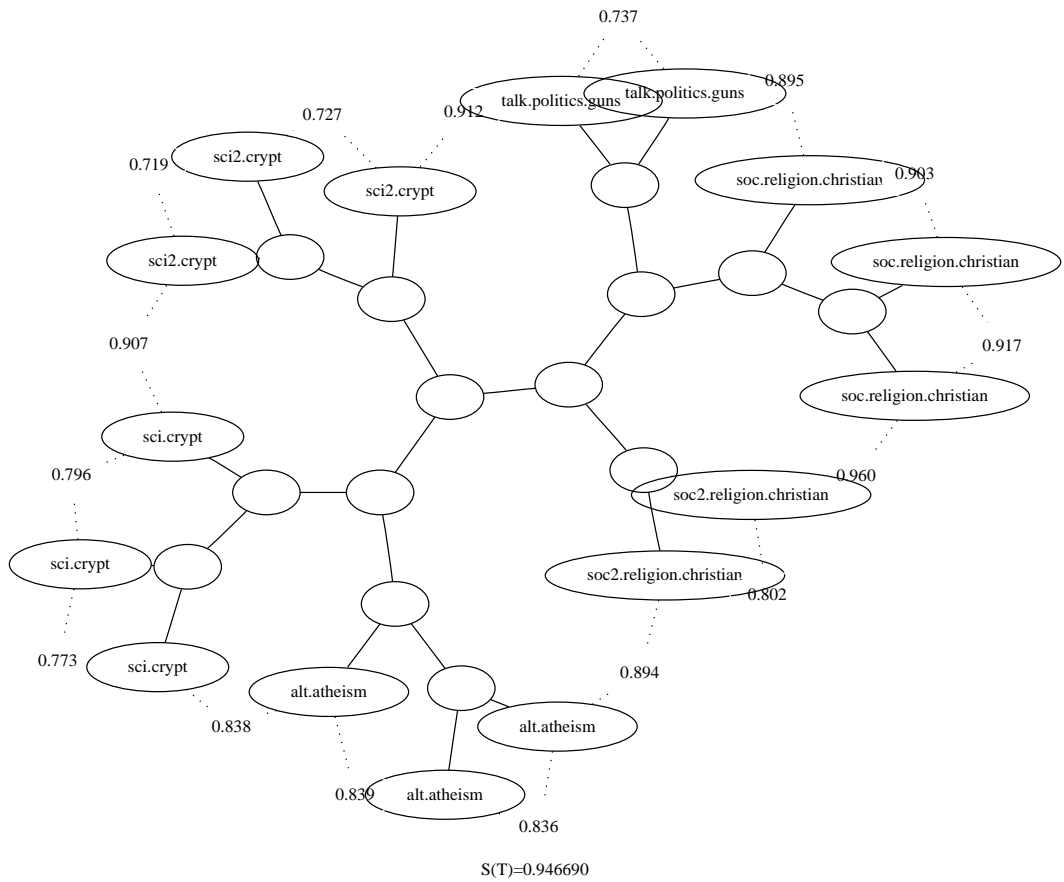}
\caption{UCI-KDD. BZIP2 compressor. Best dendrogram obtained.}
\label{APPENDIX. Fig:dendro-uci-kdd-bzip2-best}
\end{figure}

\clearpage

\begin{figure}
\centering
\includegraphics[angle=270,width=13cm]{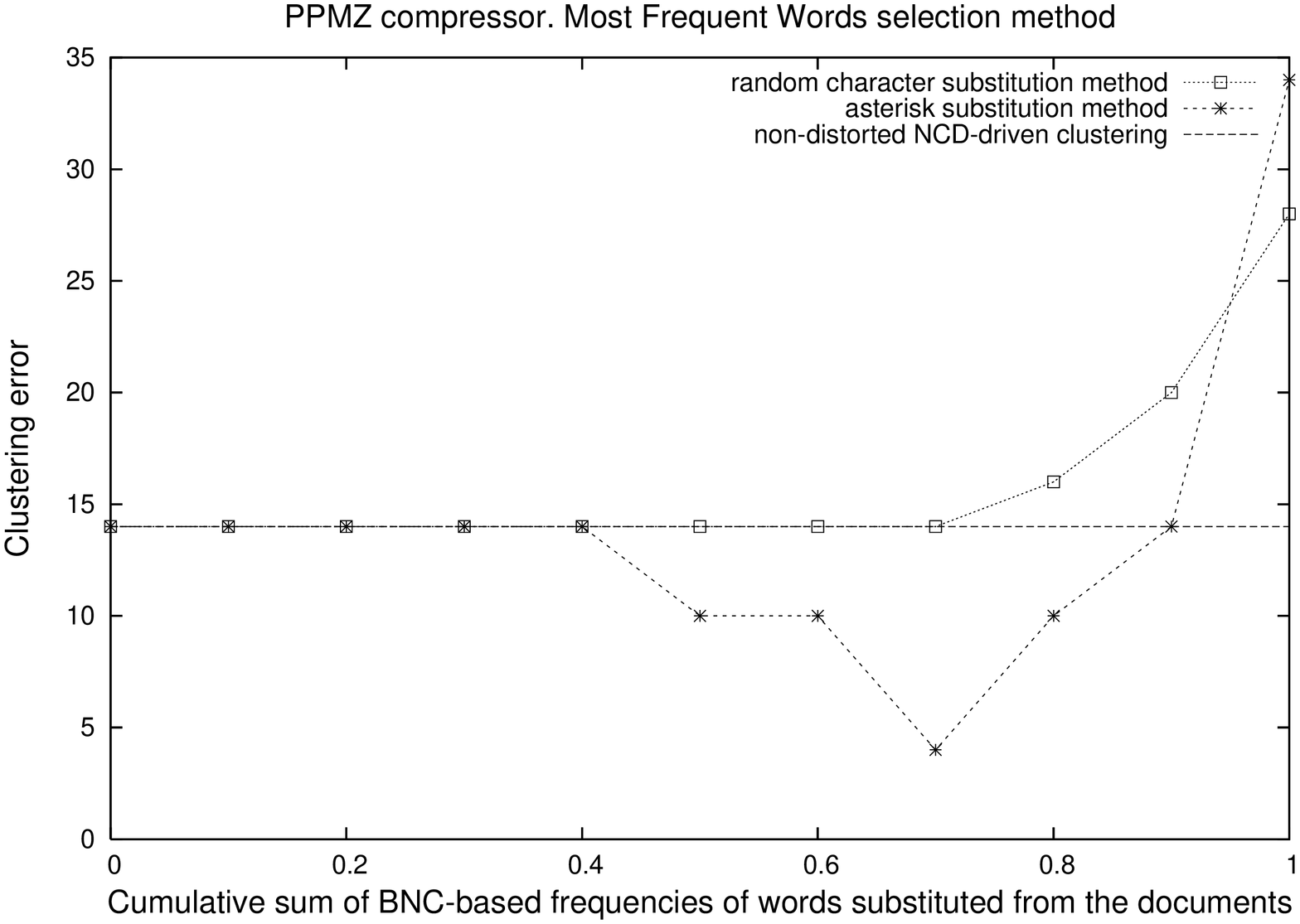}
\caption{MedlinePlus. PPMZ compressor. MFW selection method.}
\label{APPENDIX. Fig:medline-clustering-error-ppmz-mfw}
\end{figure}

\begin{figure}
\centering
\includegraphics[angle=270,width=13cm]{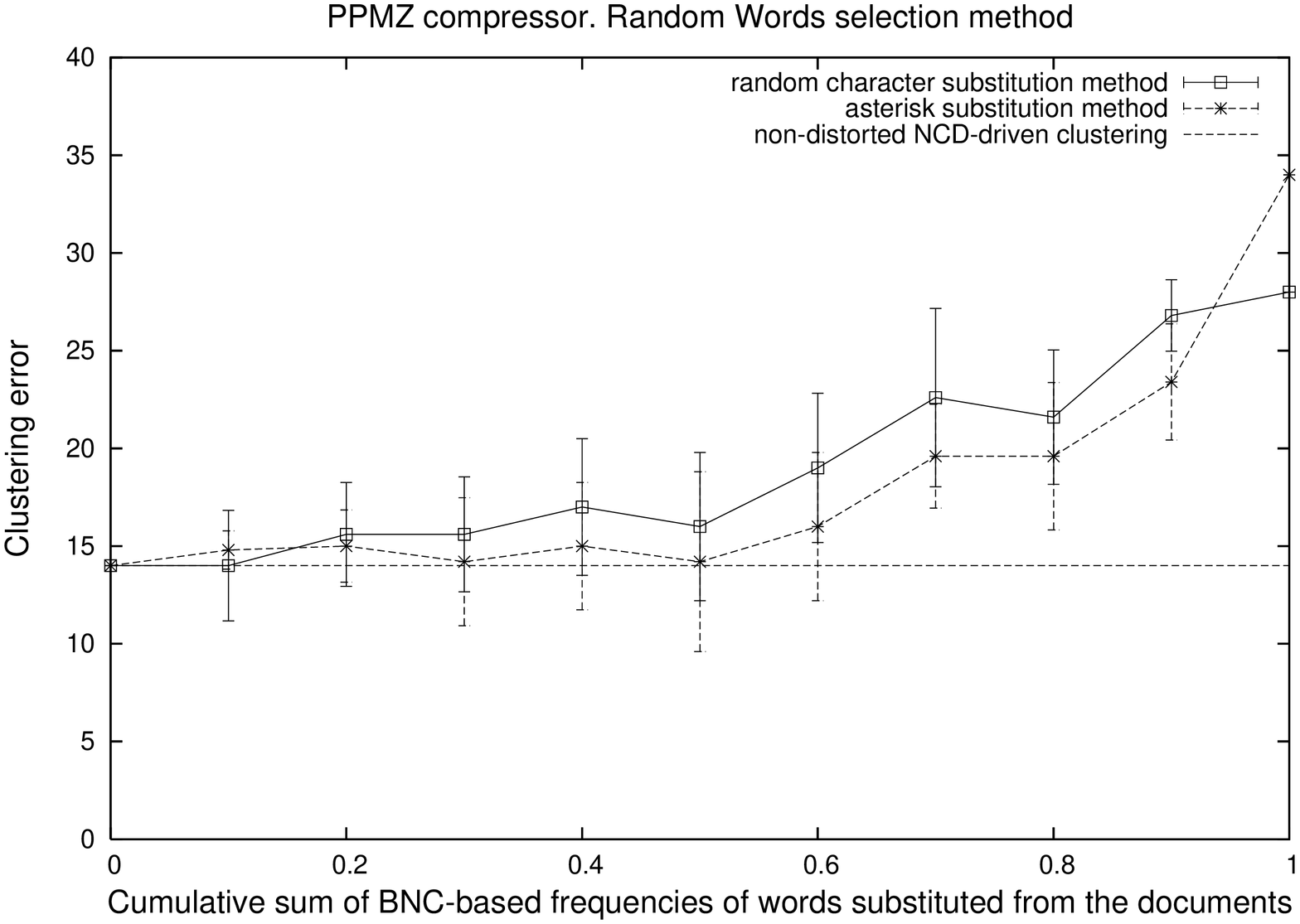}
\caption{MedlinePlus. PPMZ compressor. RW selection method.}
\label{APPENDIX. Fig:medline-clustering-error-ppmz-rw}
\end{figure}

\clearpage

\begin{figure}[ht]
\centering
\includegraphics[angle=270,width=13cm]{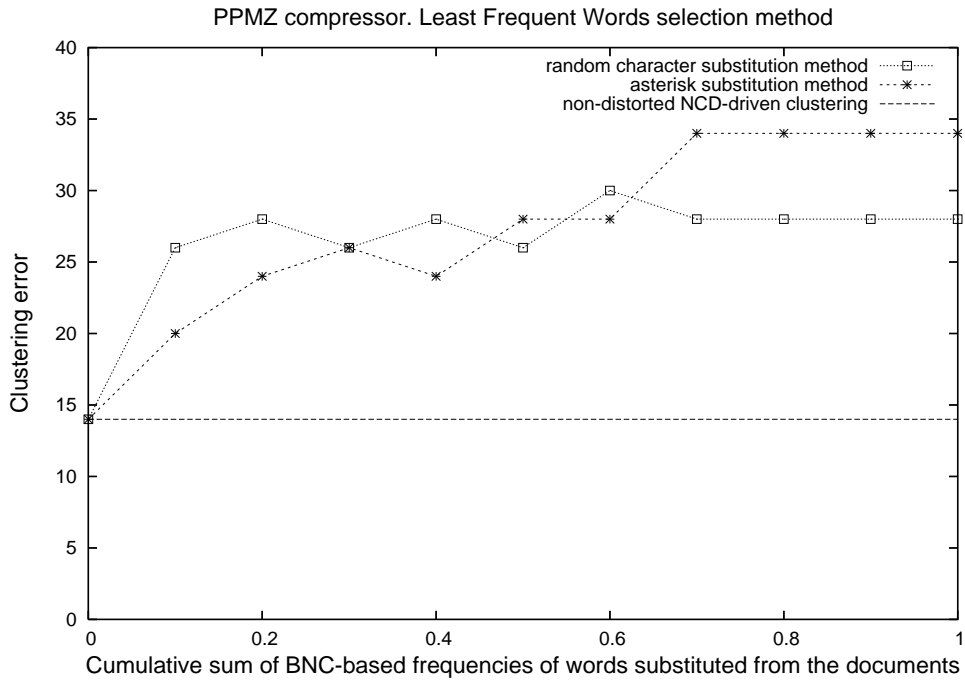}
\caption{MedlinePlus. PPMZ compressor. LFW selection method.}
\label{APPENDIX. Fig:medline-clustering-error-ppmz-lfw}
\end{figure}

The combination of the \emph{MFW selection method} and the
\emph{asterisk substitution method} improves the clustering results
as follows:

\begin{itemize}
  \item Non-distorted clustering error: 14
  \item Best clustering error: 4
\end{itemize}

This improvement can be observed by comparing Figs \ref{APPENDIX.
Fig:dendro-medline-ppmz-original} and \ref{APPENDIX.
Fig:dendro-medline-ppmz-best}. Whereas three documents are not
correctly clustered in Fig \ref{APPENDIX.
Fig:dendro-medline-ppmz-original}, only two documents are not
correctly clustered in Fig \ref{APPENDIX.
Fig:dendro-medline-ppmz-best}. Furthermore, in the best dendrogram
obtained, the documents that are not correctly clustered are
adjacent to the ones related to them.

\clearpage

\begin{figure}
\centering
\includegraphics[width=14cm]{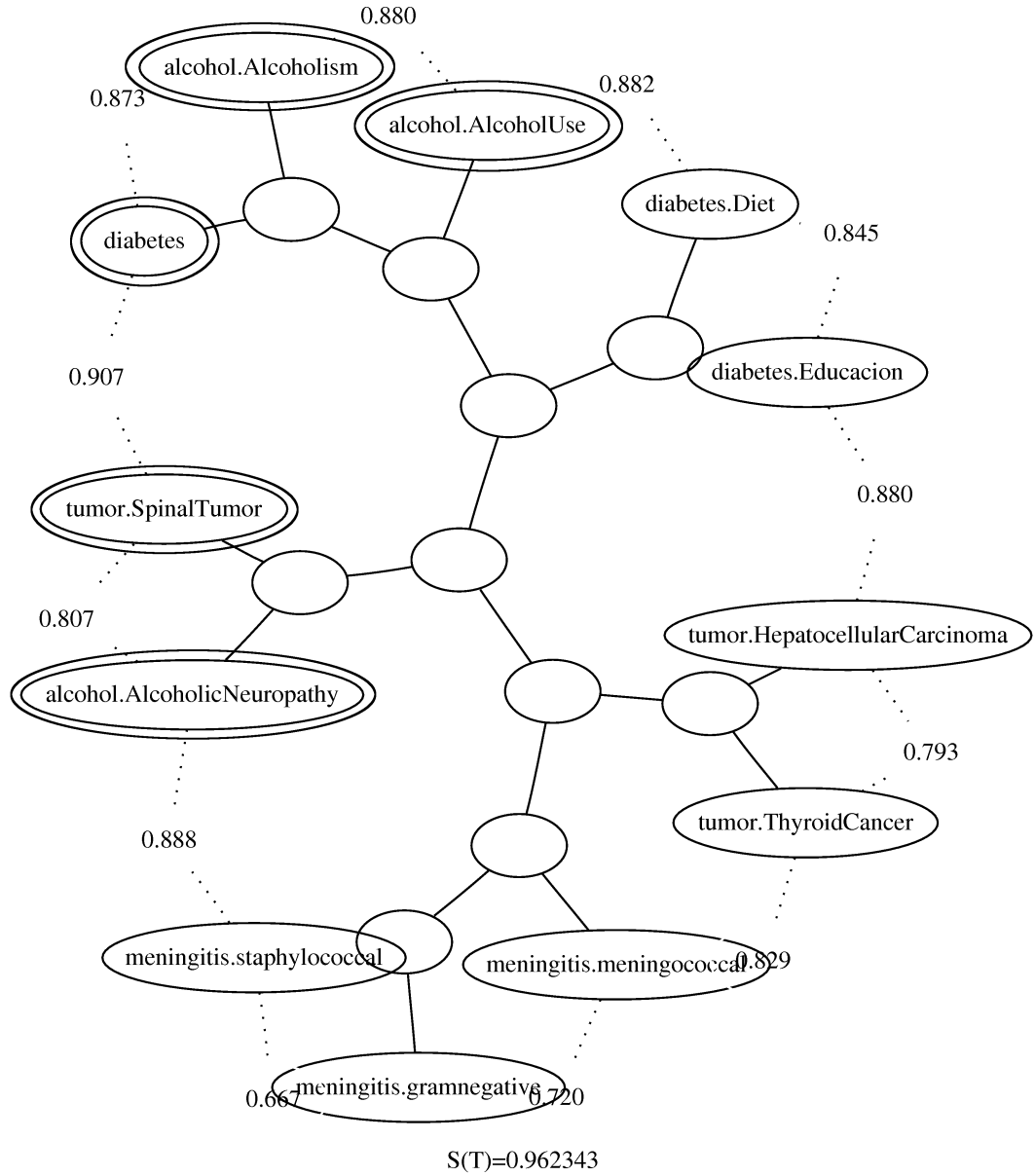}
\caption{MedlinePlus. PPMZ compressor. Dendrogram obtained with no
distortion.} \label{APPENDIX. Fig:dendro-medline-ppmz-original}
\end{figure}

\begin{figure}
\centering
\includegraphics[width=14cm]{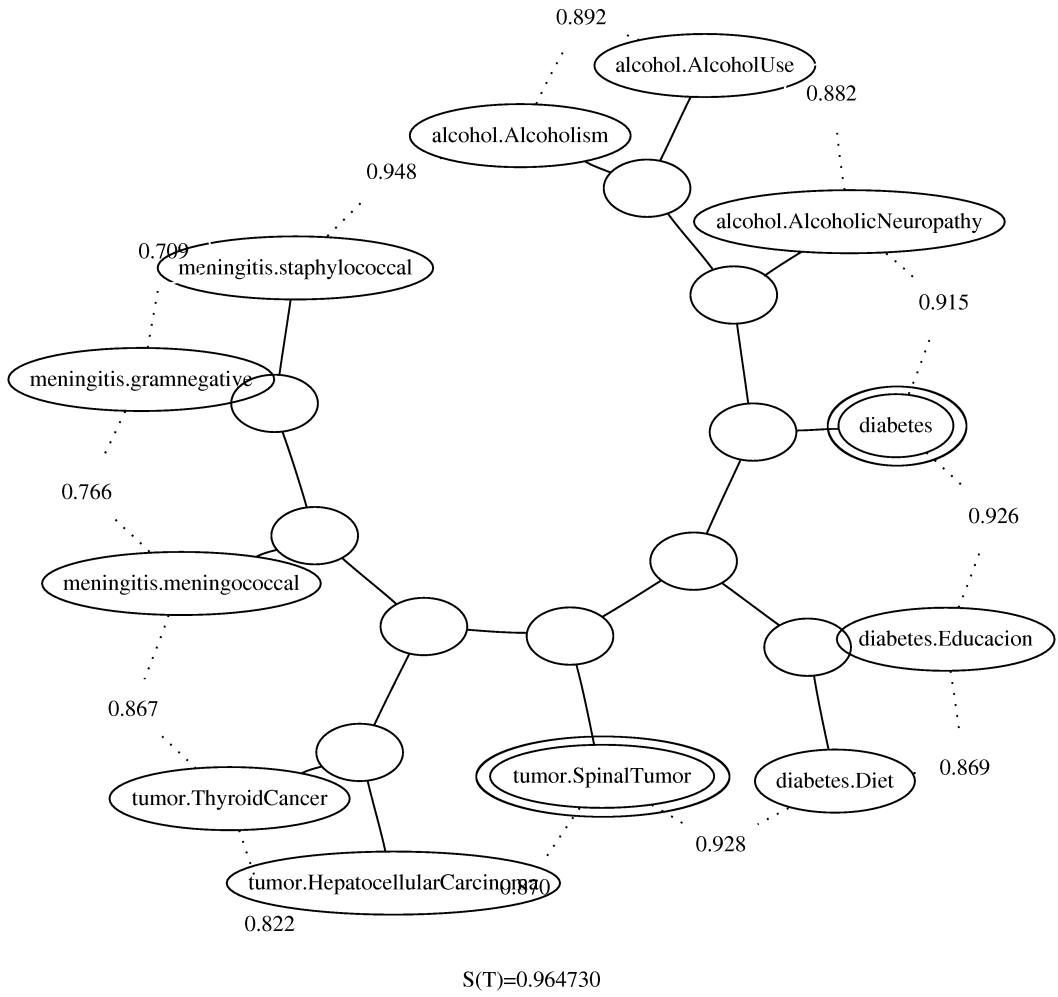}
\caption{MedlinePlus. PPMZ compressor. Best dendrogram obtained.}
\label{APPENDIX. Fig:dendro-medline-ppmz-best}
\end{figure}

\clearpage

\begin{figure}
\centering
\includegraphics[angle=270,width=13cm]{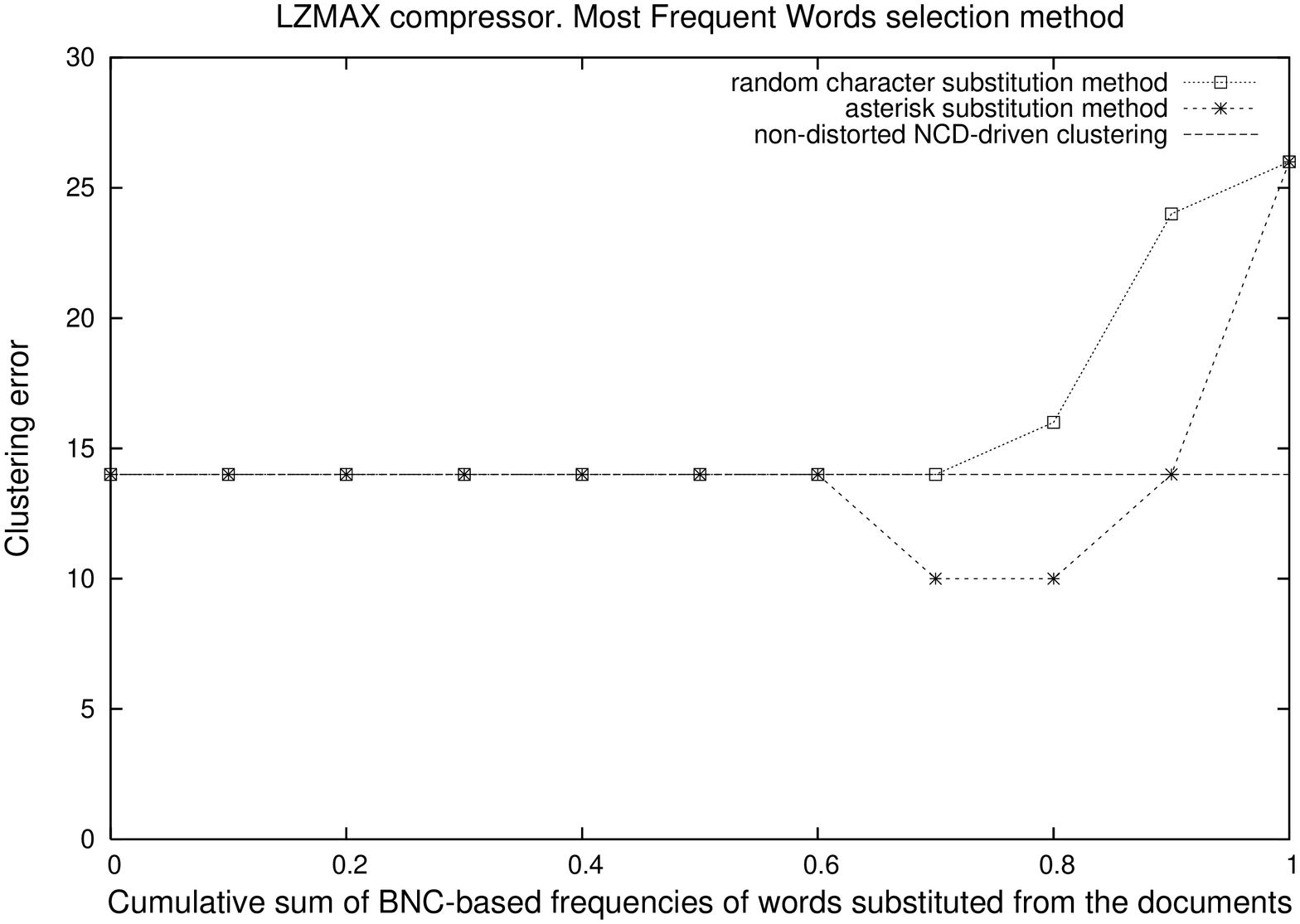}
\caption{MedlinePlus. LZMA compressor. MFW selection method.}
\label{APPENDIX. Fig:medline-clustering-error-lzma-mfw}
\end{figure}

\begin{figure}
\centering
\includegraphics[angle=270,width=13cm]{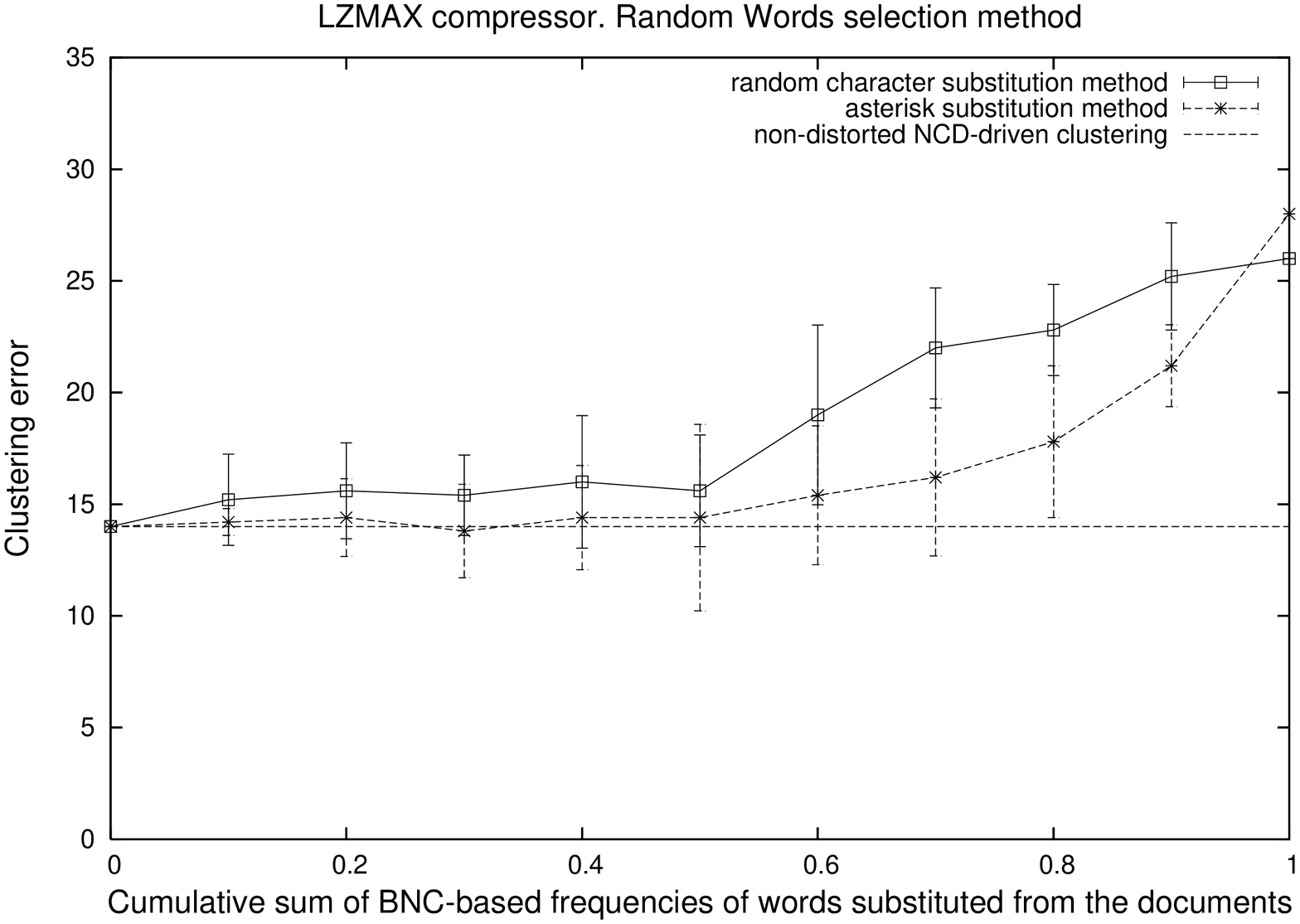}
\caption{MedlinePlus. LZMA compressor. RW selection method.}
\label{APPENDIX. Fig:medline-clustering-error-lzma-rw}
\end{figure}

\clearpage

\begin{figure}[ht]
\centering
\includegraphics[angle=270,width=13cm]{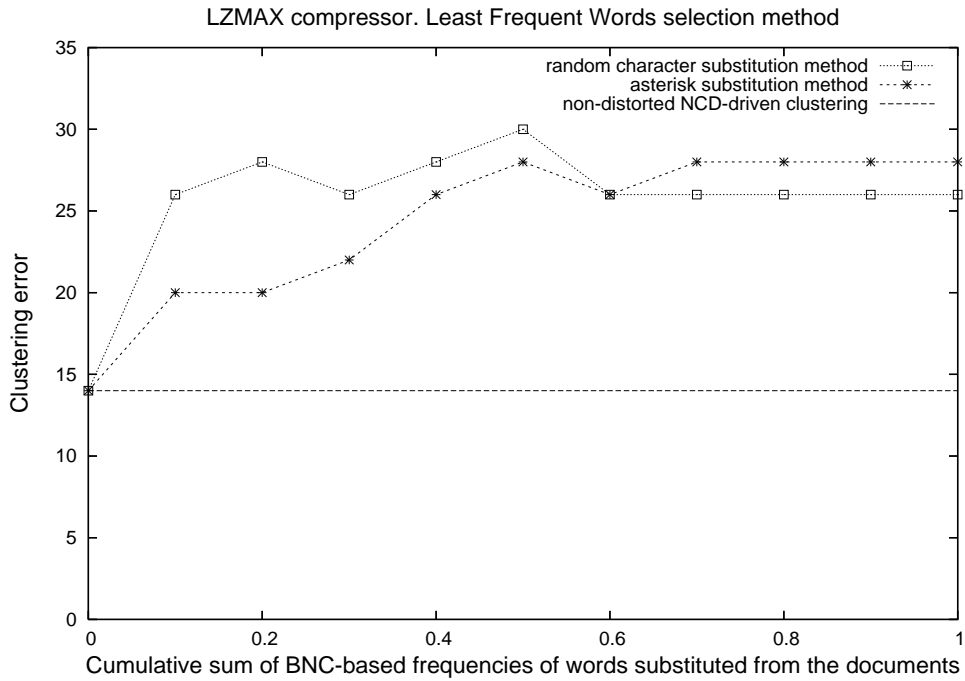}
\caption{MedlinePlus. LZMA compressor. LFW selection method.}
\label{APPENDIX. Fig:medline-clustering-error-lzma-lfw}
\end{figure}

The combination of the \emph{MFW selection method} and the
\emph{asterisk substitution method} improves the clustering results
when the texts are distorted using the set of words that accumulate
a BNC-based frequency of 0.7 and 0.8.

\begin{itemize}
  \item Non-distorted clustering error: 14
  \item Best clustering error: 10
\end{itemize}

\begin{figure}
\centering
\includegraphics[width=14cm]{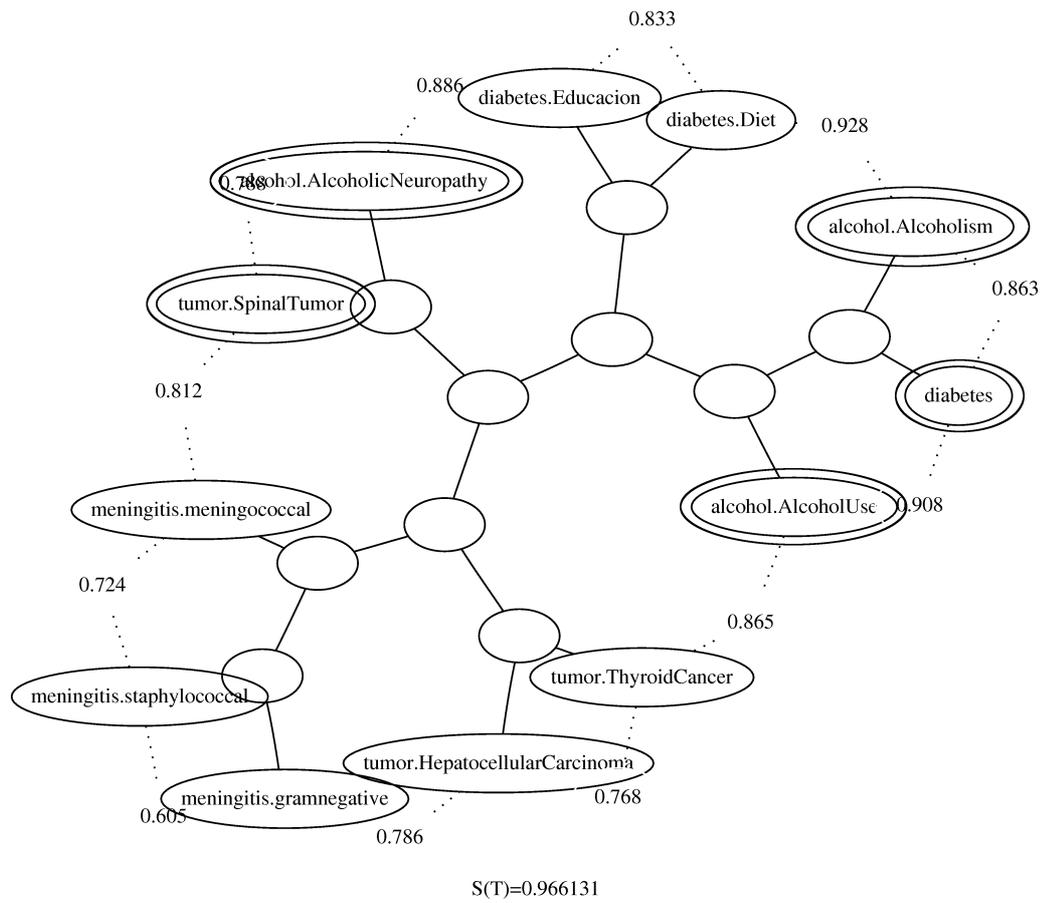}
\caption{MedlinePlus. LZMA compressor. Dendrogram obtained with no
distortion.} \label{APPENDIX. Fig:dendro-medline-lzma-original}
\end{figure}

As usual, the improvement can be observed by comparing Figs
\ref{APPENDIX. Fig:dendro-medline-lzma-original} and \ref{APPENDIX.
Fig:dendro-medline-lzma-best}. Three texts are problematic in this
case. They are about diabetes, tumor and alcohol. The difference
between the dendrogram obtained with no distortion and the best
dendrogram obtained is that, in the latter, the texts are closer to
the texts that are related to them.

\clearpage

\begin{figure}
\centering
\includegraphics[width=14cm]{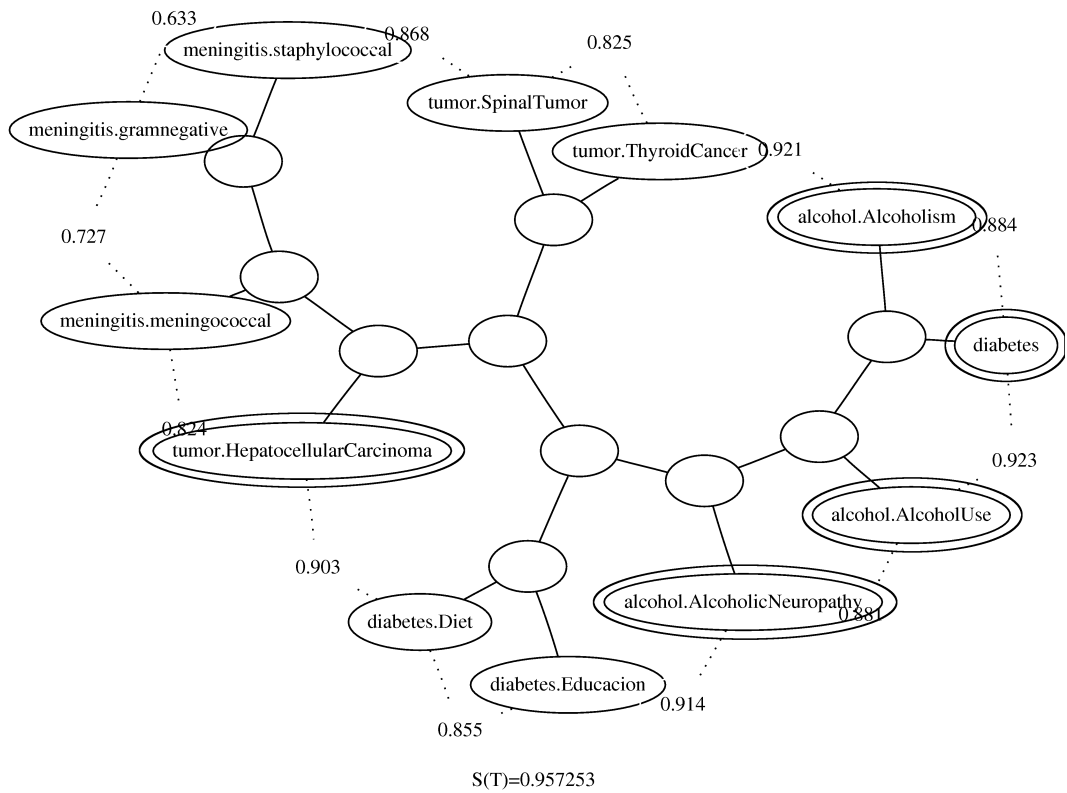}
\caption{MedlinePlus. LZMA compressor. Best dendrogram obtained.}
\label{APPENDIX. Fig:dendro-medline-lzma-best}
\end{figure}

\clearpage

\begin{figure}
\centering
\includegraphics[angle=270,width=13cm]{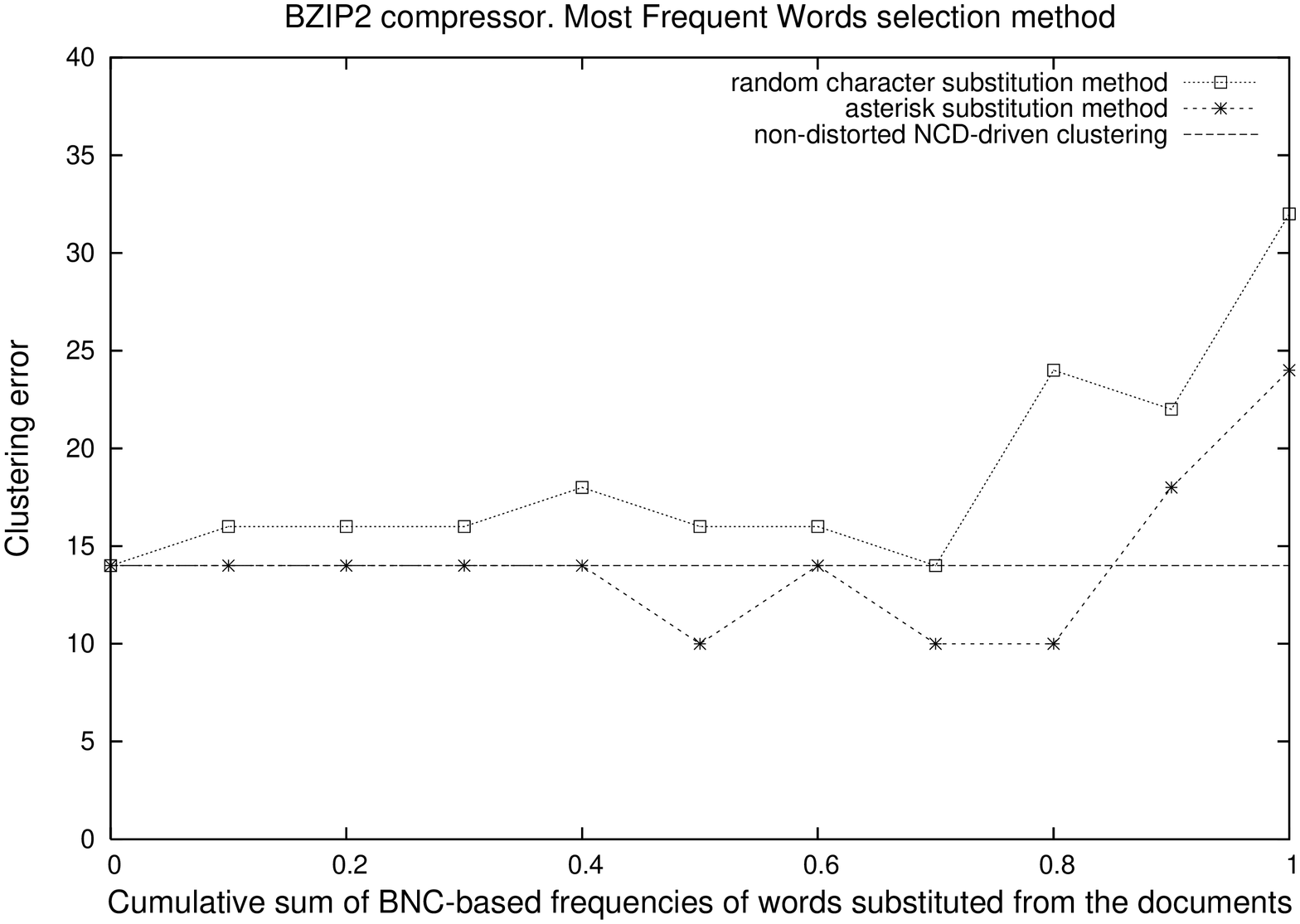}
\caption{MedlinePlus. BZIP2 compressor. MFW selection method.}
\label{APPENDIX. Fig:medline-clustering-error-bzip2-mfw}
\end{figure}

\begin{figure}
\centering
\includegraphics[angle=270,width=13cm]{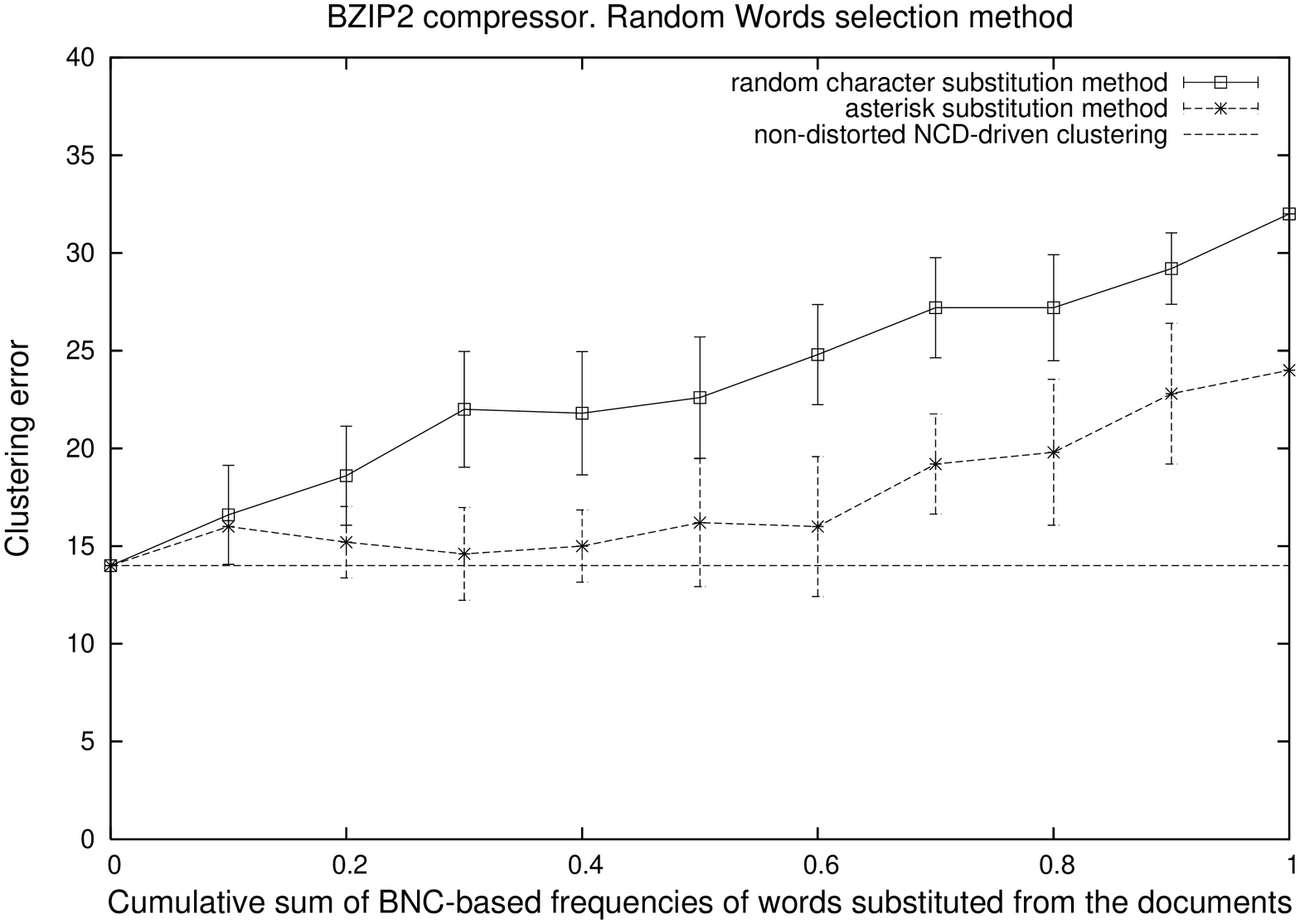}
\caption{MedlinePlus. BZIP2 compressor. RW selection method.}
\label{APPENDIX. Fig:medline-clustering-error-bzip2-rw}
\end{figure}

\clearpage

\begin{figure}[ht]
\centering
\includegraphics[angle=270,width=13cm]{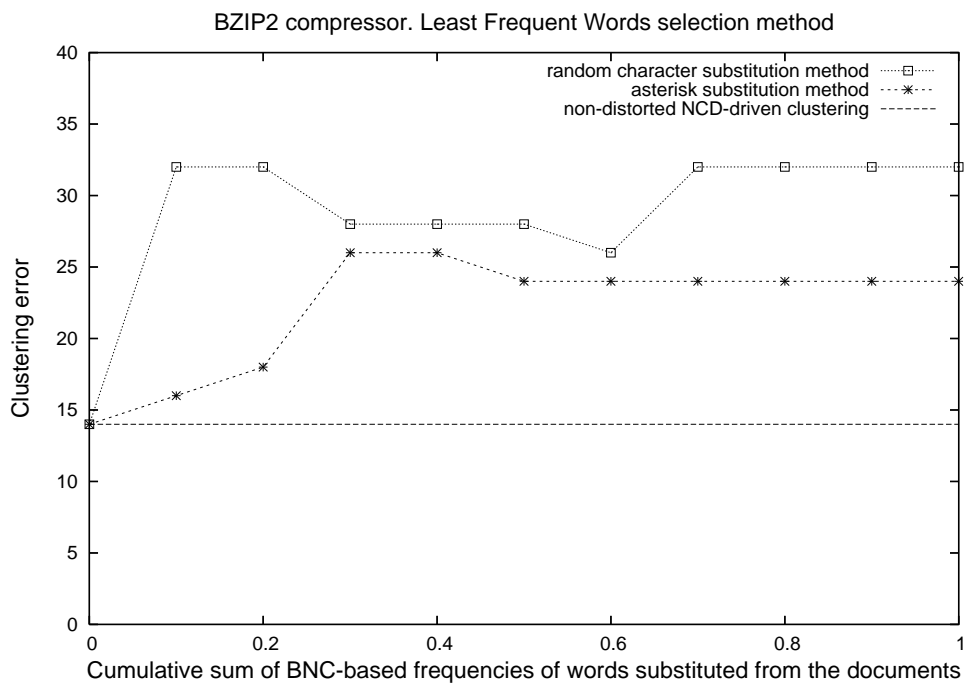}
\caption{MedlinePlus. BZIP2 compressor. LFW selection method.}
\label{APPENDIX. Fig:medline-clustering-error-bzip2-lfw}
\end{figure}

Analyzing Figs \ref{APPENDIX.
Fig:medline-clustering-error-bzip2-mfw}, \ref{APPENDIX.
Fig:medline-clustering-error-bzip2-rw}, and \ref{APPENDIX.
Fig:medline-clustering-error-bzip2-lfw} one can observe that the
best clustering results correspond to the \emph{MFW selection
method}, the worst results correspond to the \emph{LFW selection
method}, and the results corresponding to the \emph{RW selection
method} are situated in between them.

As usual, the combination of the \emph{MFW selection method} and the
\emph{asterisk substitution method} improves the clustering results.
In this case, this improvement is obtained when the texts are
distorted using the set of words that accumulate a BNC-based
frequency of 0.5, 0.7 and 0.8.

\begin{itemize}
  \item Non-distorted clustering error: 14
  \item Best clustering error: 10
\end{itemize}

Again, comparing Figs \ref{APPENDIX.
Fig:dendro-medline-lzma-original} and \ref{APPENDIX.
Fig:dendro-medline-lzma-best} one can see the clustering behavior
improvement. For this \emph{dataset-compression algorithm} pair
three texts are problematic. They are about diabetes, tumor and
alcohol. The difference between the dendrogram obtained with no
distortion and the best dendrogram obtained is that in the latter
the texts are closer to the texts to which they are related.

\clearpage

\begin{figure}
\centering
\includegraphics[width=14cm]{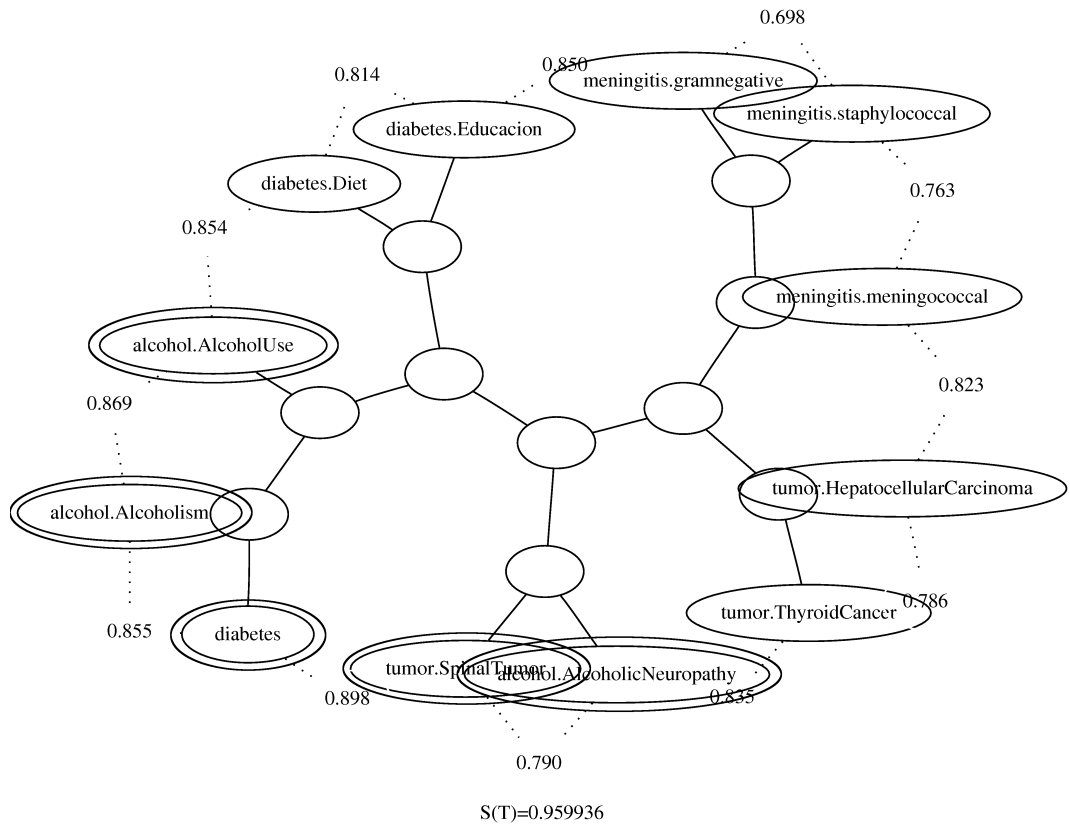}
\caption{MedlinePlus. BZIP2 compressor. Dendrogram obtained with no
distortion.} \label{APPENDIX. Fig:dendro-medline-bzip2-original}
\end{figure}

\begin{figure}
\centering
\includegraphics[width=14cm]{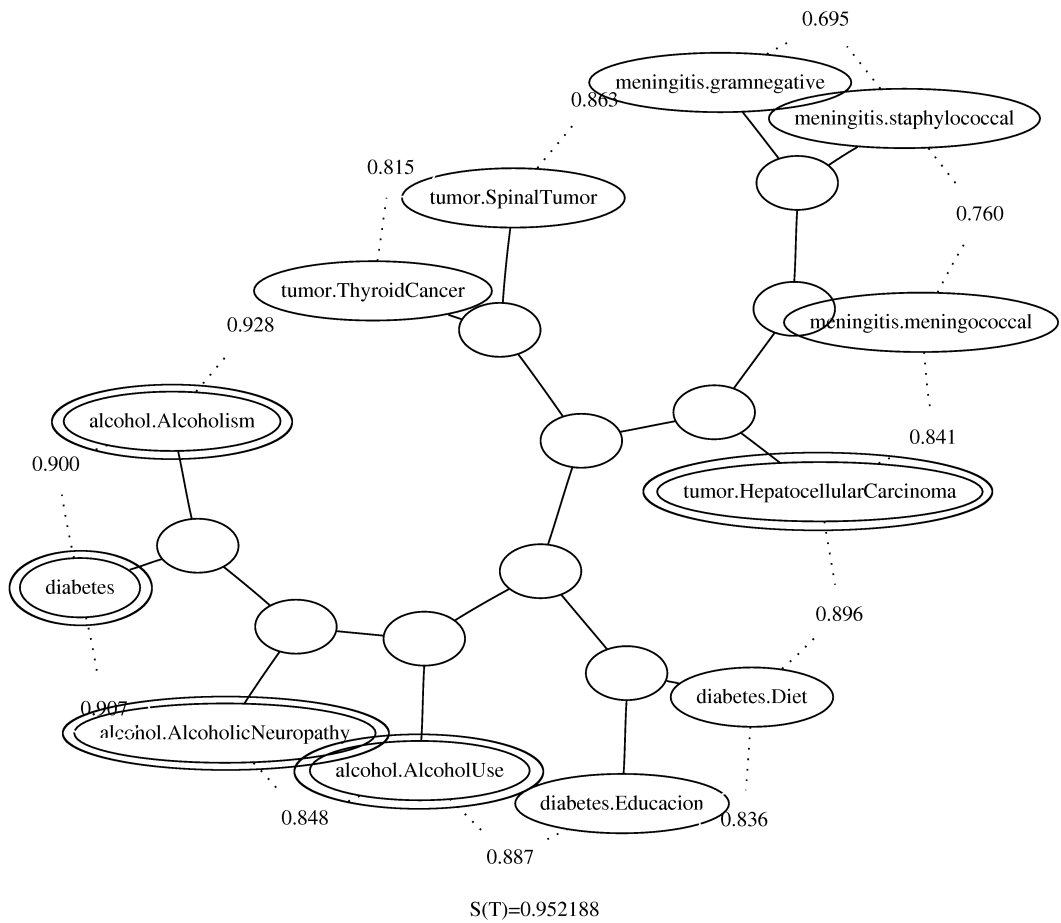}
\caption{MedlinePlus. BZIP2 compressor. Best dendrogram obtained.}
\label{APPENDIX. Fig:dendro-medline-bzip2-best}
\end{figure}

\clearpage

\begin{figure}
\centering
\includegraphics[angle=270,width=13cm]{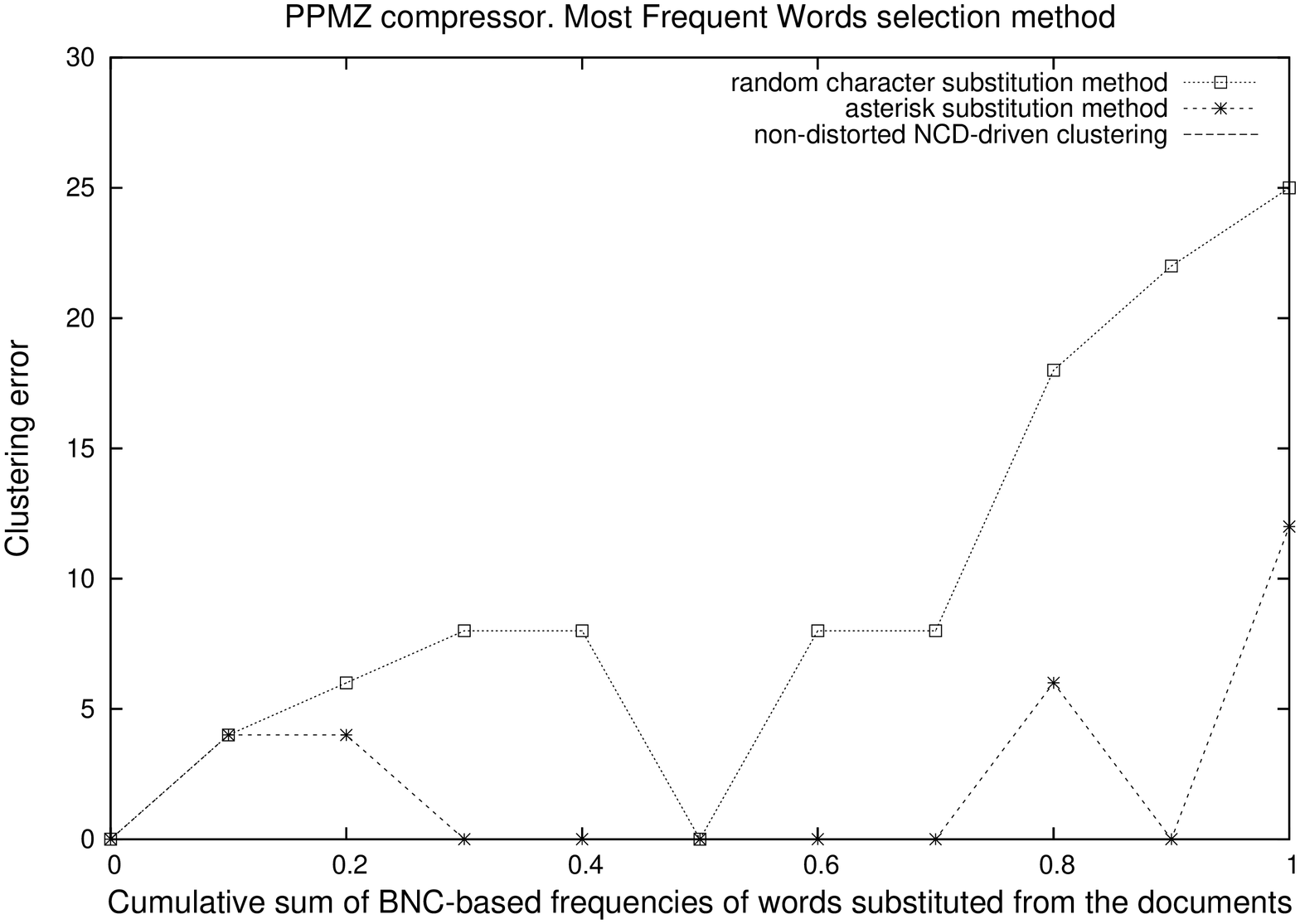}
\caption{IMDB. PPMZ compressor. MFW selection method.}
\label{APPENDIX. Fig:imdb-clustering-error-ppmz-mfw}
\end{figure}

\begin{figure}
\centering
\includegraphics[angle=270,width=13cm]{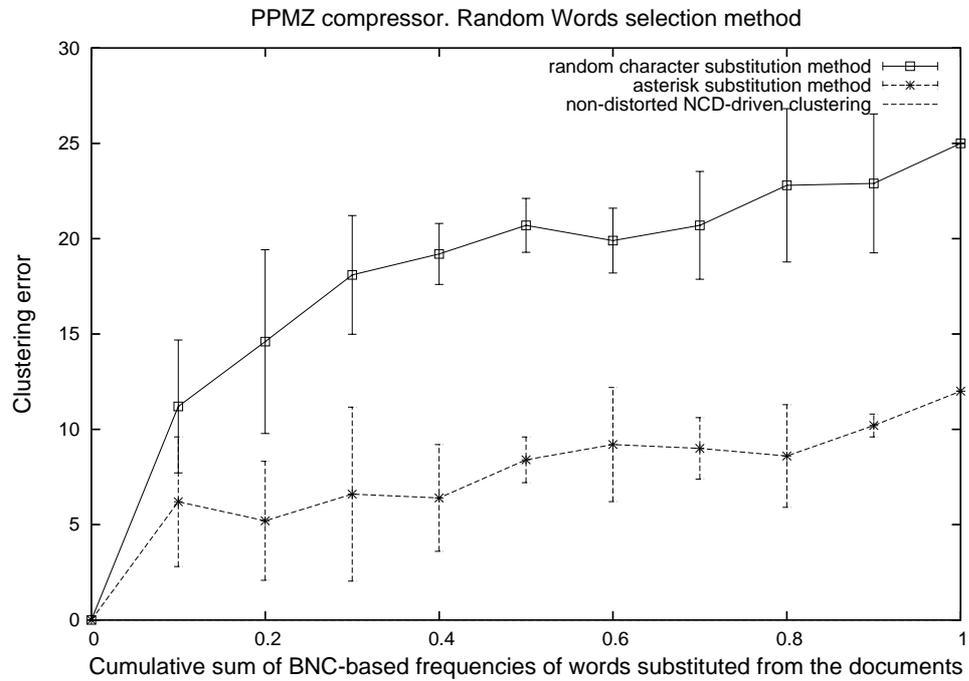}
\caption{IMDB. PPMZ compressor. RW selection method.}
\label{APPENDIX. Fig:imdb-clustering-error-ppmz-rw}
\end{figure}

\clearpage

\begin{figure}[ht]
\centering
\includegraphics[angle=270,width=13cm]{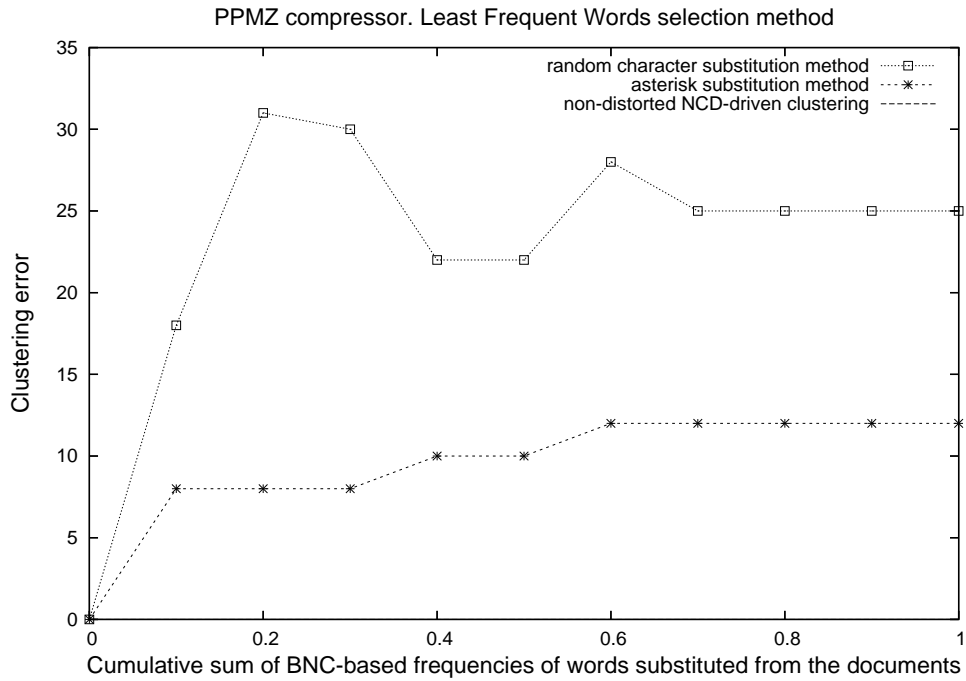}
\caption{IMDB. PPMZ compressor. LFW selection method.}
\label{APPENDIX. Fig:imdb-clustering-error-ppmz-lfw}
\end{figure}

Again, analyzing Figs \ref{APPENDIX.
Fig:imdb-clustering-error-ppmz-mfw}, \ref{APPENDIX.
Fig:imdb-clustering-error-ppmz-rw}, and \ref{APPENDIX.
Fig:imdb-clustering-error-ppmz-lfw} one can observe that the best
clustering results correspond to the \emph{MFW selection method},
the worst results correspond to the \emph{LFW selection method}, and
the results corresponding to the \emph{RW selection method} are
situated in between them.

Given that the dendrogram obtained with no distortion clusters all
the texts perfectly, only one dendrogram is shown in this case.

\clearpage

\begin{figure}
\centering
\includegraphics[width=14cm]{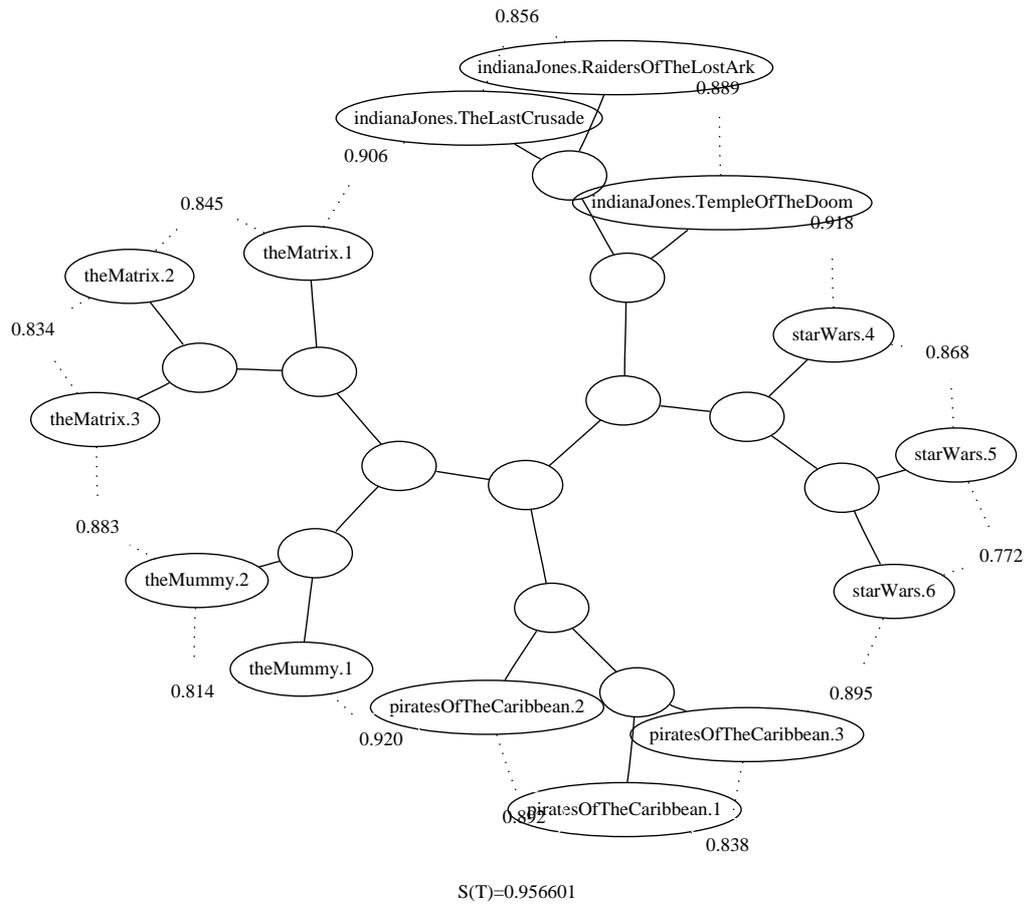}
\caption{IMDB. PPMZ compressor. Best dendrogram obtained.}
\label{APPENDIX. Fig:dendro-imdb-ppmz-best}
\end{figure}

\clearpage

\begin{figure}
\centering
\includegraphics[angle=270,width=13cm]{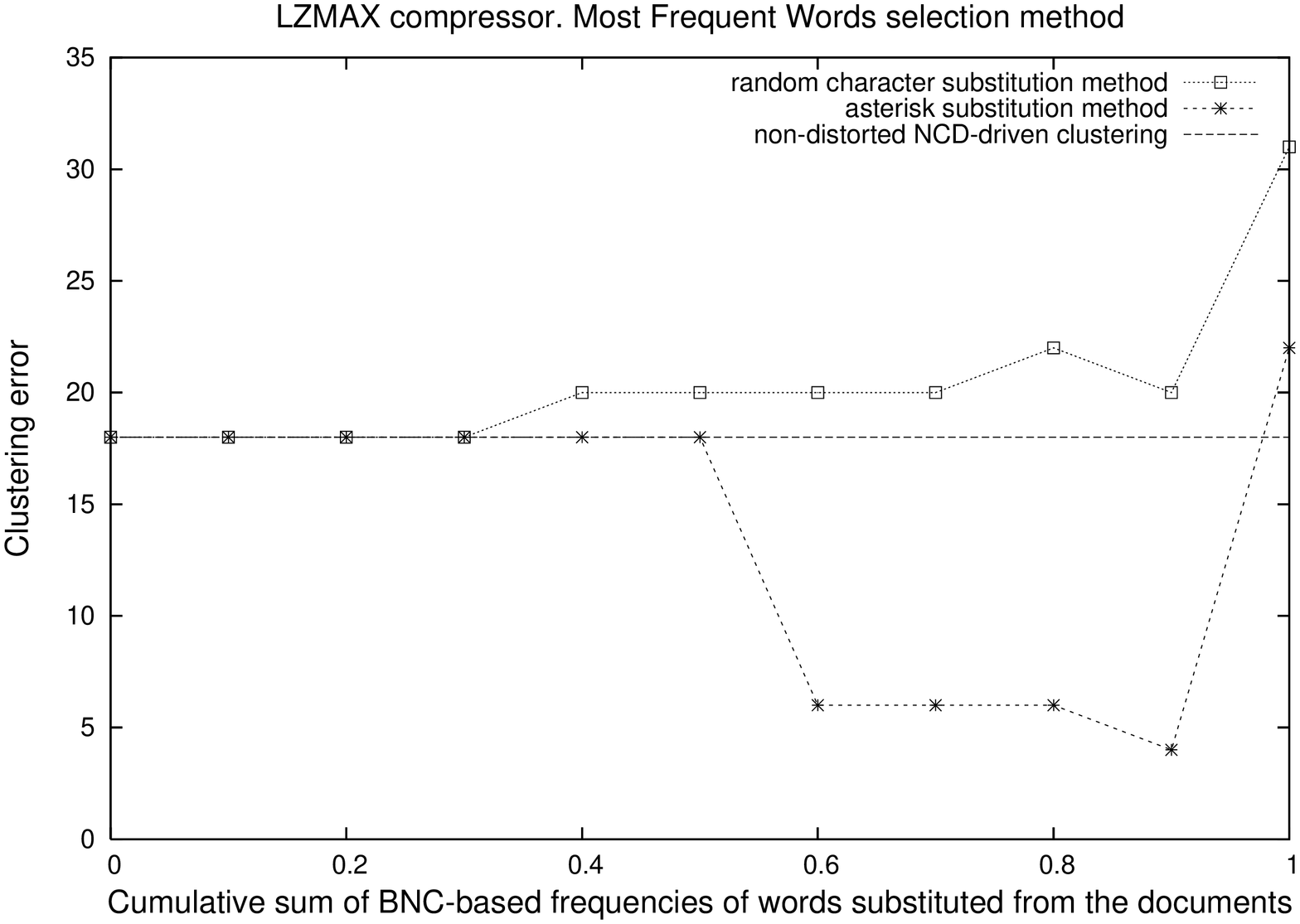}
\caption{IMDB. LZMA compressor. MFW selection method.}
\label{APPENDIX. Fig:imdb-clustering-error-lzma-mfw}
\end{figure}

\begin{figure}
\centering
\includegraphics[angle=270,width=13cm]{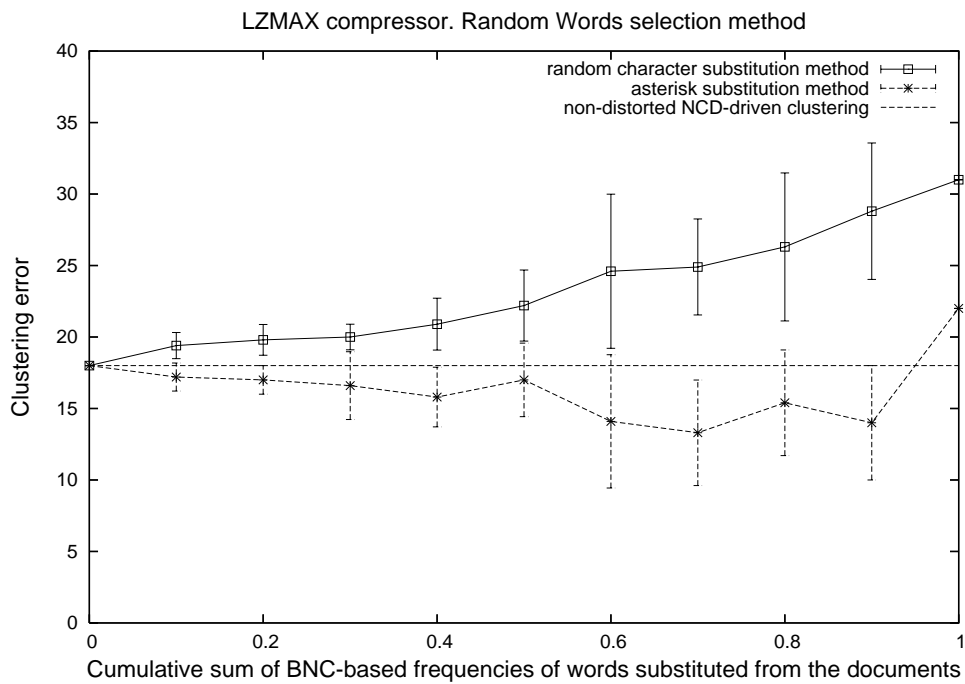}
\caption{IMDB. LZMA compressor. RW selection method.}
\label{APPENDIX. Fig:imdb-clustering-error-lzma-rw}
\end{figure}

\clearpage

\begin{figure}[ht]
\centering
\includegraphics[angle=270,width=13cm]{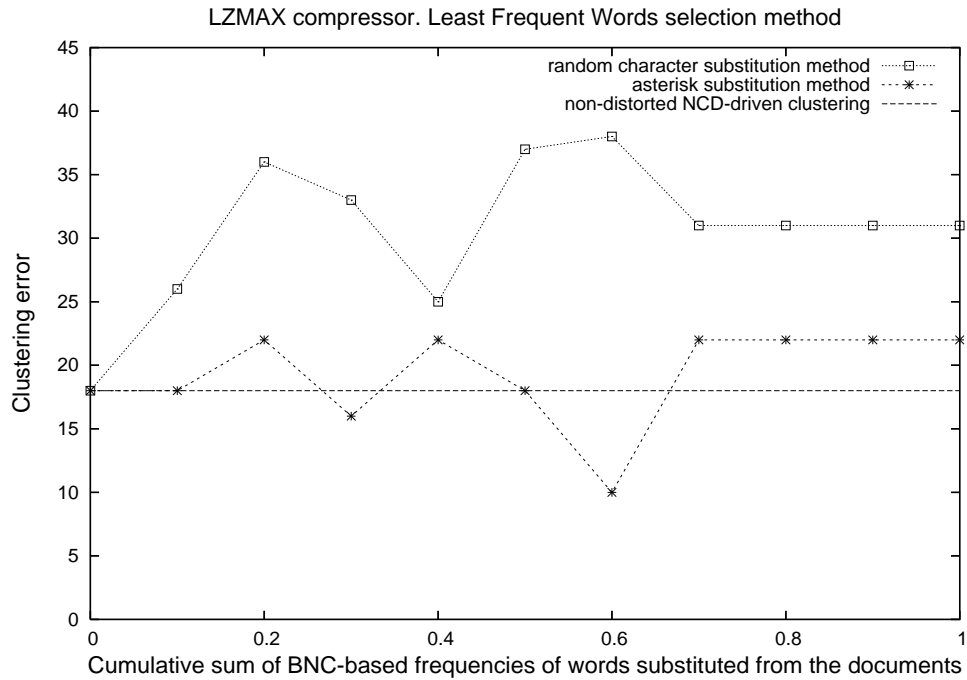}
\caption{IMDB. LZMA compressor. LFW selection method.}
\label{APPENDIX. Fig:imdb-clustering-error-lzma-lfw}
\end{figure}

The combination of the \emph{MFW selection method} and the
\emph{asterisk substitution method} improves the clustering results
when the texts are distorted using the sets of words that accumulate
a BNC-based frequency of 0.6, 0.7, 0.8 and 0.9.

\begin{itemize}
  \item Non-distorted clustering error: 18
  \item Best clustering error: 4
\end{itemize}

The improvement can be observed by comparing Figs \ref{APPENDIX.
Fig:dendro-imdb-lzma-original} and \ref{APPENDIX.
Fig:dendro-imdb-lzma-best}. The difference between both figures is
that the movies ``Pirates of the Caribbean 2'', ``Star Wars 4'' and
``Indiana Jones and the Temple of the Doom'' are incorrectly
clustered in the dendrogram depicted in Fig \ref{APPENDIX.
Fig:dendro-imdb-lzma-original}, whereas only the movie ``Pirates of
the Caribbean 2'' is incorrectly clustered in Fig \ref{APPENDIX.
Fig:dendro-imdb-lzma-best}.

\clearpage

\begin{figure}
\centering
\includegraphics[width=14cm]{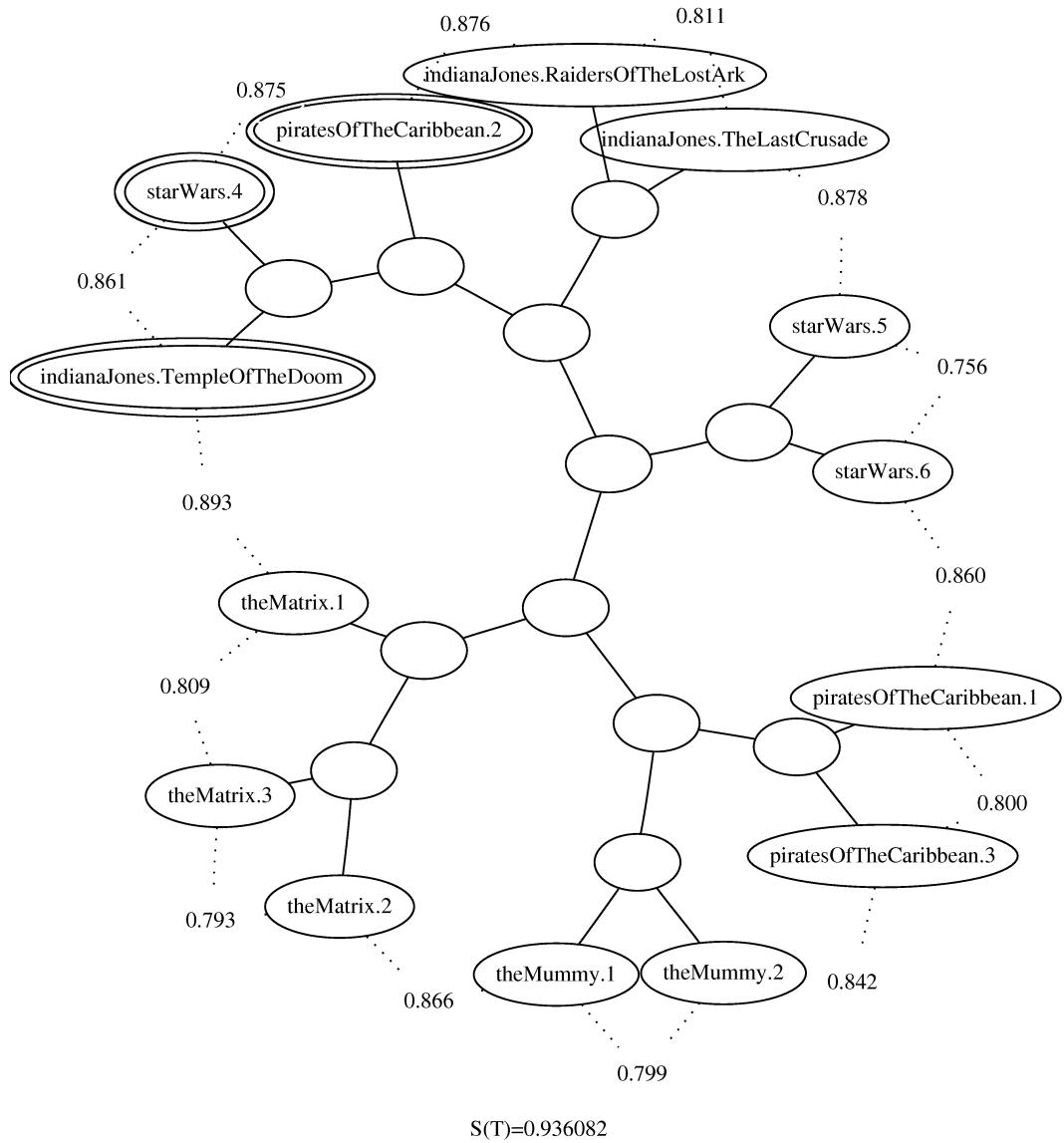}
\caption{IMDB. LZMA compressor. Dendrogram obtained with no
distortion.} \label{APPENDIX. Fig:dendro-imdb-lzma-original}
\end{figure}

\begin{figure}
\centering
\includegraphics[width=14cm]{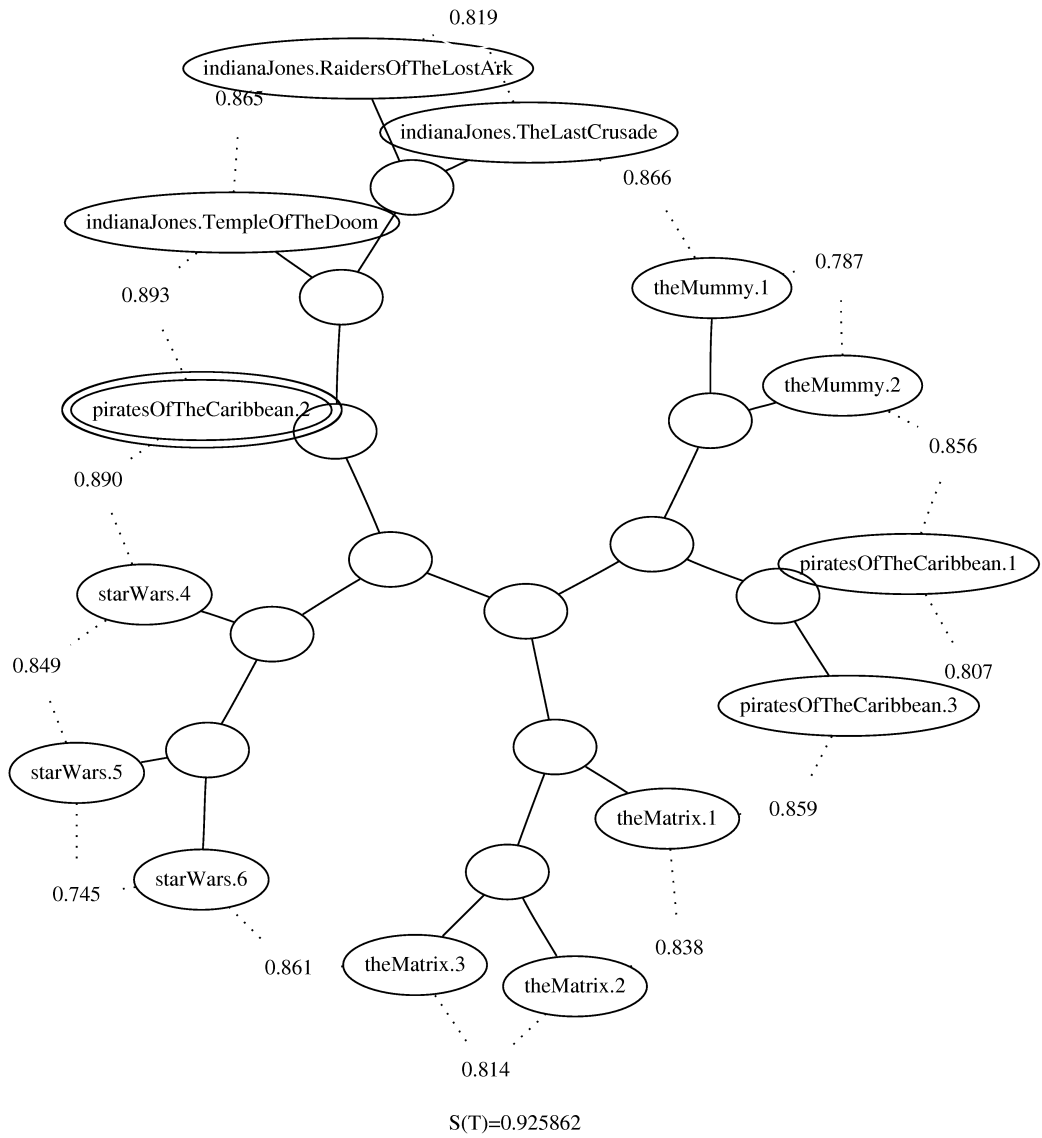}
\caption{IMDB. LZMA compressor. Best dendrogram obtained.}
\label{APPENDIX. Fig:dendro-imdb-lzma-best}
\end{figure}

\clearpage

\begin{figure}
\centering
\includegraphics[angle=270,width=13cm]{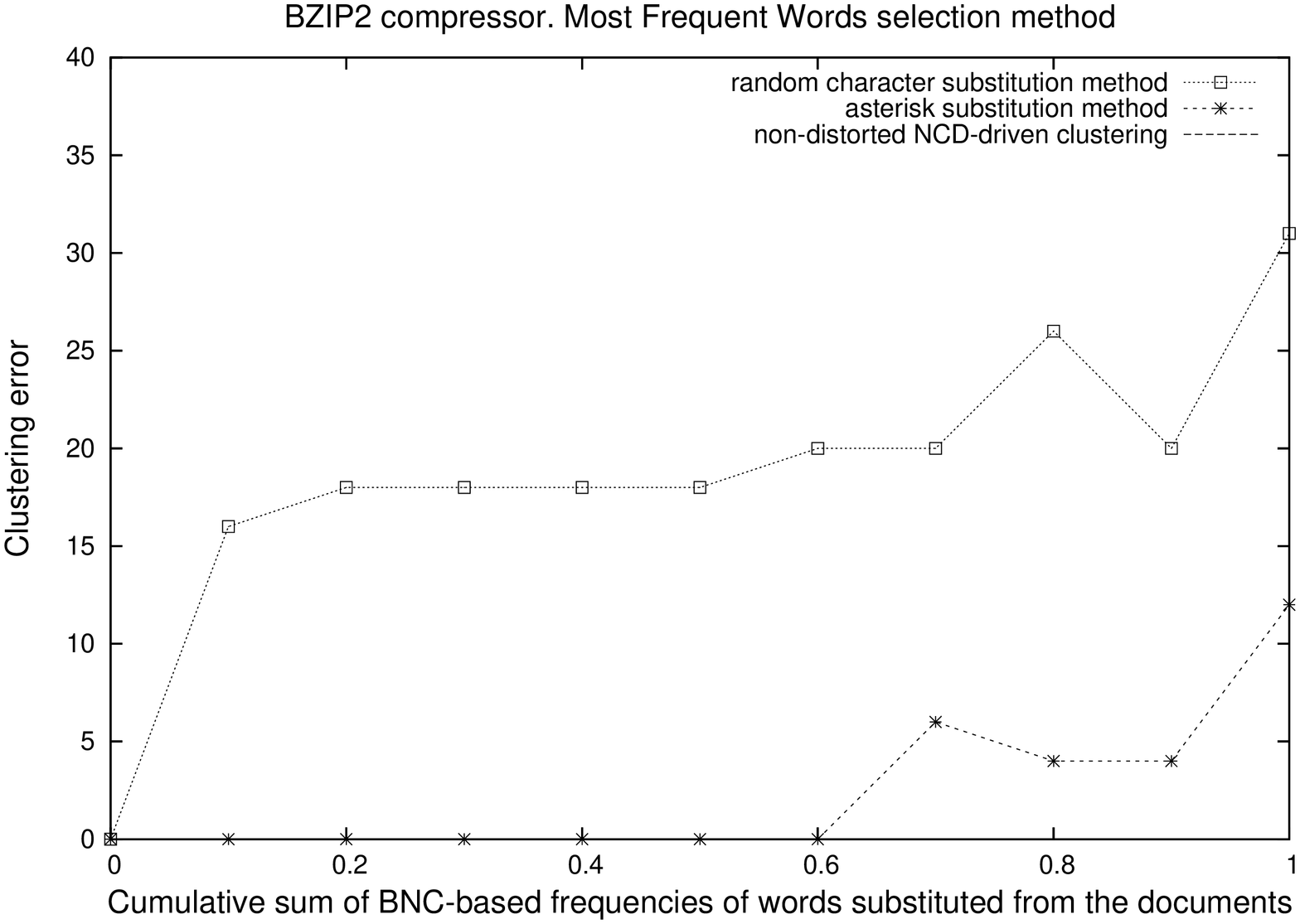}
\caption{IMDB. BZIP2 compressor. MFW selection method.}
\label{APPENDIX. Fig:imdb-clustering-error-bzip2-mfw}
\end{figure}

\begin{figure}
\centering
\includegraphics[angle=270,width=13cm]{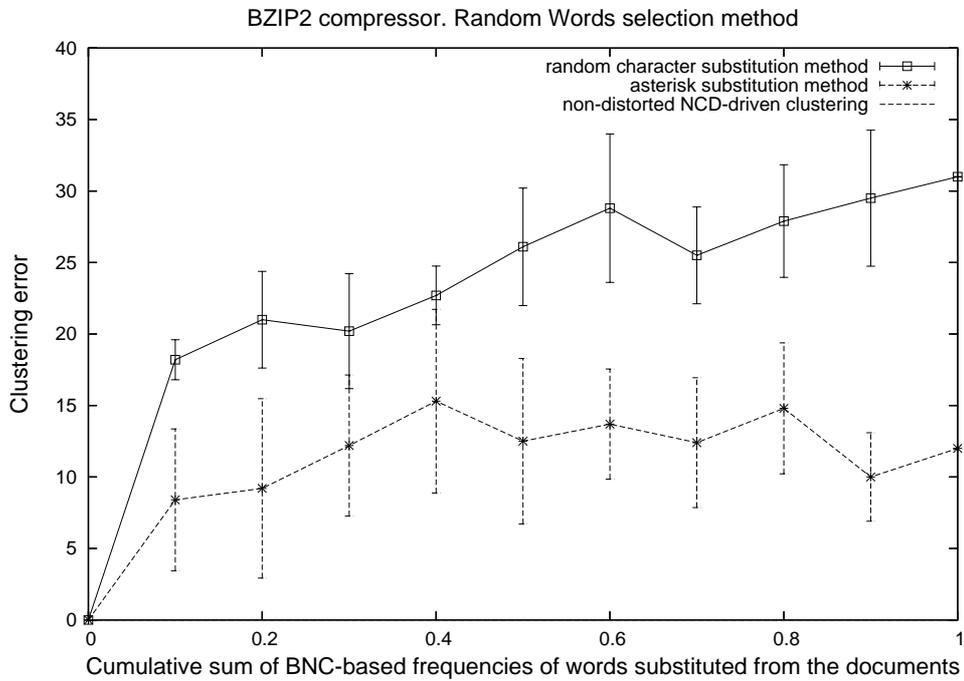}
\caption{IMDB. BZIP2 compressor. RW selection method.}
\label{APPENDIX. Fig:imdb-clustering-error-bzip2-rw}
\end{figure}

\clearpage

\begin{figure}[ht]
\centering
\includegraphics[angle=270,width=13cm]{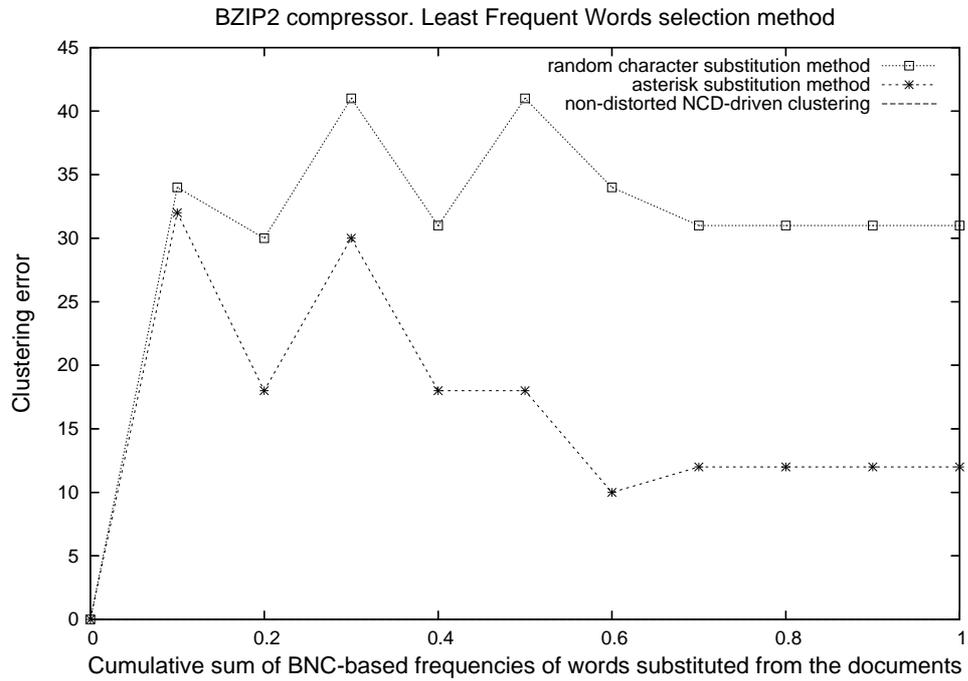}
\caption{IMDB. BZIP2 compressor. LFW selection method.}
\label{APPENDIX. Fig:imdb-clustering-error-bzip2-lfw}
\end{figure}

The clustering error obtained without distortion is 0, therefore,
only one dendrogram is depicted for this \emph{dataset-compression
algorithm} pair.

Again, analyzing Figs \ref{APPENDIX.
Fig:imdb-clustering-error-bzip2-mfw}, \ref{APPENDIX.
Fig:imdb-clustering-error-bzip2-rw}, and \ref{APPENDIX.
Fig:imdb-clustering-error-bzip2-lfw}, one can reach the conclusion
that the best clustering results correspond to the \emph{MFW
selection method}, the worst results correspond to the \emph{LFW
selection method}, and the results corresponding to the \emph{RW
selection method} are situated in between them.

\clearpage

\begin{figure}
\centering
\includegraphics[width=14cm]{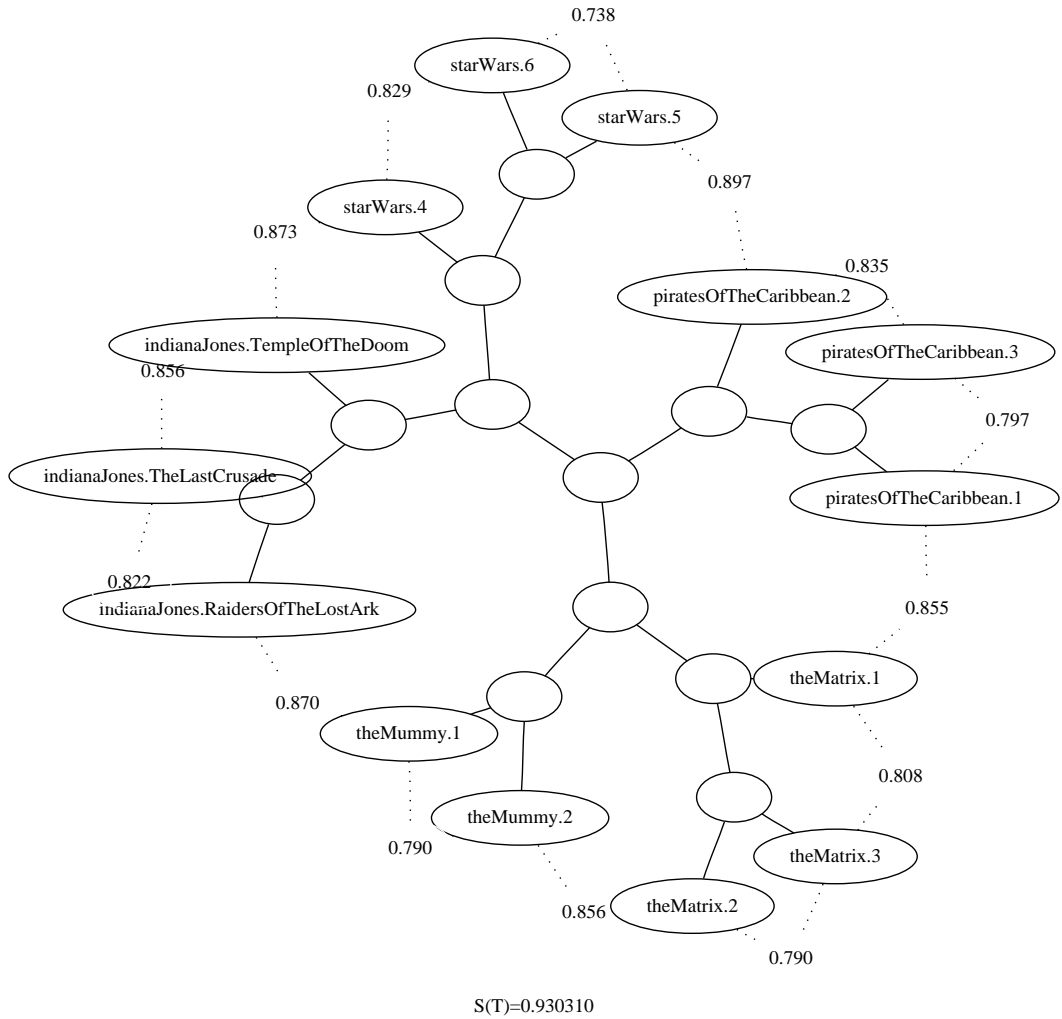}
\caption{IMDB. BZIP2 compressor. Best dendrogram obtained.}
\label{APPENDIX. Fig:dendro-imdb-bzip2-best}
\end{figure}

\bibliographystyle{is-abbrv}
\bibliography{bitacora-ordenada}

\end{document}